\documentclass[11pt,letterpaper]{article}
\usepackage{latexsym}
\usepackage{amssymb, amsmath}
\usepackage{amsfonts}
\usepackage{epsfig, epic, eepic}
\newcommand{\contract}{\lrcorner}
\newcommand{\z}{\langle z\rangle}

\def\theequation{\arabic{section}.\arabic{equation}}

\begin{document}

\begin{titlepage}
\begin{center}

\hfill hep-th/0605283\\
\hfill ZMP-HH/06-08\\

\vskip 1.5cm 
{\Large\bf Geometric Transitions on non--K\"ahler Manifolds}

\vskip 1.5cm

{\bf Anke Knauf}  \\ 

\vskip 20pt

{\em II. Institut f\"ur Theoretische Physik\\
Universit\"at Hamburg\\ 
Luruper Chaussee 149\\
22761 Hamburg, Germany}\\

\vskip 15pt

{\em Department of Physics\\University of Maryland\\
College Park, MD 20742, USA}

\vskip 20pt

{email: {\tt  anke.knauf@desy.de}} \\

\end{center}

\vskip 1.5cm

\begin{center} {\bf ABSTRACT } \end{center}

{\parindent=0.0mm

This article is based on the publications \cite{gtone, realm, gttwo} and the author's PhD--thesis. We study geometric transitions on the supergravity level using the basic idea of \cite{gtone}, where a pair of non--K\"ahler backgrounds was constructed, which are related by a geometric transition. Here we embed this idea into an orientifold setup as suggested in \cite{gttwo}.
The non--K\"ahler backgrounds we obtain in type IIA are non--trivially fibered due to their construction from IIB via T--duality with Neveu--Schwarz flux. We demonstrate that these non--K\"ahler manifolds are not half--flat and show that a symplectic structure exists on them at least locally. 

We also review the construction of new non--K\"ahler backgrounds in type I and heterotic theory as proposed in \cite{realm}. They are found by a series of
T-- and S--duality and can be argued to be related by geometric transitions as well. A local toy model is provided that fulfills the flux equations of motion in IIB and the torsional relation in heterotic theory, and that is consistent with the U--duality relating both theories. For the heterotic theory we also propose a global solution that fulfills the torsional relation because it is similar to the Maldacena--Nunez background.
\vspace*{1cm}

\today}

\end{titlepage}


\tableofcontents


\setcounter{equation}{0}
\section{Introduction and Outline}

Apart from being the most promising candidate for a theory of quantum gravity, string theory has provided many insights that are not directly related to the quest for the ultimate theory. One example are  
gauge/gravity dualities, which can bridge the divide between Planck scale and low energy physics. They provide a means to construct gauge theories from dual supergravity backgrounds.

The first example was provided by the AdS/CFT correspondence, which relates a string theory on $AdS_5\times S^5$ to a superconformal $\mathcal{N}=4$ field theory. However, we would like to be able to embed the (Minimally Supersymmetric) Standard Model (MSSM) of Particle Physics into a four--dimensional, low energy effective description of string theory, in other words we want to describe realistic gauge theories with running coupling and less or no supersymmetry. The conformal invariance and some supersymmetry can be broken in certain models that contain D--branes extended along Minkowski space as well as the compactified directions. 

Since the gauge theories obtained in this way are asymptotically free, they are accessible to perturbation theory in the high energy regime (UV). The strong coupling regime (IR), on the other hand, still poses one of the greatest challenges in field theory calculations. Quantum Chromodynamics, for example, becomes already strongly coupled at rather high energies, typically around 1GeV. It is therefore highly desirable to gain a better understanding of non--perturbative phenomena and to find tools for computing strongly coupled quantities. 

This work was motivated by the search for dualities between weakly and strongly coupled gauge theories from a string theory perspective. 
Suppose a gauge theory that has a good weakly coupled description in the UV is dual to another, strongly coupled theory. One can then analyze the strong coupling behavior by considering only the dual, weakly coupled theory. String theory is able to describe such gauge theory dualities by embedding them in string theory dualities, which are called ``geometric transitions'' \cite{ks,vafa}. 

The weak--strong duality in geometric transitions is a duality between different string theory backgrounds, in particular, between different geometries. Whereas the weakly coupled gauge theory can be described by an open string theory on D--branes, the strongly coupled theory is described by a closed string theory on a different background geometry. String theory then provides quantities relevant for the supersymmetric field theories, it can for example compute superpotentials. 

The gauge theory described by \cite{ks} and \cite{vafa} was $\mathcal{N}=1$ pure Super--Yang--Mills obtained from a geometric transition on conifold geometries.  Conifolds are non--compact Calabi--Yau manifolds. It has become apparent over recent years that string compactifications (with flux) can lead to more general (non--K\"ahler) manifolds. Such ``flux compactifications'' have some advantages over Calabi--Yau compactifications. In particular, they can address the moduli fixing problem. It is therefore the aim of this work to suggest new non--K\"ahler manifolds, that are also connected by a geometric transition. 

One would furthermore like to construct models that can describe more phenomenologically interesting gauge theories. A first step in this direction is to add matter to the Super--Yang--Mills theory, which can be achieved with certain additional D--branes. Next steps would include to replace the non--compact ``internal'' manifolds with compact ones and to break supersymmetry completely. However, the latter two topics are not part of this work.

\subsection{String Theory and Gauge Theories}

There are five different consistent superstring theories in ten dimensions: type IIA and IIB theory of oriented closed strings, type I theory of unoriented open and closed strings and two heterotic theories, which contain oriented closed strings \cite{gsw, pol, kaku, kirit}. Their massless bosonic spectra are summarized in table \ref{stringspectrum}.

Taking a closer look at the type II spectrum we see that the low energy limit  represents IIA and IIB supergravity (IIA being the trivial dimensional reduction of 11--dimensional $\mathcal{N}=1$ supergravity). The massless NS--NS spectrum consists of the metric $g_{\mu\nu}$, a scalar $\phi$ (the dilaton) and an antisymmetric tensor $B_{\mu\nu}$ (the B--field).  The fermionic sector contains two spin--3/2 and two spin--1/2 states, the former being termed gravitini, since they are the natural superpartners of the spin--2 graviton. 
Type IIA contains the RR $p$--forms of odd $p$ (a U(1) gauge field $A_\mu$ and a three-form $C_{\mu\nu\rho}$), whereas IIB contains those of even $p$ (a scalar $\chi$, a two--form $C_{\mu\nu}$ and a four--form with self--dual field strength). 
We will in the following always use the symbol $C_p$ for the RR gauge potentials and $F_{p+1}=dC_p$ for their fieldstrengths.

\begin{table}[b]
\begin{center}
  \begin{tabular}{|l|l|c|l|}\hline
    Theory & massless bosonic states & susy in d=10 & gauge group\\ \hline
    IIA    & $g_{\mu\nu}$, $\phi$, $B_{\mu\nu}$, $A_\mu$, $C_{\mu\nu\rho}$ 
      & $\mathcal{N}=2$ & $U(1)$ \\
    IIB    & $g_{\mu\nu}$, $\phi$, $B_{\mu\nu}$, $\chi$, $C_{\mu\nu}$, 
      $C_{\mu\nu\rho\sigma}$ & $\mathcal{N}=2$ & -- \\ 
    Type I & $g_{\mu\nu}$, $\phi$, $A_\mu$, $C_{\mu\nu}$ & $\mathcal{N}=1$ & 
      $SO(32)$\\
    Heterotic $SO(32)$ & $g_{\mu\nu}$, $\phi$, $A_\mu$, $B_{\mu\nu}$  & 
      $\mathcal{N}=1$ & $SO(32)$\\
    Heterotic $E_8\times E_8$ & $g_{\mu\nu}$, $\phi$, $A_\mu$, $B_{\mu\nu}$  & 
      $\mathcal{N}=1$ & $E_8\times E_8$\\ \hline
    \end{tabular}
  \caption{The massles spectra of the five superstring theories. Only type I and heterotic contain non--Abelian gauge bosons.}
  \label{stringspectrum}
\end{center}
\end{table}

One can obtain unoriented closed strings by gauging type IIB by the worldsheet parity  $\Omega$, meaning one projects out all states that are odd under orientation reversal of the worldsheet.
This theory would only contain $g_{\mu\nu}$, $\phi$ and the RR 2--form in its spectrum (as well as an even combination of fermions). Since half of the states are projected out, this unoriented version of type IIB has only $\mathcal{N}=1$ supersymmetry.
This theory is not consistent by itself, 
but it can be combined with another theory: the theory of open unoriented strings (which is also not consistent by itself) that contributes an $\mathcal{N}=1$ super--Yang--Mills multiplet (a gauge boson $A_\mu$ in the adjoint representation of SO(32) and its superpartner, a gaugino) to the massless spectrum. Together they form type I. It will become essential in section \ref{heterotic} that type I can be viewed as a ``projection'' (a so--called orientifold) of type IIB.

Heterotic string theory contains oriented closed strings, but only the right movers show supersymmetry whereas the left movers are described by bosonic string theory. The latter one requires d=26, the former one d=10, so the left movers have to be compactified on a 16--dimensional, even and self--dual lattice. There are only two such lattices, giving rise to a gauge group $E_8\times E_8$ or SO(32). 
For heterotic SO(32) the massless spectrum looks suspiciously close to type I theory. And indeed, both theories are connected by S--duality \cite{polwit} and the heterotic string theory can be viewed as a soliton of type I \cite{dabhol}. We will have more to say about this duality in section \ref{hetsec}. 

\subsubsection*{D--Branes}

Open string worldsheets have boundaries, or in other words, these strings have endpoints, which can fulfill Dirichlet or von--Neumann boundary conditions.
Dirichlet boundary conditions in $p$ spatial directions restrict the string endpoint to a $p$--dimensional hyper--surface which has been termed D$p$--brane. Dirichlet boundary conditions were disregarded for a long time, since they do not restrict momentum flow off the string.  It was discovered in \cite{branes} that these D--branes are actually dynamical objects, they can interact with strings and they are also charged under RR gauge potentials. The $(p+1)$--dimensional worldvolume $\Sigma_{p+1}$ of a D$p$--brane couples naturally to an RR $(p+1)$--form
\begin{equation}\label{rrcharge}
  \int_{\Sigma^{p+1}} C_{p+1}\,,
\end{equation}
which is called ``electric'' coupling since it has the same form as a photon coupling to the worldline of a point particle. D--branes can also be viewed as magnetic sources for RR fields. Like the photon fieldstrength $F=dA$ creates a flux through a two--dimensional (Gaussian) surface, an RR fieldstrength $F_p=dC_{p-1}$ creates a flux through a $(8-p)$--dimensional hypersurface. 
This implies that a D$p$--brane is an electric source for the same field for whom a D$(6-p)$ brane is the magnetic source. This is simply an expression of Hodge duality in $d=10$ for RR fieldstrengths. 
According to their RR spectrum type IIA contains D$p$--branes with $p=even$ and IIB with $p=odd$. 
Type I contains the RR two--form which couples to a D1--brane (or magnetically to a D5--brane). But since it is a theory of open strings (whose endpoint are free to move through the (9+1)--dimensional target space), type I also contains spacetime filling D9--branes. Note that due to the absence of RR gauge fields in the heterotic spectrum, there cannot be any D--branes in heterotic string theory. There are, however, the equivalents for the NS field, so--called NS5--branes. 

Since D--branes are electrically charged they repel each other. This force is balanced by their gravitational attraction, so a system of parallel D--branes is stable and preserves the same supersymmetry as a single D--brane. 

The concept of anti--D--branes with opposite charge has also been introduced. These would attract D--branes and eventually annihilate each other. Such systems break supersymmetry and are of particular interest for cosmology \cite{cosmo}. On compact manifolds one often seeks a mechanism to cancel D--brane charges, since there cannot be any net charge in a compact space. Apart from anti--D--branes one can consider orientifold planes (O--planes), which also carry negative charge but are non--dynamical objects. They arise in orientifolds, which combine gauging by a spacetime symmetry (orbifold) with the worldsheet parity $\Omega$, see e.g. \cite{dabholkar}. Each fixed ``point'' of such a symmetry is actually a hyperplane, this is the orientifold plane\footnote{On the orientifold plane the theory is unoriented and half the states are projected out from an oriented theory. In this sense, type I can be understood as type IIB with spacetime filling D9--branes (to introduce open strings whose endpoints can move freely in all directions) and O9--planes (to render them unoriented).}. Orientifold planes have been used successfully in compactifications with D--branes \cite{blumenhagen}--\cite{denef}, but since we only deal with {\it non--compact} manifolds, we are not required to introduce a charge--cancellation mechanism.

D--branes interact by exchanging gravitons (closed strings). But a closed string can fuse with a brane and split into an open string with both endpoints on this brane. This heuristic argument shows that a closed string theory with branes must also contain open strings. Since type II theories contain RR p--forms and their sources, D(p-1)--branes, they should also contain open strings in addition to the closed oriented strings. But we have seen above that open strings contain only half as much supersymmetry as closed oriented strings. This is explained by the D--branes themselves being BPS states that preserve half the supersymmetry. In the bulk the theory is essentially type II whereas on the brane it is (an oriented version of) type I. 

D--branes provide the possibility of obtaining interesting gauge theories from type II, because they have non--Abelian gauge theories on their worldvolume. Depending on whether the strings ending on the D--branes are oriented or unoriented, they give rise to different gauge groups.
In the simple case of open strings between N parallel D--branes (a so--called stack of D--branes) the gauge group is either U(N) or SO(N) for oriented or unoriented strings\footnote{For unoriented strings a gauge group Sp(N) is also possible \cite{pol}.}, respectively. Letting the D--branes extend along (3+1) Minkowski space results in a gauge theory in our ``observable universe''.
One can engineer phenomenologically interesting gauge theories depending on the internal compactification manifold and the orientation the D--branes take in these extra dimensions. Engineering Standard--Model like gauge groups from intersecting D--brane scenarios has also been quite successful, see for example \cite{braneworlds}--\cite{patisala}.

\subsubsection*{String Theory Compactifications}

So far we have only considered strings in flat 10--dimensional Minkowski space. To make contact with experimental observations, one needs to explain why the six extra dimensions are not detected. The usual approach is to assume them to be compactified on such small length scales that they are not visible in present day experiments. To preserve 4d Poincar\'e invariance, one assumes the 10--dimensional space to be a direct product of (warped) (3+1)--dimensional flat Minkowski space and a six--dimensional internal manifold, i.e. the metric of the internal space does not depend on external coordinates. 

We are especially interested in the case where these compactifications preserve some supersymmetry. Let us first discuss the case where all vacuum expectation values of the antisymmetric NS and RR tensors, these are called ``fluxes'', are set to zero. Then the supersymmetry condition in 4d translates into the existence of a covariantly constant spinor on the internal manifold, which characterizes a Calabi--Yau manifold.
A Calabi--Yau is a complex manifold with SU(3) holonomy, in other words it is K\"ahler and Ricci--flat (see e.g. \cite{nakahara, candel} for a review on complex geometry). In this case the external space is simply given by flat (3+1) Minkowski space. Compactification of type II theories on a six--dimensional Calabi--Yau manifold preserves $\mathcal{N}=2$ in 4d and $\mathcal{N}=1$ for type I/heterotic. The corresponding low energy effective actions (for type II) have been worked out for example in \cite{compact, mirror}.

However, a larger class of compactification manifolds is possible if one allows for vacuum expectation values of the NS and RR fieldstrengths. This idea was already raised many years ago \cite{strominger, dewit, hull}. In such ``flux compactifications'' the ten--dimensional space is a product of warped (3+1) Minkowski space and an internal manifold, which is no longer a Calabi--Yau. In contrast to the Calabi--Yau case, supersymmetry now only requires a nowhere vanishing, globally defined SU(3) invariant spinor, which characterizes a manifold with SU(3) structure, but not SU(3) holonomy, see section \ref{torsion} for details. These manifolds are in general non--K\"ahler, but they are often not even complex. See \cite{granaflux} for a comprehensive review of flux compactifications.

Flux compactifications have another major advantage: They often allow one to fix some moduli of the theory \cite{kachruori,denef}, \cite{typeiiflux}--\cite{giddings}. Moduli are scalar fields in the effective field theory that arise during compactification, they describe for example deformations of the complex structure of the compactification manifold or result from the Kaluza--Klein reduction of the ten--dimensional fields. Since their values are generally not fixed, they parameterize a continuous family of vacua. The space of all possible background values for these fields is called the moduli space. As these scalars are not observed in four--dimensional physics, they have to become sufficiently massive, for example by acquiring a vacuum expectation value. Fluxes can generate potentials for some moduli in the effective theory that fix those expectation values. 

Despite those advancements we are still faced with the problem that there might be an infinite number of vacua that can arise from string theory compactifications. It seems there is a ``landscape'' \cite{landscape} of four--dimensional vacua with a few inhabitable islands. Based on statistical analyses it has been suggested recently that the chances of finding an MSSM compatible vacuum might be one in a billion \cite{flori}.

As geometric transitions provide a duality between a background with D--branes and a background with only flux, they may also serve as a mechanism to explain the origin of the fluxes in flux compactifications that do not use D--branes. The appearance of flux can be naturally understood in the dual theory which contains their sources, the D--branes. 

\subsubsection*{M--Theory, F--Theory and Dualities}

The five superstring theories described above are related by a web of dualities. They are therefore believed to be different limits of one unique theory, $\mathcal{M}$--theory, see e.g. \cite{pol, mtheory, obers, alvarez} and references therein. We already noted that they all contain supergravity multiplets in their massless spectrum and that type IIA contains precisely the $\mathcal{N}=2$ d=10 supergravity multiplet that one obtains by trivial dimensional reduction of $\mathcal{N}=1$ d=11 supergravity. One might therefore suspect that $\mathcal{M}$--theory, whatever it may be, reduces to $\mathcal{N}=1$ d=11 supergravity in its low energy limit. This interpretation of $\mathcal{M}$--theory as an 11--dimensional theory can be made more precise by considering the strong coupling regime of type IIA, in which its BPS spectrum (of D0--branes) looks like a Kaluza--Klein tower. $\mathcal{M}$--theory can therefore also be viewed as the strong coupling limit of type IIA in which an extra dimension with radius $R\sim g_s$ opens up, $g_s$ being the string coupling in IIA.

Reducing 11--dimensional supergravity on an interval $S^1/\mathbb{Z}_2$ one can obtain heterotic $E_8 \times E_8$ similarly as a weak coupling limit. At each end of the interval (whose length is proportional to the string coupling) there are spacetime filling 9--planes with gauge group $E_8$ on them. One brane is usually called the hidden sector, the other one carries our observable world.

This establishes type IIA and heterotic $E_8 \times E_8$ as limits of the same theory. T--duality relates both type II and both heterotic theories to each other, and S--duality relates type I to heterotic SO(32). Roughly speaking, T--duality states\footnote{A more thorough discussion of this duality is relegated to appendix \ref{tduality}.} that a theory compactified on a circle with radius $R$ is dual to another theory on a circle with radius $1/R$. It also has a non--trivial action on D--branes as it exchanges Dirichlet and von--Neumann boundary conditions. T--duality along a direction parallel to the D$p$--brane turns it into a D$(p-1)$ brane; T--duality along a direction transverse to the D$p$--brane turns it into a D$(p+1)$--brane. This can consistently turn the IIA spectrum into the IIB spectrum and vice versa.

S--duality is a strong--weak coupling duality that relates one theory at coupling $g_s$ to another one at $1/g_s$. IIB is actually self--dual under this symmetry (since S--duality is part of the $SL(2,\mathbb{Z})$ symmetry of IIB), whereas heterotic SO(32) is dual to type I. It might seem peculiar that heterotic and type I can be dual, since their fundamental objects (open versus closed strings, D1 and D5--branes versus NS5--branes) are so distinct. But in the strong coupling limit of type I its D1 branes become light and its fundamental strings become heavy, so that the D1--branes of type I can actually be interpreted as the fundamental strings of the heterotic theory \cite{polwit}. It is amazing that this duality can relate an oriented to an unoriented theory.

The only missing link in the duality web is then the one between type IIB and type I, but we established that when discussing the unoriented closed string: we can obtain type I as an orientifold limit of type IIB. 

Although we still lack a precise description of $\mathcal{M}$--theory, we know that its low energy limit reduces to 11--dimensional supergravity. By dimensional reduction one finds IIA supergravity and can then follow the chain of dualities to reach the other four string theories. We will make extensive use of this ``duality chasing'' to obtain one supergravity solution from another.

There is another relation between type IIA and IIB: mirror symmetry \cite{dixon}--\cite{candelpark}. It states that compactifying IIA on a manifold X is equivalent to compactifying IIB on the {\it mirror} manifold of X. Not only does this produce agreement of the low energy effective actions, but both theories are actually equivalent on the quantum level of the SCFT. The two mirror manifolds are non--trivially related, 
their Hodge numbers are interchanged. Since IIA and IIB are already related by T--duality one might suspect that their could be a relation between T--duality and mirror symmetry. This turns out to be correct, as shown by Strominger, Yau and Zaslow \cite{syz}, and we will return to this important relation in section \ref{mirrorsym}.

F--theory is a possible 12--dimensional theory that is related to 10--dimensional type IIB theory by interpreting the two extra dimensions as a compact torus. The complex structure parameter $\tau$ of the torus is identified with the complex scalar $\lambda=\chi+i e^{-\phi}$ in IIB \cite{vafaf}. Here $\phi$ is the dilaton and $\chi$ is the RR zero--form (axion). The F--theory torus can be non--trivially fibered over the ten--dimensional base giving rise to singularities on the base. We will see in section \ref{introf} that these singular points are accompanied by orientifold planes and D--branes and determine a IIB orientifold.

F--theory is rather geometrical in nature and does not play the same role as $\mathcal{M}$--theory, since type IIB supergravity is not a Kaluza--Klein reduction of 12--dimensional supergravity. (There is no supergravity theory with 32 supercharges in d=12.) It has been suggested \cite{vafaf} as a geometrical interpretation of the $SL(2,\mathbb{Z})$ symmetry of type IIB. More details are to be found in section \ref{introf} and \cite{dabholkar,vafaf, sen}.

After reviewing the dualities that will enables us to ``chase'' backgrounds that are valid string theory solutions, we now have to explain what geometric transitions are in the context of string theory. We will do this on the well--understood example of the conifold transition.

\subsubsection*{The Conifold Transition}\label{gt}

Conifolds are non--compact Calabi--Yau threefolds. Generically, one speaks of a conical singularity if the metric takes (in some local region) the form
\begin{equation}
  ds^2 \,=\, dr^2+r^2\,ds_T^2
\end{equation}
for some base $T$. The point $r=0$ is then often called the conical point or the tip of the cone. We are interested in the case where the base $T$ is given by
\begin{equation}
  T\,=\,T^{1,1}\,=\,\left(SU(2)\times SU(2)\right)/U(1)\,.
\end{equation}
As explained in appendix \ref{coni}, this base is topologically equivalent to $S^2\times S^3$. There are two distinct ways to smooth the singularity at $r=0$, one can either blow up an $S^3$ or an $S^2$. The former manifold is then called ``deformed conifold'', the latter ``resolved conifold'' \cite{candelas}. The transition from one geometry to the other is called a ``conifold transition'' and can be pictured as shrinking the size of the $S^3$ to a point and then blowing up an $S^2$:
\begin{center}
\includegraphics[height=65pt]{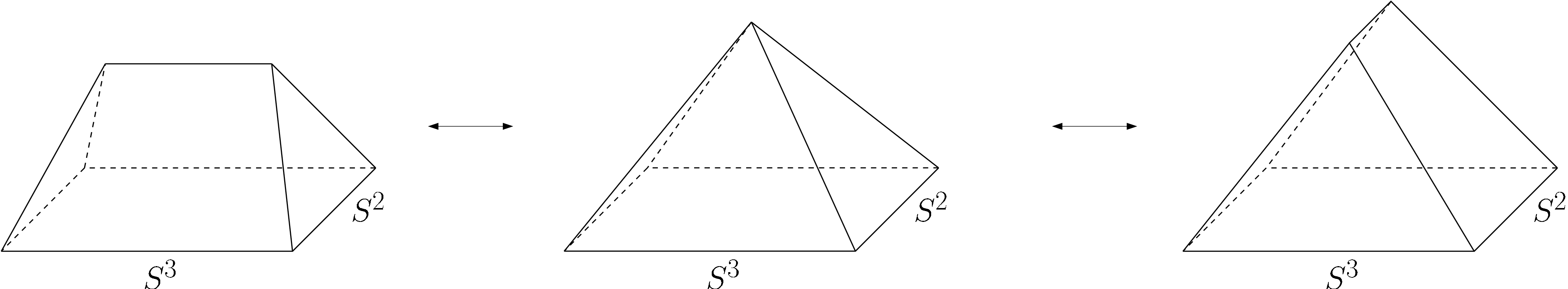}
\end{center}

Although these manifolds are non--compact and therefore not suited for string theory compactification, they provide a useful mechanism to construct gauge theories\footnote{To obtain phenomenologically relevant theories one could construct compact manifolds with conical singularities.}, as has been noted in a series of papers \cite{klebanovwitten, klebanekrasov, klebanovtseyt}. Placing $N$ parallel D3--branes at the tip of the conifold produces an $SU(N)\times SU(N)$ superconformal gauge theory on the world volume of the D--branes \cite{klebanovwitten}, which extend along external Minkowski space. This conformal invariance can be broken by ``wrapping'' branes on cycles in the internal manifold \cite{klebanekrasov}. Consider for example D6--branes which extend along three internal directions. If these internal directions are compact, one uses the term ``wrapping''. The system preserves supersymmetry when the D6--branes wrap the $S^3$ of the (deformed) conifold\footnote{Supersymmetry requires such wrapped submanifolds to be special Lagrangian.} or when D5--branes wrap the $S^2$ of the (resolved) conifold. As shown in \cite{klebanekrasov}, the resulting gauge theory exhibits a logarithmically running coupling. Since the Calabi--Yau breaks 3/4 of the supersymmetry and the D--branes another 1/2, the gauge theory in four dimensions has $\mathcal{N}=1$ supersymmetry.

The notion of a geometric transition in string theory was introduced in \cite{ks, vafa}. The basic idea is that a gauge theory constructed as in the last paragraph is dual to another theory which results from a different string theory. Let us illustrate this with the Klebanov--Strassler model \cite{ks}. They constructed a theory that flows in the IR towards a strongly coupled SU(N) Super--Yang--Mills (SYM) theory by wrapping branes on the singular conifolds (in other words the D--branes are wrapping a {\it vanishing} cycle). It has been known for a long time that SYM confines and the IR behavior is governed by the Veneziano--Yankielowicz \cite{vy} superpotential\footnote{Extensions for this superpotential have been proposed \cite{schwetz, ckl, dv}, but they shall not concern us here.}
\begin{equation}\label{venez}
  W(S)\,=\,N\,\left[\log\frac{S}{\Lambda^3}-S\right]\,,
\end{equation}
where the chiral superfield $S$ is given by
\begin{equation}
  S\,=\, Tr W^\alpha W_\alpha\,,
\end{equation}
$W^\alpha$ being the field strength of the vector multiplet, it contains the gaugino $\lambda^\alpha$ in its bottom component. Minimizing the superpotential \eqref{venez} leads to a vacuum expectation value for the gaugino bilinear in the bottom component of $S$
\begin{equation}
  \langle S\rangle \,=\,  \langle \lambda^\alpha\,\lambda_\alpha\rangle \,=\, 
    \Lambda^3\,e^{2\pi i k/N}\,,\qquad k=1,\ldots, N\,,
\end{equation}
in other words, the confining theory shows gaugino condensation. $\Lambda$ is the scale of the gauge theory. There are $N$ different vacua and the gaugino vacuum expectation value leads to chiral symmetry breaking. The original SU(N) SYM has a chiral $U(1)$ symmetry, but the vacuum breaks this to $\mathbb{Z}_2$.

This behavior of the gauge theory should somehow be visible in the underlying string theory. The string background Klebanov and Strassler considered was the singular conifold, which can be written as an embedding in four dimensional complex space as
\begin{equation}
  \sum_{i=1}^4 (z_i)^2\,=\,0\,,\qquad\qquad z_i\in\mathbb{C}^4\,.
\end{equation}
This background has an obvious U(1) symmetry under $z_i\to e^{i\alpha}z_i$, for some complex phase $\alpha$. It should be this precise U(1) that is broken by gaugino condensation. Therefore, \cite{ks} suggested that the IR limit of this theory should rather be given by a string theory on the deformed conifold
\begin{equation}
  \sum_{i=1}^4 (z_i)^2\,=\,\mu^2\,,\qquad\qquad \mu\in\mathbb{R}\,,
\end{equation}
as the residual symmetry here is a $\mathbb{Z}_2$ that acts as $z_i\to -z_i$. The constant $\mu$ is called the deformation parameter and governs the size of the blown up $S^3$.

This is one concrete example of the interplay between gauge theories and geometry. One should not think of the KS model as a flow from one geometry to the other, but it is rather a duality between the singular and the deformed background, in the UV the former is the appropriate description, whereas in the IR the latter one is relevant.

The term ``geometric transition'' in the string theory context is used for models like that developed by Vafa \cite{vafa}, who made the statement of the KS model more precise: The theory with D--branes wrapping a cycle in the resolved conifold is dual to a theory without D--branes but fluxes on the deformed conifold. The branes disappear in this ``transition'', as the cycle they wrap shrinks to zero size, but the dual cycle is blown up with the corresponding fluxes on it.

For example, in IIA one starts with N D6--branes wrapping the $S^3$ of the deformed geometry, which creates an SU(N) SYM theory on the leftover (3+1) external dimensions. In the IR this describes a confined theory, the dual string theory background is the resolved conifold with blown--up $S^2$ and RR flux (that correspond to the branes before transition) on it. Gopakumar and Vafa \cite{gopakumar} showed that this is more than just a transition on a purely geometric level. They showed that both theories (before and after transition) actually compute the same topological string amplitudes, see appendix \ref{conjecture} for details. In the following we will focus on the target space perspective of these models.

What we just described has a mirror in IIB. Resolved and deformed conifolds are (approximately) mirror to each other. Mirror symmetry on Calabi--Yaus exchanges their Hodge numbers $h^{1,1}$ and $h^{2,1}$ that represent the dimension of the cohomology classes $H^{1,1}$ and $H^{2,1}$. For a blown up $S^2$ one finds $h^{1,1}=1$ and $h^{2,1}=0$, whereas the blown up $S^3$ has $h^{1,1}=0$ and $h^{2,1}=1$. But there is one subtlety \cite{hori}: whereas the deformed conifold has only one compact 3--cycle, the resolved conifold has two compact even cycles, a 2--cycle and a 0--cycle. But
in the limit when the size of the blown up $S^2$ and $S^3$ are small, i.e. we are at the ``transition point'' from one geometry to the other, the mirror of the resolved conifold becomes effectively the deformed conifold\footnote{Strictly speaking the mirror of the resolved conifold has some variables in $\mathbb{C}^*=\mathbb{R}^+\times U(1)$ instead of $\mathbb{C}$. It is given by $x_1+x_2+x_1x_2e^{-t}+1-uv=0$, where $x_{1,2}\in\mathbb{C}^*$ and $u,v\in\mathbb{C}$, $t$ is the K\"ahler parameter or size of $S^2$ \cite{hori,aganagic}.} \cite{hori}. This means that in IIB the UV picture is given by D5--branes wrapping the $S^2$ of the resolved conifold, the IR picture is given by the deformed conifold with RR flux on the blown up $S^3$.

For obvious reasons, this duality is called open/closed duality, since after the transition there are no open strings in the theory anymore. It can also be interpreted as a large $N$ duality ($N$ is the number of D--branes), see \cite{vafa, gopakumar} or the discussion in appendix \ref{conjecture}.

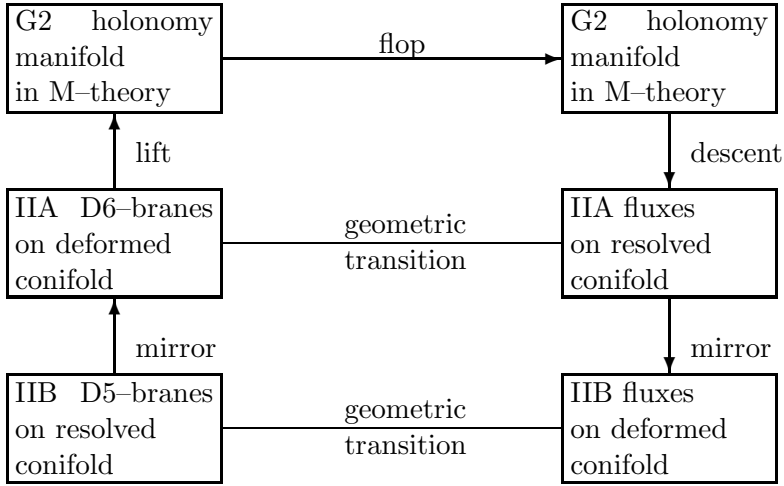
\begin{figure}[t]
\begin{center}
  \begin{picture}(300,200)\thicklines
    \put(10,0){\framebox(80,40){\begin{minipage}{75pt}
      IIB D5--branes on resolved\\ conifold
      \end{minipage}}}\thinlines
    \put(91,20){\line(1,0){128}}\thicklines
    \put(220,0){\framebox(80,40){\begin{minipage}{75pt}
      IIB fluxes\\ on deformed\\ conifold
      \end{minipage}}}\thicklines
    \put(137,18){\begin{minipage}{50pt} geometric transition \end{minipage}}
    \put(50,41){\vector(0,1){28}}
    \put(58,48){\begin{minipage}{50pt} mirror \end{minipage}}
    \put(10,70){\framebox(80,40){\begin{minipage}{75pt}
      IIA D6--branes on deformed\\conifold
      \end{minipage}}}
    \put(50,111){\vector(0,1){28}}
    \put(58,122){\begin{minipage}{50pt} lift \end{minipage}}
    \put(10,140){\framebox(80,40){\begin{minipage}{75pt}
      G2 holonomy manifold\\in M--theory 
      \end{minipage}}}
    \put(91,160){\vector(1,0){128}}
    \put(150,162){\begin{minipage}{50pt} flop \end{minipage}}
    \put(220,140){\framebox(80,40){\begin{minipage}{75pt}
      G2 holonomy manifold\\in M--theory
    \end{minipage}}}
    \put(260,139){\vector(0,-1){28}}
    \put(268,122){\begin{minipage}{50pt} descent \end{minipage}}
    \put(220,70){\framebox(80,40){\begin{minipage}{75pt}
      IIA fluxes\\ on resolved \\conifold
      \end{minipage}}}\thinlines
    \put(91,90){\line(1,0){128}}\thicklines
    \put(137,88){\begin{minipage}{50pt} geometric transition 
    \end{minipage}}
    \put(260,69){\vector(0,-1){28}}
    \put(268,48){\begin{minipage}{50pt} mirror \end{minipage}}
  \end{picture}
  \caption{Vafa's duality chain. By following the arrows through a series of mirror symmetry and flop transition we can verify the geometric transition as conjectured for IIA and IIB.} \label{vafachain}
\end{center}
\end{figure}

There is one last relation between deformed and resolved conifold that we need to exploit. They can both be obtained via dimensional reduction from a $G_2$--holonomy manifold 
that is a cone over $S^3\times \widetilde{S}^3$ \cite{flop}, where $\widetilde{S}^3$ indicates a three--sphere that remains finite at the tip of the cone, the other one has vanishing size. Basically, the deformed conifold, a cone over $S^2\times \widetilde{S}^3$, can be found by reducing on a U(1) fiber that is part of the vanishing $S^3$. The six--dimensional manifold then possesses a blown up $\widetilde{S}^3$. The resolved conifold can also be obtained by a circle reduction, but this time one reduces along a U(1) fiber belonging to $\widetilde{S}^3$, so that the geometry one obtains is a cone over $S^3\times\widetilde{S}^2$, but this time it is the two--sphere that is blown up\footnote{Taking fluxes and D--branes into account one actually reduces on twisted fibers and the $G_2$ manifold is a cone over $S^3\times \widetilde{S}^3/\mathbb{Z}_N$.}. Both geometries are related by a ``flop'' in the $G_2$ manifold, which simply exchanges $S^3\leftrightarrow\widetilde{S}^3$, we will give a more detailed description in section \ref{iia}.
All these ingredients can be connected to what we will call ``Vafa's duality chain'' throughout this article, see figure \ref{vafachain}. 

\newpage
\subsection{Outline}

It is one goal of this work to verify the duality chain in figure \ref{vafachain} for a complete supergravity background including metric and all fluxes. Although geometric transitions have been embedded in dual brane solutions \cite{branecascade, oh}, an explicit supergravity analysis has not been carried out yet. This can be done by following the arrows in figure \ref{vafachain} and requires a series of T--dualities and a flop. 

In section \ref{mirror} we discuss the first step in the duality chain, the mirror symmetry between resolved and deformed conifold. After that we will follow the whole duality chain in section \ref{chain} and we will discover non--K\"ahler backgrounds in type IIA whose torsion classes are analyzed in section \ref{torsion}.
We can furthermore use the idea of duality chasing to find new non--K\"ahler backgrounds in type I and heterotic that are also connected by a geometric transition. This will be presented in setion \ref{heterotic}.

Already the first step in this duality chain raises a puzzle: the mirror symmetry between resolved and deformed conifold. As we will explain in section \ref{mirrorsym}, mirror symmetry can be understood as three successive T--dualities, if the manifold admits a $T^3$ fibration. This is the well--known Strominger--Yau--Zaslow (SYZ) conjecture \cite{syz}. But as we will see, resolved and deformed conifold do not have the same number of isometries, it seems therefore contradictory that they should be mirrors. It is another aim of this work to resolve this puzzle and we provide one possible resolution in section \ref{mirror}. This will require some non--trivial manipulations to the metric of the resolved conifold and we can only recover a semi--flat version of the deformed conifold. In particular, we have to boost the complex structure of the resolved conifold, which is in agreement with anticipations from \cite{syz}, that the large complex structure limit can be used in the absence of proper isometry directions. The large complex structure limit we impose is in general a non--trivial action on the resolved conifold, but if we restrict ourselves to a local limit of the coordinates, this boost can be interpreted as a coordinate redefinition. In this sense, we can establish the mirror symmetry between resolved and deformed conifold only in a local limit.

Another question arises immediately when one considers the complete supergravity solution for D5--branes on the resolved conifold. It has been shown in \cite{klebanekrasov, klebanovtseyt} that these wrapped branes (also called fractional branes, since they do not only act as D5--branes but also couple to fields that seem to be created by D3--branes) give rise to NS flux. Therefore, we do not only have to find the mirror of a resolved conifold, but we have to find its mirror in the presence of NS flux. As already considered in \cite{vafa, kachru, jan}, this leads to non--K\"ahler manifolds and seems to indicate that the duality chain \ref{vafachain} needs to be modified. We will show in section \ref{mirror} that this is true and the mirror is actually a ``non--K\"ahler version'' of the deformed conifold. It differs from the deformed conifold in a very precise way, by the same ``twisting of fibers'' as advocated in \cite{jan}, but in contrast to \cite{jan} we do not find half--flat manifolds. The non--K\"ahler manifold we find only admits a symplectic, but no half--flat structure. This is not in contradiction with the literature, since the solution from \cite{jan} actually also admits a symplectic structure and we will give several arguments in section \ref{discuss} that favor the symplectic interpretation.

In section \ref{chain} we will follow all the steps linking the duality chain and we will demonstrate that it contains two non--K\"ahler backgrounds in IIA that we will argue also to be geometric transition duals. Therefore, already the duality chain as originally suggested by Vafa, gives necessarily rise to non--K\"ahler backgrounds. We will argue for the consistency of our calculation with the fact that we recover a K\"ahler background in IIB at the very end of the duality chain, which looks locally like a deformed conifold. An analysis of the global properties of these backgrounds will have to be pursued elsewhere. We will only present a local analysis, since the background we start with in type IIB is an F--theory orientifold whose global metric is unknown (it will contain singularities due to D7--branes and O7--planes). One reason we use this setup is that there is no known supergravity solution for D5--branes on the resolved conifold that would preserve supersymmetry. But as will become clear from the analysis in section \ref{mirror}, if we aim for a IIA mirror background that is close to the deformed conifold, we are restricted to the local limit anyway.

The F--theory setup has another advantage apart from providing a supersymmetric solution for D5--branes on the resolved conifold. It enables us to
suggest a generalization of Vafa's duality chain that includes additional D--branes which act as a global symmetry in the underlying gauge theory. In other words, we find a gauge theory with flavors in the fundamental representation of $SU(2)^{16}$. The emergence of this additional symmetry is a convenient by--product of the supersymmetric solution we seek from F--theory. This will be explained in section \ref{ftheorysetup}. The influence of these additional branes on the gauge theory superpotential could be determined once we know a global supergravity solution.

In section \ref{torsion} we demonstrate how the torsion classes of the IIA non--K\"ahler manifolds can be determined. The analysis remains somewhat preliminary, since we are restricted to the local limit, which does not contain any information about global properties of the manifold. But we can nevertheless show with a quite generic ansatz for the (almost) complex structure that the local metric admits a symplectic but no half--flat structure.

We can also use the concept of ``duality chasing'' (meaning to obtain one string solution from another one by applying a number of T-- or S--dualities) to find type I and heterotic string backgrounds. Starting with a IIB orientifold containing O7--planes we obtain a type I background with O9--planes after two T--dualities. Another S--duality then takes us to heterotic. We can do this with a IIB background before and after transition, so we find pairs of backgrounds in type I and heterotic.
As shown in section \ref{heterotic}, this will also lead to non--K\"ahler backgrounds because they are obtained from T--dualities with NS field. Since they are via a long duality chain related to an $\mathcal{M}$--theory flop, they can also be called geometric transition duals. We will provide a toy example in section \ref{hetsec} that is consistent with the IIB supergravity equations of motion for RR and NS flux, their change under T--duality and the torsional relation for heterotic backgrounds. 

This implies a duality between heterotic string theory with NS5 branes and another heterotic theory with only flux. It would be intersting to find out what this means at the topological string level. Since there are no open heterotic strings, the interpretation as an open/closed duality fails. In this sense, the interpretation of the geometric transition we propose for heterotic strings is not as clear as that for type II theories.

Although most analyses presented here are confined to the local limit, we will be able to propose a global solution for the heterotic background after transition by exploiting similarities with the Maldacena--Nunez \cite{mn} background, which is a valid heterotic solution. We will therefore argue that our solution is the local limit of a quite generic background that matches with MN for a specific choice of parameters. We will also verify the torsional relation for this background.

Remaining open questions and future directions will be discussed in section \ref{outlook}.  The necessary background material is provided in a variety of appendices. This article is based on \cite{gtone, realm, gttwo}, but uses the insights from \cite{gttwo} to repeat the calculations from \cite{gtone} under new assumptions. We also provide some new interpretations, especially with regard to the mirror symmetry between resolved and deformed conifold in the local limit.

\setcounter{equation}{0}
\section{Mirror Symmetry on Conifolds}\label{mirror}

We begin this section by reviewing mirror symmetry between IIA and IIB compactified on a pair of mirror Calabi--Yaus. Repeating the arguments from \cite{micu} it is demonstrated that RR flux does not alter the mirror symmetry between the two Calabi--Yaus (if the backreaction of the fluxes on the geometry is neglected), but NS flux has non--trivial consequences \cite{jan}. The mirror of a Calabi--Yau with NS flux is not a Calabi--Yau anymore. This can be understood using the Strominger--Yau--Zaslow (SYZ) conjecture which states that mirror symmetry is the same as three successive T--dualities on a supersymmetric $T^3$ fiber inside the Calabi--Yau. T--duality mixes B--field and metric components, so the mirror of a Calabi--Yau with NS flux acquires a twisting in the $T^3$ fiber due to the B--field.

After reviewing this background material, we first discuss how resolved and deformed conifold can be mirror to each other in the sense of SYZ although they do not possess the same number of isometry directions. We will show that we have to impose some non--trivial boost on the complex structure of the resolved conifold to find a mirror that resembles the deformed conifold. Furthermore, as anticipated by SYZ, we can establish the mirror symmetry only for the semi--flat limit of the metrics.
If we furthermore turn on NS flux we find a non--K\"ahler manifold as the mirror that, apart from a B--field dependent fibration, still resembles the deformed conifold. We baptize this manifold the ``non--K\"ahler deformed conifold'' and argue that it is not half--flat.

\subsection{Mirror Symmetry and Strominger--Yau--Zaslow}\label{mirrorsym}

Compactification of type II theories on a six--dimensional Calabi--Yau manifold preserves $\mathcal{N}=2$ in four dimensions and $\mathcal{N}=1$ for type I/heterotic, see e.g. \cite{gsw, kaku}. 
The fascinating aspect of these compactifications is the fact that the resulting four--dimensional theory is determined by the properties of the internal manifold. Let us consider type II theories compactified on a Calabi--Yau $X$ that is characterized by its Hodge numbers, which represent the dimensions of different cohomology classes, $h^{p,q}=dim H^{p,q}(X,\mathbb{C})$. The IIA Kaluza--Klein reduction contains $h^{1,1}$ vector multiplets, $h^{1,2}$ hypermultiplets and one tensor multiplet in the four--dimensional theory, whereas the type IIB reduction contains $h^{1,2}$ vector multiplets, $h^{1,1}$ hypermultiplets and one tensor multiplet \cite{compact, mirror}\footnote{The reason why Hodge numbers are the relevant quantities in this compactification is the fact that that all fields are expanded in harmonic forms on $X$ and harmonic p--forms ($\omega^p$ such that $\Delta \omega^p=(d+*d*)^2\omega^p=0$) are in one--to--one correspondence with the cohomology group $H^p(X)$, see e.g. \cite{nakahara, candel}.}.

Mirror symmetry is an expression of the fact that the theory obtained from IIA compactified on a Calabi--Yau 3--fold $X$ is equivalent to IIB compactified on $Y$, if $X$ and $Y$ are mirror manifolds \cite{dixon}--\cite{candelpark}. This does not only mean that one obtains the same low energy effective action in both compactified theories, but also agreement on the quantum level of the SCFT. This relates the mirror manifolds in a non--trivial way, e.g. their Hodge numbers are interchanged
\begin{equation}
  h^{1,1}(X)\,=\,h^{1,2}(Y)\,,\qquad h^{1,2}(X)\,=\,h^{1,1}(Y)\,.
\end{equation}

Mirror symmetry holds on the supergravity level when we allow for RR fluxes to be turned on (but their backreaction is neglected) \cite{micu}. In order to preserve four--dimensional Poincar\'e invariance, we only allow for vacuum expectation values of RR fields along internal directions, i.e. on the Calabi--Yau. Recall that IIA allows for even p--form fieldstrengths $F_p$ and IIB for odd p--form fieldstrengths to be turned on. There is a peculiarity about the cohomologies of Calabi--Yaus. Recall the Hodge diamond of a Calabi--Yau
\begin{equation}\label{hodgediamond}
  \begin{array}{ccccccc}
    &&&1&&&\\
    &&0&&0&&\\
    & 0 && h^{1,1} &&0&\\
    1 && h^{1,2} && h^{2,1} && 1\\
    & 0 && h^{2,2} &&0&\\
    &&0&&0&&\\
    &&&1&&&
  \end{array}
\end{equation}
where, as for all K\"ahler manifolds of complex dimension $dim_\mathbb{C} X=3$, one finds the identities $h^{1,1}=h^{2,2}$ and $h^{2,1}=h^{1,2}$.
The dimensions of certain cohomology classes vanish, in particular
\begin{equation}
  dim(H^1(X))\,=\,dim(H^5(X))\,=\,0\,,
\end{equation}
this implies that there can be no 1-- or 5--form flux turned on, since they would have to be expanded in a basis of harmonic 1-- or 5--forms on $X$. Therefore, IIB can only have 3--form flux turned on whereas IIA allows for 0, 2, 4 and 6--form fluxes\footnote{Note that for compact internal manifolds these fluxes are quantized, so they are actually governed by the integral cohomology classes $H^p(X,\mathbb{Z})$. The conifold geometries we consider a not compact, but have some compact cycles. This will cause the NS flux not to be quantized, but the RR flux still is.}. And indeed, counting dimensions one finds $2(h^{1,2}+1)$ 3--forms for type IIB and $2(h^{1,1}+1)$ even p--forms for IIA. It was shown explicitely in \cite{micu} that the low energy effective actions obtained with these RR fluxes agree. So, even and odd RR form fluxes can be mapped precisely under mirror symmetry with the interchange of $h^{1,1}(X)\leftrightarrow h^{1,2}(Y)$.

What happens to this analysis if we also allow for NS 3--form flux $H_{NS}$ to be turned on? If we follow the same reasoning as for RR fluxes (and as advocated in \cite{jan}) we encounter the following puzzle: both IIA and IIB have NS flux, which corresponds to the cohomology class $H^3(X)$ and $H^3(Y)$, respectively. But mirror symmetry maps even to odd cohomology classes and vice versa. So how can 3--form flux in IIA be mapped to any even form flux in IIB? NS flux does not get mapped to RR flux, since the RR mapping discussed above is already complete. The NS sector contains the metric and dilaton besides $H_{NS}$, but no antisymmetric tensors which could be interpreted as even degree p--forms. The only explanation seems to be that the metric and dilaton have to account for the ``missing cohomologies''. It was therefore suggested \cite{jan}, that mirror symmetry in the presence of NS flux does not lead to another Calabi--Yau manifold, but a non--K\"ahler (in fact even non--complex) manifold whose intrinsic torsion provides for the mirror of NS 3--form flux. Similar observations were made in \cite{vafa, kachru}. 

The ``geometric part'' of $H^2(X,\mathbb{C})$ in IIA is given by the fundamental two--form $J$ (which is the K\"ahler form on a complex manifold if it is closed). This is combined with $B_{NS}$ to form the complexified K\"ahler modulus $t=J +iB_{NS}$. In type IIB the corresponding quantity is the holomorphic three--form $\Omega^{3,0}\in H^3(Y,\mathbb{C})$ which is also closed on a Calabi--Yau. But it is also clear that mirror symmetry must exchange the two. If we now allow for $dB_{NS}\ne 0$, i.e. the imaginary part of the complex K\"ahler modulus to be non--closed, this should translate into the imaginary part of $\Omega^{3,0}=\Omega^++i \Omega^-$ being non--closed as well. This led the authors of \cite{jan} to suggest that the correct mirror manifold should be given by a so--called ``half--flat'' manifold, which are manifolds with SU(3) structure characterized by\footnote{We use a different notation compared to \cite{jan}, which states that $d\Omega^+\ne 0$, but the assignment of real and imaginary part is completely arbitrary.}
\begin{equation}
  d\Omega^-\ne 0\,,\qquad\mbox{but}\qquad d\Omega^+\,=\,d (J\wedge J)\,=\,0\,.
\end{equation}
It was also demonstrated to hold true in a toroidal compactification. We will return to this issue in the discussion \ref{discuss} and in section \ref{torsion}.

\subsubsection*{The Strominger--Yau--Zaslow Conjecture}

Above discussion focused on topological quantities (cohomology classes) of the compactification manifolds. Our focus is rather on the target space perspective, i.e. we are interested in the metric of internal manifolds. Fortunately, the work of Strominger, Yau and Zaslow (SYZ) \cite{syz} provides a way of finding the mirror of a large class of manifolds by simply applying T--duality, which only requires knowledge of the metric.

The SYZ conjecture states that any Calabi--Yau $X$ that has a mirror possesses a supersymmetric $T^3$--fibration (with in general singular fibers) over a base $B$. The mirror Calabi--Yau $Y$ is then given as the moduli space of the $T^3$ fibers and their flat connection. Mirror symmetry is equivalent to T--duality along these $T^3$ fibers.

Mirror symmetry can be viewed as a symmetry between BPS states. Consider D0--branes in type IIA on $X$ and D3--branes in type IIB on a $T^3$ inside the mirror manifold $Y$. The moduli space of the D0--branes is of course all of $X$, so by mirror symmetry there must be an object in IIB on $Y$ which also has moduli space $X$. The D3--brane moduli space is generated by deformations of the 3--cycle $T^3$ within $Y$ and the flat U(1) connection\footnote{Supersymmetry requires the three--cycle to be a special Lagrangian submanifold, that means the K\"ahler form restricted to this cycle as well as the imaginary part of the holomorphic 3--form vanish, and the U(1) connection on it to be flat.} on it. Both of these are generated by harmonic 1--forms 
on the three--cycle and it turns out that their moduli space (which has to match $X$) is also a $T^3$ fibration. One would actually reach the same conclusion if one would start with a generic supersymmetric three--cycle in $Y$ without assuming from the beginning that the D3--branes wrap a $T^3$.
With this logic, both $X$ and $Y$ are $T^3$ fibrations over the same base $B$. 

This led to the SYZ--conjecture: ``Mirror symmetry is three T--dualities''. The simple argument is that three T--dualities turn D0-- into D3--branes and vice versa and that such T--dualities can be performed on the supersymmetric $T^3$ fibrations without changing the moduli space. Consider a six--torus as a simple example. This is a trivial $T^3$ fibration over $T^3$. T--duality will invert the size of the $T^3$ fiber, but the mirror is again a $T^3\times T^3$. The SYZ argument is non--trivial at points where the fibers become singular. There are no isometries\footnote{The T--duality action still exists in the case without isometries \cite{greg}, although we cannot simply apply Buscher's rules from appendix \ref{tduality}.} and constructing the moduli space of D3--branes is complicated by instantons.

This problem can be avoided in the large complex structure, or semi--flat, limit where one considers the base $B$ to be large compared to the $T^3$ fibers. Semi--flat means that the metric only depends on the base coordinates $y^i$, i.e. away from the singular fibers one can write \cite{syz, jan}
\begin{equation}\label{semiflat}
  ds^2\,=\,g_{ij}\,dy^i\,dy^j+h^{\alpha\beta}\,\left(dx_\alpha+\omega_\alpha(y)\right)
    \left(dx_\beta+\omega_\beta(y)\right)
\end{equation}
where $x_\alpha$ parameterize the $T^3$ fiber and $\omega_\alpha$ are one--forms on the base that describe the twisting of the fiber as one moves along the base. In this semi--flat limit $x_\alpha$ are still isometry directions, so we can explicitely perform T--dualities. However, it is expected that the equivalence between mirror symmetry and T--duality on $T^3$ holds not only in the semi--flat limit \cite{syz}. 

The influence of NS flux on this picture was discussed in \cite{jan}. As already explained on the ground of cohomology--arguments, the mirror of a Calabi--Yau with NS--flux will no longer be a Calabi--Yau. The B--field leads to an additional twisting of the $T^3$--fibers. We will see this explicitely in section \ref{nowflux}. Let us first discuss mirror symmetry between resolved and deformed conifold in the absence of any flux.

\subsection{The Mirror of the Resolved Conifold}\label{mirrorfirst}

Recall from section \ref{gt} (see also appendix \ref{coni}) that the resolved and deformed conifold describe asymptotically a cone over $S^2\times S^3$, but the singularity at $r=0$ is smoothed out to an $S^2$ or $S^3$, respectively.
The Ricci--flat K\"ahler metric of the resolved conifold has been derived in \cite{candelas, pandozayas}
\begin{eqnarray}\label{resmetric}\nonumber
  ds^2_{\rm res} & = & \widetilde{\gamma}'\,d\widetilde{r}^2 
    + \frac{\widetilde{\gamma}'}{4}\,\widetilde{r}^2\big(d\widetilde{\psi}
    +\cos\widetilde{\theta}_1\,d\widetilde{\phi}_1 
    +\cos\widetilde{\theta}_2\,d\widetilde{\phi}_2\big)^2 \\ 
  & & + \frac{\widetilde{\gamma}}{4}\,\big(d\widetilde{\theta}_1^2
    +\sin^2\widetilde{\theta}_1\,d\widetilde{\phi}_1^2\big) 
    + \frac{\widetilde{\gamma}+4a^2}{4}\,\big(d\widetilde{\theta}_2^2
    + \sin^2\widetilde{\theta}_2\,d\widetilde{\phi}_2^2\big)\,,
\end{eqnarray}
where $(\widetilde{\phi}_i,\widetilde{\theta}_i)$ are the usual Euler angles on $S^2$, $\widetilde{\psi}=0\ldots 4\pi$ is a U(1) fiber over these two spheres and $\widetilde\gamma$ is a function of $\widetilde{r}$ that goes to zero as $\widetilde{r}\to 0$, see \eqref{defgamma} for its definitions. The constant $a$ is called resolution parameter, because it produces a finite size prefactor for the $(\widetilde{\phi}_2,\widetilde{\theta}_2)$--sphere at $\widetilde{r}=0$. This metric has clearly 3 isometries related to shift symmetries in the coordinates  $\widetilde{\psi},\,\widetilde{\phi}_1$ and $\widetilde{\phi}_2$. These are indeed the appropriate Killing directions as the metric was constructed to be invariant under $SU(2)\times SU(2)\times U(1)$ \cite{candelas}, see appendix \ref{coni} for a brief review.

The deformed conifold metric on the other hand is given by \cite{papatseyt, minasian}
\begin{eqnarray}\label{defmetric}\nonumber
  ds^2_{\rm def} &=& \hat{\Gamma}\, \left[\frac{4\,d\widetilde{r}^2}{\widetilde{r}^2
    (1-\mu^4/\widetilde{r}^4)} + \left(d\widetilde{\psi}+\cos\widetilde{\theta}_1\,
   d\widetilde{\phi}_1+\cos\widetilde{\theta}_2\,d\widetilde{\phi}_2\right)^2\right]\\
  & + & \frac{\hat{\gamma}}{4}\;\left[\left(\sin\widetilde{\theta}_1^2\,
    d\widetilde{\phi}_1^2+d\widetilde{\theta}_1^2\right) 
    +\left(\sin\widetilde{\theta}_2^2\, d\widetilde{\phi}_2^2 
    + d\widetilde{\theta}_2^2\right)\right]\\ \nonumber
  & + & \frac{\hat{\gamma}\mu^2}{2\, \widetilde{r}^2}\,\left[\cos\widetilde{\psi}\,
    \left(d\widetilde{\theta}_1\, d\widetilde{\theta}_2
    -\sin\widetilde{\theta}_1\sin\widetilde{\theta}_2\, d\widetilde{\phi}_1 \,
    d\widetilde{\phi}_2\right)\right.\\ \nonumber
  & & \left.\quad\quad+\sin\widetilde{\psi}\,\left(\sin\widetilde{\theta}_1 \,
    d\widetilde{\phi}_1\, d\widetilde{\theta}_2 + \sin\widetilde{\theta}_2\,
    d\widetilde{\phi}_2\,d\widetilde{\theta}_1\right)\right]\,,
\end{eqnarray}
with the deformation parameter $\mu$ and a similar function\footnote{The function $\hat{\gamma}$ is related to the K\"ahler potential $\hat{\mathcal{F}}$ as $\hat{\gamma}=\widetilde{r}^2
\hat{\mathcal{F}}$, and similar for $\widetilde{\gamma}$, see \eqref{defgamma}.}
$\hat{\gamma}(\widetilde{r})$, $\hat{\Gamma}$ can be read off from equation \eqref{defmetricapp}.  This metric exhibits the same structure of a $\widetilde{\psi}$--fibration over two spheres, but there are additional cross--terms in the last line. We see that $\widetilde{\psi}$ does not correspond to an isometry anymore. The U(1) symmetry associated to shifts $\widetilde{\psi}\to\widetilde{\psi}+k$ is absent. This is not a peculiarity of our coordinate choice but an inherent property of the deformed conifold. As discussed in section \ref{gt}, the deformed conifold breaks the U(1) symmetry of the singular conifold (which also the resolved conifold exhibits, compare \eqref{singbase} with \eqref{resmetricapp})\footnote{On the other hand, both singular and deformed conifold are symmetric under the exchange of the two $S^2$: $(\phi_1,\theta_1) \leftrightarrow (\phi_2,\theta_2)$, a symmetry that is broken in the resolved conifold, since one $S^2$ is blown up.}. 

How can both manifolds then be mirror in the sense of SYZ? The answer lies within the above mentioned semi--flat limit. We can still apply SYZ if the base is large compared to the $T^3$ fiber. If we identify $(\widetilde{r},\widetilde{\theta}_1,\widetilde{\theta}_2)$ as the base coordinates and $(\widetilde{\psi},\widetilde{\phi}_1,\widetilde{\phi}_2)$ as the coordinates of the $T^3$ fiber in the resolved metric, we can T--dualize along the latter. What we recover can of course not be the deformed conifold, since it lacks the $T^3$--fibration, but only a semi--flat limit that still possesses an isometry along $\widetilde{\psi}$. Moreover, as we will show now, simply T--dualizing along the 3 isometry directions is not enough. We have to impose the condition of a large base ``by hand''.

To simplify our calculation, and with some foresight to following sections, we define local coordinates. We restrict our analysis to a small neighborhood of $(r_0,\,\z,\,\langle\phi_1\rangle, \,\langle\phi_2\rangle, \,\langle\theta_1\rangle,\,\langle\theta_2\rangle)$ by introducing
\begin{eqnarray}\label{local}\nonumber
  \widetilde{r} &=& r_0+\frac{r}{\sqrt{\widetilde{\gamma}_0'}}\,,\qquad\qquad\qquad
    \;\widetilde{\psi} \,=\, \z + \frac{2 z}{\sqrt{\widetilde{\gamma}_0'}\,r_0} \\ 
    \nonumber
  \widetilde{\phi}_1 &=& \langle\phi_1\rangle + \frac{2x}
    {\sqrt{\widetilde{\gamma}_0}\,\sin\langle\theta_1\rangle}\,,\quad\quad\;
  \widetilde{\phi}_2 \,=\, \langle\phi_2\rangle + \frac{2y}{\sqrt{(\widetilde
    {\gamma}_0 + 4 a^2)}\,\sin\langle\theta_2\rangle}\\
  \widetilde{\theta}_1 &=& \langle\theta_1\rangle+\frac{2\theta_1}
    {\sqrt{\widetilde{\gamma}_0 }}\,,\qquad\qquad\qquad
  \widetilde{\theta}_2 \,=\, \langle\theta_2\rangle+\frac{2\theta_2}{\sqrt{
    (\widetilde{\gamma}_0+4 a^2)}}\,,
\end{eqnarray}
where $\widetilde{\gamma}_0$ is constant, namely $\widetilde{\gamma}(\widetilde{r})$ evaluated at $\widetilde{r}=r_0$. The coordinates $(r,\,z,\,x,\,y,\,\theta_1,\,\theta_2)$ are small fluctuations around these expectation values and we will call them ``local coordinates'' henceforth. In these local coordinates the metric on the resolved conifold takes a very simple form (in lowest order in local coordinates\footnote{After this work was complete, we extended this analysis to linear order in $r$ and we were able to resum the $\theta_i$--dependences \cite{gtthree}. It was shown that $A$ and $B$ indeed contain $\cot\theta_1$ and $\cot\theta_2$, respectively, not just their expectation values.})
\begin{equation}\label{startmetric}
  ds^2 \,=\,dr^2+(dz+A\,dx + B\, dy)^2+ (dx^2+d\theta_1^2) 
    +(dy^2+d\theta_2^2)\,,
\end{equation}
where we have defined the constants
\begin{equation}\label{defineab}
  A \,=\, \sqrt{\frac{\widetilde{\gamma}_0'}{\widetilde{\gamma}_0}}\;r_0\,\cot
    \langle\theta_1\rangle \,,\qquad\quad
  B \,=\, \sqrt{\frac{\widetilde{\gamma}_0'}{(\widetilde{\gamma}_0+4 a^2)}}\;r_0\,
    \cot\langle\theta_2\rangle\,.
\end{equation}
This is easily T--dualized along $x,y$ and $z$ (which correspond to the former isometry directions $\widetilde{\psi},\,\widetilde{\phi}_1,\,\widetilde{\phi}_2$) with the help of Buscher's rules from \eqref{buscher}. In the absence of B--field they read for T--duality along $y$
\begin{eqnarray}
  \widetilde{G}_{yy} &=& \frac{1}{G_{yy}}\,,\qquad \qquad
  \widetilde{G}_{\mu\nu} \,=\, G_{\mu\nu}-\frac{G_{\mu y}G_{\nu y}}{G_{yy}}
\end{eqnarray}
where the tilde indicates the metric after T--duality. Applying these three times one finds the mirror
\begin{eqnarray}\label{mirrormetric}\nonumber
 d\tilde{s}^2 &=& dr^2 \,+\,\alpha^{-1}\,\left(dz-\alpha A\,dx-\alpha B
    \,dy\right)^2 \,+\, \,d\theta_1^2 \,+\, d\theta_2^2 \\ 
  & & + \,\alpha(1+B^2)\, dx^2 \,+\, \alpha (1+A^2)\, dy^2  -\, 2\alpha AB\, dx\,dy\,,
\end{eqnarray}
where we have introduced
\begin{equation}\label{alpha}
  \alpha \,=\, (1+A^2+B^2)^{-1}\,.
\end{equation}
The metric \eqref{mirrormetric} does not resemble a deformed conifold, as for example the cross--term $d\theta_1\,d\theta_2$ is missing.
This can be cured by boosting the complex structure of the resolved base. Consider again \eqref{startmetric}, which can be written as
\begin{equation}\label{torimetric}
  ds^2 \,=\,dr^2+(dz+A\,dx + B\, dy)^2+ \vert d\chi_1\vert^2+\vert d\chi_2\vert^2\,,
\end{equation}
with the two tori
\begin{equation}\label{deftori}
  d\chi_1\,=\,dx+\tau_1\,d\theta_1\,,\qquad d\chi_2\,=\,dy+\tau_2\,d\theta_2\,.
\end{equation}
In \eqref{startmetric} the complex structures are simply $\tau_1=\tau_2=i$. Note, that these tori are just local versions of the spheres in \eqref{resmetric}, since locally a sphere resembles a degenerate torus\footnote{The appearance of tori instead of spheres is also consistent with dual brane pictures constructed in \cite{dasmukhi, uranga}.}. The large complex structure limit is then given by letting
\begin{equation}\label{boost}
  \tau_1\,\longrightarrow\, i-f_1\,,\qquad\qquad \tau_2\,\longrightarrow\, i-f_2
\end{equation}
with real and large $f_{1,2}$. We define them with some forsight as
\begin{equation}\label{definef}
  f_i \, =\, \frac{\beta_i}{\sqrt{\epsilon}}
\end{equation}
with finite $\beta_i$. The only other change to the metric \eqref{startmetric} that we will make is to shift the component $g_{zz}=1\,\to\, (1-\epsilon)$. Then letting $\epsilon\to 0$ in $g_{zz}$ and $f_{1,2}$ simultaneously will be our ``regularization scheme''. These transformations might seem a little ad hoc, but we will explain in more detail why we chose this particular boost of the complex structure\footnote{See \cite{gtone} for more attempts to restore the $d\theta_1\,d\theta_2$--term that did {\it not} work.}.

After three T--dualities along $x,\,y,\,z$ and letting $\epsilon \to 0$ one arrives at the local mirror metric\footnote{The attentive reader might have noticed that the complex structure boost \eqref{boost} introduces additional cross terms into the metric. Those will lead to B--field components under T--duality which have been properly taken into account when calculating the result \eqref{mirrormetricnow}. We postpone the discussion of the B--field to the next section.}
\begin{eqnarray}\label{mirrormetricnow}\nonumber
  d\tilde{s}^2 &=& dr^2 \,+\,\alpha^{-1}\,\left(dz-\alpha A\,dx-\alpha B
    \,dy\right)^2\,+ \,\alpha(1+B^2)\, dx^2  \\ \nonumber
  & & +\, \alpha (1+A^2)\, dy^2
    \, + \,(1-A^2\beta_1^2)\,d\theta_1^2 \,+\, (1-B^2\beta_2^2) 
    \,d\theta_2^2  \\ 
  & & -\, 2 AB \beta_1\beta_2\,d\theta_1 d\theta_2 \,-\, 2\alpha AB\, dx\,dy \,,
\end{eqnarray}
which we now compare to the deformed conifold metric \eqref{defmetric}. We also have to introduce local coordinates on the deformed conifold. These coordinates will be similar to \eqref{local}, but the precise coefficients will differ. We therefore leave some coefficients $a_i, b_i$ generic. The local deformed metric reads
\begin{eqnarray}\label{deflocal}\nonumber
  ds^2_{\rm def} &=& a_0\,dr^2 \,+\,a_1\,\left(dz+b_1\,dx+b_2\,dy\right)^2 
    \, +\, a_2\,(dx^2+d\theta_1^2) \,+\,a_3\,(dy^2+d\theta_2^2) \\ 
  & & +\, 2a_4\,\left[\cos\z\,(d\theta_1\,d\theta_2-dx\,dy)
    +\sin\z\,(dx\,d\theta_2+dy\,d\theta_1)\right]\,,
\end{eqnarray}
Comparing this metric to \eqref{mirrormetricnow} one makes the following observations
\begin{itemize}
  \item The semi--flat limit of the local deformed metric can be achieved by setting $\z=0$, then the isometry along $z$ is restored. In this case the $(dx\,d\theta_2+dy\,d\theta_1)$ term does not appear and both metrics have the same functional dependence. 
  \item The $d\theta_1\,d\theta_2$ cross term in \eqref{mirrormetricnow} should have the same prefactor as the $dx\,dy$ cross term apart from a minus sign.
  \item In order for the torus structure to be preserved one would expect $d\theta_1^2$ to have the same coefficient as $dx^2$ (and similarly for $d\theta_2^2$ and $dy^2$). This is identical to the statement that the two $S^2$ in the deformed metric \eqref{defmetric} are ``unsquashed''.
\end{itemize}
Let us see if all three conditions from the last two bullets can be met simultaneously by fixing the two constants $\beta_1$ and $\beta_2$. First, require
\begin{equation}
  - 2AB \beta_1\beta_2 \,=\, 2\alpha AB
\end{equation}
to match the crossterms in the last line in \eqref{mirrormetricnow}. This simply implies that 
\begin{equation}\label{condone}
  \beta_1\beta_2 \,=\, -\alpha\,.
\end{equation}
Requiring $d\theta_1^2$ to have the same coefficient as $dx^2$ (and similarly for $d\theta_2^2$ and $dy^2$) gives the following conditions
\begin{eqnarray}\nonumber
  \alpha(1+B^2) & = & (1-A^2\beta_1^2)\\
  \alpha(1+A^2) & = & (1-B^2\beta_2^2)\,.
\end{eqnarray}
Remembering that $\alpha=(1+A^2+B^2)^{-1}$ gives the surprisingly simple solutions
$\beta_1^2=\beta_2^2=\alpha$, which together with \eqref{condone} has two possible solutions
\begin{equation}\label{solbeta}
  \big\{\beta_1=\sqrt{\alpha},\;\beta_2=-\sqrt{\alpha}\,\big\}\qquad\mbox{or}\qquad 
    \big\{\beta_1=-\sqrt{\alpha},\;\beta_2=\sqrt{\alpha}\,\big\}\,.
\end{equation}
In both cases, the mirror metric in type IIA finally reads
\begin{eqnarray}\label{defmirror}\nonumber
  d\tilde{s}^2 &=& dr^2 \,+\,\alpha^{-1}\,\left(dz-\alpha A\,dx-\alpha B
    \,dy\right)^2\\ 
  & & + \,\alpha(1+B^2)\,(dx^2+d\theta_1^2)\,+\,\alpha (1+A^2)\,
    (dy^2+d\theta_2^2)\\ \nonumber
  & & +\, 2\alpha AB \,(d\theta_1 d\theta_2- dx\,dy) \,,
\end{eqnarray}
which matches indeed the $\langle\psi\rangle=0$ limit of \eqref{deflocal} with appropriate identifications of $a_i$ and $b_i$.
 So we have shown that the mirror of the local resolved metric is the semi--flat limit of a local deformed conifold, if we impose an additional boost of the complex structure to make the $T^3$ fiber small compared to the base.

In local coordinates we can restore the $dx\,d\theta_2$ and $dy\,d\theta_1$ cross terms. We can rotate the $(y,\,\theta_2)$ torus 
\begin{equation}\label{yrotation}
  \begin{pmatrix} dy \\ d\theta_2 \end{pmatrix}
  \,\longrightarrow\, \begin{pmatrix} \cos z & -\sin z \\ \sin z & \cos z \end{pmatrix}
    \,\begin{pmatrix} dy \\ d\theta_2 \end{pmatrix}\,.
\end{equation}
This does not change the term $(dy^2+d\theta_2^2)$, but the last term in \eqref{defmirror} changes as
\begin{eqnarray}\nonumber
  d\theta_1\, d\theta_2 - dx\,dy
    \longrightarrow  \cos z\,\big(d\theta_1\,d\theta_2 - dx\,dy\big) 
    +\sin z\,\big(dx\,d\theta_2+dy\,d\theta_1 \big)\,,
\end{eqnarray}
exactly as required for a deformed conifold metric! This implies of course also a change in the \linebreak $z$--fibration $(dz-\alpha A\,dx-\alpha B\,dy)$. 
This change {\it cannot} be absorbed by a shift in other coordinates\footnote{This is precisely the reason why a coordinate transformation like \eqref{yrotation} cannot produce an additional isometry on the global deformed conifold. One might be tempted to use the inverse of this coordinate transformation on the deformed global metric \eqref{defmetric} as it would eliminate the $\cos\psi$ and $\sin\psi$ terms. But it is not possible to preserve the $d\psi$--fibration at the same time and one does not recover an additional U(1) symmetry.}
{\it unless} we rotate by a constant value $\langle z\rangle$ instead of $z$. Then we can define a new $z$--coordinate as $z\to z-\alpha B(\cos\langle z\rangle-1)\,y+\alpha B\,\sin\langle z\rangle\,\theta_2$ which leaves the form of the $z$--fibration invariant. This is similar to the approach taken in \cite{ohta}, where the ``delocalized limit'' of the deformed metric is given by choosing $\psi=\z=$ constant. So we may also adopt the point of view that the mirror metric we obtained is precisely the delocalized limit of a deformed conifold with $\z=0$. We therefore conjecture the local resolved conifold \eqref{startmetric} to be mirror dual to a local deformed conifold with metric
\begin{eqnarray}\label{mirrorfinal}\nonumber
  d\tilde{s}^2 &=& dr^2 \,+\,\alpha^{-1}\,\left(dz-\alpha A\,dx-\alpha B\,
    dy\right)^2\\ 
  & & + \,\alpha(1+B^2)\,(dx^2+d\theta_1^2)\,+\,\alpha (1+A^2)\,
    (dy^2+d\theta_2^2)\\ \nonumber
  & & +\, 2\alpha AB \,\left[\cos\langle z\rangle\,(d\theta_1 d\theta_2- dx\,dy)
    +\sin \langle z\rangle\,(dx\,d\theta_2+dy\,d\theta_1)\right]\,.
\end{eqnarray}
This agrees with the local limit \eqref{deflocal}, but there is one difference: in the Ricci--flat K\"ahler metric \eqref{defmetric} both $S^2$ have the same size. This is not the case in \eqref{mirrorfinal}, since in general $1+A^2\ne 1+B^2$. This statement is of course not very meaningful in the local limit, since we could absorb these constants into a rescaling of coordinates, but it should be clear that when $A$ and $B$ are not merely understood as constants, but as functions of $r,\theta_1,\theta_2$ (recall \eqref{defineab} for their definition), the two spheres will have different size. Actually, both spheres will have the same size if we choose $\langle\theta_1\rangle=\langle\theta_2\rangle$ and let $a\to 0$, then $A=B$. This is perfectly consistent with the statement that the deformed conifold is only the mirror of the resolved conifold in the limit when the resolution parameter becomes vanishingly small.

However, some deviations from the deformed conifold metric should be expected, since, as was noted before, resolved and deformed conifold metric are not exactly mirror to each other \cite{hori, aganagic}. They become mirror when both are close to the ``transition point''. However, in this limit neither the base nor the fiber are large, on the contrary, the blown up $S^2$ or $S^3$ shrink to zero size. This explains why we are unable to recover the deformed conifold with the SYZ procedure: SYZ can only work when the base is large and we are away from the singular fiber. This is opposite to the limit when both manifolds are mirror. We therefore suggest that the ``trick'' we performed by boosting the complex structure of the local tori $(x,\theta_1)$ and $(y,\theta_2)$ can be given a more rigorous physical interpretation: our choice of local coordinates can be interpreted as one way of taking us close to the transition point\footnote{Note how in \eqref{startmetric} the resolution parameter is not visible anymore. The metric of the singular conifold \eqref{singbase} has the same local limit as the resolved conifold.}, but we also need a large base to be able to perform the SYZ procedure. Therefore, we have to boost the complex structure ``by hand'' as in \eqref{boost}. This boost is to be understood as being large only in the {\it local} framework, not on the global manifold. This discussion should also clarify that we cannot leave the local limit by re--introducing global coordinates on the mirror manifold and then claim that this is the mirror of the global resolved conifold. The assumption that we work in a small patch on the manifold entered explicitely into our calculations. 

One can, of course, apply three T--dualities to the {\it global} resolved metric \eqref{resmetric}, since it exhibits three isometries. The result has to be another Calabi--Yau that is also a $T^3$ fibration over some three--dimensional base, but it will not be the deformed conifold. We will call any metric that has in the local limit the form \eqref{mirrorfinal} a ``deformed conifold'' and specifically indicate when we mean the Calabi--Yau.

\subsection{The Mirror in the Presence of NS--Flux}\label{nowflux} 

We would like to confirm Vafa's duality chain, see figure \ref{vafachain}, by starting in the bottom left corner: D5--branes on the resolved conifold in IIB. Three T--dualities will take us to the mirror in IIA which has now D6--branes on the deformed conifold. This statement is of course to be understood in the local framework we used in the last section.

However, wrapping D5--branes on the resolved conifold \eqref{resmetric} is not trivial. First, the backreaction of the branes on the metric will create a warp factor and second, such a solution with fractional branes\footnote{The equivalence of wrapped D5--branes and fractional D3--branes was shown in \cite{klebanekrasov}.} also has to include NS flux \cite{klebanovtseyt, klebanekrasov}. 
In \cite{pandozayas} an attempt was made to find the supergravity solution describing D5--branes wrapping the (blown up) $S^2$ of the resolved conifold. Although they solved the supergravity equations of motion, it was shown in \cite{cvetic} that their solution breaks supersymmetry completely (because of non--primitive fluxes).
So far, no explicit supersymmetric solution (including metric, dilaton and all fluxes) for D5--branes on the resolved conifold exists.

We will circumvent this problem in the next section by using an F--theory setup that by construction delivers a supersymmetric IIB solution. In this section we solely focus on the effect of NS flux on mirror symmetry.

The standard brane solution describes a warped product of the flat world--volume times the flat transverse space. 
Our case is different, since we want to wrap D--branes (that also extend along Minkowski space) on a compact cycle inside a non--compact Calabi--Yau.
However, the backreaction of branes on this type of non--compact manifolds has been studied, see for example \cite{klebanekrasov,minasian, keha}. One finds again a warped product, where the warp factor is a function of the non--compact direction $r$. 
If the internal manifold is Ricci--flat with metric $g_{mn}$ and coordinates $y$ one can construct the following 10--dimensional D--brane solution
\begin{equation}
  ds_{10}^2\,=\,h^{-1/2}(y)\,ds^2_{\mathbb{R}^{1,3}} + h^{1/2}(y)\,g_{mn}(y)\,
    dy^m\,dy^n\,.
\end{equation}
In the case of regular D3--branes along Minkowski space $\mathbb{R}^{1,3}$, $h$ would be a harmonic function on the internal manifold. For fractional D3--branes $h$ is determined from the trace of the Einstein equations \cite{ks,klebanovtseyt, pandozayas}\footnote{These cases use some symmetry properties of conifolds and should not be generalized to arbitrary $g_{mn}$.}, $R=-\frac{1}{2}\Delta h$. For us, $g_{mn}$ will be the Ricci--flat K\"ahler metric of the resolved conifold \eqref{resmetric}, which is of the form $g_{mn}=dr^2+r^2\,g_{ij}\,dy^i\,dy^j$ for some 5--dimensional base $g_{ij}$. For this class of manifolds $h$ is a function of $r$ only \cite{keha}. The solution will also include RR flux (three and five--form fieldstrengths, since fractional D3--branes carry both the charge of D3--branes and of the D5--branes that they really are) and NS flux. We will neglect the RR flux for the time being since it does not affect the metric under T--duality.

The supergravity solution for metric and NS flux found in \cite{pandozayas} reads
\begin{eqnarray}\label{resoliib}\nonumber
  ds^2 & = & h(\widetilde{r})^{-1/2}\,ds_{\mathbb{R}^{1,3}}^2 +h(\widetilde{r})^{1/2}
    \left[\widetilde{\gamma}'\,d\widetilde{r}^2 +\frac{\widetilde{\gamma}'}{4}\,\widetilde{r}^2\big 
    (d\widetilde{\psi}+\cos\widetilde{\theta}_1\,d\widetilde{\phi}_1 
    +\cos\widetilde{\theta}_2\,d\widetilde{\phi}_2\big)^2 \right.\\ 
  & &+ \left.\frac{\widetilde{\gamma}}{4}\,\big(d\widetilde{\theta}_1^2+\sin^2\widetilde{\theta}_1
    \,d\widetilde{\phi}_1^2\big) +  \frac{\widetilde{\gamma}+4a^2}{4}\,\big(d\widetilde
    {\theta}_2^2+\sin^2\widetilde{\theta}_2\,d\widetilde{\phi}_2^2\big)\right]\\
  B_{NS} &=& b_1(\widetilde{r})\,d\widetilde{\theta}_1\wedge \sin\widetilde{\theta}_1
    \,d\widetilde{\phi}_1 + b_2(\widetilde{r})\,d\widetilde{\theta}_2\wedge 
    \sin\widetilde{\theta}_2\,d\widetilde{\phi}_2\,.
\end{eqnarray}

Clearly, this solution (or the accompanying RR flux) has to be modified if we want to preserve supersymmetry. But as we will show in the next section, all possible solutions for the metric of D5--branes on the resolved conifold have the same local limit. We therefore introduce local coordinates by generalizing \eqref{local} slightly as to incorporate the warpfactor $h(\widetilde{r})$, which will in the local neighborhood also be interpreted as a constant $h_0=h(r_0)$. We define
\begin{eqnarray}\nonumber
  \widetilde{r} &=& r_0+\frac{r}{\sqrt{\widetilde{\gamma}_0'\sqrt{h_0}}}\,,\qquad
    \qquad\qquad \;\widetilde{\psi} \,=\, \z + \frac{2 z}
    {\sqrt{\widetilde{\gamma}_0'\sqrt{h_0}}\,r_0}\\ \nonumber
  \widetilde{\phi}_1 &=& \langle\phi_1\rangle + \frac{2x}
    {\sqrt{\widetilde{\gamma}_0\sqrt{h_0}}\,\sin\langle\theta_1\rangle}\,,\quad\quad\;
  \widetilde{\phi}_2 \,=\, \langle\phi_2\rangle + \frac{2y}{\sqrt{
    (\widetilde{\gamma}_0+4 a^2)\sqrt{h_0}}\,\sin\langle\theta_2\rangle}\\
  \widetilde{\theta}_1 &=& \langle\theta_1\rangle+\frac{2\theta_1}{\sqrt{\widetilde
    {\gamma}_0 \sqrt{h_0}}}\,,\qquad\qquad\qquad
  \widetilde{\theta}_2 \,=\, \langle\theta_2\rangle+\frac{2\theta_2}{\sqrt{
    (\widetilde{\gamma}_0+4 a^2)\sqrt{h_0}}}\,.
\end{eqnarray}
This gives the following ansatz for the local metric of D5--branes on the resolved conifold
\begin{equation}\label{start}
  ds^2 \,=\,dr^2+(dz+A\,dx + B\, dy)^2+ (dx^2+d\theta_1^2) 
    +(dy^2+d\theta_2^2)\,,
\end{equation}
which is precisely the same as \eqref{startmetric}. $A$ and $B$ remain as defined in \eqref{defineab}.

For the NS flux we make the most generic ansatz possible, with one exception: we only allow for electric NS--flux (magnetic NS flux leads in general to non--geometrical solutions \cite{wecht, simeon, nongeomet}), i.e. B--field components that have only one leg along T--duality directions\footnote{Without loss of generality we do not include components involving $dr$ since components of the 3--form fieldstrength like $dr\wedge dx\wedge d\theta_i$ can easily be obtained from $\partial_r b_{xi}(r)\,dr\wedge dx\wedge d\theta_i$.}:
\begin{equation}\label{bstartiib}
  B_{NS}^{IIB} \,=\, b_{zi}\,dz\wedge d\theta_i + b_{xj}\,dx\wedge d\theta_j 
    + b_{yk}\,dy\wedge d\theta_k\,.
\end{equation}
In general, the coefficients $b_{zi},\,b_{xj}$ and $b_{yk}$ can depend on all base coordinates $(r,\,\theta_1,\,\theta_2)$ to preserve the background's isometries, although \cite{pandozayas} seems to indicate they should be functions of $r$ only.

This B--field has now non--trivial consequences when we perform T--dualities along $x$, $y$ and $z$. We will not merely find a local version of the deformed conifold, but a manifold with {\it twisted fibers}, that is clearly the local limit of a non--K\"ahler version of the deformed conifold.

The reason why mirror symmetry with NS field gives rise to a non--K\"ahler manifold is actually very easy to illustrate in the SYZ picture. T--duality mixes B--field and metric. In the presence of NS flux, Buscher's rules \eqref{buscher} read
\begin{eqnarray}\nonumber
  \widetilde{G}_{yy} &=& \frac{1}{G_{yy}}\,,\qquad\qquad\qquad\qquad\qquad\quad
    \widetilde{G}_{\mu y}\,=\,\frac{B_{\mu y}}{G_{yy}}\\
  \widetilde{G}_{\mu\nu} &=& G_{\mu\nu}-\frac{G_{\mu y}G_{\nu y}-B_{\mu y}
    B_{\nu y}}{G_{yy}}\\ \nonumber
  \widetilde{B}_{\mu\nu} &=& B_{\mu\nu}-\frac{B_{\mu y}G_{\nu y}-G_{\mu y}
    B_{\nu y}}{G_{yy}}\,,\qquad \widetilde {B}_{\mu y}\,=\,\frac{G_{\mu y}}{G_{yy}}\,,
\end{eqnarray}
so cross terms in the metric are traded against the corresponding components in $B_{NS}$ and vice versa.
Therefore, the $T^3$ fibers acquire a twisting by $B_{NS}$--dependent one--forms as
\begin{eqnarray}\label{fibration}\nonumber
  dx\;\longrightarrow\;d\hat{x} &=& dx-b_{xi}\,d\theta_i\\
  dy\;\longrightarrow\;d\hat{y} &=& dy-b_{yi}\,d\theta_i\\ \nonumber
  dz\;\longrightarrow\;d\hat{z} &=& dz-b_{zi}\,d\theta_i
\end{eqnarray}
under T--duality. Apart from this modification, we perform the same steps as in the last section: we boost the complex structure as in \eqref{boost} and take the limit $\epsilon\to 0$. We then fix the constants $\beta_i$ (their values do not change) and we restore the $\sin\langle z\rangle$ and $\cos\langle z\rangle$ dependence via the rotation \eqref{yrotation} or by adapting the point of view that the mirror is in the delocalized limit where $\langle z\rangle=0$.

Then we find the mirror in IIA to be
\begin{eqnarray}\label{nkdef}\nonumber
  d\widetilde{s}^2 &=& dr^2 \,+\,\alpha^{-1}\,\left[(dz-b_{zi}\,d\theta_i)-\alpha 
    A\,(dx-b_{xi}\,d\theta_i)-\alpha B\,(dy-b_{yi}\,d\theta_i)\right]^2\\ \nonumber
  & +& \alpha(1+B^2)\,\left[d\theta_1^2+(dx-b_{xi}\,d\theta_i)^2\right] \,+\, 
    \alpha (1+A^2)\,\left[d\theta_2^2 +(dy-b_{yi}\,d\theta_i)^2\right]\\ 
  & + &  2\alpha AB\,\cos \langle z\rangle \,\big[d\theta_1 d\theta_2 \,-\,  
    (dx-b_{xi}\,d\theta_i)(dy-b_{yi}\,d\theta_i)\big] \\ \nonumber
  & + &  2\alpha AB\,\sin\langle z\rangle\,\big[(dx-b_{xi}\,d\theta_i)\,d\theta_2
    +(dy-b_{yi}\,d\theta_i)\,d\theta_1 \big]\,,
\end{eqnarray}
with $\alpha$ defined in \eqref{alpha}.
We therefore conjecture the local resolved conifold to be mirror dual to a local ``non--K\"ahler deformed conifold'' with twisted fibers that make this metric inherently non--K\"ahler. 

In the absence of B--field this is clearly a K\"ahler background, since in this local version all coefficients in the metric are constants. With B--field dependent fibration the fundamental two--form will in general not be closed anymore, because it will depend on derivatives of $b_{ij}$. A more thorough analysis of this geometry will be attempted in section \ref{torsion}, but it remains somewhat incomplete because we lack the knowledge of a global background. Strictly speaking, a metric in a small patch does not carry any information about the manifold. We can, however, make some predictions on what we expect for the global solution, since supersymmetry poses restrictions on allowed non--K\"ahler manifolds.

As mentioned earlier, cross terms in the metric induce a B--field under T--duality. Due to the boost of complex structures \eqref{boost} we introduced new cross terms of the form $dx\,d\theta_1$ and $dy\,d\theta_2$. We do therefore also recover a B--field in the mirror IIA. It has a peculiar scaling\footnote{This comes from the introduction of the boost $f_{1,2}$, in other words without this boost there would be no B--field in IIA. We will comment on the ``physicality'' of this flux in the next section.} with $\epsilon$, but can nevertheless be determined to
\begin{eqnarray}\nonumber
  \sqrt{\frac{\epsilon}{\alpha}}\,\widetilde{B}_{NS}^{IIA} &=& (dx-b_{xi}\,d\theta_i)
    \wedge d\theta_1 - (dy-b_{yi}\,d\theta_i)\wedge d\theta_2\\
  & & -A\,(dz-b_{zi}\,d\theta_i)\wedge d\theta_1 + B\, (dz-b_{zi}\,d\theta_i) 
    \wedge d\theta_2\,,
\end{eqnarray}
where we have without loss of generality assumed that $\beta_1=-\beta_2=\sqrt{\alpha}$, otherwise there would be an overall minus sign. To be consistent with the metric we should also here perform the rotation \eqref{yrotation}, which changes above result to
\begin{eqnarray}\label{bnsiia}
  \sqrt{\frac{\epsilon}{\alpha}}\,\widetilde{B}_{NS}^{IIA} &=& (dx-b_{xi}\,d\theta_i)
    \wedge d\theta_1 - (dy-b_{yi}\,d\theta_i)\wedge d\theta_2
    -A\,(dz-b_{zi}\,d\theta_i)\wedge d\theta_1 \\ \nonumber
  & & + B\,\cos\langle z\rangle\, 
    (dz-b_{zi}\,d\theta_i) \wedge d\theta_2 + B\,\sin\langle z\rangle\, 
    (dz-b_{zi}\,d\theta_i) \wedge (dy-b_{yi}\,d\theta_i)\,.
\end{eqnarray}
This seems to indicate that we can have a magnetic flux (the $dx\wedge dy$ component in $\widetilde{B}_{NS}^{IIA}$), but this term vanishes if we would want to reverse the T--dualities. Because then we would have to take the semi--flat limit of the IIA background either by choosing $\langle z\rangle=0$ or by reversing above rotation. In any case, it shows that 
mirror symmetry can only be realized in the semi--flat limit where $\langle z\rangle=0$. This is of course due to the restrictions of Buscher's rules, which require isometry directions. See \cite{greg} for discussions on T--duality in the absence of isometries.

\subsection{Discussion}\label{discuss}

We would like to compare our result to the half--flat manifolds obtained in \cite{jan}. There are some obvious similarities, like the B--field dependent twist that enters into the $T^3$ fibers. But there are also some fundamental differences. 
Several arguments lead us to believe that the mirror of the resolved conifold with fluxes is not a half--flat manifold.
\begin{itemize}
\item As will be demonstrated in section \ref{torsion}, we could not find a half--flat structure for our local IIA metric, but a symplectic structure does exist (that means the fundamental two--form is closed, $dJ=0$, but the manifold is not complex). We will show in the following that the half--flat solution from \cite{jan} also possesses a symplectic structure and give some arguments that favor the symplectic interpretation.
\item If we would T--dualize the global background, we would have to take a warp factor into account that depends on an {\it internal} coordinate. In the case of \cite{jan} one would not find a half--flat structure either if the warp factor would depend on internal coordinates.
\end{itemize}

To be able to illustrate these arguments, let us briefly review the discussion from \cite{jan}.
Start with an NS5--brane in flat 10--dimensional space, this will be a warped product of the 
flat worldvolume of the brane ${\mathbb R}^{1,5}$ and the transverse four dimensional space. The wrap factor $V$ is a harmonic function on the transverse space $\mathbb{R}^4$. 
This solution can be ``smeared'' by allowing  $V$ to depend only on certain coordinates of $\mathbb{R}^4$. Consider the simple case that $V$ actually depends only on one coordinate $V=V(\xi)$ and compactify the 10--dimensional solution on a six--torus. In this way $\xi$ becomes an external coordinate. This leads to a domain--wall solution, since the 5--branes are wrapped along three internal directions and extend along three spacetime directions ${\mathbb R}^{1,2}$. The solution can be written as
\begin{eqnarray}\nonumber
  ds^2 &=& ds_{{\mathbb R}^{1,2}}^2 +V\,d\xi^2+ \left(dx_1^2 +dx_2^2+ 
    dx_3^2+V(dy_1^2+dy_2^2+dy_3^2)\right)\\ 
  H_{NS} &=& \partial_\xi V\,dy_1\wedge dy_2\wedge dy_3\,. 
\end{eqnarray}
The term in parenthesis indicates the compact part in the metric. Note that three of the internal directions are warped by $V$, but from the point of view of the internal manifold this is simply a constant, so the starting manifold X is simply a torus $T^6=T^3\times T^3$. The other three internal coordinates $x^i$ are along the brane worldvolume. The four dimensional non--compact (external) space is now given by $ds_{{\mathbb R}^{1,2}}^2 +V\,d\xi^2$. 

There are several fibers one could identify as the T--duality $T^3$. 
Choose it such that $H_{NS}$ is purely electric. Following \cite{jan}, but using slightly different notation,  
we perform T--duality along $x_1$, $x_2$ and $y_1$. Locally, the B--field can be chosen as $B=\lambda\, y_2\, dy_3\wedge dy_1$. One finds for the mirror background
\begin{eqnarray}
  ds^2 &=& ds_{\mathbb{R}^{1,2}}^2+d\xi^2\\ \nonumber
    & & +\left(dx_1^2+dx_2^2+dx_3^2 
    +V^{-1}(dy_1-\lambda y_2\,dy_3)^2 +V\,dy_2^2+V\,dy_3^2\right)\,.
\end{eqnarray}
Clearly, the internal manifold $Y$ is no longer a torus. It has the form $T^3\times Q$, where $Q$ is a non--trivial $S^1$--fibration over $T^2$.
Also note, that the B--field is ``used up'' under T--duality. It vanishes completely in the metric and since we started with a torus without any cross-terms, no B--field is generated under T--duality. 
One can still define an almost complex structure locally, in terms of three complex vielbeins
\begin{eqnarray}\label{janvielb}\nonumber
  e_1 &=& dx_1+i\sqrt{V} dy_2\\ \nonumber
  e_2 &=& dx_2+i\sqrt{V} dy_3\\ 
  e_3 &=& \frac{1}{\sqrt{V}} (dy_1-\lambda\, y_2\,dy_3)+i dx_3\,,
\end{eqnarray}
but this cannot be integrated to a complex structure on the mirror manifold. Define the fundamental 2--form as
$  J \,=\,\frac{i}{2}\delta^{ij}\,e_i\wedge \bar{e}_j$,
then it is obviously not closed, because
\begin{equation}
  dJ\,=\,\frac{2\lambda}{\sqrt{V}}\, dy_2\wedge dy_3\wedge dx_3\,,
\end{equation}
but it still satisfies
\begin{equation}\label{dJwedgeJ}
  d(J\wedge J)\,=\,0\,.
\end{equation}
One can also define a holomorphic 3--form
$  \Omega\,=\,\Omega^++i \Omega^-\,=\,e_1\wedge e_2\wedge e_3$,
which is not closed either but
\begin{equation}
  d\Omega\,=\,-\frac{\lambda}{\sqrt{V}}\,dx_1\wedge dx_2\wedge dy_2\wedge dy_3\,,
\end{equation}
which means that
\begin{equation}\label{domegaplus}
  d\Omega^+\,\ne\,0\,,\qquad d\Omega^-\,=\,0\,.
\end{equation}
Equations \eqref{dJwedgeJ} and \eqref{domegaplus} define precisely what is known as a half--flat manifold. 


However, the choice of complex structure on a six--dimensional {\it real} manifold is not unique. The metric in real coordinates does not carry any information about the complex structure.
For example, a minor change in \eqref{janvielb} given by the exchange of the imaginary parts of $e_2$ and $e_3$
\begin{eqnarray}\nonumber
  e_1 &=& dx_1+i\sqrt{V} dy_2\\ \nonumber
  e_2 &=& dx_2+i dx_3\\ 
  e_3 &=& \frac{1}{\sqrt{V}} (dy_1-\lambda\, y_2\,dy_3)+i\sqrt{V} dy_3\,,
\end{eqnarray}
leads to the conclusion $Y$ should be symplectic. One finds $J$ to be completely independ of the term containing $\lambda\,y_2$, therefore $dJ=0$, but this is not a K\"ahler manifold as it cannot be complex. For the holomorphic 3--form one finds
\begin{equation}
  d\Omega\,=\,-\frac{\lambda}{\sqrt{V}}\,(dx_1\wedge dx_2\wedge dy_2\wedge dy_3 
    +i\, dx_1\wedge dx_3\wedge dy_2\wedge dy_3)\,,
\end{equation}
so neither real nor imaginary part of $\Omega$ are closed.

However, the argument of ``cohomology counting'' presented in section \ref{mirrorsym} strongly suggests that this manifold should indeed be half--flat. It fulfills the assumption that the non--closed imaginary part of the complexified K\"ahler parameter is mapped to a part of $\Omega$ being non--closed. It turned out to be the real part of $\Omega$ in this convention, but the assignment of real or imaginary part is completely arbitrary\footnote{The authors of \cite{jan} defined $t=B_{NS}+iJ$, so that they find a mapping of non--closed real parts of $J$ and $\Omega$.}. 

But there are also more arguments that favor the interpretation as a symplectic manifold. One of them is given in the context of generalized complex geometry (GCG), as introduced by Hitchin and Gualtieri \cite{hitch, gualtieri}, see also appendix \ref{gcg} for a brief introduction. The structures they defined interpolate between usual complex and symplectic structures. The question whether GCG has relevance for string theory has been of great interest recently, see for example \cite{kapuli}--\cite{zucchini}. In particular, the B--field transform \cite{gualtieri} that takes complex or symplectic structures into generalized complex structures, can be interpreted as T--duality with magnetic NS flux \cite{kapustin}. It was argued that T--duality with only electric NS flux cannot lead to GCG, it should preserve a complex or symplectic structure\footnote{However, the work of Kapustin \cite{kapustin} differs from \cite{jan} or the analysis presented herein since he considers T--duality in {\it all directions} on an even--dimensional torus. We and \cite{jan}, on the other hand, are interested in mirror symmetry or {\it three} T--dualities.}. In a different context, it was also shown that IIA vacua should always be symplectic, whereas IIB vacua should be complex \cite{minas}. Mirror symmetry between complex and symplectic manifolds has been discussed in \cite{kachrutomas}. 


These arguments seem to suggest that mirror symmetry with NS flux leads to symplectic manifolds in IIA and only the special case discussed in \cite{jan} allows an additional half--flat structure\footnote{Twisted tori that are simultaneously symplectic and half--flat were discussed in \cite{marchesano}. They differ from \cite{jan} as they do not represent domain wall solutions. The half--flat structure is achieved by chosing the appropriate values for the flux components, i.e. the appropriate vacuum.}. As will be discussed in section \ref{torsion}, half--flat manifolds lift to manifolds with $G_2$ holonomy (which preserve $\mathcal{N}=2$ supersymmetry in d=4 upon compactification without flux). Supersymmetry arguments therefore seem to suggest that the symplectic mirror manifold admits a half--flat structure precisely if there are no additional RR or NS fluxes in IIA (and the dilaton is constant).

An interesting question in this context would be under what circumstances half--flat manifolds also admit a symplectic structure (and vice versa). One might also ask  
if half--flat manifolds can be interpreted as generalized complex manifolds and if so, of what kind (e.g. B--field transforms of complex or symplectic structures or something entirely different).
Maybe there exists a mapping between GCG and torsion classes \cite{salamon}, which describe two different approaches of classifying generalized (or non--K\"ahler) geometries. One way to achieve that might be the framework of \cite{minasfidanza} which maps quantities of GCG (pure spinors) to fundamental two--form and holomorphic three--form, which are used to determine torsion classes.


Let us now discuss what would happen if we T--dualized the global metric \eqref{resmetric} instead of its local limit \eqref{startmetric}.
F--theory tells us that flux compactifications in IIB simply lead to an overall conformal warp factor for the internal space \cite{giddings}. For IIA we do not have such a reasoning and the back--reaction of fluxes seems to be more involved, see \cite{granaflux} for an overview on flux compactifications. Thus, we will start with a space that is a conformal resolved conifold in IIB. The analysis of \cite{jan} took the back--reaction into account, but the warp factor was a function of an external coordinate (otherwise one could not compactify that direction). Our warp factor $h(r)$ is {\it not constant} on the internal manifold. Repeating the analysis following equation \eqref{janvielb} with a warp factor $V$ that depends on internal coordinates, would not produce a half--flat manifold. This is the second reason why we do not believe our result \eqref{nkdef} to be half--flat globally.

The influence of a non--trivial warp factor is already interesting for scenarios with only RR flux.
Imagine a generic conformal Calabi--Yau in IIB, let us denote it by $h\cdot X$, where $X$ is a $T^3$ fibration over some base $B$. Under T--duality, the size of the $T^3$ fibers will be inverted, but the base is not changed.
If $Y$ was the mirror of $X$ in the absence of fluxes, then $h\cdot X$ does not map to a conformal version of $Y$ under three T--dualities, as the $T^3$ fibers acquire a factor of $h^{-1}$, but the base does not. So, the mirror is locally a product $(h\cdot B)\times (h^{-1}T^3)$. One could still write this manifold as $h\cdot \hat{Y}$, but $Y$ and $\hat{Y}$ are not the same manifold as they have different $T^3$ fibers. This difference is of course trivial if the warp factor is constant on the fiber, then it can be absorbed into a coordinate redefinition. This is the case in \cite{jan}, but not if we leave the local limit of our calculation.

It would be interesting to study the influence of flux back--reactions onto the result from \cite{micu}, which states that if X is mirror to Y, then the two manifold are also mirror with RR fluxes turned on. Solutions of the type \cite{beckbeck} allow for compact internal manifolds with warp factors that do depend on internal coordinates. Their effect under T--duality should then be non--trivial, i.e. they cannot be absorbed into a rescaling of coordinates. Then the local form of the mirror $(h\cdot B)\times (h^{-1}T^3)$ suggests a more complicated backreaction of the fluxes in IIA or a modification of the mirror symmetry statement of \cite{micu}.

To summarize this section:
We find a family of non--K\"ahler manifolds in IIA that includes one K\"ahler geometry (for all NS flux switched off), which takes the form of a local deformed conifold metric. However, it does not agree with the local limit of the Ricci--flat K\"ahler metric on the deformed conifold (the two $S^2$ have different size). This was to be expected, since resolved and deformed conifold are only approximately mirror to each other. When constructing the mirror metric, we had to make explicit use of the assumption that we work in the local limit to bring the IIA metric into a form that resembles a local deformed conifold.

We also argued that
the IIA non--K\"ahler backgrounds can be more general than half--flat globally. Since T--duality preserves supersymmetry, these should be good string theory backgrounds (although we can only give their metric in the local limit), provided we start with a string theory solution in IIB. As already mentioned, the solution for wrapped D5--branes on the resolved conifold from \cite{pandozayas} breaks all supersymmetries, so we need a new idea.
We will suggest a background that is obtained from an F--theory construction and therefore inherently supersymmetric in the next section.

\setcounter{equation}{0}
\section{Geometric Transitions in Type II}\label{chain}

In this section we will use what we have learned about mirror symmetry between resolved and deformed conifold to verify ``Vafa's duality chain'', see figure \ref{vafachain}, with a concrete string theory background. We find such a background by considering an F--theory solution that is elliptically fibered over the resolved conifold. This background has the same local metric we discussed in section \ref{mirror}. Taking a generic ansatz for NS and RR fluxes we follow the duality chain from IIB to IIA, lift the resulting non--K\"ahler background to $\mathcal{M}$--theory, perform a flop, reduce again to IIA and perform a last mirror to end in IIB again. We will show that the duality chain has to be modified to include non--K\"ahler backgrounds in IIA.

The flop we perform in the $\mathcal{M}$--theory lift implies that the two non--K\"ahler backgrounds we propose in IIA are connected via a geometric transition. We will give several arguments for the consistency of this transition, in particular the ``closure'' of the duality chain as we finally end in IIB with a K\"ahler background. We will show that the fluxes do not change (much) under geometric transition, but the geometry does. This is in agreement with the interpretation that the cycles, on which we wrapped D--branes, shrink and a new geometry that contains only flux emerges after geometric transition.

To fix notation and remind the reader of the connection between F--theory and type IIB orientifolds we start with a short review, see also \cite{dabholkar} for a more detailed review article. Our particular F--theory setup will be constructed in section \ref{ftheorysetup}, the reader familiar with the concepts may want to skip ahead.

\subsection{Orientifolds and F--Theory}\label{introf}

We begin this section by reviewing some symmetries of type IIB superstrings we will need to make the connection between 12--dimensional F--theory and 10--dimensional IIB on an orientifold. Orientifolds are nothing but orbifold (manifolds gauged by a symmetry group G) that include the orientation reversal of the string worldsheet. We will denote the symmetry group by $\{G_1,G_2\}$, where $G_1$ is a pure orbifold action and only $G_2$ contains the worldsheet parity $\Omega$.

There are two perturbative $\mathbb{Z}_2$ symmetries of IIB superstrings which will be of particular interest to us. They are perturbative in the sense that they are evident at the perturbative level but believed to hold non--perturbatively, whereas non--perturbative symmetries are not apparent at the perturbative level.
\begin{itemize}
  \item Worldsheet parity $\Omega$. Type IIB is invariant under orientation reversal of the worldsheet ($\sigma\to 2\pi-\sigma$) since it is non--chiral on the worldsheet. $\Omega$ takes left movers to right movers, therefore NS--NS tensor states are even under $\Omega$ if they are symmetric (odd if they are antisymmetric). From the R--R sector the tensor $C_2$ is even (because of Fermi statistics). One combination of the fermions is even, the other is odd. To summarize,
  \begin{equation}\label{actionomega}
    \begin{array}{ll}
      g_{\mu\nu},\, \phi,\, C_2 & \mbox{even under}\; \Omega\\
      \chi,\, B_{NS},\, C_4 \quad & \mbox{odd under}\; \Omega\,.
    \end{array}
  \end{equation}
  \item Spacetime fermion number $(-1)^{F_L}$. This symmetry is only obvious in Green--Schwarz (GS) formalism, because there are no spacetime fermions in R--NS formalism. In GS formalism, the IIB superstring action reads \cite{gsw} (with worldsheet coordinates $\sigma$ and $\tau$ and light--cone coordinates $\partial_\pm=1/2(\partial_\tau\pm\partial_\sigma)$, the Regge slope is as usual denoted by $\alpha'$ and is related to the string tension)
  \begin{equation}
    S=\frac{-1}{4\pi\alpha'}\,\int d\sigma\,d\tau\,(\partial_+X^\mu \partial_-X^\mu -i S^\alpha_L\partial_- S^\alpha_L -i S^\alpha_R\partial_+ S^\alpha_R)
  \end{equation}
  where $S^\alpha_L, S^\alpha_R$ are actually spacetime spinors, both transforming as ${\bf 8_s}$ of $Spin(8)$\footnote{$Spin(8)$ is the covering group of $SO(8)$, which is the transverse rotation group if we use light cone gauge on the 10d target space. Relevant to us are the eight--dimensional representations ${\bf 8_v}$ (vector), ${\bf 8_s}$ (spinor) and ${\bf 8_c}$ (conjugate spinor).}. The type IIA action looks similar, but with $S^{\dot\alpha}_R$ transforming as a conjugate spinor ${\bf 8_c}$. This action is, amongst others, invariant under $S_L\to -S_L$, an operation which is written as $(-1)^{F_L}$. $F_L$ is the {\it spacetime} fermion number coming from left movers. Only left moving fermions are odd under this symmetry, so R--NS and R--R states are odd, whereas NS--R and NS--NS states are even. To summarize,
  \begin{equation}\label{actionfl}
    \begin{array}{ll}
      g_{\mu\nu},\, \phi,\, B_{NS}\quad & \mbox{even under}\; (-1)^{F_L}\\
      \chi,\, C_2,\, C_4 \quad & \mbox{odd under}\; (-1)^{F_L}\,.
    \end{array}
  \end{equation}
\end{itemize}

We will be particularly interested in T--duality on orientifolds, so let us see what happens to the perturbative symmetry $\Omega$ under T--duality. As explained in appendix \ref{tduality}, T--duality has a rather non--trivial action on the worldsheet. It acts as a one--sided parity transformation sending $X^i_R\to -X^i_R$, but leaving $X_L^i$ invariant ($i$ indicates the T--duality direction). A T--duality along the 9th direction yields 
\begin{eqnarray}\label{omegaundert}
  \Omega\;\mbox{in type IIB}\;&\xrightarrow{T_9}&\;I_9\,\Omega\;\mbox{in type IIA}
\end{eqnarray}
where $I_9$ symbolizes spacetime parity, i.e. inversion of the 9th coordinate $I_9:\; (X_L^9, X_R^9)\,\to\,(-X_L^9,-X_R^9)$. In other words,
\begin{equation}
  T_9\,\Omega\,T_9^{-1}\,=\,I_9\,\Omega
\end{equation}
as can be seen from
\begin{equation}
  (X_L^9,X_R^9)\,\xrightarrow{T_9^{-1}}\,(X_L^9,-X_R^9)\, \xrightarrow{\Omega}\, (-X_R^9,X_L^9) \,\xrightarrow{T_9}\, (-X_R^9,-X_L^9)\,.
\end{equation}
Acting on fermions, the parity $I_9$ flips the chirality of both left and right moving fermions. Note that $\Omega$ by itself is not a symmetry of type IIA, because starting with a left moving spinor ${\bf 8_s}$ and a right moving conjugate spinor ${\bf 8_c}$, one obtains a left moving conjugate spinor ${\bf 8_c}$ and a right moving spinor ${\bf 8_s}$. To flip the chirality the operation has to be accompanied by a parity transformation to obtain a genuine symmetry
\begin{equation}
  (\bf{8_s,8_c}) \,\xrightarrow{\Omega}\, (\bf{8_c,8_s})\,\xrightarrow{I_9}\, (\bf{8_s,8_c})\,.
\end{equation}
Thus we see that worldsheet parity is a symmetry of IIB, but IIA is only symmetric under a combination of worldsheet and spacetime parity.

Type IIB superstring theory has a non--perturbative $SL(2,\mathbb{Z})$ symmetry, 
which is given by the group represented by integer $2\times 2$ matrices with determinant one:
\begin{equation}
  \Lambda\,=\,
  \begin{pmatrix}
    a & b\\ c & d
  \end{pmatrix}
  \quad\mbox{with}\; a,b,c,d\in\mathbb{R}\qquad\mbox{and}\qquad ad-bc=1\,.
\end{equation}
Define a complex scalar field $\lambda=\chi+i e^{-\phi}$, where $\chi$ is the axion (the RR scalar) and $\phi$ the dilaton. The expectation value of the dilaton actually fixes the string coupling, $g_s=e^{\langle\phi\rangle}$. One also introduces the Einstein metric $G_{\mu\nu}=e^{-\phi/2} g_{\mu\nu}$ (with string frame metric $g_{\mu\nu}$) because it is invariant under $SL(2,\mathbb{Z})$. Recall also the other fields in the type IIB spectrum: $B_{NS}$ and the RR fields $C_2, C_4$. 
The action of $SL(2,\mathbb{Z})$ on the bosonic fields is given by
\begin{equation}
  \lambda\,\to\,\frac{a\lambda+b}{c\lambda+d}\,,\quad 
  \begin{pmatrix}
    B_{NS}\\ C_2
  \end{pmatrix} \,\to\,
  \begin{pmatrix}
    d & -c \\ -b & a
  \end{pmatrix}
  \begin{pmatrix}
    B_{NS} \\ C_2
  \end{pmatrix}\,,\quad
  C_4\,\to\,C_4\,,\quad G\,\to\, G\,.
\end{equation}
This action is generated by the elements:
\begin{eqnarray}\label{sltwoz}\nonumber
    T:\; &  \lambda\to \lambda+1 \,,\quad  &\Lambda=\begin{pmatrix}
                                                   1 & 1\\ 0 & 1
                                                 \end{pmatrix}\\[1ex]
    S: \;& \lambda\to -1/\lambda\,, \quad &  \Lambda=\begin{pmatrix}
                                                   0 & 1\\ -1 & 0
                                                 \end{pmatrix}\\[1ex] \nonumber
    R: \;& \lambda\to\lambda\,, \quad & \Lambda=\begin{pmatrix}
                                                   -1 & 0\\ 0 & -1
                                                 \end{pmatrix}\,.
\end{eqnarray}
This symmetry has far--reaching consequences for the moduli space of type IIB. It states that every $\lambda$ is non--perturbatively equivalent to any other $\lambda$ that can be reached by an $SL(2,\mathbb{Z})$ transformation This reduces the moduli space from the whole upper half plane to a much smaller subset \cite{kaku} known as the fundamental domain.

The symbol $S$ for one of the generators is not chosen by accident, it generates precisely what is known as S--duality. This becomes obvious if one considers the case where $\chi=0$ and $\lambda$ is essentially given by the string coupling. The action of $S$ then relates a theory at strong coupling ($\lambda=i/g_s$) to a theory at weak coupling ($-1/\lambda=ig_s$). Apparently, type IIB is self--dual under this symmetry, but it also connects type I and heterotic SO(32), a fact we will use in section \ref{hetsec}.
Note also that $S\,(-1)^{F_L}\,S^{-1}=\Omega$, because $S$ basically exchanges $B_{NS}$ and $C_2$ (recall \eqref{actionomega} and \eqref{actionfl}). Another important relation for the discussion of F--theory will be
\begin{equation}\label{rwithomega}
   R\,=\,\Omega (-1)^{F_L}\,,
 \end{equation}
which can be easily seen from their action on the massless spectrum. $B_{NS}$ and $C_2$ are odd, whereas all other fields are even under $R$, which is the same as the combined action $\Omega (-1)^{F_L}$, see \eqref{actionomega} and \eqref{actionfl}. 

This $SL(2,\mathbb{Z})$ symmetry has been proposed \cite{vafaf} to have a geometrical interpretation as two extra toroidal dimensions. This means there is some 12--dimensional theory, which has been termed F--theory, that gives rise to 10--dimensional IIB. The complex structure of the F--theory two--torus is hereby identified with the IIB scalar $\lambda$. We will now briefly review how this leads to orientifolds in IIB \cite{sen}, see e.g. \cite{dabholkar} for a detailed review.

Consider an elliptically fibered Calabi--Yau fourfold $K$ which is a toroidal fiber bundle over a base $B$. Even though $K$ is a smooth manifold, there will be points in the base where the fiber becomes singular and its complex structure parameter $\tau$ can have non--trivial monodromy\footnote{Monodromy on a fibration is similar to the concept of holonomy on a manifold. Instead of a parallel transport along a closed loop one considers the lift of a loop at $x$ that lies in a connected space $X$. The lift of $x$ into a fiber over $X$ is given by $c$, and going once around the loop in $X$ one finds an image $c'$ in the fiber, where in general $c\ne c'$. The monodromy is given by the right action of the fundamental group $\pi_1(X,x)$ as a permutation on the set of all $c$.} around these points. An F--theory compactification on $K$ refers to a compactification of type IIB on $B$, where the IIB scalar $\lambda=\chi+i e^{-\phi}$ is identified with the geometrical parameter $\tau$ \cite{vafaf}.
In general, $\tau$ varies over the base resulting in a non--constant field $\lambda$.

However, there are possible scenarios that allow for a constant solution of $\lambda$
\cite{sen, keshavmukhi}. 
These solutions are characterized by 24 singularities in the function describing the elliptic fibration. In the special case where these singularities appear at four different locations in a multiplicity of six, $\lambda$ is constant.
The singularities are interpreted as 24 seven--branes in F--theory, because there is a non--trivial monodromy\footnote{This monodromy indicates a magnetic source which has to be a seven--brane because seven--branes couple magnetically to scalars.} as we go around the singular point, which is precisely the one generated by $R$ in \eqref{sltwoz}: 
Moving around one singular point, the coordinate of the fiber is twisted by $R$, i.e. it is inverted, but its modular parameter $\tau$ remains unchanged. 

The base is actually an orbifold generated by the parity transformation $I_2$ that inverts both coordinates of the toroidal fiber.
If we go once around the singular point the theory comes back to itself modulo the $SL(2,\mathbb{Z})$ element $R$. We therefore have an orbifold of the base by $I_2\cdot R$, which turns into an orientifold, because $R$ is related to the worldsheet parity $\Omega$ of IIB. So, this F--theory setup gives rise to IIB on
\begin{equation}
  \frac{B}{\{1,R\,I_2\}}\,=\,\frac{B}{\{1,\Omega(-1)^{F_L}I_2\}}
\end{equation}
if one recalls the identity $R=\Omega(-1)^{F_L}$ from \eqref{rwithomega}.
Such an identification of F--theory with an orientifold is very useful and we will make extensive use of it in this thesis. Choosing the 12--dimensional manifold as a direct product\footnote{The inclusion of fluxes leads to warp factors \cite{beckbeck,sav}.} of flat Minkowski space and a Calabi--Yau fourfold with elliptic fibration preserves $\mathcal{N}=2$ supersymmetry for the resulting IIB theory (away from the fixed points of the orientifold action) \cite{sav}.

This orientifold has 4 orientifold 7--planes and 16 D7--branes that cancel their charges. This can be seen by noting that under T--duality along the $i$ and $j$ directions: $\Omega(-1)^{F_L}I_{ij}=T^{-1}_{ij}\Omega T_{ij}$, therefore this orientifold is T--dual to type IIB on $B/\{1,\Omega\}$, which has 32 D9--branes and a spacetime filling orientifold 9--plane. Under T--duality these become D7--branes and since they move in pairs (due to $I_{ij}$), there are effectively only 16 of them. The orientifold action $\Omega I_{ij}$ has four fixed points and the transverse space is 7--dimensional, so there are four O7--planes. This establishes the correspondence of an F--theory 7--brane with one O7--plane and four D7--branes in IIB. 

Also note that under T--duality, recall \eqref{omegaundert},
\begin{equation}\label{tdualori}
  \mbox{IIB on}\quad\frac{B}{\{1,\Omega (-1)^{F_L} I_{ij}\}} \,\xrightarrow{T_k}\, \mbox{IIA on}\quad 
    \begin{cases}
      B'/\{1,\Omega (-1)^{F_L} I_{ijk}\} & \mbox{if}\,k\ne i,j\\
      B'/\{1,\Omega (-1)^{F_L} I_{i}\} & \mbox{if}\,k=j
    \end{cases}\,,
\end{equation}
which means there will be orientifold 6--planes or 8--planes in IIA.
The dimension of the orientifold planes behaves under T--duality like that of D--branes: depending on whether T--duality is transverse or orthogonal to the plane its dimension is increased or decreased, respectively.

\subsection[The F--Theory Setup]{The F--Theory Setup and Orientifolds in IIA and IIB}\label{ftheorysetup}

We now want to construct a global IIB setup that is supersymmetric, contains D5--branes wrapped on the $S^2$ of the resolved conifold and allows us to add additional D--branes for global symmetries. 
Both adventures can be most conveniently undertaken at the same time within F--theory.

Consider F--theory on an elliptically fibered Calabi--Yau fourfold $X\to B$ such that the base $B$ is the resolved conifold\footnote{$B$ will not be a Calabi--Yau threefold anymore, since $X$ is a Calabi--Yau, but it is still K\"ahler \cite{gttwo}.}. Suppose $B$ contains a smooth $S^2$ 
and that there is a conifold transition from $B$ to $B'$ obtained by contracting the $S^2$ to a conical singularity and then smoothing to a deformed geometry. This gives another elliptically fibered Calabi--Yau fourfold $X'\to B'$ over a deformed base. We can introduce similar notation for the case when the cycle is contracted to a conical singularity, such that $X_0\to B_0$ describes a fibration over the singular conifold.

The Euler characteristic of $X$ can be computed from its topology. 
It was shown in \cite{gttwo} that it does not change under geometric transition, i.e. $\chi(X')=\chi(X_0)=\chi(X)$.
For the compact example in \cite{gttwo} it was computed explicitely with the {\it Maple} routine {\it schubert} \cite{schubert}, obtaining $\chi(X)=\chi(X')=19728$. This agrees with the result presented in \cite{oh} where the F--theory description for a fourfold after geometric transition was derived. Since we focus on local metrics only, our analysis does not distinguish whether the manifold is compact or not. But since we will not introduce any charge cancellation mechanisms for the wrapped D5--branes, let us assume a non--compact manifold henceforth.

As explained in section \ref{introf}, this setup provides us with an orientifold in type IIB with D7--branes and O7--planes. When we wrap D5--branes on the $S^2$ of the resolved geometry, we obtain an intersecting D5/D7--brane scenario on an orientifold IIB, which preserves supersymmetry \cite{realm}. 
The metric of the base $B$ has to resemble the resolved conifold locally, but globally it will also contain singularities that correspond to the 7--branes. Adding D5--branes creates warp factors. To incorporate these effects we make the following generic ansatz for the base
\begin{eqnarray}\label{basemetric}\nonumber
  ds^2 & = & h_0(\widetilde{r})\,d\widetilde{r}^2 + h_1(\widetilde{r})\,\big 
    (d\widetilde{\psi}+\cos\widetilde{\theta}_1\,d\widetilde{\phi}_1 
    +\cos\widetilde{\theta}_2\,d\widetilde{\phi}_2\big)^2\\ 
  & &+ \big(h_2(\widetilde{r})\,d\widetilde{\theta}_1^2
    +h_3(\widetilde{r})\,\sin^2\widetilde{\theta}_1\,d\widetilde{\phi}_1^2\big) +  
    \big(h_4(\widetilde{r})\,d\widetilde{\theta}_2^2+ h_5(\widetilde{r})\,
    \sin^2\widetilde{\theta}_2\,d\widetilde{\phi}_2^2\big)\,,
\end{eqnarray}
which allows in particular for the two spheres to be asymmetric and squashed. This ansatz is motivated by the idea that in the absence of D--branes and fluxes we should recover the K\"ahler metric. Also, for $\widetilde{r}\to\infty$ the warp factors should approach 1, so we will suppress any $\theta_{1,2}$--dependence in the functions $h_i$ although it would not influence the following local analysis. 

Let us compare this to the local metric in IIB \eqref{startmetric} we T--dualized in the last section. We define again local coordinates
\begin{eqnarray}\label{baselocal}\nonumber
  \widetilde{r} &=& r_0+\frac{r}{\sqrt{h_0(r_0)}}\,,\qquad\qquad\qquad
    \;\widetilde{\psi} \,=\, \z + \frac{z}{\sqrt{h_1(r_0)}}\\ \nonumber
  \widetilde{\phi}_1 &=& \langle\phi_1\rangle + \frac{x}
    {\sqrt{h_3(r_0)}\,\sin\langle\theta_1\rangle}\,,\quad\quad\;
  \widetilde{\phi}_2 \,=\, \langle\phi_2\rangle + \frac{y}{\sqrt{ h_5(r_0)}\,
    \sin\langle\theta_2\rangle}\\
  \widetilde{\theta}_1 &=& \langle\theta_1\rangle+\frac{\theta_1}{\sqrt{
    h_2(r_0)}}\,,\qquad\qquad\qquad
  \widetilde{\theta}_2 \,=\, \langle\theta_2\rangle+\frac{\theta_2}{\sqrt{
    h_4(r_0)}}\,,
\end{eqnarray}
which gives the same simple form of the local metric 
\begin{equation}\label{startiib}
  ds^2 \,=\,dr^2+(dz+A\,dx + B\, dy)^2+ (dx^2+d\theta_1^2) 
    +(dy^2+d\theta_2^2)\,,
\end{equation}
where we have with a slight abuse of notation kept the name $A$ and $B$ for the constants, but they are now more generically given by
\begin{equation}\label{defineabtwo}
  A\,=\,\sqrt{\frac{h_1(r_0)}{h_3(r_0)}}\;\cot\langle\theta_1\rangle\,,\qquad
  B\,=\,\sqrt{\frac{h_1(r_0)}{h_5(r_0)}}\;\cot\langle\theta_2\rangle\,.
\end{equation}
Apart from this redefinition of $A$ and $B$, the mirror symmetry analysis will be completely unchanged from section \ref{mirror}. The mirror is then given by \eqref{nkdef}, which is the local limit of D6--branes on the deformed metric. We will show shortly the consistency of this construction with an orientifold in IIA.
It would, of course, be interesting to find a Ricci--flat K\"ahler metric on this Calabi--Yau fourfold and the global supersymmetric type IIB solution with it, but this is beyond the scope of this thesis. We will therefore restrict ourselves to the local limit henceforth.

We do, however, need some global information, in particular the orientation of the D7 and D5--branes on the base $B$.
The choice is up to us. We have to specify over which directions the F--theory fibration degenerates, that determines the position of the orientifold planes and the D7--branes with them. We will consider the version discussed in \ref{introf}, where the fiber degenerates over a two--dimensional subspace\footnote{See \cite{gttwo} for explanations why a degeneration over 0, 4 and 6--cycles actually does not work.}, giving rise to four fixed points with one orientifold plane each and there are four D7--branes on top of each O7--plane. This translates into a constant complex structure $\tau$ on the fiber which also implies a constant dilaton $\phi_B$ in type IIB.

We need our IIB background to be invariant under this orientifold action, which is given by $\Omega\,(-1)^{F_L}\,I_{ij}$.
Since the IIB background is invariant under $ \Omega\,(-1)^{F_L}$, we require the metric to be invariant under spacetime parity $I_{ij}$ of the two coordinates $x_i$ and $x_j$ over which the fiber degenerates. There are many choices for $x_i$ and $x_j$, but recall from \eqref{tdualori} that under T--duality this orientifold becomes
\begin{equation}
  \mbox{IIB on}\quad\frac{B}{\{1,\Omega (-1)^{F_L} I_{ij}\}} \,\xrightarrow{T_{xyz}}
    \,\mbox{IIA on}\quad \frac{B'}{\{1,\Omega (-1)^{F_L} I_{ij}\cdot I_{xyz}\}}\,.
\end{equation}
We want the final metric in IIA after 3 T--dualities to resemble some version of a deformed conifold, so the parity operator in IIA is also severely restricted. In other words, we need to choose $x_i$ and $x_j$ such that $I_{ij}$ is a symmetry of the resolved conifold and $I_{ij}\cdot I_{xyz}$ (which one finds after three T--dualities along $x$, $y$ and $z$) is a symmetry of the deformed metric. Of course, the reasoning in IIA is different as $\Omega (-1)^{F_L}$ is not a symmetry of IIA, but we have to make sure that that under $I_{ij}\cdot I_{xyz}$ no components inherent to the deformed conifold metric are projected out. This will become clearer when we discuss our example.

The choice we adapt is that the F--theory torus is fibered non--trivially over the two--torus $(x,\,\theta_1)$. This is actually the only choice that preserves all the symmetries we require \cite{gttwo}. The D5--branes are wrapped on the two--torus (or sphere) given by $(y,\,\theta_2)$ (recall from \eqref{resmetric} that this is the sphere that remains finite as $\widetilde{r}\to 0$). This means that under three T--dualities 
\begin{equation}
  \mbox{IIB on}\quad\frac{B}{\{1,\Omega (-1)^{F_L} I_{x\theta_1}\}} 
    \,\xrightarrow{T_{xyz}}\, \mbox{IIA on}\quad \frac{B'}{\{1,\Omega (-1)^{F_L} 
    I_{yz\theta_1}\}}\,.
\end{equation}
In IIB the D5--branes extend along Minkowski space and $(y,\,\theta_2)$, whereas the D7/O7--system extends along Minkowski space and $(r,\,y,\,z,\,\theta_2)$. After three T--dualities the D7/O7 system has turned into D6/O6 which extend along Minkowski space and $(r,\,x,\,\theta_2)$, whereas the original D5--branes become D6--branes on the three-cycle $(x,\,z,\,\theta_2)$.

Or schematically, in type IIB
\begin{equation}\nonumber
  \begin{array}{lllllllllll}
    D5:\quad & 0&1&2&3&-&-&-&-&y&\theta_2\\
    D7/O7:\quad & 0&1&2&3&r&z&-&-&y&\theta_2\,,
  \end{array}
\end{equation}
which turns after three T--dualities along $x,\,y$ and $z$ into IIA with
\begin{equation}\nonumber
  \begin{array}{lllllllllll}
    D6:\quad & 0&1&2&3&-&z&x&-&-&\theta_2\\
    D6/O6:\quad & 0&1&2&3&r&-&x&-&-&\theta_2\,.
  \end{array}
\end{equation}
It is easy to see that the metric \eqref{startiib} is indeed invariant under $I_{x\theta_1}$ (remember that $A$ contains $\cot\langle \theta_1\rangle$, so it is odd under this parity\footnote{The parity operation acts on the global manifold, so it does not merely send $\theta_1\to -\theta_1$, but acts on the global coordinate $\widetilde{\theta_1}$. Comparison to \eqref{baselocal} shows that this also implies $\langle\theta_1\rangle\to -\langle\theta_1\rangle$.}) and the mirror \eqref{nkdef} will be symmetric under $I_{yz\theta_1}$ after we impose some restrictions on the B--field components (more in the next section).

Note that the D7--branes extend along the non--compact direction $r$. A similar brane configuration on the resolved conifold has been considered by \cite{ouyang}, but it was not constructed from F--theory. It was shown there how strings stretching between D7 and D5--branes (or D6 and D6) give rise to a global symmetry. It is not a gauge symmetry because of the large volume factor associated with the D7--branes extending along the non--compact direction $r$. We will call these D7 or D6 that originate from F--theory ``flavor branes'' to distinguish them from the D5 or D6 that carry the gauge theory.

Before moving on with our duality chain let us comment on the gauge theory that results from this brane setup. As demonstrated in \cite{vafa}, the gauge theory on the D5 or D6 branes gives rise to $\mathcal{N}=1$ SYM in d=4. In IIB there are additionally $4\times4$ D7--branes at four fixed points. Each stack of four D7--branes gives rise to an $SO(8)$ symmetry (not SO(4) because the D7--branes also have a ``mirror image'' on the ``other side'' of the orientifold plane, so there are effectively 8 branes between which the strings can stretch). So the global symmetry in this setup is $SO(8)^4$ which can be broken by Wilson lines to $(SO(4)\times SO(4))^4\simeq SU(2)^{16}$. In IIA there are now eight fixed points of the orientifold action $\Omega\, (-1)^{F_L}\, I_{yz\theta_1}$. Therefore, there are eight O6--planes, each accompanied by two D6--branes for charge cancellation. The symmetry group generated by eight stacks of D6 is therefore $SO(4)^8\simeq SU(2)^{16}$. So in both IIA and IIB we consider a generalization of pure $\mathcal{N}=1$ SYM to a symmetry with flavors in the fundamental representation of $SU(2)^{16}$. If we are far away from the flavor branes (far away from the orientifold points), those flavors will be heavy and integrated out, so that in the low energy limit the effective field theory reduces to that discussed by \cite{vafa}. See \cite{gttwo, ouyang} for more details.

\subsection{Non--K\"ahler Transitions in IIA}\label{iia}

Let us now turn to the ``duality chain''. We will show that there are two non--K\"ahler backgrounds in IIA that resemble deformed and resolved conifold apart from B--field dependent fibrations and are related by a flop in $\mathcal{M}$--theory. In this way they are geometric transition duals, because one could start with the deformed geometry, shrink a three--cycle and blow up a two--cycle in the resolved version of our non--K\"ahler conifold. The only difference is, that our two-- and three--cycle contain non--trivial B--field fibrations. 

From the F--theory setup in the last section we know that every metric for D5--branes on the resolved conifold takes the local form \eqref{startiib}:
\begin{equation}\nonumber
  ds^2 \,=\,dr^2+(dz+A\,dx + B\, dy)^2+ (dx^2+d\theta_1^2) 
    +(dy^2+d\theta_2^2)\,.
\end{equation}
We also have to re--evaluate the assumptions about NS and RR flux. We cannot use the solution from \cite{pandozayas} because it breaks supersymmetry and our background contains additional D7--branes on an orientifold. 

We keep the assumption that there is only electric NS flux. Recall that $B_{NS}$ is odd under the combined symmetry $\Omega (-1)^{F_L}$, see \eqref{actionomega} and \eqref{actionfl}, so it also has to be odd under parity $I_{x\theta_1}$ to be invariant under the orientifold action $\Omega (-1)^{F_L} I_{x\theta_1}$. This means that only flux components with precisely one leg along the directions of the degenerating fiber $(x,\,\theta_1)$ survive. This restricts our ansatz \eqref{bstartiib} to\footnote{This is not the most generic ansatz, since we did not include $b_{rx},\, b_{r\theta_1}$ or any magnetic flux that might still be invariant under the orientifold action. We still trust our ansatz to be generic enough for our purposes. It would be useful to find a supergravity solution that confirms that.}
\begin{equation}\label{bnsstartiib}
  B_{NS}^{IIB}\,=\, b_{x\theta_2}\,dx\wedge d\theta_2 + b_{y\theta_1}\,dy\wedge 
    d\theta_1 + b_{z\theta_1}\,dz\wedge d\theta_1\,.
\end{equation}
The same symmetry arguments apply to RR fields, we therefore
make a generic choice for the RR two--form gauge potential
\begin{eqnarray}\label{crrstartiib}\nonumber
  C_2 &=& c_1\,dx\wedge dz + c_2\,dx\wedge dy + c_3\, dx\wedge d\theta_2\\
  & &  + c_4\,dy\wedge d\theta_1 + c_5\,dz\wedge d\theta_1 + c_6\,d\theta_1\wedge 
    d\theta_2\,,
\end{eqnarray}
where the components $c_i$ as well as $b_{ij}$ are in general allowed to depend on $(r,\,\theta_1,\,\theta_2)$ (to preserve the isometries of the background). The orientifold action also restricts them to be even under $\theta_1\to-\theta_1$. Note that $c_2$ and $c_6$ did not appear in the solution of \cite{pandozayas}, but they are allowed in our orientifold setup.
Since wrapped D5--branes act as fractional branes, there will also be an RR four--form potential. $C_4$ is even under $\Omega (-1)^{F_L}$, so we only allow for components that are even under parity as well
\begin{equation}\label{startcfour}
  C_4\,=\,c_7\,dx\wedge dy\wedge dz \wedge d\theta_1 + c_8 \,dx\wedge dy\wedge
     d\theta_1\wedge d\theta_2 + c_9\,dx\wedge dz\wedge d\theta_1\wedge d\theta_2\,.
\end{equation}
The self duality of its fieldstrength is realized by taking $F_5=(1+*_{10})\,dC_4$.

This is of course a specific toy example. One could furthermore restrict $C_2$ to be along the space transverse to the D5--brane only, i.e. along $(x,\,z,\,\theta_1)$ or allow for components containing $dr$ (the $r$--dependence of the RR fieldstrengths is taken care of by the $r$--dependence of the coefficients $c_i$). We should also note that away from the orientifold point more types of fluxes are allowed. However, as long as we do not know the full supergravity background we can very well demonstrate our calculation with this toy model.

\subsubsection*{The IIA non--K\"ahler background before transition}

The three T--dualities are performed as in section \ref{nowflux} and the result is a special version of \eqref{nkdef}, which becomes under the specific choice of B--field we made
\begin{eqnarray}\label{finaliia}\nonumber
  d\widetilde{s}^2 &=& dr^2 \,+\,\alpha^{-1}\,\left[(dz-b_{z\theta_1}\,d\theta_1)
    -\alpha A\,(dx-b_{x\theta_2}\,d\theta_2)-\alpha B\,(dy-b_{y\theta_1}\,d\theta_1)
    \right]^2\\ \nonumber
  & +& \alpha(1+B^2)\,\left[d\theta_1^2+(dx-b_{x\theta_2}\,d\theta_2)^2\right] \,+\, 
    \alpha (1+A^2)\,\left[d\theta_2^2 +(dy-b_{y\theta_1}\,d\theta_1)^2\right]\\ 
  & + &  2\alpha AB\,\cos \langle z\rangle \,\left[d\theta_1 d\theta_2 \,-\,  
    (dx-b_{x\theta_2}\,d\theta_2)(dy-b_{y\theta_1}\,d\theta_1)\right] \\ \nonumber
  & + &  2\alpha AB\,\sin\langle z\rangle\,\left[(dx-b_{x\theta_2}\,d\theta_2)
    \,d\theta_2 +(dy-b_{y\theta_1}\,d\theta_1)\,d\theta_1 \right]\,.
\end{eqnarray}
Note that this is indeed precisely the correct choice of B--field components that makes \eqref{finaliia} symmetric under $I_{yz\theta_1}$ (with $A$ and $\sin\langle z\rangle$ odd, $B$ and $\cos\langle z\rangle$ even\footnote{As explained in footnote 7 of section \ref{ftheorysetup}, this is to be understood as a parity of the global background.} under the parity  $I_{yz\theta_1}$).

To simplify notation in the following analysis, let us define coordinates (or rather one--forms) that include the B--field dependent fibration
\begin{eqnarray}\nonumber
  d\hat{x} &=& dx-b_{x\theta_2}\,d\theta_2\\ 
  d\hat{y} &=& dy-b_{y\theta_1}\,d\theta_1\\ \nonumber
  d\hat{z} &=& dz-b_{z\theta_1}\,d\theta_1\,.
\end{eqnarray}

The RR fields in the mirror IIA are also found by applying Buscher's rules \eqref{rrtdual}.
The resulting RR one--form which corresponds to the intersecting D6--branes is
\begin{equation}\label{rroneform}
  C_1^{IIA}\,=\, c_1\,d\hat{y} - c_2\,d\hat{z} + c_7\,d\theta_1\,.
\end{equation}
The RR three--form field is found to be
\begin{eqnarray}\label{rrthreeiia}\nonumber
  C_3^{IIA}&=& C_{xy1}\,d\hat{x}\wedge d\hat{y} \wedge d\theta_1 
    + C_{xz1}\,d\hat{x}\wedge d\hat{z} \wedge d\theta_1 + C_{yz1}\,d\hat{y}\wedge 
    d\hat{z} \wedge d\theta_1 \\ 
  & + & C_{yz2}\,d\hat{y}\wedge 
    d\hat{z} \wedge d\theta_2 +C_{y12}\,d\hat{y}\wedge d\theta_1 \wedge d\theta_2
    +C_{z12}\,d\hat{z}\wedge d\theta_1 \wedge d\theta_2
\end{eqnarray}
with components defined as
\begin{eqnarray}\label{rrcomp}\nonumber
  C_{xy1} &=& -c_5+f_1 c_1\,, \qquad\qquad\quad  C_{xz1} \,=\, c_4-f_1c_2\\ 
  C_{yz1} &=& Af_1c_1\,\qquad\qquad\qquad\quad\;\;\, C_{yz2} \,=\, 
    -c_3+f_2(Bc_1-c_2)\\ \nonumber
  C_{y12} &=& c_9+f_2c_7\,,\qquad\qquad\qquad C_{z12} \,=\, c_8+Bf_2c_7\,.
\end{eqnarray}
The appearance of $f_{1,2}$ needs some explanation. Recall that these constants were fixed by the metric to $\pm \sqrt{\alpha/\epsilon}$. The question is if these fields are unphysical because they become infinitely large in the limit $\epsilon \to 0$.  The approach taken in \cite{gtone} was to rescale the metric by a conformal factor $\sqrt{\epsilon}$, such that the B--field (which has an overall factor of $1/\sqrt{\epsilon}\,$) becomes finite. This would on the other hand imply that all components in the RR--fields {\it not} containing $f_{1,2}$ scale with some positive power of $\epsilon$ and vanish in the $\epsilon\to 0$ limit. This is particularly unphysical for the 1--form \eqref{rroneform}, since its absence would indicate the absence of D6--branes. 

Another 
approach taken in \cite{gtone, gttwo} is to make explicit use of the local limit, where $f_{1,2}$ are constant. Note that $f_1$ appears in terms with $d\theta_1$ and $f_2$ in terms with $d\theta_2$. If we define new coordinates
\begin{equation}\label{thetahat}
  d\hat{\theta}_i \,=\, d(f_i\,\theta_i)
\end{equation}
then all terms containing $f_i$ can be absorbed into these new coordinates.
This interpretation is completely consistent if $\alpha$ is treated as a constant. If we wanted to leave local coordinates, we would have to define $d\hat{\theta}_i \,=\, d(F_i\,\theta_i)$ with $\partial_{\theta_i} F_i=f_i$ and restrict $f_i=f_i(\theta_i)$. 

Let it suffice to say that the problem of the unphysicality of some background fields can be cured in the local limit. We will henceforth keep these terms and ask the reader to keep in mind that their divergence for $\epsilon\to 0$ is not severe. In section \ref{iib} we will argue that these terms might actually be ``large complex structure artefacts'' that should vanish when we want to leave the large complex structure limit.

Finally, let us note that there is also a 5--form field
\begin{equation}\label{rrfive}
  C_5^{IIA} \,=\, \big[c_6+c_3\,f_1-(c_4-Bc_5)\,f_2+(c_2-Bc_1)\,f_1f_2\big] 
    \,d\hat{x} \wedge d\hat{y} \wedge d\hat{z}\wedge d\theta_1\wedge d\theta_2
\end{equation}
and furthermore the dilaton $\phi_A$, which gives rise to a string coupling $g_A$
\begin{equation}
  e^{\phi_A}\,=\,g_A \,=\,\frac{g_B}{\sqrt{1-\frac{\epsilon}{\alpha}}}
    \quad \xrightarrow {\epsilon\to 0}\quad g_B\,=\,e^{\phi_B}\,.
\end{equation}
Apparently, the dilaton remains constant under T--duality if we take the limit that $\epsilon\to 0$, so $\phi_A=\phi_B=\phi$. For completeness, let us also quote the B--field, which was already evaluated in \eqref{bnsiia}. It now has a slightly different fibration structure in the coordinates, but remains as
\begin{eqnarray}\label{bnsfinaliia}\nonumber
  \sqrt{\frac{\epsilon}{\alpha}}\,B_{NS}^{IIA} &=& d\hat{x}\wedge d\theta_1 -
    d\hat{y}\wedge d\theta_2 -A\,d\hat{z}\wedge d\theta_1 \\
  & & + B\,\cos\langle z\rangle\, 
    d\hat{z} \wedge d\theta_2 + B\,\sin\langle z\rangle\, 
    d\hat{y} \wedge d\hat{z}\,.
\end{eqnarray}

We have already commented on the properties of this non--K\"ahler manifold in section \ref{discuss}. Let us now focus on finding the background it is dual to. 
Both geometries should be related by a flop transition in $\mathcal{M}$--theory just as in the Calabi--Yau case discussed in \cite{flop}. Let us therefore discuss how this background lifts to $\mathcal{M}$--theory.

\subsubsection*{The M--theory Flop}

In order to not overload this section with details, let us
make a simple choice for the background fields in \eqref{bnsstartiib} and \eqref{crrstartiib}. Let us assume
\begin{equation}\label{ansatz}
  b_{z\theta_1}\,=\,0\,\qquad c_i\,=\,0\quad\mbox{except}\quad c_1\,=c_1(r,\,\theta_1,\,\theta_2)\,.
\end{equation}
This simplifies the RR one--form in the mirror IIA. It will become useful to write \eqref{rroneform} under this assumption as
\begin{equation}\label{oneform}
  C_1\,=\,\Delta_1\,d\hat{x} - \Delta_2\,d\hat{y}\,, 
\end{equation}
where $\Delta_1$ and $\Delta_2$ are not necessarily given by zero and $-c_1$, respectively, if we allow for an extra gauge degree of freedom in the one--form potentials.
As usual in the presence of a gauge field $C_1$ and dilaton $\phi$, type IIA on a manifold $X$ is lifted to $\mathcal{M}$--theory on a twisted circle via
\begin{equation}
  ds_{\mathcal{M}}^2 \,=\, e^{-2\phi/3}\,ds_X^2 + e^{4\phi/3}\,(dx_{11}+C_1)^2
\end{equation}
with $x_{11}$ parameterizing the extra dimension with radius $R$, $x_{11}=0\ldots 2\pi R$. In the limit $R\to 0$ we recover 10--dimensional IIA theory. The gauge fields in our case enter into the metric so that it becomes
\begin{eqnarray}\label{liftbefore}\nonumber
  ds_{\mathcal{M}}^2 &=& e^{-2\phi/3}\,dr^2 + e^{-2\phi/3}\,\alpha^{-1}\, 
    \big(dz-\alpha A\,d\hat{x} -\alpha B\, d\hat{y}\big)^2 
    +  e^{4\phi/3}\,\big(dx_{11}+\Delta_1\,d\hat{x}-\Delta_2\,d\hat{y}\big)^2\\ 
    \nonumber
  & & + e^{-2\phi/3}\,\left[\alpha(1+B^2)\,(d\theta_1^2+d\hat{x}^2) + 
    \alpha(1+A^2)\,(d\theta_2^2+d\hat{y}^2)\right]\\ 
  & & + e^{-2\phi/3}\,2\alpha AB\,\left[\cos\langle z\rangle\, (d\theta_1\, 
    d\theta_2-d\hat{x}\,d\hat{y}) + \sin\langle z\rangle \,
    (d\hat{x}\,d\theta_2 + d\hat{y}\,d\theta_1)\right]\,.
\end{eqnarray}
The two fibration terms in the first line are of special interest. They are very similar in structure, even more so if one introduces new coordinates $\psi_1$ and $\psi_2$ via
\begin{equation}\label{xeleven}
  dz\,=\,d\psi_1-d\psi_2\qquad\mbox{and}\qquad dx_{11}\,=\,d\psi_1+d\psi_2.
\end{equation}
This happens, of course, with some forsight. To explain why this choice is particular convenient to perform the flop, we need to discuss similarities and differences compared to \cite{flop}, which discussed the flop of Vafa's scenario. 

It was argued in \cite{flop} that deformed and resolved conifold both lift to a $G_2$--holonomy manifold with symmetry group $SU(2)\times SU(2)\times U(1)$.
Moreover, it was shown in \cite{cvetlupope}, that there is a whole family of $G_2$--holonomy metrics (that includes the lift of resolved and deformed conifold) of the form\footnote{A similar ansatz was discussed in \cite{brand}, which corresponds to $g_3=1$ and $a^2(1-\xi^2)=b^2$, so there are only four free parameters instead of six. Of course, the requirement of $G_2$ holonomy restricts these parameters, only one of the six is actually free, so that the solutions from \cite{cvetlupope} correspond to a one--parameter family of $G_2$ metrics.}
\begin{eqnarray}\label{gtwoansatz}\nonumber
  ds^2 &=& dr^2 + a^2\,\big[(\Sigma_1+\xi \sigma_1)^2 + (\Sigma_2+\xi \sigma_2)^2
    \big] + b^2(\sigma_1^2+\sigma_2^2)\\
  & &  + c^2 (\Sigma_3-\sigma_3) + f^2 (\Sigma_3+g_3\,\sigma_3)^2
\end{eqnarray}
where $\sigma_i$ and $\Sigma_i$ are two sets of SU(2) left--invariant one--forms, because all these $G_2$--holonomy metrics have $S^3\times S^3$ principal orbits, i.e. $SU(2)\times SU(2)$ symmetry. This is of course inspired by the usual notation for conifold geometries, see appendix \ref{coni}. In terms of Euler angles on the two $S^3$ these left--invariant one--forms are given as\footnote{We use slightly different notation than \cite{cvetlupope}, in particular we use $-\phi_2$ instead of $\phi_2$.}
\begin{equation}\label{leftforms}
  \begin{array}{ll}
    \sigma_1\,=\,\cos\psi_1\,d\theta_1+\sin\psi_1\,\sin\theta_1\,d\phi_1 & 
      \Sigma_1\,=\,\cos\psi_2\,d\theta_2-\sin\psi_2\,\sin\theta_2\,d\phi_2 \\
    \sigma_2\,=\,-\sin\psi_1\,d\theta_1+\cos\psi_1\,\sin\theta_1\,d\phi_1\quad & 
      \Sigma_2\,=\,-\sin\psi_2\,d\theta_2-\cos\psi_2\,\sin\theta_2\,d\phi_2\\
    \sigma_3\,=\,d\psi_1+\cos\theta_1\,d\phi_1 & 
      \Sigma_3\,=\,d\psi_2-\cos\theta_2\,d\phi_2
  \end{array}
\end{equation}
and satisfy $d\sigma_i=-1/2\,\epsilon_{ijk}\,\sigma^j\wedge \sigma^k$ and $d\Sigma_i=-1/2\,\epsilon_{ijk}\,\Sigma^j\wedge \Sigma^k$. 
The metric in these vielbeins is obviously invariant under a left acting $SU(2)\times SU(2)$ and there is a U(1) symmetry generated by the shift symmetries $\psi_1\to\psi_1+k$ and $\psi_2\to\psi_2+k$, which is why $\psi_1-\psi_2$ was identified as the 11th direction in \cite{cvetlupope}. 

The general setup \eqref{gtwoansatz} that we adopted from \cite{cvetlupope} has less symmetry than the metric in \cite{flop} for which the flop was discussed. 
In particular, it allows for a solution that looks like the lift of a deformed conifold, but with two $S^2$ of different size (so it includes not only the Calabi--Yau deformed conifold). This becomes obvious if the Calabi--Yau metrics for resolved and deformed conifold are written in vielbeins \eqref{leftforms} as
\begin{eqnarray}
  ds_{\rm def}^2 &=& A^2\,\sum_{i=1}^2(\sigma_i-\Sigma_i)^2 + B^2\,\sum_{i=1}^2
    (\sigma_i+\Sigma_i)^2 + C^2 \,(\sigma_3-\Sigma_3)^2 +D^2\,dr^2\\
  ds_{\rm res}^2 &=& \widetilde{A}^2\,\sum_{i=1}^2(\sigma_i)^2 + \widetilde{B}^2 
    \,\sum_{i=1}^2(\Sigma_i)^2 + \widetilde{C}^2\,(\sigma_3-\Sigma_3)^2 + 
    \widetilde{D}^2\,dr^2
\end{eqnarray}
with the coefficients $A,B$ etc. determined by K\"ahler and Ricci flatness condition, see \eqref{resmetricapp} and \eqref{defmetricapp}.
This clearly shows that the deformed conifold is completely symmetric under $\mathbb{Z}_2:$ $\sigma_i\leftrightarrow \Sigma_i$, whereas the resolved conifold does not have this symmetry, due to $\widetilde{A}\ne \widetilde{B}$. This is precisely the statement that the two $S^2$ do not have the same size in the resolved geometry, but they do in the deformed. To see this consider
\begin{equation}\nonumber
  \sigma_1^2+\sigma_2^2\,=\,d\theta_1^2+\sin^2\theta_1\,d\phi_1^2\,,\qquad
    \Sigma_1^2+\Sigma_2^2\,=\,d\theta_2^2+\sin^2\theta_2\,d\phi_2^2\,,
\end{equation}
which implies for the metric describing the two $S^2$
\begin{eqnarray}\nonumber
  ds_{\rm def}^2 &=& (A^2+B^2)[\sin^2\theta_1\,d\phi_1^2+d\theta_1^2
    +\sin^2\theta_2\,d\phi_2^2+d\theta_2^2]+\ldots \\ \nonumber
  ds_{\rm res}^2 &=& \widetilde{A}^2\,(\sin^2\theta_1\,d\phi_1^2+d\theta_1^2)
    +\widetilde{B}^2\,(\sin^2\theta_2\,d\phi_2^2+d\theta_2^2)+\ldots
\end{eqnarray}
for deformed and resolved conifold, respectively.
Note that the parameter $\xi$ in \eqref{gtwoansatz} controls the asymmetry between the two $S^2$. On the other hand, the deformed metric has cross--terms $\sigma_i\Sigma_i$ that the resolved conifold does not exhibit. This is the reason why the resolved metric has a U(1) that the deformed does not have.

If deformed and resolved conifold have such different symmetry properties, how can they be reductions of the same $\mathcal{M}$--theory manifold?

The answer to this question as given by \cite{flop} is that a $G_2$--holonomy metric with symmetry $SU(2)\times SU(2)\times U(1)$ can be reduced to six dimensions in two different ways. Topologically, the manifold in question is equivalent to a cone over $S^3\times \widetilde{S}^3$ that has a $U(1)$ fiber on which one can reduce to d=6. One can either reduce on a fiber that belongs to an $S^3$ of vanishing size (this yields a six--dimensional manifold with blown--up $\widetilde{S}^3$, the deformed conifold) or on a fiber that belongs to an $\widetilde{S}^3$ of finite size (this gives a finite size $\widetilde{S}^2$ in six dimensions, the resolved geometry)\footnote{Furthermore modding out by a $\mathbb{Z}_N$ in both cases gives a singularity corresponding to N D6--branes or a non--singular solution with N units of flux, respectively \cite{flop}.}. In other words, both scenarios are related by an exchange of the finite size $\widetilde{S}^3$ with the vanishing $S^3$ which is called a ``flop transition''.

A cone over $S^3\times \widetilde{S}^3$ is given by $\mathbb{R}^+\times S^3\times \widetilde{S}^3$ which is equivalent to $\mathbb{R}^4\times \widetilde{S}^3$. The topology of this manifold can be viewed as \cite{flop}
\begin{equation}
  (u_1^2+u_2^2+u_3^2+u_4^2)-(v_1^2+v_2^2+v_3^2+v_4^2)\,=\,V\,,\qquad\quad\mbox{with}
    \quad u_i,v_i\in \mathbb{R}\,.
\end{equation}
For $V>0$ the blown up $\widetilde{S}^3$ is described by $u_i$ and $v_i$ correspond to $\mathbb{R}^4$. For $V<0$ their roles are exchanged. The flop transition can then be viewed as a sign flip in $V$ or as an exchange of the two $S^3$. Since each $S^3$ is described by a set of SU(2) left invariant one--forms, this amounts to an exchange $\sigma_i\leftrightarrow\Sigma_i$. But note that this also implies that the U(1) fiber along which one reduces to d=6 is contained either in $\sigma_3$ or $\Sigma_3$, i.e. it is given either by $\psi_1$ or $\psi_2$, but not by $x_{11}=\psi_1+\psi_2$ as we would like to define it.

This discussion was for the Calabi--Yau metrics. The ``non--K\"ahler deformed conifold'' we found in section \ref{mirror} does {\it not} have two $S^2$ of same size. We therefore need to use the more general ansatz \eqref{gtwoansatz} from \cite{cvetlupope} and adopt a flop different from the one suggested for the Calabi--Yaus in \cite{flop}.


It was established in \cite{cvetlupope} that the limit $c=0$ of the $G_2$ metric \eqref{gtwoansatz} contains resolved and deformed conifold in different regions of the parameter space\footnote{In particular, \cite{cvetlupope} solved the differential equations for the $r$--dependent coefficients $a,b,c,f,g_3$ and $\xi$ and showed that the resulting K\"ahler form looks like that for the resolved conifold. It was not considered how a flop between resolved and deformed conifold can be performed.}. They chose $x_{11}=\psi_1-\psi_2$, which is close to what we attempt to do. But it becomes obvious from the crossterms
\begin{eqnarray}\nonumber
  \sigma_1 \Sigma_1+\sigma_2\Sigma_2 &=& \cos(\psi_1-\psi_2)\,[d\theta_1
    \,d\theta_2 -\sin\theta_1\sin\theta_2\,d\phi_1\,d\phi_2] \\
  & + & \sin(\psi_1-\psi_2)\,\,[\sin\theta_1\,d\phi_1\,d\theta_2
    + \sin\theta_2\,d\phi_2\,d\theta_1]
\end{eqnarray}
that $\psi_1-\psi_2$ has to be identified with $\psi$ (or $z$ in local coordinates) to produce the typical $\cos\psi$ and $\sin\psi$ terms for a deformed conifold, recall for example \eqref{defmetric}. We therefore choose $z=\psi_1-\psi_2$ and $x_{11}=\psi_1+\psi_2$ in \eqref{xeleven}.

After this excursion into the literature, let us now discuss our IIA background.
Our metric \eqref{liftbefore} does of course not describe $S^3\times S^3$ principal orbits. Recall that our coordinates $x,y,z$ are non--trivially fibered due to the B--field components which entered into the metric.
We can nevertheless adopt the ansatz \eqref{gtwoansatz} with a different definition of vielbeins
\begin{equation}\label{sigmas}
  \begin{array}{ll}
    \sigma_1\,=\,\cos\psi_1\,d\theta_1 +\sin\psi_1\,d\hat{x} &
      \Sigma_1\,=\,\cos\psi_2\,d\theta_2 -\sin\psi_2\,d\hat{y} \\
    \sigma_2\,=\,-\sin\psi_1\,d\theta_1 +\cos\psi_1\,d\hat{x}\quad &
      \Sigma_2\,=\,-\sin\psi_2\,d\theta_2 -\cos\psi_2\,d\hat{y} \\
    \sigma_3\,=\,d\psi_1-\alpha A\,d\hat{x} &
      \Sigma_3\,=\,d\psi_2+\alpha B\,d\hat{y}\,. \\
  \end{array}
\end{equation}
Then we can write \eqref{liftbefore} in terms of these vielbeins as in \eqref{gtwoansatz}. But our metric does not have $G_2$ holonomy. It only possesses a $G_2$ structure, as will be discussed in section \ref{torsion}.

Let us make one last simplifying assumption. Consider the term $(\Sigma_3+g_3\,\sigma_3)$, which becomes in our vielbeins
\begin{equation}\label{simplerroneform}
  (\Sigma_3+g_3\,\sigma_3)\,=\,(d\psi_1+g_3\,d\psi_2-\alpha Ad\hat{x} + g_3\,\alpha Bd\hat{y})\,.
\end{equation}
We would like to match this with the twisted $\mathcal{M}$--theory circle $(dx_{11}+C_1)$.
Since we want to identify $\psi_1+\psi_2$ with $x_{11}$, we also have to identify the terms in the fibration with the one--form gauge field 
\eqref{oneform}. Assume we can use a gauge choice for $C_1$ such that $\Delta_1=-\alpha A$, then we can use the freedom in $g_3$ to bring also the other term in the required form $g_3 \alpha B=-\Delta_2$. Let us therefore assume right from the start that we can set $g_3=1$ and choose the one--form to be
\begin{equation}\label{gaugechoice}
  C_1\,=\,-\alpha A \,d\hat{x}+\alpha B\,d\hat{y}\,.
\end{equation}
Then we can bring our metric \eqref{liftbefore} into the form \eqref{gtwoansatz} using the one--forms \eqref{sigmas}. After a little rearrangement, this takes the form\footnote{Here we have ignored that our metric does not contain $\cos z$ and $\sin z$, but only their expectation values. This can be taken into account by defining $\psi_1-\psi_2=\langle z\rangle + z$ and keeping only the lowest order in this local coordinate, but it does not influence the following statements.}:
\begin{eqnarray}
  ds^2 &=& e^{-2\phi/3}\,dr^2+e^{-2\phi/3}\,\alpha(1+A^2)\,(\Sigma_1^2+\Sigma_2^2) 
    + e^{-2\phi/3}\,\alpha(1+B^2)\,(\sigma_1^2+\sigma_2^2)\\ \nonumber
  & + & 2e^{-2\phi/3}\,\alpha
    AB \,(\sigma_1\Sigma_1+\sigma_2\Sigma_2) + e^{-2\phi/3}\,\alpha^{-1}\,
    (\sigma_3-\Sigma_3)^2 + e^{4\phi/3}\,(\sigma_3+\Sigma_3)^2 \,.
\end{eqnarray}
The identification of parameters with \eqref{gtwoansatz} is as follows:
\begin{equation}
\begin{array}{ll}
  a^2\,=\,e^{-2\phi/3}\,\alpha(1+A^2)\,,\quad & c^2\,\,=\,e^{-2\phi/3}\,\alpha^{-1} \\
  b^2\,=\,e^{-2\phi/3}\,(1+A^2)^{-1}\,, & f^2\,=\,e^{4\phi/3}\,\\
  \xi\; \,=\,AB\,(1+A^2)^{-1}\,, & g_3\,\,=\,1\,,
\end{array}
\end{equation}
the only difference being that we consider the limit $f=0$ as the reduction to ten dimensions instead of $c=0$ as in \cite{cvetlupope}, i.e. we reduce along $x_{11}=\psi_1+\psi_2$.

The flop has to be different from the case considered in \cite{flop}, since we do not want to exchange the role of $\psi_1$ and $\psi_2$, but we want to exchange $x_{11}$ and $z$ as these are the naturally fibered coordinates in \eqref{liftbefore}. Furthermore, we have the asymmetry factor $\xi$, so that our metric does not exhibit the $\mathbb{Z}_2$ symmetry $\sigma_i\leftrightarrow\Sigma_i$ as the lift of the Calabi--Yau deformed conifold does. We define our flop transition by the assumption that after flop a reduction along $x_{11}$ should produce a resolved geometry. This means in particular that the cross terms $\sigma_1\Sigma_1$ and $\sigma_2\Sigma_2$ in \eqref{gtwoansatz} have to vanish. This, together with $x_{11}\leftrightarrow z$, can be achieved by
\begin{eqnarray}\label{flipflop}\nonumber
  \sigma_3-\Sigma_3 &\leftrightarrow&\sigma_3+\Sigma_3\\
  \sigma_i &\to& \Sigma_i\\ \nonumber
  \Sigma_i &\to& \xi\,(\sigma_i-\Sigma_i)\qquad\mbox{with}\quad i=1,2\,.
\end{eqnarray}
This results in the following metric after flop
\begin{eqnarray}\label{liftafter}\nonumber
   ds^2 &=& e^{-2\phi/3}\,dr^2+e^{-2\phi/3}\,\frac{\alpha A^2B^2}{1+A^2}\,\big
    (d\theta_1^2+d\hat{x}^2\big) + e^{-2\phi/3}\,\frac{1}{1+A^2}\,\big
    (d\theta_2^2+d\hat{y}^2\big) \\
  & & + e^{-2\phi/3}\,\alpha^{-1}\,\big(dx_{11}-\alpha A\,d\hat{x}+\alpha B
    \,d\hat{y}\big)^2+e^{4\phi/3}\,\big(dz-\alpha A\,d\hat{x}-\alpha B\,
    d\hat{y}\big)^2\,,
\end{eqnarray}
which can now be reduced along the same $x_{11}$ to the IIA background after transition.

\subsubsection*{The IIA non--K\"ahler background after transition}

Dimensional reduction on the same $x_{11}$ does clearly not give the same metric as before flop. Instead, we find
\begin{eqnarray}\label{iiaafter}\nonumber
   ds^2 &=& dr^2+\alpha A^2B^2(1+A^2)^{-1}\,\big
    (d\theta_1^2+d\hat{x}^2\big) + (1+A^2)^{-1}\,\big
    (d\theta_2^2+d\hat{y}^2\big) \\
  & & +e^{2\phi}\,\big(dz-\alpha A\,d\hat{x}-\alpha B\,
    d\hat{y}\big)^2\,.
\end{eqnarray}
with one--form gauge field
\begin{equation}\label{newoneform}
  \widetilde{C}_1 \,=\, \sqrt{\alpha}\,\big(- A\,d\hat{x}+ B\,d\hat{y}\big)
\end{equation}
where we rescaled $x_{11}$ with $1/\sqrt{\alpha}$. Recall that the coordinates $d\hat{x}, d\hat{y}$ describe B--field dependent circle fibrations over $x,y$, so this manifold is non--K\"ahler in precisely the same spirit as the ``non--K\"ahler deformed conifold'' before flop \eqref{finaliia}. Comparing it to \eqref{startiib} shows that it also possesses the characteristic metric of a resolved conifold (locally). Note that the dilaton is the same as before flop, $\phi_A=\phi_B=\phi=$ const.

To summarize: we claim the metric \eqref{iiaafter}, which we call ``non--K\"ahler resolved conifold'' to be transition dual to the metric \eqref{finaliia}, the ``non--K\"ahler deformed conifold''. The latter one was a manifold with D6--branes wrapping a 3--cycle, whereas the former describes a blown--up 2--cycle with fluxes on it. We have not considered all the fluxes yet.

In particular, $B_{NS}^{IIA}$ from \eqref{bnsfinaliia} lifts to M--theory as a 3--form field $\mathcal{C}=B_{NS}^{IIA}\wedge dx_{11}$. Since we reduce along the same 11th coordinate after flop, this field is reproduced exactly as before and remains a passive spectator.
The RR three--form fields from \eqref{rrthreeiia} lift directly to three--form flux in $\mathcal{M}$--theory, so it remains unchanged under flop as well. The components \eqref{rrcomp} are simplified by our ansatz \eqref{ansatz}, they now amount to
\begin{equation}\label{rriiasimple}
  C_{xy1} \,=\, f_1 c_1\,, \qquad C_{yz1} \,=\, Af_1c_1\,,\qquad C_{yz2} \,=\, 
    Bf_2c_1\,,
\end{equation}
all others vanish. The five--form does not change either\footnote{There is no five form in $\mathcal{M}$--theory, but the IIA five form is dual to a three form that can be lifted to 11d.}, it is still  given by \eqref{rrfive} but it is simplified by the ansatz \eqref{ansatz} to
\begin{equation}\label{rrfiveiiasimple}
  C_5^{IIA} \,=\, -Bc_1\,f_1f_2 
    \,d\hat{x} \wedge d\hat{y} \wedge d\hat{z}\wedge d\theta_1\wedge d\theta_2\,.
\end{equation}
So, in conclusion, all fields except the RR one--form remain unchanged under flop transition. This should of course be expected, since the effect of a geometric transition is to remove the D--branes, but not the fluxes. In fact, one would expect all fluxes to remain unchanged under this transition. The changed one--form is only due to the gauge choice we employed in \eqref{gaugechoice}. This choice was by no means {\it necessary} to perform the flop, but tremendously convenient.

There is another consistency check for our background that involves a relation between NS and RR three--form fieldstrength. 
The fluxes have to satisfy a linearized supergravity equation of motion \cite{ks, oh}
\begin{equation}\label{eom}
  F_3\,=\,*_6\,H_{NS}\,,
\end{equation}
where $F_3=dC_2$ is the RR fieldstrength, $H_{NS}=dB_{NS}$ is the NS fieldstrength.
We started with an ansatz for the B--field $B_{NS}^{IIB}=b_{x\theta_2}\,dx\wedge d\theta_2+b_{y\theta_1}\,dy\wedge d\theta_1$ which is allowed under orientifold action. A particular simple choice would be to allow the coefficients to depend only on $r$, as in \cite{pandozayas}. The corresponding field strength would then be
\begin{equation}
  H^{IIB}_{NS}\,=\,\partial_r b_{x\theta_2}\,dr\wedge dx\wedge d\theta_2  +\partial_r 
    b_{y\theta_1}\,dr\wedge dy\wedge d\theta_1.
\end{equation}
This would be consistent with the following RR fieldstrength 
\begin{eqnarray}\nonumber
  F_3^{IIB} &=& F_{xz2}\,dx\wedge dz\wedge d\theta_2 
    + F_{yz1}\,dy\wedge dz\wedge d\theta_1\\
   &=&*_6 (H_{ry1}\,dr\wedge dy\wedge d\theta_1) 
    + *_6 (H_{rx2} \,dr\wedge dx\wedge d\theta_2)\,,
\end{eqnarray}
where $H_{ijk}$ indicates the corresponding component of $H_{NS}^{IIB}$.
(The precise relation between $F_{xz2}$ and $H_{ry1}$ involves a numerical factor from the Hodge operator on this curved manifold.)
Can this be realized with the simple ansatz \eqref{ansatz}? The answer is yes, if we consider
\begin{equation}
  C_2^{IIB}\,=\,c_1(\theta_2)\,dx\wedge dz\,.
\end{equation}
This will have a fieldstrength $F_3^{IIB}\,=\,\partial_{\theta_2} c_1(\theta_2)\,dx\wedge dz\wedge d\theta_2$, but no $dy\wedge dz\wedge d\theta_1$ component. This means that for the equation of motion \eqref{eom} to be satisfied we also need the $dr\wedge dx\wedge d\theta_2$ component of $H_{NS}^{IIB}$ to vanish, so that
\begin{equation}
  F_3^{IIB}\,=\,\partial_{\theta_2} c_1(\theta_2)\,dx\wedge dz\wedge d\theta_2
    \,=\, *_6 H_{NS}^{IIB} \,=\, *_6\,(\partial_r b_{y\theta_1}\, dr\wedge 
    dy\wedge d\theta_1)\,.
\end{equation}
This can be achieved by letting $b_{x\theta_2}=$ constant. 
Allowing for more RR components than only $c_1$ to be switched on will also allow for more generic B--field components.

One can actually show that the most generic ansatz\footnote{This includes $dr\wedge dx$ or $dr\wedge d\theta_1$ components for $B_{NS}$ and $C_2$.} for IIB 2--form fluxes, that are allowed under orientifold action, will always yield $b_{x\theta_2}=$ constant and $c_3=c_4=c_5=c_6=$ constant. Otherwise we cannot fulfill the equation of motion if all background fields only depend on $r,\theta_1,\theta_2$ and we do not allow for magnetic NS flux. One also finds $c_1=c_1(r,\theta_2)$ and $c_2=c_2(r,\theta_2)$, only $b_{z\theta_1}$ and $b_{y\theta_1}$ can depend on all base coordinates.

One comment is in order: in the discussion above we have always restricted the fluxes to be symmetric under orientifold operation. If we want to consider the full IIB theory with unbroken $\mathcal{N}=2$ supersymmetry, we actually have to move away from the orientifold planes, i.e. we restrict our local coordinates to a patch that does not contain any orientifold point. 
In that case we do not have to follow the restrictions imposed on the fluxes under orientifold symmetry, but the equation of motion \eqref{eom} still restricts the RR fluxes in terms of NS fluxes. This will allow for much more generic fluxes.

One particularly interesting example would be to introduce a IIB RR two--form component
\begin{equation}
  C_2^{IIB}\,=\,c_{10}\,dy\wedge d\theta_2
\end{equation}
which is allowed away from the orientifold point. We still require $c_{10}$ to be independent of $x,y,z$ to preserve the isometries of the background. Under three T--dualities this creates a new term in the IIA three--form
\begin{equation}
  C_3^{IIA}\,=\,c_{10}\,d\hat{x}\wedge d\hat{z}\wedge d\theta_2\,.
\end{equation}
This term is interesting, because it describes flux along the three-cycle on which the branes are wrapped. Its existence implies that we can define a complexified volume of the blown up three--cycle before flop in $\mathcal{M}$--theory as
\begin{equation}
  V\,=\,\sqrt{det G}+i\,\vert c_{10}\vert\,,
\end{equation}
as anticipated in \cite{flop} ($G$ being the metric on the three--cycle). This helps us to avoid the singularity in the flop transition, because even when the three--cycle shrinks to zero, there is still a finite imaginary part in $V$. This means one can smoothly transform from the deformed to the resolved geometry in the $\mathcal{M}$--theory lift. This imaginary part was interpreted as a gauge theory $\theta$--angle in \cite{flop}. That means, it would have to be closed (recall that the $\theta$--angle in supersymmetric field theories is constant). To answer the question if this is possible in our setup we can again consider the IIB equations of motion.

The fieldstrength of this new term would be 
\begin{equation}
  F_3^{IIB}\,=\,\partial_r c_{10}(r,\theta_1,\theta_2)\,dr\wedge dy\wedge 
    d\theta_2 - \partial_{\theta_1} c_{10}(r,\theta_1,\theta_2)\,dy\wedge d\theta_1
    \wedge d\theta_2\,.
\end{equation}
The problem with this term is that the linearized equation of motion \eqref{eom} would imply a $dx\wedge dz\wedge d\theta_1$ or $dr\wedge dx\wedge dz$ term for $H_{NS}$. But this is magnetic flux which we do not allow for. So, the only solution is that the RR three--form fieldstrength in IIB has to vanish as well which can be achieved by setting $c_{10}$ to a constant, this implies $dC_2^{IIB}=0$.
In the mirror IIA this implies indeed that $dC_3^{IIA}=0$, which justifies its interpretation as a gauge theory $\theta$--angle.

In conclusion, we have shown that we can construct a new pair of string theory backgrounds that are non--K\"ahler and deviate from deformed and resolved conifold in a very precise manner: the $T^3$ fibers are twisted by the B--field. They are related by a geometric transition, because their respective lifts to $\mathcal{M}$--theory are related by a flop. We will comment on possible implications for gauge theories in section \ref{outlook}.

This concludes our discussion of the geometric transition in IIA (an analysis of the SU(3) structure is relegated to section \ref{torsion}). We can now ``close the duality chain'' by performing another mirror which takes us back to IIB. We should recover a K\"ahler background similar to the Klebanov--Strassler model \cite{ks}, since we started with a K\"ahler manifold in IIB.

\subsection{K\"ahler Transitions in IIB}\label{iib}

In principle the analysis follows the same steps as laid out when T--dualizing the resolved conifold from IIB to IIA with NS and RR flux in section \ref{iia}. Only now, our starting background is the non--K\"ahler version of the resolved conifold in IIA, the complete background is described in the end of the last section. Again, we can only recover a local (semi--flat) version of the deformed conifold, this time in IIB. We will make the fascinating observation that the same mechanism that converted B--field into metric cross terms and vice versa will now serve to restore $b_{x\theta_2}$ and $b_{y\theta_1}$ as B--field and the metric will be completely free of any B--field dependent fibration. The IIA B--field will of course enter into the metric again, but looking at \eqref{bnsfinaliia} and taking the semi--flat limit $\langle z\rangle=0$ we find
\begin{eqnarray}\nonumber
  \sqrt{\frac{\epsilon}{\alpha}}\,B_{NS}^{IIA} &=& d\hat{x}\wedge d\theta_1 -
    d\hat{y}\wedge d\theta_2 -A\,d\hat{z}\wedge d\theta_1 
    + B\, d\hat{z} \wedge d\theta_2\,.
\end{eqnarray}
The IIA B--field components are entirely given by $A$ and $B$ --- the original metric components. The original IIB B--field components $b_{x\theta_2}$ and $b_{y\theta_1}$ contained in $d\hat{x}=dx-b_{x\theta_2}\,d\theta_2$ and $d\hat{y}=dy-b_{y\theta_1}\,d\theta_1$ will be converted into B--field again and do not contribute to the metric after three T--dualities.

We could take the background \eqref{iiaafter} and RR fluxes \eqref{rriiasimple}, \eqref{rrfiveiiasimple} and T--dualize them to IIB, but it should be obvious that all the steps performed during the flop do not rely on our simplifying assumption \eqref{ansatz}. In particular, we could have taken the full B--field and 3--form fluxes, as they do not participate in the flop transition at all. We could also replace $dz\to d\hat{z}=d\hat{\psi}_1-d\hat{\psi}_2$ anywhere in the $\sigma_i$ and $\Sigma_i$ by introducing $d\hat{\psi}_1=d\psi_1-\frac{1}{2} b_{z\theta_1}\,d\theta_1$ and $d\hat{\psi}_2=d\psi_1+\frac{1}{2} b_{z\theta_1}\,d\theta_1$, then still $d\hat{x}_{11}=d\hat{\psi}_1+d\hat{\psi}_2=d\psi_1+d\psi_2$ and we would reduce on an untwisted fiber. A little harder is the question whether we could use the full RR one--form gauge potential before flop
\begin{equation}\label{fullrrone}
   C_1^{IIA}\,=\, c_1\,d\hat{y} - c_2\,d\hat{z} + c_7\,d\theta_1
\end{equation}
and use a gauge choice and the freedom in the parameter $g_3$ to bring this into the form \eqref{gaugechoice}. 
The term $-c_2 \,d\hat{z} + c_7\,d\theta_1$ would have to be gauged away or absorbed into a redefinition of $dx_{11}$. Locally that should be no problem, but if we want to allow for a coordinate dependence of $c_i$ this might not always be possible\footnote{We can of course always perform a flop transition, no matter what the 1--form gauge field looks like. But if we want to use the analysis based on the SU(2) invariant one--forms, we require this particular choice of $C_1$.}. The only simplifying assumption we will therefore make in this section is
\begin{equation}\label{assumption}
  c_2 (r,\,\theta_1,\,\theta_2)\,=\,c_7 (r,\,\theta_1,\,\theta_2)\,=\,0\,.
\end{equation}

Otherwise we will take the full IIA background after transition as our starting point for the last piece of the duality chain. The metric then reads
\begin{eqnarray}\label{startiia}\nonumber
   ds^2_{IIA} &=& dr^2 +e^{2\phi}\,\big[(dz-b_{z\theta_1}\,d\theta_1)-\alpha A\,
    (dx-b_{x\theta_2}\,d\theta_2)-\alpha B\,(dy-b_{y\theta_1}\,d\theta_1)\big]^2\\
  & & +\frac{\alpha A^2B^2}{1+A^2}\,\big[d\theta_1^2+(dx-b_{x\theta_2}\,d\theta_2)^2 
    \big] + \frac{1}{1+A^2}\,\big [d\theta_2^2+(dy-b_{y\theta_1}\,d\theta_1)^2\big] \,,
\end{eqnarray}
where we have written the fibration structure explicitely as a reminder that the original IIB B--field is contained in this metric. We would like to stress the readers patience with introducing another set of symbols for the metric components giving the spheres:
\begin{equation}\label{candd}
  C\,=\,\frac{\alpha A^2B^2}{1+A^2}\,,\qquad D\,=\,\frac{1}{1+A^2}\qquad\mbox{and} \qquad \alpha_0^{-1}\,=\,CD+\alpha^2\,e^{2\phi}\,(CB^2+DA^2)
\end{equation}
analogous to the definition of $A,\,B$ and $\alpha$ in \eqref{defineab} and \eqref{alpha}. The IIA B--field is given by 
\begin{eqnarray}\nonumber
  B_{NS}^{IIA} &=& f_1\,(dx-b_{x\theta_2}\,d\theta_2)
    \wedge d\theta_1 +f_2\, (dy-b_{y\theta_1}\,d\theta_1)\wedge d\theta_2\\
  & & -f_1\,A\,(dz-b_{z\theta_1}\,d\theta_1)\wedge d\theta_1 -f_2\, B\, 
    (dz-b_{z\theta_1}\, d\theta_1) \wedge d\theta_2\,,
\end{eqnarray}
where we have reversed the coordinate transformation \eqref{yrotation} to obtain a background with isometry in $z$--direction, i.e. we take the semi--flat limit again. Recall that $f_{1,2}$ were found to be $f_1=-f_2=\sqrt{\frac{\alpha}{\epsilon}}$, but we will ``forget'' their infinity for the moment and come back to it later. The RR fields are given by the one--form \eqref{newoneform}
and three--form \eqref{rrthreeiia} whose components under the assumption \eqref{assumption} become
\begin{eqnarray}\label{definee}\nonumber
  C_{xy1} &=& -c_5+f_1 c_1\,, \qquad\qquad  C_{xz1} \,=\, c_4\,,\qquad\qquad
    C_{yz1} \,=\, Af_1c_1\\
  C_{yz2} &=& -c_3+Bf_2c_1\,, \qquad\quad
    C_{y12} \,=\, c_9\,,\qquad \qquad C_{z12} \,=\, c_8
\end{eqnarray}
as well as five--form \eqref{rrfive} which is now given by
\begin{equation}
  C_5^{IIA} \,=\, \big[c_6+c_3\,f_1-(c_4-Bc_5)\,f_2-Bc_1\,f_1f_2\big] 
    \,dx \wedge dy \wedge dz\wedge d\theta_1\wedge d\theta_2\,.
\end{equation}

T--dualizing this background along $x,\,y$ and $z$ again fails to produce the $d\theta_1\, d\theta_2$--crossterm typical for a deformed conifold. We have to use the same ``trick'' as when going from IIB to IIA in section \ref{mirror}, we have to boost the complex structure of the $(\hat{x},\,\theta_1)$ and $(\hat{y},\,\theta_2)$ tori:
\begin{equation}
  d\hat{\chi}_1\,=\,d\hat{x}+i\,d\theta_1\,\to\, 
    d\hat{x}+(i-\tilde{f}_1)\,d\theta_1\,,\qquad
  d\hat{\chi}_2\,=\,d\hat{y}+i\,d\theta_2\,\to\, 
    d\hat{y}+(i-\tilde{f}_2)\,d\theta_2\,.
\end{equation}
Then performing three T--dualities is tedious but nevertheless straight forward. Again, $\tilde{f}_1$ and $\tilde{f}_2$ have to be very large, so we set them to $\tilde{f}_i\,=\,\tilde{\beta_i}/\sqrt{\tilde{\epsilon}}$
and change the $g_{zz}$ component in the metric as $g_{zz}\to g_{zz}-\tilde{\epsilon}$. Taking the limit $\tilde{\epsilon}\to 0$ we recover a metric that has some resemblance with a deformed conifold
\begin{eqnarray}\label{finaliib}\nonumber
  ds_{IIB}^2 &=& dr^2 + \frac{e^{-2\phi}}{\alpha_0 CD}\,\Big[dz+A f_1\,d\theta_1 + 
    Bf_2\,d\theta_2 +\alpha_0\alpha ADe^{2\phi}\,(dx-f_1\,d\theta_1)\\ 
  & & +\alpha_0\alpha BC e^{2\phi}\,(dy-f_2\,d\theta_2)\Big]^2 + 
    \alpha_0\left(D+\alpha^2B^2e^{2\phi}\right)\,(dx-f_1\,d\theta_1)^2\\ \nonumber
  & + & \left(C-\alpha^2 A^2\tilde{\beta}_1^2\right)\,d\theta_1^2 
    +\alpha_0\left(C+\alpha^2A^2e^{2\phi}\right)\,(dy-f_2\,d\theta_2)^2 
    + \left(D-\alpha^2 B^2 \tilde{\beta}_2^2\right)\,d\theta_2^2\\ \nonumber
  & -& \alpha^2\tilde{\beta}_1\tilde{\beta}_2 AB\,d\theta_1\,d\theta_2 - 
    \alpha_0\alpha^2 AB e^{2\phi}\,(dx-f_1\,d\theta_1)(dy-f_2\,d\theta_2)\,.
\end{eqnarray}
A few comments are in order.
\begin{itemize}
  \item We took the limit $\tilde{\epsilon}\to 0$ without taking into account that $f_i$ actually scales like $\epsilon^{-1/2}$. These are two different limits, but interchanging their order or letting $\epsilon=\tilde{\epsilon}$ does not influence the result. 
  \item The metric still has a fibration structure, now in terms of $- f_i\,d\theta_i$. Did we not promise to recover a K\"ahler background? The key is that this fibration over $x$, $y$ and $z$ does not depend on the B--field anymore. What was B--field in IIA actually stems from metric cross terms in IIB. So we reversed the entanglement of metric and $B_{NS}$ that dominated the IIA backgrounds, both before and after transition.
\end{itemize}

The $f_i$--dependent fibration can actually be removed in several ways. We could again take the approach from \cite{gtone, gttwo} and introduce $d\hat{\theta}_i=d(f_i\,\theta_i)$ as in \eqref{thetahat}, which is exact. In the local limit this is justified and all $f_i\,d\theta_i$ terms can be absorbed into the coordinates $x,y$ and $z$.

But we can also argue for the removal of the unwanted $f_i$--terms on more physical grounds.
What we recover in \eqref{finaliib} is naturally the large complex structure and semi--flat limit of the deformed conifold. Not only does it have the $z$--isometry we will break by introducing the cross terms accompanied by $\cos\langle z\rangle$ and $\sin\langle z\rangle$, but it also has a large scaling of the base coordinates $\theta_i$ ($r$ is non--compact to start with). This is different from the situation we encountered in section \ref{nowflux}, where the boost of complex structures on the resolved conifold did not appear explicitely in the deformed mirror metric, but only in the flux (it became part of a large B--field, from where it now comes back to haunt us). The fibration with large $f_i$ we find here is the reverse transformation to the large complex structure boost we did to the IIB background {\it before} transition. In the local limit this is nothing but a coordinate transformation, which we can now reverse to get back to our original coordinates. We can therefore argue that leaving the large complex structure limit is synonym with removing the $f_i$--dependent terms.
The postulated mirror metric (not in the large complex structure limit anymore) is then
\begin{eqnarray}\label{postulatefinaliib}\nonumber
  ds_{IIB}^2 &=& dr^2 + \frac{e^{-2\phi}}{\alpha_0 CD}\,\left[dz+\alpha_0\alpha AD
    e^{2\phi}\,dx+\alpha_0\alpha BC e^{2\phi}\,dy\right]^2\\\nonumber
  & & +\alpha_0\left(D+\alpha^2B^2e^{2\phi}\right)\,dx^2 + \left(C-\alpha^2 A^2
    \tilde{\beta}_1^2\right)\,d\theta_1^2\\
  & & +\alpha_0\left(C+\alpha^2A^2e^{2\phi}\right)\,dy^2 + \left(D-\alpha^2 B^2
    \tilde{\beta}_2^2\right)\,d\theta_2^2\\\nonumber
  & & -2\alpha^2\tilde{\beta}_1\tilde{\beta_2} AB\,d\theta_1\,d\theta_2 - 
    2\alpha_0\alpha^2 AB e^{2\phi}\,dx\,dy\,.
\end{eqnarray}
We would have found the same result if we had adopted the point of view that the IIA B--field was only a large complex structure artefact or if we had gauged it away. Similar remarks hold true for the RR fields that could also be freed from the ``unphysical'' $f_i$--terms.

We then proceed as in section \ref{mirror}. We require the $d\theta_1\,d\theta_2$ term to have the same coefficient as the $dx\,dy$ term, implying
\begin{equation}
  \tilde{\beta}_1\tilde{\beta}_2\,=\,-\alpha_0e^{2\phi}\,.
\end{equation}
We would also like to combine $x$ and $\theta_1$ as well as $y$ and $\theta_2$ into spheres/tori. This will in general not be possible for both. We therefore require it for $y$ and $\theta_2$, since we want to rotate these coordinates to restore the deformed conifold metric. The $(x,\,\theta_1)$ sphere will remain squashed.
Solving for the $dy^2$ and $d\theta_2^2$ coefficients to be equal gives
\begin{equation}\label{betatwo}
  \tilde{\beta}_2\,=\,\frac{\sqrt{D-\alpha_0(C+\alpha^2A^2e^{2\phi})}}{\alpha B}
\end{equation}
which determines $\tilde{\beta}_1$ to be
\begin{equation}\label{betaone}
  \tilde{\beta}_1\,=\,\frac{-\alpha_0\alpha Be^{2\phi}}{\sqrt{D-\alpha_0(C 
    +\alpha^2A^2e^{2\phi})}}\,.
\end{equation}
If we then also use the coordinate rotation \eqref{yrotation}, we obtain as the final IIB metric after transition
\begin{eqnarray}\label{finaliibmetric}
  d\tilde{s}_{IIB}^2 &=& dr^2 + \frac{e^{-2\phi}}{\alpha_0 CD}\,\left[dz
    +\alpha_0\alpha AD e^{2\phi}\,dx+\alpha_0\alpha BC e^{2\phi}\,dy\right]^2\\ 
    \nonumber
  & & +\alpha_0\left(D+\alpha^2B^2e^{2\phi}\right)\,(dx^2 + \zeta\, d\theta_1^2)  
    +\alpha_0\left(C+\alpha^2A^2e^{2\phi}\right)\,(dy^2 + d\theta_2^2)\\\nonumber
  & & + 2\alpha_0\alpha^2 AB e^{2\phi}\,\big[\cos\langle z\rangle 
    (d\theta_1\,d\theta_2-dx\,dy)+\sin\langle z\rangle (dx\,d\theta_2+dy\,d\theta_1) 
    \big]\,,
\end{eqnarray}
where we have introduced the ``squashing factor''
\begin{equation}\label{squash}
  \zeta \,=\,\frac{C-\alpha^2 A^2\tilde{\beta}_1^2}{\alpha_0\,(D+\alpha^2 B^2 
    e^{2\phi})}\,.
\end{equation}

One might ask if we could have obtained a metric that comes closer to the deformed conifold (without the squashing factor). The only step in the calculation that allows for a deviation from the derivation presented here would be the flop \eqref{flipflop}. The way we define this flop determines the coefficients $C$ and $D$ (the size of the two spheres after transition). One could entertain the idea of leaving $D$ as a free parameter and determine its value by requiring $\zeta=1$, i.e. the $(x,\,\theta_1)$ sphere to be unsquashed. This seems like a worthwhile idea, but for the ansatz \eqref{gtwoansatz} there does not exist any flop transition that would allow to fix $D$ independently of $C$. All flops that produce the right metric (i.e. close to a resolved conifold after transition) allow only for an overall factor which would be the same for both spheres, i.e. $C$ and $D$ would both be scaled by the same factor. That does not solve the squashing problem and we therefore adhere to the choice \eqref{flipflop}, which seems most natural (since it sends $\sigma_{1,2} \to \Sigma_{1,2}$).

We therefore find that the final IIB metric after flop \eqref{finaliibmetric} is not quite a deformed conifold due to the asymmetry in the $(x,\,\theta_1)$ sphere/torus. In the local version presented above it is of course K\"ahler (all coefficients are constant), but we cannot make any statement about the global behavior. Remember that we do not have the global metric for our starting background with D7/O7 and D5 branes. Thus, we conclude that the local metric \eqref{startiib} is transition dual to the local metric \eqref{finaliibmetric}. Both of them are K\"ahler, in contrast to the pair in IIA which contained a fibration depending on the B--field of IIB.

After having determined the metric we should now pay attention to the NS and RR fields of this background. 
The B--field splits in two parts, one ``physical'' and one large complex structure artifact:
\begin{eqnarray}
  \widetilde{B}_{NS}^{IIB} &=& b_{y\theta_1}\,dy\wedge d\theta_1 + 
     b_{x\theta_2}\,dx\wedge d\theta_2 + b_{z\theta_1}\,dz\wedge d\theta_1\\ \nonumber
  & & + \tilde{f}_1\,\left[\alpha A\,d\hat{\tilde{z}}\wedge d\theta_1 + \alpha 
    B\,d\hat{\tilde{z}} \wedge d\theta_2 + d\hat{\tilde{x}}\wedge d\theta_1 
    + d\hat{\tilde{y}}\wedge d\theta_2 \right]
\end{eqnarray}
where we encounter precisely the same fibration as we have seen in the metric
\begin{eqnarray}\label{hattildecoord}\nonumber
  d\hat{\tilde{z}} &=& dz+Af_1\,d\theta_1+Bf_2\,d\theta_2\\
  d\hat{\tilde{x}} &=& dx-f_1\,d\theta_1\\ \nonumber
  d\hat{\tilde{y}} &=& dy-f_2\,d\theta_y\,.
\end{eqnarray}
Again we take the point of view that this is simply a large complex structure artifact and should not be present in the IIB background after we leave this limit. Moreover, the whole second line in the last equation scales with this large prefactor, so we will argue that after leaving the large complex structure limit the B--field should read
\begin{equation}\label{finaliibbns}
  \widetilde{B}_{NS}^{IIB} \,=\, b_{y\theta_1}\,dy\wedge d\theta_1 + 
     b_{x\theta_2}\,dx\wedge d\theta_2 + b_{z\theta_1}\,dz\wedge d\theta_1\,.
\end{equation}
In other words, we recover the B--field we started with in \eqref{bnsstartiib}.

We now turn to the RR fields, there are one, three and five--form in IIA. It turns out that there will be no RR zero--form (axion), but only two-- and four--form in the mirror IIB. Omitting terms that scale with $f_i$ we find
\begin{eqnarray}\label{ctwoafter}
  \widetilde{C}_2^{IIB} &=& -\sqrt{\alpha}B\,dx\wedge dz-c_3\,dx\wedge d\theta_2 
    -\sqrt{\alpha}A\,dy\wedge dz-c_4\,dy\wedge d\theta_1\\ \nonumber
  & & -c_5\,dz\wedge d\theta_1-c_6\,d\theta_1\wedge d\theta_2\\
  \widetilde{C}_4^{IIB} &=& -c_8\,dx\wedge dy\wedge d\theta_1\wedge d\theta_2
    -c_9\,dx\wedge dz\wedge d\theta_1\wedge d\theta_2\,.
\end{eqnarray}
As expected, this is very close to the two-- and four--form we started with in IIB before transition, see \eqref{crrstartiib} and \eqref{startcfour}. As already mentioned, we should recover the fluxes that correspond to the D--brane setup from before transition, but the cycle on which the branes were wrapped has shrunk and we have a completely different geometry. The only term that looks out of place in \eqref{ctwoafter} is the $dy\wedge dz$ term. This was not part of our ansatz \eqref{crrstartiib} as it is not invariant under the orientifold action. The reason this term appears is the gauge choice we made in \eqref{gaugechoice}. One can check explicitely that the reverse T--dualities with the original IIA one--form \eqref{fullrrone} do indeed reproduce the fluxes we started with before transition. There is only one minor difference: all RR fields have an overall minus sign. This is not worrysome, since the orientation of the cycles, which enter into the quantization condition for the fluxes, can contribute an overall sign.

Something very peculiar happens to the dilaton. According to T--duality rules \eqref{buscher} it is given by
\begin{equation}
  e^{2\tilde{\phi}}\,=\, \tilde{G}_{zz}^{-1} G_{yy}^{-1} g_{xx}^{-1}\,e^{2\phi}
\end{equation}
where $G$ and $\tilde{G}$ indicate the metric after T--duality along $x$ and $y$, respectively, $g$ and $\phi$ are the metric and dilaton of the starting background in IIA. Plugging in the corresponding values this becomes
\begin{equation}
   e^{2\tilde{\phi}}\,=\,(\alpha_0 CD e^{2\phi})^{-1} (\alpha_0(C+\alpha^2 A^2e^{2\phi})) (C+\alpha^2 A^2e^{2\phi})^{-1}\,e^{2\phi}\,=\, (CD)^{-1}\,.
\end{equation}
Since the IIA dilaton $\phi$ became part of the metric during the flop, it now cancels in this equation. The IIB dilaton is not given by the IIA dilaton anymore. In the local background the final dilaton in IIB is still a constant and its vacuum expectation value could be fixed to zero by fixing the expectation values of $C$ and $D$. But it depends in principle on the size of the two spheres/tori in the IIA background after flop. If we leave the local limit it would not be constant anymore, in contrast to the other IIB and both IIA backgrounds.

It would be interesting to compare the background we found to other known IIB backgrounds with D5--branes on the resolved/singular conifolds that flow to a deformed conifold geometry in the IR. In particular, Klebanov--Strassler \cite{ks} and Maldacena--Nunez \cite{mn} constructed such backgrounds and it was shown by Minasian et al. \cite{grana} that there is a one--parameter family interpolating between these backgrounds. Ours might be a member of that family. We will return to this in section \ref{hetglobal} and use those similarities to postulate a global solution.

In conclusion, we have shown that the ``duality chain'' needs to be modified due to fluxes. The NS field enters into the metric giving rise to non--K\"ahler backgrounds in IIA. We constructed a pair of such backgrounds that we call ``non--K\"ahler deformed'' and ``non--K\"ahler resolved conifold'' and which are related by a flop in $\mathcal{M}$--theory. That we can close the duality chain and return to a K\"ahler background in IIB provides a consistency check of our calculation.

In Vafa's original paper \cite{vafa}, it was already anticipated that fluxes will modify the background. However, the assumption there was that fluxes will only lead to warp factors. Therefore, they calculated topological string amplitudes for the Calabi--Yau backgrounds and considered fluxes to be turned on as a perturbation. This is not the picture we find here. The IIB B--field enters into the IIA metric as a non--trivial fibration. Turning off IIA flux does not remove the impact of the IIB B--field which is now part of the metric rather than flux. We therefore suggest the impact of fluxes to be not as trivial as anticipated in \cite{vafa}. We will comment on possible ways to modify the arguments from \cite{vafa} accordingly in section \ref{outlook} .

\setcounter{equation}{0}
\section{Classification of IIA non--K\"ahler Manifolds}\label{torsion}

In this section we attempt to classify the IIA non--K\"ahler manifolds we constructed in section \ref{iia}. As already mentioned, we do not find a half--flat manifold after performing three T--dualities with fluxes. We will show that this does not contradict supersymmetry requirements. First, it has been
shown that lifting a 10d manifold on a twisted circle (i.e. with
gauge field and dilaton as in our case) can still give a supersymmetric M--theory
background (a $G_2$ holonomy manifold), even if the 10d manifold was not half--flat \cite{minakas}. 
Furthermore, we actually do not expect an 11d manifold with $G_2$ holonomy,
since our M--theory background has flux turned on.
\enlargethispage{\baselineskip}

Let us set the stage and explain how six and seven dimensional manifolds are classified in terms of their intrinsic torsion. This is interesting for string theory, because these manifolds have $SU(3)$ or $G_2$ structure, which occur naturally in string theory compactifications if one requires supersymmetry in d=4 \cite{minas},  \cite{gauntlett}--\cite{behrndt}. See \cite{gstructure, sala} for a rather mathematical exposure to manifolds with $G$--structure, we will in the following rely on \cite{salamon,lust}. 
Then we will illustrate the calculation of torsion classes for the local IIA backgrounds. We will show that with a very generic choice of complex structure we can find a symplectic, but no half--flat structure on these metrics. This local statement can of course not be assumed to hold true for a global background.

\subsection{SU(3) and $G_2$ Structure Manifolds}

We are interested in compactifications that leave some supersymmetry and Poincar\'e invariance unbroken in d=4. The latter requires a 10--dimensional metric of the form
\begin{equation}
  ds_{10}^2 \,=\, e^{A(y)}\,h_{\mu\nu}\,dx^\mu\,dx^\nu + g_{ab}\,dy^a\,dy^b
\end{equation}
with flat Minkowski space parameterized by the noncompact coordinates 
$x^\mu$ ($\mu=0\ldots 3$) and flat metric $h_{\mu\nu}$. The warp factor $e^{A(y)}$ depends only on the internal coordinates $y^a$ ($a=1\ldots 6$). If all background fluxes are set to zero, supersymmetry requires the external space to be flat ($A(y)=0$) and the existence of a covariantly constant spinor $\eta$ on the internal manifold, see e.g. \cite{gsw}. For each such spinor (that makes the supersymmetry variations of the fermions vanish) there is one copy of the minimal supersymmetry algebra in d=4. But a covariantly constant spinor also implies that the internal manifold has SU(3) holonomy, thus it must be a Calabi--Yau.

This strong condition can be relaxed if we allow for non--vanishing vacuum expectation values of the fields, i.e. fluxes \cite{strominger}--\cite{giddings}. The 10--dimensional solution will then be a warped product of Minkowski space and some internal manifold which does no longer possess SU(3) holonomy, but only SU(3) structure.
The 4d supersymmetry
condition of a covariantly constant spinor on the internal space is relaxed to the existence of a globally
defined, nowhere--vanishing spinor that is constant with respect to a {\sl torsional} connection \cite{strominger}, i.e.
\begin{equation}
  \nabla^T\,\eta\,=\,(\nabla+T)\,\eta\,=\,0\,,
\end{equation}
where $\nabla$ is the Levi--Civita connection and $T$ is the torsion.  This reduces the structure group of the 6d manifold from $SO(6)$ (the rotation group) to $SU(3)$. If the torsion vanishes, the manifold has SU(3) {\it holonomy} and is therefore Ricci--flat and K\"ahler, i.e. a Calabi--Yau. These types of compactifications preserve $\mathcal{N}=2$ for type II or $\mathcal{N}=1$ for heterotic and type I theories, in other words they leave 1/4 of the supercharges unbroken.

The existence of an SU(3) invariant spinor is equivalent to the statement that the manifold has SU(3) struture.
But SU(3) structures are also characterized by a 2--form $J$ and a three--form $\Omega$ \cite{hitchin}. Since two--forms are in the adjoint representation ${\bf 15}$ of SO(6) and three--forms in the ${\bf 20}$ of SO(6), they decompose under SU(3) as
\begin{eqnarray}\nonumber
  {\bf 15} &=& {\bf 1+8+3+\bar{3}}\\ \nonumber
  {\bf 20} &=& {\bf 1+1+3+\bar{3}+6+\bar{6}}\,,
\end{eqnarray}
so there is one singlet under SU(3) corresponding to $J$ and two singlets for $\Omega$, which correspond to its real and imaginary parts $\Omega=\Omega_++i\Omega_-$. $J$ and $\Omega$ fulfill the compatibility relations
\begin{equation}
  J\wedge \Omega_+ \,=\, J\wedge \Omega_-=0\qquad\mbox{and}\qquad
  \Omega_+ \wedge \Omega_- \,=\, \frac{2}{3}\,J\wedge J\wedge J\,.
\end{equation}
\enlargethispage{\baselineskip}

The torsion\footnote{We are a bit sloppy here and do not distinguish between contorsion and torsion.} can be viewed as a one--form with values in the Lie--Algebra $\mathfrak{so(6)}$, which decomposes into $\mathfrak{su(3)}$ and its orthogonal complement, $\mathfrak{so(6)}=\mathfrak{su(3)}\oplus\mathfrak{su(3)}^\perp$. Only the $\mathfrak{su(3)}^\perp$ part has a non--trivial action on the SU(3) invariant tensors (or spinors), this is called the {\it intrinsic} torsion with values in {\small $\bigwedge^1$}$\otimes \mathfrak{su(3)}^\perp$ \cite{sala}.
It also decomposes under representations of $SU(3)$. In particular, 
\begin{eqnarray}\nonumber
  \mbox{\small $\bigwedge^1$} \otimes \mathfrak{su(3)}^\perp &=& ({\bf 3\oplus
    \bar{3}})\otimes({\bf 1\oplus 3\oplus \bar{3}})\\
  &=& ({\bf 1\oplus 1})\oplus({\bf 8\oplus 8})\oplus({\bf 6\oplus \bar{6}})
    \oplus({\bf 3\oplus \bar{3}})\oplus({\bf 3\oplus \bar{3}})\,,
\end{eqnarray}
where {\small $\bigwedge^1$}$\sim{\bf 3\oplus \bar{3}}$, $\mathfrak{su(3)}\sim{\bf 8}$ and $\mathfrak{su(3)}^\perp\sim{\bf 1\oplus 3\oplus\bar{3}}$.
This implies that the intrinsic torsion $T_0$ lies in 5 classes \cite{salamon, lust}: $T_0\in{\cal W}_1\oplus{\cal W}_2\oplus{\cal W}_3\oplus{\cal W}_4\oplus{\cal W}_5$, which match precisely the decomposition under SU(3) as given above. Each of these torsion classes can be given by a component of the SU(3) decomposition of $dJ$ and $d\Omega$.

The obstruction for the torsional connection to be the Levi--Civita connection is measured in the
failure of fundamental 2--form and holomorphic 3--form to be closed. Defining a set of real
vielbeins $\{e_i\}$ one can define an almost complex structure as
\begin{eqnarray}\label{complstruc}\nonumber
  E_1 &=& e_1+i\,e_2\\
  E_2 &=& e_3+i\,e_4\\ \nonumber
  E_3 &=& e_5+i\,e_6\,,
\end{eqnarray}
which gives rise to a (1,1)--form w.r.t. this almost complex structure
\begin{equation}
  J\,=\,e_1\wedge e_2+e_3\wedge e_4+e_5\wedge e_6\,.
\end{equation}
Similarly, one defines a holomorphic 3--form w.r.t. this almost complex structure
\begin{equation}
  \Omega \,=\,\Omega_+ + i\,\Omega_-
  \,=\,(e_1+i\,e_2)\wedge(e_3+i\,e_4)\wedge(e_5+i\,e_6)\,,
\end{equation}
where $\Omega_\pm$ are the real and imaginary part of $\Omega$, respectively.
The torsion classes are then determined by the following forms:
\begin{eqnarray}\nonumber
  {\mathcal W}_1 &\leftrightarrow& dJ^{(3,0)} \,,\qquad\qquad\qquad\;\;\;
  {\mathcal W}_2 \,\leftrightarrow\, (d\Omega)^{(2,2)}_0 \\
  {\mathcal W}_3 &\leftrightarrow& (dJ)^{(2,1)}_0 \,,\qquad\qquad\qquad
  {\mathcal W}_4 \,\leftrightarrow\, J\wedge dJ \\ \nonumber
  {\mathcal W}_5 &\leftrightarrow& d\Omega^{(3,1)}\,,
\end{eqnarray}
where the subscript $0$ refers to the primitive part. A $(p,q)$--form $\beta$ is primitive, i.e. $\beta \in \bigwedge^{p,q}_0$ if $\beta\wedge J=0$.
It is immediately obvious that complex manifolds have to have vanishing
${\mathcal W}_1$ and ${\mathcal W}_2$ and K\"ahler manifolds are determined by $T_0\in {\mathcal W}_5$. A symplectic structure has torsion contained in $\mathcal{W}_2$ and $\mathcal{W}_5$.

Decomposing $\Omega=\Omega_+ +i\,\Omega_-$ we can write
more precisely \cite{salamon}
\begin{eqnarray}\nonumber
  d\Omega_\pm \wedge J &=& \Omega_\pm \wedge dJ \,=\, {\mathcal W}_1^\pm
    \, J\wedge J\wedge J\\
  d\Omega_\pm^{(2,2)} &=& {\mathcal W}_1^\pm\,J\wedge J+{\mathcal W}_2^\pm\wedge J \\ 
    \nonumber
  dJ^{(2,1)} &=& (J\wedge{\mathcal W}_4)^{(2,1)}+{\mathcal W}_3\,,
\end{eqnarray}
so ${\mathcal W}_1$ is given by two real numbers, ${\mathcal W}_1={\mathcal W}_1^+ +{\mathcal W}_1^-$,
${\mathcal W}_2$ is a (1,1) form and ${\mathcal W}_3$ is a (2,1) form.
With the definition of the contraction
\begin{equation}
  \contract\,:\,\mbox{\small $\bigwedge$}{}^k \,T^*\otimes \mbox{\small 
    $\bigwedge$}{}^n \,T^* \longrightarrow \,\mbox{\small $\bigwedge$}{}^{n-k} \,T^*
\end{equation}
and the convention $(e_1\wedge e_2)\,\contract\,(e_1\wedge e_2\wedge e_3\wedge e_4)
\,=\,e_3\wedge e_4$ we can define \cite{salamon}
\begin{eqnarray}
   {\mathcal W}_4 &=& \frac{1}{2}\, J \,\contract\, dJ \,,\qquad\qquad\qquad
   {\mathcal W}_5 \,=\, \frac{1}{2}\, \Omega_+ \,\contract\, d\Omega_+\,.
\end{eqnarray}
A half--flat manifold is specified by $T_0\in{\mathcal W}_1^-\oplus{\mathcal W}_2^-
\oplus{\mathcal W}_3$, which follows from $J\wedge dJ=0$ and $d\Omega_+=0$, but $d\Omega_-\ne 0$ (this lead to the terminology ``half--flat''). This implies it can be complex or non--complex. Note that the assignment of $\Omega_-$ and $\Omega_+$ may be switched by simply exchanging real and imaginary parts in the complex vielbeins $E_i$ in \eqref{complstruc}.

Similar statements hold true for M--theory on 7--manifolds, which would require $G_2$
holonomy to preserve 1/4 supersymmetry in d=4 in the absence of flux. Turning on fluxes relaxes this condition to the existence of a globally defined $G_2$ invariant spinor. In terms of torsion classes, the fundamental object now is a $G_2$ singlet which is a nowhere vanishing 3--form $\Phi$ and its failure to be closed and/or co--closed determines the torsion. The relevant structure
group is $G_2$ and the intrinsic torsion  decomposes under this group. This results in four torsion classes for the 7--manifold:
$\tau_0\in{\mathcal X}_1\oplus{\mathcal X}_2\oplus{\mathcal X}_3\oplus{\mathcal X}_4$. They are given by \cite{salamon}
\begin{eqnarray}\nonumber
   d\Phi &=& {\mathcal X}_1\,(*\Phi)+{\mathcal X}_4\wedge\Phi+{\mathcal X}_3 \\
   d(*\Phi) &=& \frac{4}{3}\,{\mathcal X}_4\wedge(*\Phi)+{\mathcal X}_2\wedge\Phi \,.
\end{eqnarray}

In \cite{salamon,hitchin} it was demonstrated how a manifold with $SU(3)$ structure can be lifted to a
$G_2$ holonomy\footnote{The lift of a half--flat manifold, the so--called Iwasawa manifold, to seven--manifolds of $G_2$ holonomy or $SU(3)$ structure as well as to eightdimensional $Spin(7)$ manifolds was discussed in \cite{misra}.}. One defines the $G_2$ invariant 3--form as
\begin{equation}\label{defPhi}
  \Phi\,=\,\Omega_+ + J\wedge e^7
\end{equation}
where $e^7$ parameterizes the 7th direction, such that the resulting 7--manifold is
a (warped) product $M\times I$ with $I\subset \mathbb{R}$. This produces Hitchin's flow equations \cite{hitchin} if the 6--manifold is half--flat. Hitchin concluded that every half--flat manifold can be lifted to a $G_2$ holonomy metric and conversely that if the seven--dimensional manifold has holonomy in $G_2$ then there exists a half--flat structure on the six--dimensional manifold for all $t\in I$. Hitchin's flow equations
\begin{equation}
  dJ\,=\,\frac{\partial\Omega_+}{\partial t}\,,\qquad\qquad\qquad
    d\Omega_-\,=\,-J\wedge\frac{\partial J}{\partial t}
\end{equation}
describe the evolution of $J$ and $\Omega$ with $t$. Of course, on a (trivial) product manifold $M\times I$ a manifold $M$ with SU(3) {\it holonomy} can always be lifted to a $G_2$ {\it holonomy}.

This is not the case if the 7d manifold is non--trivially fibered over the 6d manifold \cite{salamon}. Such a non--trivial fibration occurs naturally if we lift a 6d background with RR one--form to 7d. The fiber is twisted by the gauge field whose field strength enters into the $G_2$ torsion classes. This implies that SU(3) holonomy does not necessarily lead to $G_2$ holonomy when lifting on a twisted fiber.

In contrast, we are more interested in the reverse case discussed in \cite{minakas}. Starting with an $SU(3)$--structure manifold $X$ they constructed a $G_2$--structure manifold $Y$ as a lift over a twisted circle with dilaton $\phi$ and gauge field $A$:
\begin{equation}\label{Ymetric}
  ds_Y^2\,=\,e^{-2\alpha\phi}\,ds_X^2\,+\,e^{2\beta\phi}\,(dz+A)^2\,.
\end{equation}
We will adopt the string frame in which $\alpha=1/3$ and $\beta=2/3$. One now defines the 3--form on the 7--manifold not like in \eqref{defPhi} but rather as
\begin{equation}\label{defPhinew}
  \Phi\,=\,e^{-\phi}\,\Omega_+\,+\,e^{-\frac{2}{3}\phi}\,J\wedge e^7\,.
\end{equation}
This gives straightforward relations between the torsion classes ${\mathcal W}_i$ and
${\mathcal X}_j$ that generally involve the field strength $F=dA$ and the derivative of the dilaton $d\phi$. 
It was shown in \cite{minakas} that requiring $G_2$ holonomy (i.e. $d\Phi=d(*\Phi)=0$ or equivalently ${\mathcal X}_i=0$) leads to the following constraints on the $SU(3)$ torsion classes\footnote{The case without dilaton was already discussed in \cite{salamon}. Then the resulting 6d manifold is still half--flat, or more precisely  half--flat and {\it almost K\"ahler} with torsion $T_0 \in {\mathcal W}_2^+$. They also anticipated that allowing for a circle fibration with non--constant size would turn on ${\mathcal W}_5$.}:
\begin{eqnarray}\label{minastorsion}\nonumber
  {\mathcal W}_1^\pm &=& {\mathcal W}_2^-\,=\, {\mathcal W}_3\,=\, {\mathcal 
    W}_4\,=0\,\\
  {\mathcal W}_2^+ &=& -e^\phi\,F_0^{(1,1)}\,,\qquad\qquad {\mathcal 
    W}_5\,=\,\frac{1}{3}\,d\phi\,.
\end{eqnarray}
Note, that only in the string frame ${\mathcal W}_4=0$, otherwise it is also proportional to $d\phi$.
This shows that the 6--manifold does not need to be K\"ahler (if $F_0^{(1,1)}\ne 0$),
but it does not need to be half-flat either (it still could be if $d\phi=0$).

This short discussion was intended to clarify that half--flat manifolds are not the
only manifolds that can be lifted to a $G_2$ holonomy, resulting in a supersymmetric compactification. One has to be specific about which type of lift is chosen. It is immediately clear that our scenario requires the 7th direction to be a twisted circle, since the IIA background has a gauge field $A$. But since we have also other background fluxes turned on, we obtain a torsional M--theory background after the lift.
Therefore, the manifold we propose in IIA is neither half--flat nor has it torsion restricted to ${\mathcal W}_2^+ \oplus {\mathcal W}_5$. For the local metric constructed in section \ref{iia} we find a symplectic structure, but without knowledge of a full supersymmetric background we cannot make any assertions about the global structure of the manifold.
It is nevertheless instructive to discuss the local type IIA backgrounds as examples.

\subsection{Torsion Classes before Geometric Transition}\label{torsionbefore}

We will only discuss the IIA case, since the local IIB backgrounds are trivially K\"ahler. It will be shown that with a quite generic ansatz for the almost complex structure we can find a symplectic structure on the local metric, but no half--flat structure.

The metric for the ``non--K\"ahler deformed conifold'' was given in \eqref{finaliia}. Since we set $b_{z\theta_1}=0$ throughout the flop analysis, we will also employ this choice here. The metric then reads
\begin{eqnarray}\nonumber
  d\widetilde{s}^2 &=& dr^2 \,+\,\alpha^{-1}\,\left[dz-\alpha A\,(dx-b_{x\theta_2}
    \,d\theta_2)-\alpha B\,(dy-b_{y\theta_1}\,d\theta_1)\right]^2\\ \nonumber
  & +& \alpha(1+B^2)\,\left[d\theta_1^2+(dx-b_{x\theta_2}\,d\theta_2)^2\right] \,+\, 
    \alpha (1+A^2)\,\left[d\theta_2^2 +(dy-b_{y\theta_1}\,d\theta_1)^2\right]\\ 
  & + &  2\alpha AB\,\cos \langle z\rangle \,\left[d\theta_1 d\theta_2 \,-\,  
    (dx-b_{x\theta_2}\,d\theta_2)(dy-b_{y\theta_1}\,d\theta_1)\right] \\ \nonumber
  & + &  2\alpha AB\,\sin\langle z\rangle\,\big[(dx-b_{x\theta_2}\,d\theta_2)
    \,d\theta_2 +(dy-b_{y\theta_1}\,d\theta_1)\,d\theta_1 \big]\,.
\end{eqnarray}
Let us furthermore assume that $b_{x\theta_2}$ and $b_{y\theta_1}$ are functions of $r$ only, we showed in section \ref{iia} that this simple choice can give a consistent background that fulfills the supergravity equations of motion \eqref{eom}.

To define an almost complex structure, let us first choose a set of six real vielbeins that encapsulates this metric. We follow the choice for the Maldacena--Nunez solution \cite{mn}, since it also describes a deformed conifold that has two $S^2$ of different size. The vielbeins for this solution were given in \cite{grana}, we will discuss it in more detail in section \ref{heterotic}. The vielbeins we choose are
\begin{eqnarray}\label{iiavielb} \nonumber
  {\bf e}^1 &=&dr\,,\qquad\qquad\qquad
    {\bf e}^2\,=\,\alpha^{-1/2}\,\big(dz +A(dx-b_{x\theta_1} d\theta_1)
    +B(dy-b_{y\theta_2}d\theta_2)\big) \\ 
  {\bf e}^3 &=& \frac{1}{\sqrt{1+B^2}}\,d\theta_2\,,\qquad\quad
    {\bf e}^5 \,=\, \frac{1}{\sqrt{1+B^2}}\,(dy-b_{y\theta_2}
    d\theta_2) \\ \nonumber
  {\bf e}^4 &=& \sqrt{\alpha(1+B^2)}\left(\sin{\langle z\rangle } (dx-b_{x\theta_1} 
    d\theta_1) + \cos{\langle z\rangle }\, d\theta_1 
    + \frac{AB}{1+A^2}\,d\theta_2\right) \\ \nonumber
  {\bf e}^6 &=& \sqrt{\alpha(1+B^2)}\left(\cos{\langle z\rangle } (dx-b_{x\theta_1} 
    d\theta_1) - \sin{\langle z\rangle }\, d\theta_1 
    - \frac{AB}{1+A^2}\,(dy-b_{y\theta_2}d\theta_2)\right)\,.
\end{eqnarray}
Following the discussion in \cite{papatseyt} we write down the most generic candidate for a fundamental 2--form
\begin{equation}
  J\,=\,\sum_{i<j}\,a_{ij}\, {\bf e}^i\wedge {\bf e}^j\,,
\end{equation}
where the coefficients $a_{ij}$ could in principle depend on all coordinates.
This has to be compatible with an almost complex structure
${\mathcal J}^A_B=\delta^{AC} J_{CB}$, i.e. we require ${\mathcal J}^2=-1$. 
The resulting 12 equations for the 15 coefficients $a_{ij}$ are solved in the appendix of \cite{papatseyt}.
If we furthermore make the assumption that $a_{13}=a_{14}=0$, the almost complex structure takes a particularly simple form \cite{papatseyt}\footnote{This assumption might seem very restrictive, but we also considered permutations of the vielbeins $({\bf e}^3,{\bf e}^4,{\bf e}^5,{\bf e}^6)$. All scenarios have $\mathcal{W}_4=0$ in common and in some $\mathcal{W}_1^-$ or $\mathcal{W}_1^+$ can be zero as well. The only case with more vanishing torsion classes is the symplectic one.}.
The complex vielbeins can be written as
\begin{eqnarray}\label{defcomplvielb} \nonumber
  E^1 &=& {\bf e}^1+i\,{\bf e}^2 \\
  E^2 &=& {\bf e}^3+i\,(X\,{\bf e}^4-P\,{\bf e}^6)\\ \nonumber
  E^3 &=& {\bf e}^5+i\,(X\,{\bf e}^6+P\,{\bf e}^4)\,,
\end{eqnarray}
where $X^2=1-P^2$ and $P=a_{34}$. In the following we will make the simplifying assumption that $P$ (and therefore $X$) is a function of $r$ only.

With this setup, $J$ and $\Omega$ are defined as
\begin{eqnarray} 
  J &=& \frac{i}{2}\,\sum_{i=1}^3 E^i \wedge \overline{E}^i\,,\qquad \qquad
  \Omega \,=\, E^1\wedge E^2\wedge E^3
\end{eqnarray}
and one can calculate $dJ$ and $d\Omega$ and all five torison classes with them. One immediately notices that ${\mathcal W}_4=0$, because
\begin{equation}
  J\wedge dJ\,=\,2\alpha\,\left(P(r)P'(r)+X(r)X'(r)\right)
  \,dr\wedge dx \wedge dy\wedge d\theta_1 \wedge d\theta_2\,,
\end{equation}
which is identically zero because of $P(r)^2+X(r)^2=1$ (the prime denotes derivative w.r.t. $r$). However, the metric does not allow for a half--flat structure, because there is no choice of $P(r)$ that makes either $d\Omega_+=0$ or $d\Omega_-=0$.
We can nevertheless choose $P(r)$ to give a symplectic structure. Consider ${\mathcal W}_1^\pm$ given by $d\Omega_\pm\wedge J={\mathcal W}_1^\pm\,J\wedge J\wedge J $:
\begin{eqnarray}
  d\Omega_+\wedge J &=& \frac{2\sqrt{\alpha}}{\sqrt{1-P(r)^2}}\,P'(r)\,
  dr\wedge dx\wedge dy\wedge dz\wedge d\theta_1\wedge d\theta_2 \\ \nonumber
  d\Omega_-\wedge J &=& -\alpha\,\left(\cos\z \sqrt{1-P(r)^2} + \sin\z P(r)\right)\times \\
  & & \left((1+B^2)b_{y\theta_1}'(r)+(1+A^2)b_{x\theta_2}'(r)\right)\, dr\wedge dx\wedge dy\wedge dz\wedge d\theta_1\wedge d\theta_2\,.
\end{eqnarray}
Note that in our local background the coefficients $A,\,B$ and $\alpha$ are simply constants and we have assumed the IIB $B_{\rm NS}$--field components to have $r$--dependence only. Obviously, ${\mathcal W}_1^+$ vanishes if $P(r)$ is constant. ${\mathcal W}_1^-$ vanishes if
\begin{equation}\label{solP}
  P\,=\,-\cos\z\,=\,\mbox{constant}.
\end{equation}
It turns out, that for this value also ${\mathcal W}_3$ vanishes, and in this case $dJ=0$. Let us stress again, that there is no choice for $P(r)$ that would give ${\mathcal W}_5=0$ or ${\mathcal W}_2^\pm=0$.
The remaining torsion classes could only vanish if the IIB $B_{\rm NS}$ field was constant. In that case we would trivially recover a closed two and three--form, since then all metric components would be constant.
For completeness, let us also give ${\mathcal W}_5$ and ${\mathcal W}_2^\pm$ with the choice \eqref{solP} for $P$:
\begin{eqnarray}\nonumber
  \mathcal{W}_2^+ &=& -\sqrt{\frac{1}{1+B^2}}\,\frac{B b_{y\theta_1}'}{2}\,\big(
    {\bf e}^2\wedge {\bf e}^3-\cos\z\, {\bf e}^1\wedge {\bf e}^6-\sin\z\, {\bf e}^1
    \wedge {\bf e}^4\big)\\ 
  & & -\sqrt{\frac{\alpha}{1+B^2}}\,\frac{AB^2 b_{y\theta_1}'}{2}\,\big(
    {\bf e}^2\wedge {\bf e}^4-\cos\z\, {\bf e}^1\wedge {\bf e}^5+\sin\z\, {\bf e}^1
    \wedge {\bf e}^3\big)\\ \nonumber
  & & -\sqrt{\alpha(1+B^2)}\,\frac{A b_{x\theta_2}'}{2}\,\big(
    {\bf e}^1\wedge {\bf e}^5-\cos\z\, {\bf e}^2\wedge {\bf e}^4+\sin\z\, {\bf e}^2
    \wedge {\bf e}^6\big)\\ \nonumber
  & & -\frac{AB b_{y\theta_1}'}{1+B^2}\,\sin\z\,\big[
    \cos\z\, ({\bf e}^3\wedge {\bf e}^4-{\bf e}^5\wedge {\bf e}^6)+\sin\z\, ({\bf e}^4
    \wedge {\bf e}^5-{\bf e}^3\wedge {\bf e}^6)\big]\\ \nonumber
  & & -\left(\frac{b_{y\theta_1}'}{\sqrt{\alpha}(1+B^2)}-\sqrt{\alpha}(1+B^2)
    b_{x\theta_2}'\right)\,({\bf e}^3\wedge {\bf e}^5+{\bf e}^4\wedge {\bf e}^6) 
\end{eqnarray}
\begin{eqnarray}\nonumber
  \mathcal{W}_2^- &=& -\sqrt{\frac{1}{1+B^2}}\,\frac{B b_{y\theta_1}'}{2}\,\big(
    {\bf e}^1\wedge {\bf e}^3+\cos\z\, {\bf e}^2\wedge {\bf e}^6+\sin\z\, {\bf e}^2
    \wedge {\bf e}^4\big)\\ 
  & & -\sqrt{\frac{\alpha}{1+B^2}}\,\frac{AB^2 b_{y\theta_1}'}{2}\,\big(
    {\bf e}^1\wedge {\bf e}^4+\cos\z\, {\bf e}^2\wedge {\bf e}^5-\sin\z\, {\bf e}^2
    \wedge {\bf e}^3\big)\\ \nonumber
  & & +\sqrt{\alpha(1+B^2)}\,\frac{A b_{x\theta_2}'}{2}\,\big(
    {\bf e}^2\wedge {\bf e}^5+\cos\z\, {\bf e}^1\wedge {\bf e}^4-\sin\z\, {\bf e}^1
    \wedge {\bf e}^6\big)\\ \nonumber
  & & +\frac{2AB b_{y\theta_1}'}{1+B^2}\,\sin\z\,\big(
    {\bf e}^3\wedge {\bf e}^5+{\bf e}^4\wedge {\bf e}^6\big)\,
    -\, \frac{(1+\alpha A^2B^2\cos 2\z)b_{y\theta_1}'+\alpha(1+B^2)b_{x\theta_2}'}
    {2\sqrt{\alpha}(1+B^2)}\times \\ \nonumber
  & & \quad\big[\cos\z\, ({\bf e}^3\wedge {\bf e}^4-{\bf e}^5\wedge {\bf e}^6)
    +\sin\z\,({\bf e}^4\wedge {\bf e}^5-{\bf e}^3\wedge {\bf e}^6)\big]
    \\[2ex] \nonumber
  \mathcal{W}_5 &=& \frac{1}{2}\,\sqrt{\alpha}\,\left((1+A^2)b_{y\theta_1}'+(1+B^2)
    b_{x\theta_2}'\right)\,{\bf e}^2 + \frac{1}{2}\,\sqrt{\alpha(1+B^2)}\,A 
    b_{x\theta_2}'\,{\bf e}^4\\
  & & +\frac{1}{2}\,\frac{Bb_{y\theta_1}'}{1+B^2}\,{\bf e}^5 
    -\frac{1}{2}\,\sqrt{\frac{\alpha}{1+B^2}}\,AB^2 b_{y\theta_1}'\,\left(
    \cos\z\,{\bf e}^4+\sin\z\,{\bf e}^6\right)\,.
\end{eqnarray}
The vielbeins ${\bf e}_i$ are defined in \eqref{iiavielb}. This completely specifies a symplectic structure on the ``non--K\"ahler deformed conifold''.

\subsection{Torsion Classes after Geometric Transition}

Very similar remarks hold true for the local IIA metric after transition which we termed ``non--K\"ahler resolved conifold''.
We can find a symplectic structure, but ${\mathcal W}_2^\pm$ and ${\mathcal W}_5$ are nonzero. 
The metric after transition was obtained in \eqref{iiaafter} to be:
\begin{eqnarray}
   ds^2 &=& dr^2 + e^{2\phi}\,\big(dz-\alpha A\,(dx-b_{x\theta_2}\,d\theta_2)
    -\alpha B\,(dy-b_{y\theta_1}\,d\theta_2)\big)^2\\ \nonumber
  & + & \alpha A^2B^2(1+A^2)^{-1}\,\big (d\theta_1^2+(dx-b_{x\theta_2}\,
    d\theta_2)^2\big) 
    + (1+A^2)^{-1}\,\big(d\theta_2^2+(dy-b_{y\theta_1}\,d\theta_1)^2\big)\,.
\end{eqnarray}
We take again the ansatz \eqref{defcomplvielb} for the complex structure but now with real vielbeins
\begin{eqnarray}\nonumber
  {\bf e}^1 &=& dr\,,\qquad\qquad\qquad{\bf e}^2\,=\,e^\phi\,\big(dz
    -\alpha A(dx-b_{x\theta_2} d\theta_2)-\alpha B(dy-b_{y\theta_1}d\theta_1)\big) \\
  {\bf e}^3 &=& \frac{1}{\sqrt{1+A^2}}\,d\theta_2\,, \qquad\quad
    {\bf e}^5\,=\,\frac{1}{\sqrt{1+A^2}}\,(dy-b_{y\theta_1}d\theta_1) \\ \nonumber
  {\bf e}^4 &=&\sqrt{\frac{\alpha}{1+A^2}}\,AB\,\left(\sin\z\, (dx-b_{x\theta_2} 
    d\theta_2) + \cos\z\,d\theta_1\right) \\ \nonumber
  {\bf e}^6 &=& \sqrt{\frac{\alpha}{1+A^2}}\,AB\,\left(-\cos\z\, (dx-b_{x\theta_2} 
    d\theta_2) - \sin\z\,  d\theta_1 \right)\,.
\end{eqnarray}
This different choice of vielbeins is of course inspired by the resolved conifold \cite{papatseyt}. One also finds ${\mathcal W}_4=0$ automatically. Again, ${\mathcal W}_1^+$ can only vanish if $P(r)$ is constant and solving ${\mathcal W}_1^-=0$ has the same solution $P(r)=-\cos\z$. There is no choice of $P(r)$ that would allow for ${\mathcal W}_5=0$ or ${\mathcal W}_2^\pm=0$. With the choice $P=-\cos\z$ the remaining torsion classes are\pagebreak
\begin{eqnarray}\nonumber
  \mathcal{W}_2^+ &=& -\frac{1}{2}\,\sqrt{\alpha(1+A^2)}\,e^\phi\,b_{y\theta_1}'\,
    \left({\bf e}^2\wedge {\bf e}^3-\cos\z\, {\bf e}^1\wedge {\bf e}^6-\sin\z\, 
    {\bf e}^1 \wedge {\bf e}^4\right)\\
  & & -\frac{1}{2}\,\alpha\sqrt{1+A^2}\,Ae^\phi\,b_{x\theta_2}'\,\big(
    {\bf e}^1\wedge {\bf e}^5-\cos\z\, {\bf e}^2\wedge {\bf e}^4+\sin\z\, {\bf e}^2
    \wedge {\bf e}^6\big)\\ \nonumber
  & & -\frac{b_{y\theta_1}'-\alpha A^2B^2\,b_{x\theta_2}'}{2\sqrt{\alpha}\,AB}\,
    \left({\bf e}^3\wedge {\bf e}^5+{\bf e}^4\wedge {\bf e}^6\right)
    \\[2ex]\nonumber
  \mathcal{W}_2^- &=& -\frac{1}{2}\,\sqrt{\alpha(1+A^2)}\,e^\phi\,b_{y\theta_1}'\,
    \left({\bf e}^1\wedge {\bf e}^3+\cos\z\, {\bf e}^2\wedge {\bf e}^6+\sin\z\, 
    {\bf e}^2 \wedge {\bf e}^4\right)\\
  & & +\frac{1}{2}\,\alpha\sqrt{1+A^2}\,Ae^\phi\,b_{x\theta_2}'\,\big(
    {\bf e}^2\wedge {\bf e}^5+\cos\z\, {\bf e}^1\wedge {\bf e}^4-\sin\z\, {\bf e}^1
    \wedge {\bf e}^6\big)\\ \nonumber
   & & -\frac{b_{y\theta_1}'+\alpha A^2B^2\,b_{x\theta_2}'}{2\sqrt{\alpha}\,AB}\,
    \left[\cos\z\,({\bf e}^3\wedge {\bf e}^4-{\bf e}^5\wedge {\bf e}^6)-
    \sin\z\,({\bf e}^3\wedge {\bf e}^6-{\bf e}^4\wedge {\bf e}^5)\right]
    \\[2ex]
  \mathcal{W}_5 &=& \frac{\sqrt{\alpha(1+A^2)}}{2A}\,e^\phi\,b_{y\theta_1}'\;{\bf e}^5
    +\frac{1}{2}\,\alpha\sqrt{1+A^2}\,Ae^\phi\,b_{x\theta_2}'\;{\bf e}^4
    +\frac{b_{y\theta_1}'+\alpha A^2B^2\,b_{x\theta_2}'}{2\sqrt{\alpha}\,AB}\;
    {\bf e}^2\,,
\end{eqnarray}
where $\phi$ is the IIA dilaton which we found to be exactly the same as the
IIB dilaton before transition and constant.

We see that the geometric transition maps the torsion classes $W_2^{\pm}$ and
$W_5$ into themselves. This can be translated into a statement about $G_2$ torsion classes, using the definition of the three--form \eqref{defPhinew}. So, also the $G_2$ torsion classes $\mathcal{X}_i$ are mapped into themselves. But we know that
the flop just replaces the usual $x^{11}$ direction with the $z$--fibration.
These two circles are used to lift $SU(3)$ torsion classes to $G_2$ torsion classes and this implies that the $G_2$ torsion classes should not change during the flop.

In conclusion, we have argued that on grounds of supersymmetry we do not expect a half--flat manifold. Our lift includes a constant dilaton, one might therefore expect the torsion classes \eqref{minastorsion} to reduce to $\mathcal{W}_2^+\ne 0$, leading to a half--flat structure. But we also lift other RR fluxes to $G$--fluxes in $\mathcal{M}$--theory, which means that supersymmetry does not require $G_2$ holonomy on the 7d manifold.
Apart from that, we only have a local metric which does not show supersymmetry (all components and warp factors are approximated by constants). We could, however, find a symplectic structure on this local background which is in agreement with arguments from section \ref{discuss}.

\setcounter{equation}{0}
\section{Geometric Transitions in Type I and Heterotic}\label{heterotic}

\begin{figure}[ht]
\begin{center}
  \begin{picture}(300,200)\thicklines
    \put(10,0){\framebox(80,40){\begin{minipage}{75pt}
      Heterotic SO(32) NS5 on non--K\"ahler
      \end{minipage}}}\thinlines
    \put(91,20){\line(1,0){128}}\thicklines
    \put(220,0){\framebox(80,40){\begin{minipage}{75pt}
      Heterotic SO(32) fluxes\\ on non--K\"ahler
      \end{minipage}}}
    \put(137,18){\begin{minipage}{55pt} geometric transition? \end{minipage}}
    \put(50,69){\vector(0,-1){28}}
    \put(58,48){\begin{minipage}{50pt} S--duality \end{minipage}}
    \put(10,70){\framebox(80,40){\begin{minipage}{75pt}
      Type I D5 on\\ non--K\"ahler
      \end{minipage}}}
    \put(50,139){\vector(0,-1){28}}
    \put(58,122){\begin{minipage}{50pt} $T^2$ \end{minipage}}
    \put(10,140){\framebox(80,40){\begin{minipage}{75pt}
      IIB D5--branes on orientifold of resolved
      \end{minipage}}}\thinlines
    \put(91,160){\line(1,0){128}}\thicklines
    \put(137,158){\begin{minipage}{55pt} geometric transition \end{minipage}}
    \put(220,140){\framebox(80,40){\begin{minipage}{75pt}
      IIB fluxes on\\ orientifold of\\ deformed
    \end{minipage}}}
    \put(260,139){\vector(0,-1){28}}
    \put(268,122){\begin{minipage}{50pt} $T^2$ \end{minipage}}
    \put(220,70){\framebox(80,40){\begin{minipage}{75pt}
      Type I fluxes on non--K\"ahler
      \end{minipage}}}\thinlines
    \put(91,90){\line(1,0){128}}\thicklines
    \put(137,88){\begin{minipage}{55pt} geometric transition? \end{minipage}}
    \put(260,69){\vector(0,-1){28}}
    \put(268,48){\begin{minipage}{50pt} S--duality \end{minipage}}
  \end{picture}
  \caption{The heterotic duality chain. Following the arrows we can construct non--K\"ahler backgrounds in type I and heterotic theory that are dual to the type IIB backgrounds before and after transition. This implies that also the new backgrounds are in a sense transition duals.} \label{hetenchain}
\end{center}
\end{figure}
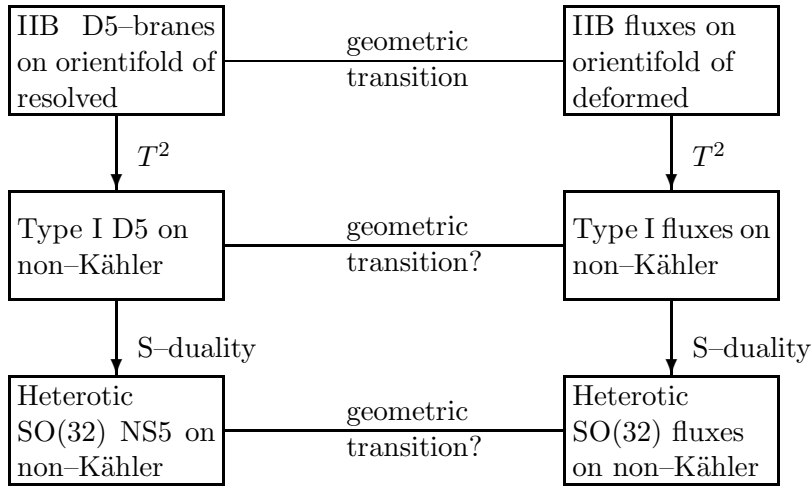

The same mechanism as that discussed in section \ref{chain} can be used to go beyond Vafa's duality chain and construct new transition dual backgrounds in type I and heterotic theory. The F--theory setup takes us naturally to the orientifold corner of type IIB which is basically type I. We only need to perform 2 T--dualities that convert the D7/O7 system into spacetime filling D9/O9. This gives rise to open and closed unoriented strings --- type I. From there we can perform another S--duality and obtain heterotic backgrounds, see figure \ref{hetenchain}. These new backgrounds will also be non--K\"ahler, since the B--field enters into the metric when we T--dualize from the IIB orientifold to type I similar to the analysis in section \ref{chain}.

Following these dualities on both sides of the geometric transition will give us backgrounds that are connected to a flop in $\mathcal{M}$--theory (as performed in section \ref{iia}) via a very long duality chain. Therefore, we claim these backgrounds are also transition duals. Supergravity equations of motion and the torsional constraint \cite{strominger} for heterotic strings pose severe restrictions on the allowed type of fluxes. We provide a toy example that is consistent with the IIB orientifold action, the IIB linearized supergravity equation of motion and the torsional relation in the U--dual\footnote{U--duality is the combined action of T-- and S--duality.} heterotic background.

We can also exploit the fact that the local heterotic metric we find after transition has a similar structure as the solution constructed by Maldacena--Nunez \cite{mn}. This enables us, for the first time in this work, to leave the local limit and propose a {\it global} solution in heterotic theory that is consistent with our IIB orientifold setup. For the construction of vector bundles on heterotic and type I backgrounds we refer the reader to \cite{gttwo}, where also their behavior under geometric transition has been studied.

\subsection[Another F--Theory Setup]{Another F--Theory Setup and IIB Orientifold}

Many of the considerations in this section are very similar to the F--theory setup constructed in section \ref{ftheorysetup}, but it should be immediately clear that we cannot use the same four--fold. What we constructed in section \ref{ftheorysetup} was an elliptic fibration over a resolved conifold base with the torus fibers degenerating over $(x,\,\theta_1)$, which means the D7/O7 system extends along $(r,y,\,z,\,\theta_2)$. 
To convert this into a spacetime filling D9/O9 system, we would have to T--dualize along $(x, \theta_1)$, but $\theta_1$ does not correspond to an isometry of conifold geometries. We therefore need a different orientation of the F--theory torus. 

First we need to define the two directions along which we want to T--dualize.
The logical candidates are among the directions $(x,y,z)$, for the same reason we chose them in section \ref{chain}: they are the isometry direction of the resolved conifold. But $z$ is not an isometry direction of the manifold after transition (being a deformed conifold). So we would like to avoid T--duality along $z$
and will T--dualize along $x$ and $y$. We then need the D7/O7 to extend orthogonal to the T--duality directions, otherwise they do not lead us to type I.

In summary, we start again with a fourfold that is elliptically fibered over the resolved conifold base, but the fiber degenerates over $(x,y)$. This has of course consequences for the IIB orientifold. We now consider
\begin{equation}
  \mbox{IIB on}\quad\frac{B}{\{1,\Omega (-1)^{F_L} I_{xy}\}} \,,
\end{equation}
where $B$ is the base that looks locally like a resolved conifold.
This means, the branes are oriented as follows:
\begin{equation}\nonumber
  \begin{array}{lllllllllll}
    D5:\quad & 0&1&2&3&-&-&-&-&y&\theta_2\\
    D7/O7:\quad & 0&1&2&3&r&z&-&\theta_1&-&\theta_2\,.
  \end{array}
\end{equation}
After T--duality along $x$ and $y$ this turns into
\begin{equation}\nonumber
  \begin{array}{lllllllllll}
    D5:\quad & 0&1&2&3&-&-&x&-&-&\theta_2\\
    D9/O9:\quad & 0&1&2&3&r&z&x&\theta_1&y&\theta_2\,,
  \end{array}
\end{equation}
which is consistent with a type I scenario.

There is a slight problem with this orientifold. The resolved conifold metric is not invariant under $I_{xy}$! Therefore, we have to project out certain components of the metric. Recalling the local metric of the resolved conifold base \eqref{startiib}
\begin{equation}\label{startresol}\nonumber
  ds^2 \,=\,dr^2+(dz+A\,dx + B\, dy)^2+ (dx^2+d\theta_1^2) 
    +(dy^2+d\theta_2^2)
\end{equation}
we see that the $dx\,dz$ and $dy\,dz$ cross terms spoil the invariance under $I_{xy}$. To eliminate them we have to ``untwist'' the $z$--fiber. 
However, the orientifold action does not require us to eliminate all terms from the $z$--fibration, we can keep those that are invariant under $I_{xy}$, like $dx^2$ for example. We therefore make the generic ansatz
\begin{equation}\label{oriansatz}
  ds^2_{IIB}\,=\,dr^2+dz^2+d_1\,|dz_1|^2+d_2\,|dz_2|^2
\end{equation}
with the two tori defined as (note that these are different tori than the ones in \eqref{deftori})
\begin{equation}
  dz_1\,=\,dx+\tau_1\,dy\,,\qquad\quad dz_2\,=\,d\theta_1+\tau_2\,d\theta_2\,.
\end{equation}
This construction in terms of tori is especially convenient since it will preserve  supersymmetry (such toroidal orbifold models have already been considered in \cite{daskat}). We see that when we define (recall that $\alpha^{-1}=1+A^2+B^2$)
\begin{equation}
  d_1\,=\,(1+A)^2\,,\quad \tau_1\,=\,\frac{1}{1+A^2}\,\left(AB+i\,\alpha^{-1/2}\right)
    \,,\qquad d_2\,=\,1\,,\quad \tau_2\,=\,i 
\end{equation}
we obtain a metric that is precisely \eqref{startresol} without the unwanted cross-terms
\begin{equation}\label{resolori}
  ds^2\,=\,dr^2+dz^2+(1+A^2)\,dx^2+(1+B)^2\,dy^2+2AB\,dx\,dy+d\theta_1^2
    +d\theta_2^2\,.
\end{equation}
We could generate a larger class of metrics that are related to this orientifold version of the resolved conifold by allowing more generic complex structures on the tori\footnote{Supersymmetry would then have to be restored by an appropriate choice of fluxes.}. The only choice we have to require for all of them is
\begin{equation}
  {\rm Re}\,\tau_2\,=\,0
\end{equation}
because the resolved conifold does not have any $d\theta_1\,d\theta_2$ cross term and we want our starting background before transition to be ``close'' to a resolved conifold. This will enable us to argue for the existence of a contractible 2--cycle. We could, however, also have taken the point of view that ``untwisting'' the $z$--fiber should remove all crossterms, also the $dx\,dy$ term that comes from the $z$--fibration. This can be achieved by setting ${\rm Re}\,\tau_1\,=\,0$ and we will also allow for this case, but keep in mind that $\tau_1$ can in principle have both real and imaginary part.

Note that the setup we chose is again a model with four O7--planes each with six D7--branes on top and we have a constant complex structure on the F--theory torus, or in other words a constant axion--dilaton in IIB. Not only is it constant, but actually zero, because D7 and O7 charges cancel exactly, so we set as in \cite{realm}
\begin{equation}
  \chi^{IIB}\,=\,0\,,\qquad\phi^{IIB}\,=\,0\,.
\end{equation}

Adding D5--branes to this background will simply act as a warp factor in IIB \cite{giddings}, i.e. a harmonic function $H(r)$. Since we work in the local limit anyway, we can absorb this into the coordinate differentials as we did in section \ref{chain}.

At the orientifold point we can also make an ansatz for the B--field, which is invariant under $\Omega(-1)^{F_L}\,I_{xy}$ if all its components have precisely one leg along the T--duality directions. We did not allow for any magnetic NS flux when we constructed the background in \ref{ftheorysetup}, so let us keep the assumption that there are no $dx\,dz$ or $dy\,dz$ components\footnote{The analysis of this section would not be influenced by allowing $dx\,dz$ or $dy\,dz$ components in the B--field. We would simply acquire also a $z$--dependent twisting of the T--duality fibers $x$ and $y$.}. Our ansatz will therefore be
\begin{equation}\label{oribns}
  B_{NS}^{IIB}\,=\,b_{xi}\,dx\wedge d\theta_i + b_{yj}\,dy\wedge d\theta_j\,,
\end{equation}
where the coefficients can now depend on $(r,z,\theta_1,\theta_2)$, since we do not want to T--dualize along $z$ anymore. 
With the same reasoning we also have to make an ansatz for the RR two--form:
\begin{equation}\label{oribrr}
  C_2^{IIB}\,=\,c_1\,dx\wedge dz+ c_2\,dx\wedge d\theta_1 + c_3\,dx\wedge d\theta_2 + 
    c_4\,dy\wedge dz+c_5\,dy\wedge d\theta_1 +c_6\,dy\wedge d\theta_2\,. 
\end{equation}
The coefficients $c_i$ are in general allowed to depend on $(r,z,\theta_1,\theta_2)$.  This implies an RR three--form fieldstrength
\begin{eqnarray}\label{iibthree}\nonumber
  F_3^{IIB} &=& F_{xz1}\,dx\wedge dz\wedge d\theta_1 +  F_{xz2}\,dx\wedge dz\wedge 
    d\theta_2 +  F_{yz1}\,dy\wedge dz\wedge d\theta_1 \\ \nonumber
  & & + F_{yz2}\,dy\wedge dz\wedge d\theta_2 + F_{rxz}\,dr\wedge dx\wedge dz + 
    F_{rx1}\,dr\wedge dx\wedge d\theta_1 \\ 
  & & + F_{rx2}\,dr\wedge dx\wedge d\theta_2 + F_{ryz}\,dr\wedge dy\wedge dz + 
    F_{ry1}\,dr\wedge dy\wedge d\theta_1 \\ \nonumber
  & & + F_{ry2}\,dr\wedge dy\wedge d\theta_2\,.
\end{eqnarray}
As in \cite{realm} we ignore the RR four--form for simplicity.

In conclusion, we make the generic ansatz \eqref{oriansatz} for the metric and \eqref{oribns} and \eqref{iibthree} for the fluxes for the IIB background {\it at} the orientifold point before transition.
We will mostly focus on the theory at the orientifold point, which takes us to type I. But it will be interesting to compare the type I theory that we find after two T--dualities to the IIB theory we would have obtained if we had T--dualized the IIB background {\sl away} from the orientifold point. They will have many similarities. But note that away from the orientifold point there are more allowed flux components.

Similar remarks hold true for the background after transition, but here we have an F--theory fourfold that is fibered over a base which resembles the deformed conifold. We now want to find an orientifold version of this, too. We will start with the semi--flat limit we obtained from T--duality, because it does not have any $dx\,d\theta_2$ or $dy\,d\theta_1$ crossterms that would not be invariant under $I_{xy}$. In other words, we start with \eqref{postulatefinaliib}, but impose the values for $\tilde{\beta_i}$ as in \eqref{betatwo}, \eqref{betaone}:
\begin{eqnarray}\label{goodiib}\nonumber
  d\tilde{s}_{IIB}^2 &=& dr^2 + \frac{e^{-2\phi}}{\alpha_0 CD}\,\left[dz
    +\alpha_0\alpha AD e^{2\phi}\,dx+\alpha_0\alpha BC e^{2\phi}\,dy\right]^2\\
  & & +\alpha_0\,D_1\,(dx^2 + \zeta\, d\theta_1^2)  
    +\alpha_0\,C_1\,(dy^2 + d\theta_2^2)\\\nonumber
  & & + 2\alpha_0\alpha^2 AB e^{2\phi}\,(d\theta_1\,d\theta_2-dx\,dy)\,,
\end{eqnarray}
with the squashing factor $\zeta$ defined in \eqref{squash}. 
At the danger of overloading notation we have introduced two other abbreviations
\begin{equation}
  C_1 \,=\, C+\alpha^2A^2e^{2\phi}\,,\qquad\qquad\mbox{and}\qquad\qquad
  D_1 \,=\, D+\alpha^2B^2e^{2\phi}
\end{equation}
where $C$ and $D$ were defined in \eqref{candd} and govern the size of the two $S^2$ in the non--K\"ahler resolved metric we found in IIA after transition. With this definition the constant $\alpha_0$ from \eqref{candd} becomes
\begin{equation}
  \alpha_0\,=\,C_1 D_1-\alpha^4 e^{4\phi}\,A^2B^2.
\end{equation}

Again, this metric has unwanted cross terms in the form $dx\,dz$ and $dy\,dz$, but it can also be brought into the form \eqref{oriansatz} after untwisting the z--fiber. However, the complex structures of the two tori will now be very different, in particular we find
\begin{equation}
  {\rm Re}\,\tau_2\,\ne\,0\qquad\mbox{but}\qquad {\rm Re}\,\tau_1\,=\,0\,,
\end{equation}
which is exactly opposite as for the background before transition. The second of these relations might not be immediately obvious, but is simply due to the cancellation of the $dx\,dy$ cross terms from the $dz$--fibration with those from the last line in \eqref{goodiib}. Again, we allow for quite generic values for $\tau_i$ (supersymmetry will fix them depending on the fluxes we choose), but we will always require ${\rm Re}\,\tau_1\,=\,0$.
Requiring \eqref{goodiib} without $dx\,dz$ or $dy\,dz$ cross terms to match with
\begin{equation}\label{iiboriafter}
  d\tilde{s}^2\,=\,dr^2+\tilde{d}_1\,|dz_1|^2+\tilde{d}_2\,|dz_2|^2 
    +\tilde{d_3}\,dz^2\,
\end{equation}
gives the following values for the constants $\tilde{d}_i$ and complex structures 
\begin{eqnarray}
  \tilde{d}_1 &=& C^{-1}\,,\qquad \tilde{d}_2\,=\,\zeta\alpha_0\, D_1\,,\qquad 
    \tilde{d}_3\,=\,(\alpha_0CD\,e^{2\phi})^{-1}\\[1ex]\nonumber
  \tau_1 &=& i\,\sqrt{\frac{C}{D}}\,,\qquad 
    \tau_2 \,=\,\frac{\alpha^2 A B e^{2\phi}
    +i\,\sqrt{C_1D_1\zeta-\alpha^4A^2B^2e^{4\phi}}}{\zeta\,D_1}\,.
\end{eqnarray}
The B--field does not change under geometric transition, as we showed on our walk through the duality chain, see \eqref{finaliibbns}, so also after transition we have \eqref{oribns}. The orientifold action also allows for the same type of RR fields as in \eqref{oribrr}. We will also assume a vanishing axion--dilaton for the orientifold background after transition.

Both of these backgrounds will now be T--dualized to type I, because the D7/O7 system turns into spacetime filling D9/O9 under T--duality along $x$ and $y$, as explained above. This gives rise to non--K\"ahler backgrounds in type I due to the same mechanism that mixes B--field and metric as encountered in section \ref{chain}.

\subsection{Non--K\"ahler Backgrounds in Type I Theory}

Before geometric transition we find the metric after performing two T--dualities along $x$ and $y$ on \eqref{oriansatz} to be
\begin{eqnarray}\label{typibefore}\nonumber
  ds^2_{I} &=& dr^2+dz^2+|dz_2|^2-2\alpha AB\,(dx-b{x\theta_i}\,d\theta_i)
    (dy-b_{y\theta_j}\,d\theta_j)\\
  & & +\alpha(1+B^2)\,(dx-b{x\theta_i}\,d\theta_i)^2 + \alpha(1+A^2)\, 
    (dy-b_{y\theta_j}\,d\theta_j)^2\,.
\end{eqnarray}
This metric has some by now familiar properties: we encounter the usual fibration structure, i.e. the fibers corresponding to the T--duality directions are twisted by the B--field. Therefore, this background will in general not be K\"ahler, since the derivative of these B--fields make the fundamental two--form non--closed. But there is one fundamental difference: we did not boost the complex structure of the $(x,\theta_1)$ and $(y,\theta_2)$ tori since this would lead to cross terms that are projected out by $I_{xy}$ and, moreover, we do not aim at regaining a metric that looks like a deformed conifold. We are not attempting to find the mirror manifold in IIA, therefore we do not expect to find a deformed conifold (or something close to it). So there is simply no reason why we should have to impose this non--trivial boost of complex structures. Furthermore, since we do not T--dualize along $z$ we do not have to worry about isometries for this direction. This is a simpler scenario than the one considered in the type II theories in section \ref{chain}. It is rather closer to the case considered in \cite{jan} in the sense that the B--field is completely ``used up'', which it has to be since $B_{NS}$ is not part of the type I spectrum.

Therefore, we find an explicit realization of a non--K\"ahler manifold for string theory backgrounds in type I. 
This should still be a complex manifold \cite{daskat, dasgreenbeck}\footnote{Non--complex manifolds in heterotic theory have been considered in \cite{andrei}, for example. Our orientifold construction is similar to models considered in \cite{daskat}, therefore we would expect it to yield complex manifolds as well.}, though, and we need to determine the precise value of the fluxes to ensure that the supergravity equations of motion are solved. So, let us finish the analysis of the type I background before transition by evaluating the RR fields and the dilaton. The type of background we consider, a toroidal orbifold with non--trivial complex structure, has already been studied in \cite{daskat}. The complex structure will be fixed by the fluxes and one particular simple choice is the assumption that the RR and NS field strengths are constant (although $B_{NS}$ and $C_2$ are not). This assumption was shown to be consistent with a metric of type \eqref{oriansatz} where ${\rm Re}\,\tau_i\,=\,0$, a choice which is possible for our orientifold setup. Under this condition, the following constraint is imposed on the fields \cite{daskat, dasgreenbeck}
\begin{equation}\label{fluxconstraint}
  B^{NS}_{[x\mu}\,C^{RR}_{\nu y]}\,=\,0\,,
\end{equation}
where $[\,,\,]$ indicates antisymmetrization of all enclosed indices.
This has important consequences for the RR field we find after T--duality. 
As derived in \cite{daskat}, only those RR fields with one leg along the T--duality direction survive and we find for the RR three-form fieldstrength and the string coupling
\begin{eqnarray}\label{typirr}\nonumber
  F_3^I &=& F_{xz1}^{IIB}\,dy\wedge dz \wedge d\theta_2 + F_{xz2}^{IIB}\,dy\wedge dz 
    \wedge d\theta_1  - F_{yz2}^{IIB}\,dx\wedge dz\wedge d\theta_1\\ 
  & & - F_{yz1}^{IIB}\,dx\wedge dz\wedge d\theta_2+ F_{rxz}^{IIB}\,dr\wedge dy\wedge 
    dz + F_{rx1}^{IIB}\,dr\wedge dy\wedge d\theta_2\\ \nonumber
  & & + F_{rx2}^{IIB}\,dr\wedge dy\wedge d\theta_1 - F_{ryz}^{IIB}\,dr\wedge dx\wedge 
    dz - F_{ry1}^{IIB}\,dr\wedge dx\wedge d\theta_2\\ \nonumber
  & & - F_{ry2}^{IIB}\,dr\wedge dx\wedge d\theta_1\\ \label{idilbefore}
  g^I &=& e^{\phi^I} \,=\,\sqrt{\alpha} \,,
\end{eqnarray}
where $F^{IIB}_{ijk}$ are the components of the RR field strength we started with in IIB, see \eqref{iibthree}. 
Note that the string coupling is still a constant in our local limit, but it could in principle depend on $(r,\theta_1,\theta_2)$ through $\alpha$, if we leave the local limit. In any case, the dilaton does not vanish anymore. It is interesting that there is no B--field dependent fibration in the RR three form. This agrees with observations made in \cite{daskat, dasgreenbeck} and is due to the constraint \eqref{fluxconstraint}. Note that this was not the case for the IIA mirror, see equation \eqref{rrthreeiia} for example, where the fibration structure is encrypted in the hatted coordinates.

In type IIB, the RR fieldstrength $F_3^{IIB}$ and the NS fieldstrength $H_{NS}^{IIB}$ are related due to the linearized equation of motion \cite{ks, oh}
\begin{equation}\label{lineom}
  F^{IIB}_3 \,=\, *H_{NS}^{IIB}\,.
\end{equation}
In other words we can fix
\begin{eqnarray}\label{sugraeom}
  F^{IIB}_{ryz} dr\wedge dy\wedge dz &=& \partial_r c_4(r,z,\theta_1,\theta_2)\,dr\wedge 
    dy \wedge dz\\ \nonumber
  &=& *\big[(\partial_{\theta_2}\,b_{x\theta_1} -  \partial_{\theta_1}\,b_{x\theta_2})\,dx\wedge d\theta_1 \wedge d\theta_2\big]
\end{eqnarray}
and similarly for the other components, where we use the convention
$\epsilon^{rxyz\theta_1\theta_2}\,=\,+1$. This is completely analogous to the discussion following equation \eqref{eom}, we will therefore not repeat it for all components. Let us simply state that this orientifold setup is much less restrictive than the one considered in section \ref{chain} and we can have more components of $F_3$ and $H_{NS}$ turned on. This is due to the fact that we have more B--field components that are consistent with the orientifold action and that the coefficients are now also allowed to depend on $z$. Even requiring $b_{x\theta_i}$ and $b_{y\theta_j}$ to be functions of $r$ only will not result in any of the $c_i$ to be forced to a constant.


We would like to address the question if the background we derived here can indeed show any geometric transition, i.e. can we shrink the two--cycle the D5 branes are wrapped on and blow up a dual three-cycle with fluxes on it? To answer this question we can consider the type IIB metric away from the orientifold point. T--dualizing this gives another IIB background which turns out to be surprisingly similar to the type I we just derived. Since we know that IIB away from the orientifold point shows geometric transition (this is the original Vafa model), we can infer that the type I background does, too, since it is dual to this IIB background.

Starting with the full IIB metric before orientifolding \eqref{startresol} and the same ansatz for $B_{NS}$ as in \eqref{oribns} we would have found
\begin{eqnarray}\label{iibaway}\nonumber
  d\hat{s}^2_{IIB} &=& dr^2+\alpha\,dz^2+|dz_2|^2-2\alpha AB\,(dx-b{x\theta_i}\,d\theta_i)
    (dy-b_{y\theta_j}\,d\theta_j)\\
  & & +\alpha(1+B^2)\,(dx-b{x\theta_i}\,d\theta_i)^2 + \alpha(1+A^2)\, 
    (dy-b_{y\theta_j}\,d\theta_j)^2\,,
\end{eqnarray}
but now with non--vanishing $B_{NS}$
\begin{equation}
  B_{NS}\,=\,-\alpha A\,(dx-b{x\theta_i}\,d\theta_i)\wedge dz -\alpha B 
    (dy-b_{y\theta_j}\,d\theta_j)\wedge dz\,,
\end{equation}
which was to be expected because these are precisely the $dx\,dz$ and $dy\,dz$ cross--terms from the starting metric, so they turn into B--field components via Buscher's rules \eqref{buscher}. We see that the only difference in the metric is a warp factor for $dz^2$. In our local limit this is simply a constant and we can rescale
\begin{equation}
  z\,\longrightarrow\,z'\,=\,\sqrt{\alpha}\,z\,,
\end{equation}
then the type I and type IIB metrics after T--duality agree completely. Since \eqref{iibaway} is dual to the IIB background that shows transition, we can infer that also the type I background \eqref{typibefore} has a two--cycle that can be shrunk and be exchanged for a blown--up three cycle.

The dual background with blown up three--cycle is found by T--dualizing the orientifold ansatz after transition \eqref{iiboriafter}. The steps are the same as for the background before transition and pretty straightforward. There is no extra boost of the complex structures required. This brings us to the type I metric after transition
\begin{eqnarray}\label{typiafter}\nonumber
  d\tilde{s}_I^2 &=& dr^2 + (\alpha_0CD e^{2\phi})^{-1}\,dz^2 + \zeta\,\alpha_0 
    D_1\,|dz_2|^2\\
  & & +C\,(dx-b_{xi}\,d\theta_i)^2 + D\,(dy-b_{yj}\,d\theta_j)^2\,.
\end{eqnarray}
Again, the B--field is completely used up under T--duality. The RR fluxes will take the same form as in \eqref{typirr}, although the precise coefficients may differ. The string coupling is evaluated to be
\begin{equation}
  g^I\,=\,e^{\phi^I}\,=\,\sqrt{CD}\,,
\end{equation}
which is not the same value we found before transition \eqref{idilbefore}, but in the local limit both are constant.

Again we want to compare this to the IIB background away from the orientifold point that we would have obtained after two T--dualities. This is done by T--dualizing \eqref{goodiib}, which still has the $dx\,dz$ and $dy\,dz$ cross--terms. After a little algebra one finds almost exactly the same metric apart from a warp factor for $dz^2$, which turns out to be surprisingly simple
\begin{eqnarray}\nonumber
  d\hat{\tilde{s}}_{IIB}^2 &=& dr^2 + e^{-2\phi}\,dz^2 + \zeta\,\alpha_0 
    (D+\alpha^2B^2 e^{2\phi})\,|dz_2|^2\\
  & & +C\,(dx-b_{xi}\,d\theta_i)^2 + D\,(dy-b_{yj}\,d\theta_j)^2\,.
\end{eqnarray}
After a rescaling
\begin{equation}
  z\,\longrightarrow\,z'\,=\,(\alpha_0 CD)^{-1/2}\,z
\end{equation}
this agrees with the type I metric. The only difference is, as before transition, that this IIB background has a non--vanishing $B_{NS}$ field and we do not have to restrict the ans\"atze for the fluxes to be invariant under orientifold operation. But note that we have established a connection between the semi--flat IIB background after transition \eqref{goodiib} with the type I background after transition \eqref{typiafter}, instead of considering the full IIB background (with restored $\cos\z$ and $\sin\z$ terms). 

We can therefore conclude that the type I backgrounds constructed in \eqref{typibefore} and \eqref{typiafter} are transition dual. Let us repeat the argument: Each of the type I metrics is essentially identical to a IIB background which is T--dual to one of the IIB backgrounds discussed in section \ref{iib}. The backgrounds in section \ref{iib} are transition duals because we found them by following the duality chain. If now one type I background is T--dual to the IIB background before and the other to the IIB background after transition, this implies that they are also transition duals. They are connected via an even longer duality chain than the one we followed in section \ref{chain}. 

Both type I backgrounds are non--K\"ahler, because they are T--dual to IIB backgrounds with NS field. It would be interesting to confirm that they are really complex as anticipated in \cite{realm, daskat}, but we cannot show this conclusively if we only know the local metric. Again, we would require knowledge of the global metric of the F--theory fourfold to be able to extend this analysis to global backgrounds.

\subsection{Non--K\"ahler Backgrounds in Heterotic Theory}\label{hetsec}

As noted in the introduction, type I and heterotic string theory are related via a weak--strong coupling duality, so--called S--duality. It does not only exchange weak and strong coupling constant, but also RR and NS two--forms. This also means, that the sources for the RR two--form have to be turned into sources for $B_{NS}$, i.e. D5--branes become so--called NS5--branes in heterotic string theory. Recall that heterotic strings cannot couple to D--branes.

S--duality is generated by the $SL(2,\mathbb{Z})$ element $S$ in \eqref{sltwoz}. 
Recall that it leaves the Einstein metric $G_{\mu\nu}=e^{-\phi/2} g_{\mu\nu}$ invariant and acts on the complex scalar field $\lambda=\chi+i e^{-\phi}$ as $\lambda \to -1/\lambda$. For $\chi=0$, S--duality simply relates $e^\phi$ to $e^{-\phi}$ (in other words strong to weak string coupling) and exchanges NS and RR 2--forms. Since the axion $\chi$ vanishes in our case and there is no NS field in type I, the S--duality rules read
\begin{equation}
  g^{het}_{\mu\nu}\,=\, e^{-\phi_I} g_{\mu\nu}^I\,,\qquad \phi_{het}\,=\,-\phi_I\,,
    \qquad\qquad B_{NS}^{het}\,=\,C_2^I\,,
\end{equation}
where $g_{\mu\nu}$ indicates the string frame metric we have been working with so far.
It is evident that S--duality maps the type I spectrum (which does not contain $B_{NS}$ since it gets projected out by $\Omega$ when constructing an unoriented closed string) to the heterotic SO(32) spectrum (which does not contain $C_2^{RR}$ since it cannot couple consistently to D--branes).
One can also show that under these replacements the type I and heterotic action are really identical \cite{polwit}.

We will now S--dualize the type I backgrounds \eqref{typibefore} before and \eqref{typiafter} after transition to obtain heterotic backgrounds. These will also be non--K\"ahler, so we construct heterotic string backgrounds with torsion. Those have been the subject of intensive study, see e.g. \cite{strominger, hull,dasgreenbeck, hullwit} and references in \cite{granaflux}, and it will be interesting to see how the new, non--compact models we discuss here, fit into the existing literature.

Before transition, we find the S--dual of the metric \eqref{typibefore}
\begin{eqnarray}\label{hetbefore}\nonumber
  ds^2_{het} &=& \alpha^{-1/2}\,\big(dr^2+dz^2+|d\chi_2|^2)-2\sqrt{\alpha}\, AB\,
    (dx-b{x\theta_i}\,d\theta_i)(dy-b_{y\theta_j}\,d\theta_j)\\
  & & +\sqrt{\alpha}\,(1+B^2)\,(dx-b{x\theta_i}\,d\theta_i)^2 + \sqrt{\alpha}\, 
    (1+A^2)\, (dy-b_{y\theta_j}\,d\theta_j)^2\,.
\end{eqnarray}
The torsion three--form and string coupling are found from \eqref{typirr}
\begin{eqnarray}\label{torsionform}\nonumber
  H^{het}\,=\,F_3^I &=& F_{xz1}^{IIB}\,dy\wedge dz \wedge d\theta_2 +
    F_{xz2}^{IIB}\,dy\wedge dz \wedge d\theta_1  - F_{yz2}^{IIB}\,dx\wedge dz\wedge 
    d\theta_1\\ 
  & & - F_{yz1}^{IIB}\,dx\wedge dz\wedge d\theta_2 + F_{rxz}^{IIB}\,dr\wedge dy\wedge 
    dz + F_{rx1}^{IIB}\,dr\wedge dy\wedge d\theta_2\\ \nonumber
  & & + F_{rx2}^{IIB}\,dr\wedge dy\wedge d\theta_1 - F_{ryz}^{IIB}\,dr\wedge dx\wedge 
    dz - F_{ry1}^{IIB}\,dr\wedge dx\wedge d\theta_2\\ \nonumber
  & & - F_{ry2}^{IIB}\,dr\wedge dx\wedge d\theta_1\\
  g^{het} &=& \frac{1}{\sqrt{\alpha}} \,,
\end{eqnarray}
After transition, taking the S--dual of \eqref{typiafter}, we find
\begin{eqnarray}\label{hetafter}\nonumber
  d\tilde{s}_{het}^2 &=& \frac{1}{\sqrt{CD}}\,\left[dr^2 + (\alpha_0CD e^{2\phi})^{-1}\,
    dz^2 + \zeta\,\alpha_0 D_1\,|d\chi_2|^2\right]\\
  & & +\sqrt{\frac{D}{C}}\,(dx-b_{xi}\,d\theta_i)^2 + \sqrt{\frac{C}{D}}\,
    (dy-b_{yj}\,d\theta_j)^2\\ \label{hetdilatafter}
  g^{het}&=&\frac{1}{\sqrt{CD}}\,.
\end{eqnarray}
This, together with a torsion three--form of the same type as \eqref{torsionform}, specifies the heterotic background we claim to be transition dual to the background obtained as the S--dual of the type I background before transition in \eqref{hetbefore}. 

Let us repeat the reasoning why we claim these backgrounds to be transition dual. We verified Vafa's duality chain to the extend that we found a IIB background that has the local structure of a deformed conifold after a series of T--dualities and an $\mathcal{M}$--theory flop. Trusting this duality chain means, both IIB backgrounds are actually transition dual. We then constructed type I backgrounds that are T--dual to an orientifold version of these IIB backgrounds. We also verified that the type I metrics are actually very close to IIB after two T--dualities {\sl away} from the orientifold limit. Therefore, we can trust that those metrics in type I contain a contractible two-- and three--cycle, respectively.
Since the heterotic backgrounds possess the same metric as the type I backgrounds (apart from an overall factor given by the coupling) they should also be transition duals.

Both heterotic backgrounds have to fulfill a torsional relation to preserve supersymmetry \cite{strominger}. With constant dilaton (as we have in the local limit) this torsional relation reads
\begin{equation}
  H^{het}\,=\,*dJ
\end{equation}
with fundamental two--form $J$. But the fluxes were already constrained in IIB by the linearized equation of motion \eqref{eom}. This implies the following chain of reasoning for the mapping of the
fluxes from IIB to heterotic
\begin{equation}\label{fluxlogic}
  *(H_{NS}^{IIB}) \,=\, F_3^{IIB} \quad\xrightarrow{T{xy}}\quad F_3^I \;=\; H^{het} \,=\, *(dJ)\,,
\end{equation}
where T--duality $T_{xy}$ along $x$ and $y$ imposes relations between the type IIB RR flux $F_3^{IIB}$ and the type I three--form $F_3^I$ that can be read of from \eqref{typirr}
\begin{eqnarray} \nonumber
  F_{rx(z,1,2)}^{IIB} &=& -F_{ry(z,2,1)}^I\,,\qquad\qquad\quad
  F_{ry(z,1,2)}^{IIB} \,=\, F_{rx(z,2,1)}^I\\
  F_{xz(1,2)}^{IIB} &=& -F_{yz(2,1)}^I\,,\qquad\qquad\qquad
  F_{yz(1,2)}^{IIB} \,=\, F_{xz(2,1)}^I\,.
\end{eqnarray}
Note that the B--field components $b_{x\theta_i}$ and $b_{y\theta_j}$ we start with in IIB appear at the end of the chain in heterotic theory in the metric and are contained in $dJ$. Therefore, this connection is highly non--trivial and might not always be consistent for an arbitrary choice of background fluxes. It means that there has to exist a complex structure on the heterotic metric that is compatible with the T--duality action on the RR forms.

We can demonstrate this for a simple toy example\footnote{This differs from the example considered in \cite{realm}.}. We will make a quite restrictive ansatz for the fluxes and work strictly in the local limit where $A,B,C,D=$constant (and so are $\alpha, \alpha_0$). Let us choose for the IIB RR two--form
\begin{equation}
  C_2^{IIB} \,=\, c_1(r)\,dx\wedge dz + c_4(r)\,dy\wedge dz\,,
\end{equation}
which means there will be only two components in the RR fieldstrength. They are related to the IIB NS--field via the linearized equation of motion \eqref{lineom} and we find
\begin{eqnarray}\label{iibthreeform}\nonumber
  F_{rxz}^{IIB}\,dr\wedge dx\wedge dz &=& *[(\partial_{\theta_2} b_{y\theta_1}-
    \partial_{\theta_1} b_{y\theta_2})\, dy\wedge d\theta_1\wedge d\theta_2] \\ 
    \nonumber
  &=& a_1 \,(\partial_{\theta_2} b_{y\theta_1}-\partial_{\theta_1} b_{y\theta_2})
    \,dr\wedge dx\wedge dz\\
  F_{ryz}^{IIB}\,dr\wedge dy\wedge dz &=& *[(\partial_{\theta_2} b_{x\theta_1}
    -\partial_{\theta_1} b_{x\theta_2})\, dx\wedge d\theta_1\wedge d\theta_2]\\ 
    \nonumber
  &=& a_2\, (\partial_{\theta_2} b_{x\theta_1}
    -\partial_{\theta_1} b_{x\theta_2})\,dr\wedge dy\wedge dz\,,
\end{eqnarray}
where the constants $a_i$ contain the numerical factor due to the Hodge star operator $*$ on the six--dimensional IIB metric \eqref{resolori}.
In order to fulfill the supergravity equation of motion we also have to ensure that the IIB NS field strength does not have any other components than those appearing in \eqref{iibthreeform}. This imposes the requirement that the components $b_{x\theta_i}$ and $b_{y\theta_j}$ are functions of $(\theta_1,\theta_2)$ only and not of $r$ or $z$.
Under T--duality these fluxes turn into an RR three--form in type I:
\begin{equation}\label{ithree}
  F_3^I \,=\, -F_{rxz}^{IIB}\,dr\wedge dy\wedge dz + F_{ryz}^{IIB}\,dr\wedge dx \wedge dz
\end{equation}
which becomes $H^{het}$ after S--duality. The question now is: does a complex structure (or rather fundamental two--form) exist for the heterotic background that is compatible with this torsion three--form?

There are of course many complex structures on the real six--manifold that is described by the metric \eqref{hetbefore}. One possible choice is to take the real vielbeins
\begin{eqnarray}\label{hetvielb}\nonumber
  e^1 &=& \alpha^{-1/4}\,dr\,,\qquad\qquad\qquad e^2\,=\, \alpha^{-1/4}\,dz\\
  e^3 &=& \alpha^{-1/4}\,d\theta_1\,,\qquad\qquad\quad\; e^4 \,=\, 
    \alpha^{-1/4}\,|\tau_2|\,d\theta_2\\ \nonumber
  e^5 &=& \alpha^{-1/4}\,\sqrt{\frac{1+A^2}{2}}\,\left((dy-b_{yi}\,d\theta_i) 
    +\gamma_2\, (dx-b_{xi}d\theta_i)\right)\\ \nonumber
  e^6 &=& \alpha^{-1/4}\,\sqrt{\frac{1+A^2}{2}}\,\left((dy-b_{yi}\,d\theta_i) 
    +\gamma_3\, (dx-b_{xi}d\theta_i)\right)
\end{eqnarray}
with the coefficients $\gamma_i$ being determined by the metric to
\begin{equation}\label{gammas}
  \gamma_2\,=\,\frac{-AB\pm\alpha^{-1/2}}{1+A^2}\,,\qquad 
    \gamma_3\,=\,\frac{-AB\mp\alpha^{-1/2}}{1+A^2}\,.
\end{equation}
With the canonical choice of complex structure as in \eqref{complstruc} the fundamental two--form becomes
\begin{equation}
  J\,=\,e^1\wedge e^2+ e^3\wedge e^4+e^5\wedge e^6\,=\,\{J_1\}_{b_{ij}=0} + \{J_2\}
\end{equation}
where we have explicitely separated $J$ into a B--field independent part $J_1$ and a part that contains the IIB B--field components $b_{x\theta_i}$ and $b_{y\theta_j}$, given by $J_2$. Since we work in the local limit
\begin{equation}
  dJ_1\,=\,0
\end{equation}
trivially. One might expect such a splitting to be always possible, since in the absence of any flux a K\"ahler background maps to another K\"ahler background under T--duality and only switching on NS flux creates torsion. This is of course correct, but a splitting of the fundamental two--form is only possible if we know the ``right'' complex structure on the K\"ahler manifold, in other words if we know the K\"ahler form. Not any choice of real vielbeins $e_i$ will lead to a closed $J_1$. These issues have been discussed in section \ref{torsion}.

For the local limit this splitting is trivially always possible. But keep in mind that the choice \eqref{hetvielb} with the complex structure imposed by $J$ is by no means unique. We view this choice as an illustrative example.
For the non--closed part we find
\begin{equation}
  J_2\,=\,b_{yi}\,dx\wedge d\theta_i-b_{xj}\,dy\wedge d\theta_j - (b_{x\theta_1} b_{y\theta_2}-b_{x\theta_2} b_{y\theta_1})\,d\theta_1\wedge d\theta_2
\end{equation}
up to an overall minus sign related to the sign ambiguity in $\gamma_i$ in \eqref{gammas}. The torsional relation $H^{het}=*(dJ_2)$ then implies
\begin{eqnarray}\label{hetthree}\nonumber
  H^{het}_{rxz}\,dr\wedge dx\wedge dz &=& *[-(\partial_{\theta_2} b_{x\theta_1}-
    \partial_{\theta_1} b_{x\theta_2})\, dy\wedge d\theta_1\wedge d\theta_2]\\ 
    \nonumber
  &=& -a_1\,(\partial_{\theta_2} b_{x\theta_1}-
    \partial_{\theta_1} b_{x\theta_2})dr\wedge dx\wedge dz\\
  H^{het}_{ryz}\,dr\wedge dy\wedge dz &=& *[(\partial_{\theta_2} b_{y\theta_1}-
    \partial_{\theta_1} b_{y\theta_2})\, dx\wedge d\theta_1\wedge d\theta_2]\\ 
    \nonumber
  &=& a_2\,(\partial_{\theta_2} b_{y\theta_1}-
    \partial_{\theta_1} b_{y\theta_2})\,dr\wedge dy\wedge dz
\end{eqnarray}
with all other components vanishing because $b_{ij}$ does not depend on $r$ or $z$. We have again included numerical factors $a_i$ to incorporate the Hodge star operator, they are exactly the same as in \eqref{iibthreeform}.
We would like to match this torsion form with the type I three--form \eqref{ithree}, which requires
\begin{equation}
  -F^{IIB}_{rxz} \,=\, H^{het}_{ryz}\,,\qquad\qquad F^{IIB}_{ryz} \,=\, H^{het}_{rxz}\,.
\end{equation}
Comparing $F^{IIB}$ given by the supergravity equation of motion in \eqref{iibthreeform} and $H^{het}$ given through the torsional relation in \eqref{hetthree}, we see that the first and second identity require
\begin{eqnarray}\label{condia}
  -a_1 \,=\, a_2\qquad\qquad\mbox{and}\qquad\qquad a_2 \,=\, -a_1\,,
\end{eqnarray}
respectively. 
The constants $a_1$ and $a_2$ are determined by the Hodge star operator, which is given on a six--manifold by
\begin{equation}
  *(dx^{\mu_1}\wedge dx^{\mu_2}\wedge dx^{\mu_3}) \,=\, \frac{1}{3!}\,\sqrt{|g|}\,
    \epsilon^{\mu_1\mu_2\mu_3}_{\phantom{\mu_1\mu_2\mu_3}\nu_1\nu_2\nu_3}\,
    dx^{\nu_1}\wedge dx^{\nu_2}\wedge dx^{\nu_3}\,.
\end{equation}
Since $a_1$ is determined by $\epsilon^{y\theta_1\theta_2}_{\phantom{y\theta_1\theta_2}rxz}$, $a_2$ is determined by $\epsilon^{x\theta_1\theta_2}_{\phantom{x\theta_1\theta_2}ryz}$ and $\epsilon^{y\theta_1\theta_2\,rxz}\,=\,-\epsilon^{x\theta_1\theta_2\,ryz}$, this seems perfectly consistent. We conclude that the choice of flux and complex structure in our toy example is consistent with the duality chain \eqref{fluxlogic} when $a_1\,=\, -a_2$.
The precise value of $a_1$ could be found from the metric \eqref{hetbefore}, but we will not do so here.

In summary, we found new non--compact, non--K\"ahler manifolds with local metric \eqref{hetbefore} and \eqref{hetafter}, that are related via S--duality to the type I backgrounds constructed in the last section. We argued the type I backgrounds to be transition duals, therefore also the heterotic non--K\"ahler backgrounds should show geometric transition. We demonstrated for a specific choice of fluxes that this background fulfills the torsional relation with torsion three--form \eqref{torsionform}, which was in turn related to the RR three form flux in the IIB orientifold. This IIB flux was also shown to fulfill the linearized supergravity equation of motion.

We now turn to the question if we can find a {\it global} background that reduces in the local limit to the ones we constructed here.

\subsection{A Global Heterotic Solution}\label{hetglobal}

One would, of course, like to leave the local limit. We cannot simply let the coordinates that we kept fixed vary, but if we find a global supergravity solution
that reduces to the one we found in the local limit, then we can safely assume this as one possible solution for our global background.

Let us therefore take a closer look at
backgrounds that have some similarity with ours. There are two solutions that come to mind in IIB: The Klebanov--Strassler (KS) model \cite{ks} and the Maldacena--Nunez solution (MN) \cite{mn}\footnote{The supersymmetry of these backgrounds has been shown in \cite{cvetic} and \cite{papatseyt}.}. The full MN solution is only known after transition, so let us focus on that region and compare the two. 

Both models describe a gravity dual of the far IR limit of a gauge theory. In the usual gauge/gravity duality, the IR of the supergravity theory corresponds to small radial coordinate $r$, the UV to large $r$. Although the gauge theory after transition is already in the far IR (the confining phase), the dual supergravity solution is nevertheless valid at all scales\footnote{This is not quite true for the MN background, which is only valid at small $r$ \cite{grana}, we will return to this issue later.}. We want to alert the reader to the fact that there are two different concepts of IR and UV: on the gauge theory side the UV is described by D--branes wrapped on a large $S^2$, whereas the (far) IR is described by fluxes on a large $S^3$. Both gauge theory phases have a supergravity dual, in which UV corresponds to large $r$ and IR to small $r$. We will in the following only discuss a solution after transition, so when we distinguish between IR and UV it will always be in the supergravity dual side. The gauge theory is already in the confining phase.

KS and MN both ``flow'' towards some version of the deformed conifold after transition. The geometry is different, in particular the MN solution does not have two spheres of the same size as KS does. Both models also have fluxes turned on, but in the MN case there is only RR flux (the reason being that they only start with D5 branes before transition, whereas KS start with D5 and D3, resulting in different solutions for the fluxes). So it seems that the MN background is well suited for S--dualizing it to a heterotic theory. We cannot do the same to the KS solution because its NS flux could turn into RR flux, which the heterotic theory does not contain. Also, the MN background seems closer to what we found after geometric transition in IIB, we also recovered two $S^2$ of different size.

Let us therefore quote the MN background \cite{mn}, which is after S--duality in heterotic
\begin{eqnarray}\label{mnglobal}
  ds^2_{\rm MN} & = & N\,d\widetilde{r}^2 + \frac{N}{4}\, \left(d\psi + {\rm cos}\,
    \tilde\theta_1 \,d\widetilde{\phi}_1 + {\rm cos}\,\widetilde{\theta}_2
    \,d\widetilde{\phi}_2\right)^2  \\ \nonumber
 & + & \frac{N}{4}\, \left(e^{2g} + a^2\right) \left(d\widetilde{\theta}_2^2 
    + {\rm sin}^2 \widetilde{\theta}_2\,d\widetilde{\phi}_2^2\right) + \frac{N}{4}
    \,\left(d\widetilde{\theta}_1^2 + {\rm sin}^2 \widetilde{\theta}_1\,
    d\widetilde{\phi}_1^2\right) \\ \nonumber
  & - & \frac{Na}{2}\Big[{\rm cos}\,\psi (d\widetilde{\theta}_1 
    d\widetilde{\theta}_2 - {\rm sin}\,\widetilde{\theta}_1  {\rm sin}\,
    \widetilde{\theta}_2\,d\widetilde{\phi}_1 
    d\widetilde{\phi}_2) + {\rm sin}\,\psi ({\rm sin}\,\widetilde{\theta}_1
    \,d\widetilde{\phi}_1\,d\widetilde{\theta}_2 + {\rm sin}
    \,\widetilde{\theta}_2\,d\widetilde{\phi}_2\,d\widetilde{\theta}_1)\Big]
\end{eqnarray}
with the definition (following the notation of \cite{grana})
\begin{equation}
  a(\widetilde{r}) \,= \, -\frac{2\widetilde{r}}{{\rm sinh}~2\widetilde{r}}\,, 
    \qquad e^{2g} \, =\,4 \widetilde{r}~{\rm coth}~2\widetilde{r} - \frac{4 
    \widetilde{r}^2}{{\rm sinh}^2~2\widetilde{r}} - 1\,.
\end{equation}
The dilaton is given by
\begin{equation}\label{mndilaton}
  e^{2\phi}\,=\,\frac{e^{g+2\Phi_0}}{{\rm sinh}\,2\widetilde{r}}
\end{equation}
where $\Phi_0$ is some constant value that could for example be fixed from the U--dual background. The NS three--form for this heterotic background was found in \cite{mn} to be
\begin{equation}\label{mnhns}
  H^{NS}_{MN} \,=\, -\frac{N}{4}\,\left[(\omega^1-A^1)\wedge(\omega^2-A^2)\wedge
    (\omega^3-A^3)-\sum_a\,F^a\wedge(\omega^4-A^4)\right]
\end{equation}
with one--forms
\begin{eqnarray}\nonumber
  \omega^1 &=& \cos\psi\, d\widetilde{\theta}_1+\sin\psi\,\sin\widetilde{\theta}_1
    \, d\widetilde{\phi}_1\,\qquad\quad\quad\; A^1\,=\,a\,d\widetilde{\theta}_2\\
  \omega^2 &=& -\sin\psi\, d\widetilde{\theta}_1+\cos\psi\,\sin\widetilde{\theta}_1
    \,d\widetilde{\phi}_1\,,\qquad\quad A^2\,=\,-a\,\sin\widetilde{\theta}_2\, 
    d\widetilde{\phi}_2\\ \nonumber
  \omega^3 &=& d\psi+\cos\widetilde{\theta}_1\,d\widetilde{\phi}_1\,,
    \qquad\qquad\qquad\qquad\quad
    A^3\,=\,-\cos\widetilde{\theta}_2\,d\widetilde{\phi}_2
\end{eqnarray}
and the fieldstrength $F$ is defined as $F^a=dA^a+\epsilon^{abc}\,A_bA_c$. These one--forms are not quite the right vielbeins for observing the SU(3) structure of this background \cite{grana}, since it is not a conformal Calabi--Yau. (For the Ricci flat K\"ahler metric on the deformed conifold these vielbeins with the canonical complex structure would give closed two and three form, but not for a background with different size of the $(\widetilde{\phi}_1,\widetilde{\theta}_1)$ and $(\widetilde{\phi}_2,\widetilde{\theta}_2)$ spheres. Therefore, they will not produce the correct SU(3) structure for this non--K\"ahler background either.) We will return to the issue of the right complex structure when we discuss the torsional relation of the global heterotic background.
 
If we want to compare this to our local background, we should also introduce local coordinates here. Let us define them as 
\begin{eqnarray}\label{mnlocalcoord}\nonumber
  \widetilde{r} &=& r_0 + r\\ 
  \psi &=& \z+z\,,\qquad\qquad\quad\; \widetilde{\theta}_i\,=\, \langle 
    \theta_i\rangle + \theta_i\\ \nonumber
  \widetilde{\phi}_1 &=& \langle\phi_1\rangle+\frac{x}{\sin\langle\theta_1\rangle}\,, 
    \qquad \widetilde{\phi}_2 \,=\, \langle\phi_2\rangle
    +\frac{y}{\sin\langle\theta_2\rangle}\,.
\end{eqnarray}
The local MN background is then found to be
\begin{eqnarray}\label{mnlocal} \nonumber
  ds^2_{\rm MN} & = & N\,dr^2 + \frac{N}{4}\, \left(dz + {\rm cot}\,\langle\theta_1
    \rangle\,dx + {\rm cot}\,\langle\theta_2\rangle\,dy\right)^2  \\ 
 & & +  \frac{N}{4}\, \left(e^{2g} + a^2\right) \left(d\theta_2^2 + dy^2\right) + 
    \frac{N}{4}\,\left(d\theta_1^2 + dx^2\right) \\ \nonumber
& & -  \frac{Na}{2}\Big[{\rm cos}\,\z (d\theta_1 d\theta_2
    - dx\,dy) + {\rm sin}\,\z (dx\,d\theta_2 + dy\,d\theta_1)\Big]\,.
\end{eqnarray}

We now want to compare this to the heterotic background we found after transition
\eqref{hetafter}. It was given by
\begin{eqnarray}\label{ourlocal}
  d\tilde{s}_{het}^2 &=& \mathcal{A}_1\,dz^2 + \mathcal{A}_2\,(dy-b_{yi}\,
    d\theta_i)^2 +\mathcal{A}_3 \,(dx-b_{xj}\,d\theta_j)^2 + \mathcal{A}_4\,
    |dz_2|^2 + \mathcal{A}_5\,dr^2\,.
\end{eqnarray}
The coefficients could be read off from \eqref{hetafter}, but we can also leave them arbitrary to allow for a larger class of backgrounds. Recall that $dz_2=d\theta_1+\tau_2\,d\theta_2$ and $dz_1=dx+\tau_1\,dy$, where we had found that the IIB background before transition was characterized by Re$\,\tau_2=0$, whereas after transition Re$\,\tau_1=0$. We will now assume that we can consistently deform the background after transition in a way that converts both tori to square ones, i.e. also Re$\,\tau_2=0$, together with a choice of B--field (these are the components of the IIB B--field {\sl before} transition)
\begin{equation}
  B_{NS}^{IIB} \,=\, b_{x\theta_2}\,dx\wedge d\theta_2 + b_{y\theta_1}\,dy\wedge 
    d\theta_1\,.
\end{equation}
This is a special choice of \eqref{oribns}, which is consistent with our IIB orientifold setup.
Supersymmetry will be restored by an appropriate choice of RR fluxes. The effect on the metric \eqref{ourlocal} is, after a little rearrangement,
\begin{eqnarray}\label{ourlocali}\nonumber
  d\tilde{s}_{het}^2 &=& \mathcal{A}_1\,dz^2 + \mathcal{A}_2\,dy^2 +\mathcal{A}_3 
    \,dx^2 + \mathcal{A}_4\,|dz_2|^2 +\mathcal{A}_5\,dr^2\\
  & & +\mathcal{A}_2\,b_{y\theta_1}^2\,d\theta_1^2 + \mathcal{A}_3\,
    b_{x\theta_2}^2\,d\theta_2^2 -2 \left(\mathcal{A}_2\,b_{y\theta_1}\,dy\,d\theta_1 
    + \mathcal{A}_3\,b_{x\theta_2}\,dx\,d\theta_2\right)\,. 
\end{eqnarray}
If this is to coincide with the local MN background \eqref{mnlocal}, we have to impose a few requirements: we want the cross terms $dx\,d\theta_2$ and $dy\,d\theta_1$ to have the same prefactor and we want the $(x,\,\theta_1)$ as well as the $(y,\theta_2)$ spheres to be unsquashed. This gives the following constraints on the coefficients and B--field
\begin{equation}\label{consti}
  \mathcal{A}_3\,=\,\frac{\mathcal{A}_2}{|\tau_2|^2}\,,\qquad 
    \mathcal{A}_4\,=\,\mathcal{A}_2\,\left(\frac{1}{|\tau_2|^2}-b_{y\theta_1}^2\right)
    \,,\qquad b_{x\theta_2}\,=\,|\tau_2|^2\,b_{y\theta_1}
\end{equation}
and converts our local metric \eqref{ourlocali} to
\begin{eqnarray}\label{ourlocaltwo}\nonumber
  d\tilde{s}_{het}^2 &=& \mathcal{A}_2 \left[\left(dy^2+
    d\theta_2^2\right)+\frac{1}{|\tau_2|^2}\,\left(dx^2+d\theta_1^2\right) 
    -2 b_{y\theta_1}\,\left(dy\,d\theta_1+dx\,d\theta_2\right)\right]\\
  & & +\mathcal{A}_5\,dr^2+\mathcal{A}_1\,dz^2 \,.
\end{eqnarray}
We now perform a local coordinate transformation
\begin{eqnarray}\label{localyrot}\nonumber
  y &\longrightarrow & \sin\z\,y+\cos\z\,\theta_2\\
  \theta_2 &\longrightarrow & -\cos\z\,y
    +\sin\z\,\theta_2\\ \nonumber
  z &\longrightarrow & z+ \cot\langle\theta_1\rangle\,x 
    + \cot\langle\theta_2\rangle \,y\,,
\end{eqnarray}
which might remind the reader of a similar transformation in section \ref{mirror}, in particular \eqref{yrotation}.
Then \eqref{ourlocaltwo} becomes
\begin{eqnarray}\nonumber
   d\tilde{s}_{het}^2 &=& \mathcal{A}_5\,dr^2+\mathcal{A}_1\,\left(dz+ \cot\langle\theta_1
    \rangle\,dx + \cot\langle\theta_2\rangle\,dy\right)^2\\ 
  & & + \mathcal{A}_2 \left[\left(dy^2+ d\theta_2^2\right)
    +\frac{1}{|\tau_2|^2}\,\left(dx^2+d\theta_1^2\right)\right]\\ \nonumber
  & & -2\mathcal{A}_2 b_{y\theta_1}\,\left[\sin\z\left(dy\,d\theta_1
    +dx\,d\theta_2\right)+\cos\z\left(d\theta_1\,d\theta_2-dx\,dy
    \right)\right]\,.
\end{eqnarray}
Comparing this to the local MN background \eqref{mnlocal}, we see that we can exactly match the two backgrounds with the following choice for the coefficients
\begin{equation}\label{fixcoef}
  \mathcal{A}_1 \,=\,\mathcal{A}_3\,=\,\frac{\mathcal{A}_5}{4}\,=\,\frac{N}{4}\,, 
    \qquad \mathcal{A}_2\,=\,\frac{N(e^{2g}+a^2)}{4}\,,\qquad \mathcal{A}_4\,=\,\frac{N e^{2g}}{e^{2g}+a^2}\,. 
\end{equation}
This has consequences for the IIB B--field and the complex structure of the tori (since they are related to the coefficients via \eqref{consti})
\begin{eqnarray}
  B_{NS}^{IIB} &=& a \,dx\wedge d\theta_2 +\frac{a}{e^{2g}+a^2}\,dy\wedge d\theta_1\\
  dz_1 &=& dx + i\,dy\,,\qquad\qquad dz_2 \,=\,d\theta_1 +i\,\sqrt{e^{2g}+a^2}
    \,d\theta_2\,.
\end{eqnarray}
Note that the complex struture of the $z_1$--torus was not fixed during the considerations here but remains the same as in \eqref{iiboriafter}, the IIB orientifold ansatz after geometric transition.

Thus, we have shown that with an appropriate choice of IIB B--field before transition and complex structure of the $z_2$--torus after transition in IIB, our solution coincides with the local limit of the MN background. 
Reversing this argument, we can also claim that the choice \eqref{fixcoef} gives a valid {\sl global} solution if we leave the local limit and allow our coordinates to vary, since then we recover the MN background, which has been shown to be supersymmetric \cite{papatseyt}. But we can even go beyond that and claim that the global heterotic background we find after transition is given in terms of generic coefficients $\mathcal{A}_i$
\begin{eqnarray}\label{hetfinal}
  &d\tilde{s}^2_{het} & = \mathcal{A}_5\,d\widetilde{r}^2 + \mathcal{A}_1\, \left(d\psi 
    +a_1 {\rm cos}\,\widetilde{\theta}_1 \,d\widetilde{\phi}_1 + b_1{\rm cos}\,\widetilde{\theta}_2\,d\widetilde{\phi}_2\right)^2  \\ 
    \nonumber
 & & +  \mathcal{A}_2\left(d\widetilde{\theta}_2^2 + {\rm sin}^2
    \widetilde{\theta}_2\,d\widetilde{\phi}_2^2\right) +\mathcal{A}_3\,\left(d\widetilde{\theta}_1^2 + {\rm 
    sin}^2 \widetilde{\theta}_1\,d\widetilde{\phi}_1^2\right) \\ \nonumber
& & -  2\mathcal{A}_2b_{y\widetilde{\theta}_1}\Big[{\rm cos}\,\psi (d\widetilde{\theta}_1 d\widetilde{\theta}_2
    - {\rm sin}\,\widetilde{\theta}_1  {\rm sin}\,\widetilde{\theta}_2\,d\widetilde{\phi}_1 d\widetilde{\phi}_2) 
    + {\rm sin}\,\psi ({\rm sin}\,\widetilde{\theta}_1\,d\widetilde{\phi}_1\,d\widetilde{\theta}_2 + {\rm sin}
    \,\widetilde{\theta}_2\,d\widetilde{\phi}_2\,d\widetilde{\theta}_1)\Big]\,,
\end{eqnarray}
where we have re--introduced global coordinates by reversing \eqref{mnlocalcoord}.
Although the MN background was derived for the IR (small $r$ limit) only \cite{grana}, our global solution should be valid in the UV (large $r$ limit) as well, but we cannot use the identification \eqref{fixcoef} there. The UV limit of MN was derived in \cite{grana}, and we will return to this issue shortly. The dilaton $\phi$ for this background can be determined from the warp factors in the metric. The NS three--form (or torsion three--form) $H$ would be given by the torsional relation \cite{dasbecksq, carcudallu}
\begin{equation}\label{torsrelat}
  H\,=\,e^{2\phi}\,*\,d\left(e^{-2\phi}\,J\right)
\end{equation}
with fundamental two--form $J$. Note that the dilaton is not constant anymore, as anticipated in our local analysis, where it became obvious how metric components would give rise to a coordinate--dependent dilaton if we leave the local limit, see \eqref{hetdilatafter}.

We could in principle now evaluate \eqref{torsrelat} to find the generic three--form for our postulated global background. We will illustrate this in the example where the coefficients $\mathcal{A}_i$ match indeed the MN solution. As pointed out in \cite{grana}, the appropriate vielbeins are
\begin{eqnarray}\label{mnvielb}\nonumber
 e^1 &=& \sqrt{N}\,d\widetilde{r}\,, \quad e^5 \,=\, \frac{\sqrt{N}}{2}\,e^g\,d\widetilde{\theta}_2\,, \quad
   e^2\,=\,\frac{\sqrt{N}}{2}\,\big(d\psi+\cos\widetilde{\theta}_1\,d\widetilde{\phi}_1+\cos\widetilde{\theta}_2\,
   d\widetilde{\phi}_2\big)\\
 e^3 &=& \frac{\sqrt{N}}{2}\,\big(\sin\psi \sin\widetilde{\theta}_1\,d\widetilde{\phi}_1
   +\cos\psi\, d\widetilde{\theta}_1-a\,d\widetilde{\theta}_2\big)\\ \nonumber
 e^4 &=& -\frac{\sqrt{N}}{2}\,\Big[{\mathcal B}\,e^g\sin\widetilde{\theta}_2\,d\widetilde{\phi}_2
   + {\mathcal A} (\cos\psi\sin\widetilde{\theta}_1\,d\widetilde{\phi}_1-\sin\psi\,d\widetilde{\theta}_1
   + a\sin\widetilde{\theta}_2\,d\widetilde{\phi}_2)\Big]\\ \nonumber
 e^6 &=& -\frac{\sqrt{N}}{2}\,\Big[{\mathcal A}\,e^g\sin\widetilde{\theta}_2\,d\widetilde{\phi}_2
   - {\mathcal B} (\cos\psi\sin\widetilde{\theta}_1\,d\widetilde{\phi}_1-\sin\psi\,d\widetilde{\theta}_1
   +a\sin\widetilde{\theta}_2\,d\widetilde{\phi}_2)\Big]
\end{eqnarray}
which give rise to the metric \eqref{hetfinal} with identification \eqref{fixcoef}. The coefficients $\mathcal{A}$ and $\mathcal{B}$ satisfy $\mathcal{A}^2+\mathcal{B}^2=1$ and are given as\footnote{They play the same role as $P$ and $X$ introduced in section \ref{torsionbefore} and stem from the generic ansatz made for the complex structure. Note that they also carry $r$--dependence only, the same assumption we used in section \ref{torsionbefore}.}
\begin{equation}
  {\mathcal A}\,=\,\coth \,2\widetilde{r}-2\widetilde{r} \,{\rm csch}^2 2
    \widetilde{r}\,,\qquad
  {\mathcal B}\,=\,{\rm csch}\, 2\widetilde{r}\,\sqrt{-1+4\widetilde{r}\,\coth\,
    2\widetilde{r} -4\widetilde{r}^2\,{\rm csch}^2\,2\widetilde{r}}\,.
\end{equation}
We then make the canonical choice of complex structure where the fundamental two--form is given by $J=e^1\wedge e^2+ e^3\wedge e^4+e^5\wedge e^6$. 

\noindent This amounts to
\begin{eqnarray}\nonumber
  J &=& \frac{N}{2}\,d\widetilde{r}\wedge (d\psi+\cos\widetilde{\theta}_1
    \,d\widetilde{\phi}_1 +\cos\widetilde{\theta}_2\,d\widetilde{\phi}_2)\,  \\ 
    \nonumber
  & - & \frac{N}{4}\,\,{\cal A}\sin\widetilde{\theta}_1\,d\widetilde{\theta}_1
    \wedge d\widetilde{\phi}_1 - \frac{N}{4}\,\,\left(-{\cal A}^2a+{\cal A} e^{2g}
    -2{\cal B} ae^g\right) \sin\widetilde{\theta}_2 \,d\widetilde{\theta}_2\wedge
    d\widetilde{\phi}_2  \\
  & + & \frac{N}{4}\,\left({\cal A}a+{\cal B}e^g\right)\,\left[\sin\psi
    (d\widetilde{\theta}_1\wedge d\widetilde{\theta}_2-\sin\widetilde{\theta}_1
    \sin\widetilde{\theta}_2\, d\widetilde{\phi}_1\wedge d\widetilde{\phi}_2)
    \right.\\ \nonumber
 & & \left.\qquad\qquad\qquad +\cos\psi(\sin\widetilde{\theta}_1\,
    d\widetilde{\theta}_2\wedge d\widetilde{\phi}_1-\sin\widetilde{\theta}_2\,
    d\widetilde{\theta}_1\wedge d\widetilde{\phi}_2)\right]\,.
\end{eqnarray}
The background dilaton can be extracted from the warped metric
or from \cite{mn,papatseyt}, and is given by \eqref{mndilaton}.
With this dilaton one computes 
\begin{equation}
   d(e^{-2\phi}J)\,=\,e^{-2\phi}\left(-2\frac{\partial\phi}{\partial \widetilde{r}}\,d\widetilde{r}\wedge
   J+dJ\right)\,.
\end{equation}
The Hodge dual of this expression is most easily found in terms of
vielbeins, since then 
\begin{equation}
  *\,(e^{\alpha_1}\wedge e^{\alpha_2}\wedge e^{\alpha_3}) \,=\, \frac{1}{3!}\;
    \epsilon^{\alpha_1 \alpha_2\alpha_3}_{\phantom{\alpha_1 \alpha_2\alpha_3}
    \mu_1\mu_2\mu_3} \; e^{\mu_1}\wedge e^{\mu_2}\wedge e^{\mu_3}\,.
\end{equation}
We choose the orientation so that $\epsilon^{123456}=1$.
Inverting \eqref{mnvielb} and replacing the coordinate differentials by
vielbeins one finds
\begin{eqnarray}\nonumber
  e^{2\phi}&*& d(e^{-2\phi}J)\,=\,\frac{1}{ \sqrt{N}\,F_2(\widetilde{r})}\,\left[\frac{F_2(\widetilde{r})\,
    (1+8\widetilde{r}^2-\cosh 4\widetilde{r})\,(4\widetilde{r}-{\rm sinh} 4\widetilde{r})}{F_1(\widetilde{r})\,{\rm sinh}^2 2\widetilde{r}}\, e^1\wedge 
    e^2\wedge e^6 \right.\\ \nonumber
 & &\left.+\frac{2\,(-1+2\widetilde{r}\coth 2\widetilde{r})}{\sinh 2\widetilde{r}}\, e^1\wedge e^3\wedge e^5 + 
    \frac{(1+8\widetilde{r}^2-\cosh 4\widetilde{r})}{\sinh^3 2\widetilde{r}}\,e^1\wedge e^4\wedge e^6 \,  \right. \\
 & &\left. + \left(-\frac{\widetilde{r}}{{\rm sinh}^2 \widetilde{r}}+\frac{1}{{\rm sinh} \widetilde{r}\,\cosh \widetilde{r}}
    -\frac{\widetilde{r}}{\cosh^2 \widetilde{r}}\right)\,e^2\wedge e^4\wedge e^5 \,\right.  \\ \nonumber
 & &\left.+\frac{F_2^2(\widetilde{r})}{{\rm sinh} \widetilde{r}\,\cosh \widetilde{r}}\,e^2\wedge e^3\wedge e^6 
    + \,\frac{(-4\widetilde{r}+{\rm sinh} 4\widetilde{r})}{\sinh^2 2\widetilde{r}}\,e^3\wedge e^4\wedge e^6\right]
\end{eqnarray}
with $F_1(\widetilde{r})$ and $F_2(\widetilde{r})$ defined by
\begin{equation}
  F_1(\widetilde{r})\,=\,-1+8\widetilde{r}^2+\cosh\, 4\widetilde{r}
    -4\widetilde{r}\,{\rm sinh}\, 4\widetilde{r} \,,\quad
  F_2(\widetilde{r})\,=\,\sqrt{-1+4\widetilde{r}\,(\coth\, 2\widetilde{r}
    -\widetilde{r}\,{\rm csch}^2\,2\widetilde{r})}\,.
\end{equation}
This three--form is the torsion for our background \eqref{hetfinal} with dilaton \eqref{mndilaton} and coefficients \eqref{fixcoef}.
In terms of global coordinates ($\widetilde{r}, \widetilde{\theta}_i, \widetilde{\phi}_i, \psi$) the torsion $H$ is given by
\begin{eqnarray}\nonumber
  H &=& \,-\frac{Na'}{4}\,\cos\psi\,d\widetilde{r}\wedge(d\widetilde{\theta}_1\wedge
    d\widetilde{\theta}_2-\sin\widetilde{\theta}_1\sin\widetilde{\theta}_2\,
    d\widetilde{\phi}_1\wedge d\widetilde{\phi}_2)\\ \nonumber
  & &-\, \frac{Na'}{4}\,\sin\psi\,d\widetilde{r}\wedge 
    (\sin\widetilde{\theta}_2\,d\widetilde{\theta}_1\wedge
    d\widetilde{\phi}_2-\sin\widetilde{\theta}_1\,d\widetilde{\theta}_2\wedge 
    d\widetilde{\phi}_1)\\ \nonumber
  & & +\,\frac{Na}{4}\,\sin\psi\,d\widetilde{\theta}_1\wedge d\widetilde
    {\theta}_2\wedge  (d\psi+\cos\widetilde{\theta}_1\,d\widetilde{\phi}_1
    +\cos\widetilde{\theta}_2\,d\widetilde{\phi}_2)\\ \nonumber
  & & -\,\frac{N}{4}\,\,(\sin\widetilde{\theta}_1\cos\widetilde{\theta}_2 
    - a\cos\psi\cos\widetilde{\theta}_1 \sin\widetilde{\theta}_2) \,
    d\widetilde{\theta}_1\wedge d\widetilde{\phi}_1\wedge d\widetilde{\phi}_2
\end{eqnarray}
\begin{eqnarray}\nonumber
  & & -\,\frac{N}{4}\,\,(\sin\widetilde{\theta}_2\cos\widetilde{\theta}_1 
    - a\cos\psi\cos\widetilde{\theta}_2 \sin\widetilde{\theta}_1) \,
    d\widetilde{\theta}_2\wedge d\widetilde{\phi}_1\wedge d\widetilde{\phi}_2 \\ 
  & & -\,\frac{N}{4}\,\,\sin\widetilde{\theta}_1\,d\widetilde{\theta}_1\wedge 
    d\widetilde{\phi}_1\wedge d\psi +\frac{N}{4}\,\,\sin\widetilde{\theta}_2\,
    d\widetilde{\theta}_2\wedge d\widetilde{\phi}_2\wedge d\psi \\ \nonumber
  & & -\,\frac{Na}{4}\cos\psi\,(\sin\widetilde{\theta}_2\,d\widetilde{\theta}_1\wedge 
    d\widetilde{\phi}_2\wedge d\psi  -\sin\widetilde{\theta}_1\,d\widetilde{\theta}_2
    \wedge d\widetilde{\phi}_1\wedge d\psi)\\ \nonumber
  & & -\, \frac{Na}{4}\,\sin\psi\sin\widetilde{\theta}_1\sin\widetilde{\theta}_2\,
    d\widetilde{\phi}_1\wedge d\widetilde{\phi}_2\wedge d\psi
\end{eqnarray}
with $a'=\partial a/\partial \widetilde{r}$. It is easy to check that this matches precisely the MN three--form \eqref{mnhns} and therefore confirms our background to be a valid superstring solution. Moreover, in \cite{gttwo} it was shown how one can construct vector bundles for this type of backgrounds that are derived from F--theory. Their behavior under conifold transition was also studied there and we will not repeat the arguments here.

%

The knowledge of a global heterotic solution that is consistent with our IIB orientifold setup now enables us to make some predictions for the global behavior of the IIB B--field $b_{y\theta_1}$ and
$b_{x\theta_2}$ and the complex structure of the $z_2$ torus as well.
The global heterotic metric \eqref{hetfinal}
contains of course also the global IIB B--fields. It was obtained by
connecting the local pictures in both theories and then using the
similarity of the heterotic metric with Maldacena--Nunez \cite{mn} to
obtain the global picture. In our case of interest, a background
with only NS flux, we know MN to be a valid solution in the IR (for small $r$).
Comparing \eqref{hetfinal} with the MN metric \eqref{mnglobal} determines $B_{NS}$\footnote{One
could as well solve the heterotic equations of motion (see also
\cite{grana})}. For small $\widetilde{r}$  
\begin{eqnarray}\nonumber
  b_{x\theta_2} & =& -1 + \frac{2}{3}\,\widetilde{r}^2 - \frac{14}{45}\,\widetilde{r}^4 + 
    {\cal O}(\widetilde{r}^{6}) \\ 
  b_{y\theta_1} & =& -1 + \frac{10}{3}\,\widetilde{r}^2-\frac{446}{45}\,\widetilde{r}^4+ {\cal O}(\widetilde{r}^{6})
\end{eqnarray}
Near $\widetilde{r} \to 0$ both B--field components are constant as one
might have expected. Having determined the B--field we can also
fix the $z_2$--torus, since they are related via \eqref{consti}. The complex structure is given as 
\begin{equation}
  \vert\tau_2\vert^2 \, = \, 1 + \frac{8}{3} \widetilde{r}^2 - \frac{32}{45}\, \widetilde{r}^4 +{\cal 
    O}(\widetilde{r}^{6})
\end{equation}
which tells us how the ($\theta_1,\theta_2$) torus varies as we move along the radial direction. In fact, near $\widetilde{r}\to 0$: $\tau_2 \equiv i\vert\tau_2\vert = i$ which, along with $\tau_1 = i$, completely specifies the IR (small $r$) behavior in
IIB.

The discussion in \cite{mn} does not extend to the UV regime. Here we
can rely on the analysis of \cite{grana} which embeds the MN background
in a class of interpolating solutions between MN and KS. Using
their results we can obtain the large $\widetilde{r}$ behavior of the
B--fields (the small $\widetilde{r}$ behavior agrees with that from MN):
\begin{eqnarray}\nonumber
  b_{x\theta_2} & = & -2\,e^{-2\widetilde{r}} + a_{\rm uv}\,(2\widetilde{r}-1)\,e^{-\frac{10\widetilde{r}}{3}} 
    - \frac{1}{2}\, a_{\rm uv}^2\,(2\widetilde{r}-1)^2\,e^{-\frac{14\widetilde{r}}{3}}+ {\cal O}(e^{-6\widetilde{r}}) \\
  b_{y\theta_1} & =&  -2\,e^{-2\widetilde{r}} - a_{\rm uv}\,(2\widetilde{r}-1)\,e^{-\frac{10\widetilde{r}}{3}} 
    -\frac{1}{2}\, a_{\rm uv}^2\,(2\widetilde{r}-1)^2\,e^{-\frac{14\widetilde{r}}{3}}+ {\cal O}(e^{-6\widetilde{r}})
\end{eqnarray}
 where $a_{\rm uv}=-\infty$ corresponds to MN in the interpolating
scenario. The complex structure then results in
\begin{equation}
  \vert\tau_2\vert^2 \,=\, \,1+2\,e^{-4\widetilde{r}}- a_{\rm uv}\,(2\widetilde{r}-1)\,e^{-\frac{4\widetilde{r}}{3}} 
    + \frac{1}{2}\,a_{\rm uv}^2 \,(2\widetilde{r}-1)^2\,e^{-\frac{8\widetilde{r}}{3}}+ {\cal O}
    (e^{-\frac{16\widetilde{r}}{3}})\,.
\end{equation}
Notice that for $\widetilde{r}\to\infty$ the IIB 
$B$--fields vanish and the complex structure approaches again
$\tau_2=i$ and $\tau_1=i$.

This finalizes the study of the global heterotic background \eqref{hetfinal}. With a different deformation of our local background \eqref{ourlocal} we can also propose a global IIB metric. The heterotic metric is essentially the same as that for the IIB orientifold after two T--dualities.
In \cite{gttwo} it was shown that we can also obtain
the same local limit as the $\mathcal{N}=2$ background studied in \cite{ntwofivebrane}. This strongly suggests that the manifold we obtained admits an $\mathcal{N}=2$ supersymmetric solution and only fluxes break supersymmetry to $\mathcal{N}=1$. (Recall that in IIB fluxes only act as an overall conformal warp factor, which is not visible in local coordinates, there we can always absorb a warp factor into the coordinate differentials.) This is precisely the scenario discussed by Gopakumar and Vafa \cite{gopakumar, vafa}. Therefore, although we do not exactly recover a (conformal) conifold in the strict sense\footnote{Recall that our ``walk through the duality chain'' in section \ref{chain} led us to a metric that resembles a deformed conifold, but had two different sized $S^2$ and one of them was ``squashed'', for example.}, we still seem to have recovered a valid string theory background at the end of the duality chain in IIB.

Let us summarize the accomplishments of this section. We found non--K\"ahler backgrounds in type I and heterotic theory that are T-- and U--dual, respectively, to (an orientifold of) the K\"ahler IIB backgrounds constructed in section \ref{chain}. This means, they are part of a long duality chain that eventually relates them via a flop in $\mathcal{M}$--theory. To our knowledge, these backgrounds provide the first attempt of constructing geometric transitions in heterotic (or type I) theory. It would be very interesting to study the effects on the underlying gauge theory and to find an interpretation in topological string theory. We will comment in the next section on the challenges that non--K\"ahler manifolds pose in this context.

Furthermore, although most of this work only describes backgrounds in the local limit, we were able to propose a global extension for the heterotic background after transition by using a similarity with the Maldacena--Nunez solution. We suggest that a larger class of heterotic solutions is possible by leaving the coefficients $\mathcal{A}_i$ in \eqref{hetfinal} generic, but we confirmed the torsional relation for the case where they match with the MN solution. In general, they will be determined by the torsional relation, where the torsion three--form is given in terms of the U--dual IIB RR form, as illustrated in \eqref{fluxlogic}. We provided a (local) toy example in which all fluxes were consistent with this logic.

\setcounter{equation}{0}
\section{Conclusion and Outlook}

\subsection{Summary}

The purpose of this work was to verify Vafa's duality chain, to discuss mirror symmetry with NS flux on conifold geometries and to propose new non--K\"ahler backgrounds that are also related by a geometric transition in IIA, type I and heterotic SO(32).

We started in section \ref{mirrorfirst} by showing how resolved and deformed conifold are approximately mirror to each other, although they do not possess the same number of isometries. As anticipated by Strominger, Yau and Zaslow, we had to employ a ``large complex structure limit'' and could only recover a semi--flat version of the deformed conifold, i.e. one that does indeed have the same number of isometries as the resolved conifold. We argued that the typical deformed conifold metric could be restored with a special coordinate transformation. This assumption relied on our use of local coordinates and might not hold true in a global scenario. The use of local coordinates also enabled us to hold the $T^3$ fiber coordinates fixed and boost the complex structure ``by hand'', such that the T--duality fiber becomes small compared to the base. The fact that resolved and deformed conifold are only mirror to each other in the limit when both resolution and deformation parameter become small could also be observed in this local limit. But we do not believe this to hold globally, since the large complex structure boost we perform alters the manifold non--trivially. In the local limit, however, this boost can be interpreted as a trivial coordinate redefinition.

It should nevertheless be clear that globally the mirror of the Calabi--Yau resolved conifold is not the Calabi--Yau deformed conifold, as was pointed out in \cite{aganagic, vafa}, but the metric we found resembles the local limit of a deformed conifold. 

Equipped with the established mirror symmetry between a local resolved and a local deformed conifold we determined the influence of NS--flux on this picture. In accordance with the literature on this topic we found a non--K\"ahler manifold as the mirror of the local resolved conifold. This mirror manifold has nevertheless close resemblance to the local deformed conifold. The only difference is that the $T^3$ fibers acquire a ``twist'' by the B--field. Although this seems very close in spirit to the half--flat manifolds found in \cite{jan}, we argued in section \ref{discuss} that this manifold which we call ``non--K\"ahler deformed conifold'' is not half--flat. There are several resolutions to this discrepancy. First, the half--flat manifolds from \cite{jan} do not only admit a half--flat but also a symplectic structure, which is in agreement with other observations \cite{minas} that IIA backgrounds should always be symplectic.
This fits perfectly with the torsion classes for our backgrounds that were computed in section \ref{torsion}. We demonstrated that the local IIA non--K\"ahler backgrounds we constructed do not admit a half--flat but a symplectic structure.

Another difference to the half--flat models is that our mirror background does not lift to a purely geometric solution in $\mathcal{M}$--theory and does therefore not exhibit $G_2$ holonomy, but only a $G_2$ structure. There are additional fluxes turned on, as we showed by explicitely T--dualizing a IIB background which looks locally like a resolved conifold, but was constructed from F--theory. This F--theory setup implies that we actually T--dualize an orientifold. Nevertheless, the resolved conifold metric is completely invariant under the orientifold operation we constructed in section \ref{ftheorysetup}. We also made a generic ansatz for the fluxes which are allowed on this orientifold and used this background to ``walk along Vafa's duality chain'' in sections \ref{iia} and \ref{iib}. It turned out that the mirror metric was also perfectly consistent with an orientifold in IIA. The ``non--K\"ahler deformed conifold'' we find under mirror symmetry with NS flux was invariant under the combined action of the IIB orientifold and three T--dualities. 

\begin{figure}[ht]
\begin{center}
  \begin{picture}(300,210)\thicklines
    \put(10,0){\framebox(80,40){\begin{minipage}{75pt}
      IIB D5 and\\ D7/O7 on\\ K\"ahler resolved
      \end{minipage}}}\thinlines
    \put(91,20){\line(1,0){128}}\thicklines
    \put(220,0){\framebox(80,40){\begin{minipage}{75pt}
      IIB fluxes\\ on K\"ahler\\ deformed
      \end{minipage}}}
    \put(137,18){\begin{minipage}{50pt} geometric transition \end{minipage}}
    \put(50,41){\vector(0,1){28}}
    \put(58,48){\begin{minipage}{50pt} mirror \end{minipage}}
    \put(10,70){\framebox(80,40){\begin{minipage}{75pt}
      IIA D6 and O6 on ``non--K\"ahler deformed''
      \end{minipage}}}
    \put(50,111){\vector(0,1){28}}
    \put(58,122){\begin{minipage}{50pt} lift \end{minipage}}
    \put(10,140){\framebox(80,40){\begin{minipage}{75pt}
      G2 structure\\ with G--flux in M--theory 
      \end{minipage}}}
    \put(91,160){\vector(1,0){128}}
    \put(150,162){\begin{minipage}{50pt} flop \end{minipage}}
    \put(220,140){\framebox(80,40){\begin{minipage}{75pt}
      G2 structure\\ with G--flux in M--theory
    \end{minipage}}}
    \put(260,139){\vector(0,-1){28}}
    \put(268,122){\begin{minipage}{50pt} descent \end{minipage}}
    \put(220,70){\framebox(80,40){\begin{minipage}{75pt}
      IIA fluxes on ``non--K\"ahler resolved''
      \end{minipage}}}\thinlines
    \put(91,90){\line(1,0){128}}\thicklines
    \put(137,88){\begin{minipage}{50pt} geometric transition \end{minipage}}
    \put(260,69){\vector(0,-1){28}}
    \put(268,48){\begin{minipage}{50pt} mirror \end{minipage}}
  \end{picture}
  \caption{The duality chain proposed in section \ref{chain}. The backgrounds in IIA have to be replaced by non--K\"ahler versions of deformed and resolved conifold and the $\mathcal{M}$--theory lift does not possess $G_2$ holonomy anymore.} \label{ourchain}
\end{center}
\end{figure}
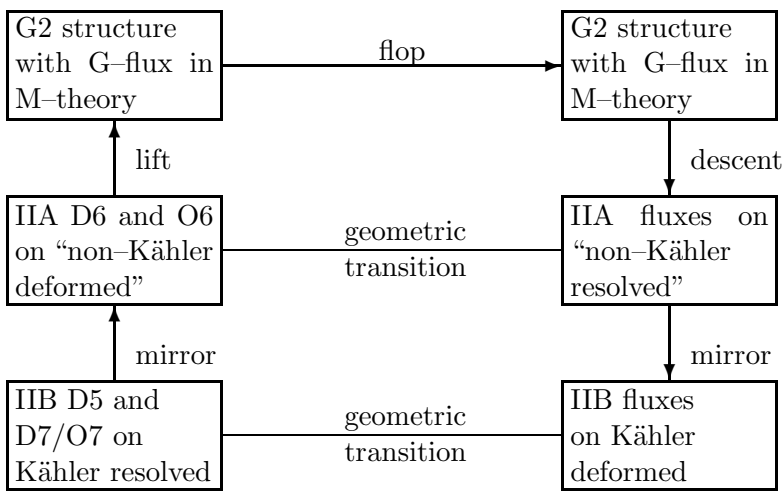

We showed beyond any reasonable doubt that Vafa's duality chain figure \ref{vafachain} should be modified for a full supergravity background that will necessarily include NS and RR flux. NS flux modifies the mirror symmetry between two Calabi--Yaus. Even if we do not start with a Calabi--Yau (since the base of the F--theory setup constructed in section \ref{ftheorysetup} is only conformally K\"ahler) we find a mirror geometry that has a $T^3$ fibration which is twisted due to the $B_{NS}$ field. We therefore proposed a modification of Vafa's duality chain in section \ref{chain} that can be pictured as in figure \ref{ourchain}.

The two non--K\"ahler backgrounds are related via a flop in $\mathcal{M}$--theory and therefore transition duals. We also showed that away from the orientifold point we can have flux components turned on that lift to a closed 3--form in $\mathcal{M}$--theory which is oriented along the 3--cycle that shrinks under the transition. This 3--form  makes the flop a smooth transition, since it can be interpreted as the imaginary part of the complexified volume of the three cycle. Even if the cycle shrinks to zero, there is no singularity because the imaginary part remains finite. 

The consistency of our calculations can be argued from the fact that we do indeed recover a K\"ahler background at the end of the duality chain in IIB. It resembles the local version of a deformed conifold. We furthermore showed that the fluxes do not change during the flop tansition, but the two--cycle that the D5--branes wrapped before transition was shrunk and a dual three--cycle blown up. This has to be true since the flop in ${\mathcal M}$--theory was performed such that resolved and deformed geometry are exchanged (which implies the exchange of blown--up two and three cycle, respectively).

Moreover, our F--theory setup also allowed us to introduce additional D7--branes in IIB (D6--branes in IIA) that lead to an additional global symmetry for the underlying gauge theory. We constructed flavor groups in $SU(2)^{16}$. But as long as we are far away from the orientifold fixed points, these flavors are heavy and the effective low energy theory is pure $\mathcal{N}=1$ SU(N) Super--Yang--Mills, as in Vafa's scenario. Nevertheless, it would be interesting to determine the effect of the additional branes on the superpotentials.

In section \ref{heterotic} we left Vafa's duality chain and considered a new duality chain that took us to type I and heterotic SO(32), see figure \ref{hetenchain}. We had to consider a different F--theory setup, because T--duality of a IIB orientifold can only lead to type I if the T--duality directions are orthogonal to the orientifold planes, such that one obtains spacetime filling orientifold planes. This second F--theory setup led us to consider an orientifold action that does not leave conifold geometries invariant. We had to project out certain components of the metric. What we constructed was essentially a toroidal orbifold. We could nevertheless argue that the type I and heterotic metrics also possess contractible two-- and three--cycles and should therefore be transition dual. The reason we believe this to be true is that performing two T--dualities on the full IIB background (without projecting out certain components of the metric) produces almost exactly the same metric. Since the IIB backgrounds are related via a geometric transition (as shown in section \ref{chain}), also the type I and heterotic backgrounds from section \ref{heterotic} should be transition duals. They are connected via a very long duality chain to a flop in $\mathcal{M}$--theory.

This interpretation is of course not as rigorous as the one for the type II backgrounds, since we considered different orientifolds in section \ref{chain} and \ref{heterotic}. The gauge theory interpretation and topological string analysis still remain unclear.

Let us also mention that the IIB supergravity equations of motion and the torsional constraint in heterotic pose serious constraints on the IIB RR and NS flux. We constructed a toy example in section \ref{hetsec} for which these relations can be fulfilled {\it and} be consistent with the relation imposed by T--duality. 

Finally, we were also able to leave the local limit which was imposed on us throughout this work by the lack of the metric on the F--theory fourfold. We only know the local metrics on the bases\footnote{It should also be clear after the discussion in section \ref{mirror} that we have to work in the local limit if we want to follow Vafa's duality chain. We can only find a mirror in IIA that has some resemblance with a deformed conifold if we work with local coordinates in the semi--flat and large complex structure limit. For the heterotic duality chain there is no such restriction.}. However, there are known globally valid backgrounds on conifold geometries in IIB. We exploited the fact that the Maldacena--Nunez (MN) \cite{mn} background can be S--dualized to heterotic theory and fulfills the torsional relation. We showed that their metric reduces in the local limit to the one we found in heterotic after transition (with an appropriate choice of B--field and complex structure in our IIB backgrounds). Therefore, we argued that we can reverse this argument and postulate the MN background as one valid {\it global} solution for our heterotic scenario. We proposed a more generic background in which the coefficients would have to be fixed by imposing the torsional relation.

\subsection{Outlook and Remaining Open Questions}\label{outlook}

The calculations presented in this article raise (at least) four immanent questions.
Can one find a global solution in IIB and repeat the analysis for the type II duality chain (and for the heterotic chain as well)?
How does the open/closed duality argument based on topological string amplitudes that Gopakumar and Vafa \cite{gopakumar} gave for Calabi--Yaus need to be modified for the non--K\"ahler backgrounds we constructed in IIA?
What is the interpretation of the geometric transition in heterotic theory?
What are the implications for the resulting gauge theory, can one for example compute its superpotential?
We will address these issues and a few other remaining problems in this section.

\subsubsection*{Global Solutions}

As already pointed out several times, we lack the global metric for two reasons. First, there is no known supersymmetric solution of D5--branes on the resolved conifold. Attempts on solving the supergravity equations of motion \cite{pandozayas} have not been successful (the solution violates the primitivity condition for the fluxes \cite{cvetic}). Second, we use an F--theory fourfold that contains the resolved conifold as its base to gain a supersymmetric setup and to introduce ``flavor branes''. We would need to find the global metric of this fourfold, which is in general very hard.
The F--theory base will furthermore contain singularities at the points where the F--theory torus degenerates. We also used the local metric to ensure we work in a patch that does not have these singularities.

But let us assume we had a global supersymmetric solution of D5--branes on the resolved conifold. If we wanted to find the global mirror of this manifold we would want to apply SYZ and perform three T--dualities. The discussion in section \ref{mirrorfirst} should have made it clear that simply T--dualizing the resolved metric does not produce the deformed conifold. We needed to impose a non--trivial boost of the complex structure of the resolved conifold metric. In a global setup, this boost (which was given by the large functions $f_{1,2}$ in \eqref{boost}) will be a function of the base--coordinates (the solutions found in \eqref{solbeta} contain $\alpha$, which would in general depend on the coordinates $(r,\theta_1,\theta_2)$, if we had not worked in the local limit). This means, it cannot be interpreted as a coordinate transformation, but it has a non--trivial influence on the complex structure of the manifold. Effectively, we want to shift (compare to \eqref{boost})
\begin{equation}
  d\theta_i\;\to\;(1+i\,f_i)\,d\theta_i\,,
\end{equation}
which might not always be integrable, depending on the specific value $f_i$ takes in a global framework. One would therefore have to make sure that the background after this non--trivial boost of the complex structure still fulfills the supergravity equations of motion\footnote{In a global setup one would also see $(x,\theta_1)$ and $(y,\theta_2)$ as  spheres and not tori. This would require a different approach to make the $T^3$ fiber large, but a boost of the $\theta_i$ directions should still be possible. One just has to ensure that this boost also produces $dx\,d\theta_1$ and $dy\,d\theta_2$ crossterms.}. 

As already pointed out towards the end of section \ref{mirrorfirst}, one could T--dualize the global resolved conifold metric without any non--trivial boost, since it possesses three isometry directions. The result will be another Calabi--Yau with $T^3$ fiber, but it will not be the deformed conifold. If we want to confirm Vafa's duality chain we have to take the limit where resolved and deformed conifold are actually mirror, meaning where both resolution and deformation parameter are small. We argued in section \ref{mirrorfirst} that one possible way to do that was given by using local coordinates. It would be interesting to find out if one can also realize the mirror symmetry in this limit with a {\it global} setup, but the simple fact that the mirror of the resolved conifold lacks the $d\theta_1\,d\theta_2$--cross term, which is typical for the deformed conifold, seems to forbid that.

So, not only do we lack a global solution from a supergravity viewpoint, but it is also unclear how one could verify Vafa's duality chain with a global metric. Three T--dualities on a global resolved conifold will not take us to a global deformed conifold unless we find some other method to stay close to the transition point than the large complex structure limit proposed here. In our analysis, to establish the mirror symmetry between resolved and deformed conifold, we were literally forced into the local limit.

If we want to follow the heterotic duality chain we do not aim at finding the deformed conifold metric. Therefore, we can T--dualize a IIB metric without any non--trivial manipulations to its complex structure. This means, if we knew a global IIB solution we could find a global heterotic solution under U--duality. There is one problem, though: we need an orientifold of IIB, otherwise two T--dualities will only return us to IIB and not lead to type I. In the local setup we simply projected out metric components that were not invariant under the orientifold action. The resulting metric took the form of a toroidal orbifold, which seems promising in terms of admitting supersymmetric solutions \cite{daskat}. Nevertheless, for a global solution we should carefully check whether supersymmetry is preserved under this projection. But this side of the story seems to face less difficulties that the type II duality chain.

\subsubsection*{Topological Models}

Let us assume we had global solutions in type II and heterotic, or at least we had some global information about the manifolds. Which global information do we need precisely to define a topological sigma model (see appendix \ref{sigma}) and topological string amplitudes on these backgrounds? 

We discuss the IIA case first. Our observations here will be closely linked to generalized complex geometries (GCG) \cite{hitch, gualtieri}. The goal would be to repeat the arguments from \cite{gopakumar} that are summarized in appendix \ref{conjecture}. One would like to show that the open string partition function (describing D6--branes on the ``non--K\"ahler deformed conifold'') agrees with the closed string partition function (describing flux on the ``non--K\"ahler resolved conifold'').

Our case differs from \cite{gopakumar} because we consider non--K\"ahler manifolds. As shown in \cite{gates} and reviewed in appendix \ref{sigma}, one can construct a (2,2) supersymmetric sigma model on a non--K\"ahler target if $H\ne 0$ and the target manifold is bi--Hermitian, i.e. it admits two complex structures and the metric is Hermitian w.r.t. both of them. This is also the data required to define a {\it generalized K\"ahler} structure \cite{gualtieri}, see appendix \ref{gcg} for a brief introduction to generalized complex geometry. It has furthermore been shown that any other extension of (2,2) sigma models, e.g. with semi--twisted chiral superfields, also requires a generalized K\"ahler target \cite{lindstrom}. We therefore come to the first conclusion that the global IIA manifolds need to possess a bi--Hermitian or generalized K\"ahler structure. In GCG a generalized K\"ahler structure is defined by two {\it commuting} generalized complex structures (this does not mean that the two complex structures of the bi--Hermitian structure commute).


It has then been shown by Kapustin and Li \cite{kapuli}, see also appendix \ref{gensigma}, that one can define a topological sigma model for generalized K\"ahler targets, if the manifold is generalized Calabi--Yau w.r.t. to one of the generalized complex structures. Then the topological observables depend only on the cohomology of this generalized complex structure, but not on both.

If we wanted to use their results to compute the string partition functions we would have to identify a generalized K\"ahler structure on the global IIA manifolds. With the knowledge of the global background one could repeat the torsion class analysis from section \ref{torsion} to find out what type of non--K\"ahler manifold we constructed. If it turns out to be symplectic also {\it globally}, it cannot be generalized K\"ahler, because there exists only one generalized complex structure on a symplectic manifold, it is of the type \eqref{symplec}.
So far, there is no direct relation between the torsion classes generally used to classify non--K\"ahler backgrounds and generalized complex structures. It would be fascinating to discover a relation between both formalisms. Promising attempts in this direction seem to be to relate the fundamental two and three--form (the objects used for torsion classes) to pure spinors (objects in GCG) \cite{minas}. Both approaches have been used to derive supersymmetry conditions, either in terms of torsion classes \cite{minakas,lust} or in terms of pure spinors \cite{minas}. It should be possible to find a mapping between those criteria.

The question we have to answer is: Can T--duality with B--field convert our IIB background into a generalized K\"ahler background in IIA? This should be the case if our IIB background was K\"ahler with $H=0$ and described by a (2,2) supersymmetric sigma model (the assumptions made by Gopakumar and Vafa \cite{vafa,gopakumar}). If we then find a mirror that is non--K\"ahler and has $H\ne0$, there should still exist a (2,2) supersymmetric sigma model, because T--duality preserves supersymmetry. Therefore, this manifold would have to be generalized K\"ahler.

But this is not the case we constructed. First, our IIB starting background has non--trivial $H$. For a (2,2) sigma model to exist, the manifold would have to be non--K\"ahler with torsion $H$ precisely as outlined in \cite{gates}. There would have be two Hermitian complex structures that are covariantly constant w.r.t. to two different torsional connections. The arguments in \cite{gttwo} suggest that the base of the F--theory fourfold is actually K\"ahler and only acquires a conformal warp factor due to the D5--branes. We would have to find out if a conformal factor changes the metric in such a way that it allows for a bi--Hermitian structure. The impact of a conformal rescaling on the torsion classes has been considered in \cite{lust}. For a conformal K\"ahler manifold only $\mathcal{W}_4$ and $\mathcal{W}_5$ can be non--zero. This tells us the intrinsic torsion of the manifold, it remains to be seen if this gives rise to two complex structures (invariant under two torsional connections) such that the metric is Hermitian w.r.t. both of them.

Furthermore, consider the mirror we find in IIA. On a global manifold we do not want to impose the non--trivial boost of the complex structure, because we do not know if it preserves supersymmetry. If we did not boost the complex structure, we would not find a B--field in the mirror\footnote{This is a feature for all mirror symmetries with only {\it electric} NS flux. If the B--field has one leg along the T--duality directions it will be converted into metric. If the metric does not possess any ``electric'' (in the same sense as above) cross terms, then no new B--field will be generated. This is the same as the case discussed in \cite{jan}.}. Recalling equation \eqref{bnsiia}, we see that the B--field scales with the large function $f_i$, without the complex structure boost there would not have been any B--field (we can also adopt the point of view that this B--field is merely a large complex structure artefact which vanishes when we leave the large complex structure limit). But that means we would find a non--K\"ahler target with $H=0$. There is no (2,2) sigma model for this case.

There seem to be only two explanation for this puzzle: either the manifold will still admit a K\"ahler structure although it has a non--trivial B--field fibration (but that seems unlikely, if we were not able to find one in the local limit, it will in general be even harder on a global metric) or we have to modify our arguments due to RR flux.

A type II string theory background in the absence of any flux requires a Calabi--Yau manifold to preserve $\mathcal{N}=2$ supersymmetry in four dimensions\footnote{The D--branes or fluxes break another 1/2 of the supersymmetry.}. Turning on only NS flux will make the manifold in general non--K\"ahler and we are in the sigma model framework of \cite{gates}. If we have RR flux turned on, the manifold can still be non--K\"ahler even if the NS flux vanishes. But this does not fit into any sigma model framework. It is still completely unclear how to incorporate RR flux into a sigma model. The argument by Gopakumar and Vafa \cite{vafa,gopakumar} rested on the assumption that the Calabi--Yau would be modified when fluxes are turned on but that the topological analysis remained unchanged. They treated the fluxes as a perturbation that created the superpotential, but did not influence the topological string amplitudes. They even argued that the topological amplitudes do not depend on RR flux.

This seems to clash with our considerations above: turning off the NS field should give a K\"ahler target based on sigma model arguments, but it will not in general do so as RR flux can still warp the manifold in a non--trivial way. It might be true that RR fluxes do not influence the topological amplitudes, but since they influence the target manifold, they should still dictate the sigma model.

To bring all these different observations together and resolve above puzzles will have to be left for future work. Independent of the questions we raised here, it would be of great interest to find a way to include RR flux into sigma models.

Our interest into GCG goes beyond its connection to sigma models. For example, 
one would like to make a connection between B--field transforms in GCG and T--duality with B--field. An ordinary complex structure gives rise to a generalized complex structure on the direct sum of tangent and cotangent bundle $T\oplus T^*$, which is of type \eqref{complex}
\begin{equation}
  \mathcal{J}_J\,=\,\begin{pmatrix} -J & 0 \\ 0 & J^* \end{pmatrix}
\end{equation}
where $J$ is a complex structure in the usual sense on $T$. The B--field transform of this generalized complex structure is found to be \eqref{btransformj}
\begin{equation}
  \mathcal{J}\,=\,e^{-B}\mathcal{J}_J e^{B} \,=\,
    \begin{pmatrix} -J & 0 \\ BJ+J^*B & J^* 
    \end{pmatrix}\,.
\end{equation}
Note that in this case only the $(0,2)$ component of the real two--form $B$ has any effect. B--field transforms of complex structures are always block--lower--diagonal, an observation used in \cite{kapustin}. This actually seems to indicate that only a B--field transform a.k.a. {\sl T--duality} with non--vanishing {\sl magnetic} part of the NS flux leads to GCG (generalized K\"ahler) but {\sl not} T--duality with only electric NS flux (i.e. (1,1) part) as in our case or in \cite{jan}. 

The analysis in \cite{kapustin} dealt with T--duality ``in all directions'' on a six torus which leads from IIB to IIB. The situation might be different when one considers an odd number of T--dualities going from IIA to IIB. Furthermore, it has been noted that T--duality with magnetic NS flux leads to {\sl non--geometric} solutions \cite{wecht,simeon,nongeomet}. Usually, one thinks of geometric quantities as sections of bundles associated with the frame bundle that transform from chart to chart under diffeomorphisms. In string theory, the symmetry group is larger than diffeomorphisms. There may exist a more general $SO(d,d)$ valued transition, this will for example mix metric and B--field, making them not well--defined separately. This is what \cite{wecht, simeon, nongeomet} mean by ``non--geometric''. It seems, Kapustin \cite{kapustin} claims these non--geometric backgrounds are realized as GCG under T--duality with magnetic B--field, but T--duality with only electric NS field would not lead to GCG.

As elaborated in appendix \ref{gcg}, holomorphic 3--forms and symplectic 2--forms play an important role in GCG (they are related to pure spinor line bundles, see \eqref{purespinor}). This is also reminiscent of topological string theory and might hint to a deeper connection. There has been a series of papers which connects string theory backgrounds to the notation of pure spinors developed in the mathematical framework \cite{minas}. It has further been noted that mirror symmetry seems to have an interpretation as the exchange of these two pure spinors \cite{minasfidanza} and that GCG provide a good framework for compactifications on manifolds with $SU(3)\times SU(3)$ structure \cite{jangrana}.
It is therefore our believe that GCG will help us to understand many properties of string theory that we were so far not able to describe properly.


\subsubsection*{Heterotic Interpretation}

Heterotic string theories are based on closed strings only. So, how would one interpret a geometric transition, which is an open/closed duality in type II, for heterotic strings? 

We already pointed towards one interpretation as a brane/flux duality. Since under S--duality D5--branes become NS5--branes, we can consider the heterotic geometric transition to be a duality between a background with NS5--branes and a background with only NS flux, see figure \ref{hetenchain}. This is of course a target space perspective. To be able to say anything about topological heterotic strings one would need to find a sigma model that describes supersymmetry for only the right movers, i.e. with (0,2) supersymmetry, and find a way to perform a twist that renders the theory topological. Such ``half twists'' (of the right moving sector only) have been considered \cite{wittentop, wittwo, katz, kapu}, but it is not clear under what circumstances one can obtain a topological theory.

In \cite{gttwo}  we suggested a (0,2) sigma model by starting with the (2,2) supersymmetric sigma model \eqref{sigmaaction} and breaking supersymmetry for the left movers explicitely.
We added non--interacting fields {\it only} in the left--moving sector. This breaks the left moving supersymmetry, and one might therefore hope
to obtain an action for (0,2) models from the (2,2) model, at least {\it
classically}. On the other hand, a possible (0,2) action is also restricted
because this will be the action for heterotic string. It turns out,
there are few allowed changes one can do to find the
classical (0,2) action from a given (2,2) action, see section 3.1 in \cite{gttwo} for details.

The half--twist is performed
only to the right moving sector of the sigma model.
This was presented first in \cite{wittentop}. 
In the future we would like to understand the Chiral de Rham complex CDR and the chiral differential operators CDO
\cite{malikov} in more detail, since they have emerged as the relevant mathematical objects for (half--)twisted conformal field theories \cite{wittwo, kapu}. Constructing a (0,2) topological sigma model with and without NS5--branes would be very fascinating, not only under the considerations we raised within this article.


\subsubsection*{Gauge Theories and Superpotentials}

We already commented briefly on the dual gauge theory of our supergravity backgrounds in the end of section \ref{ftheorysetup}. Since we introduced additional D7--branes in IIB with our F--theory setup, there will be an additional global symmetry for the gauge theory. We argued that we find a global symmetry group $SU(2)^{16}$ both in IIA and IIB. This was due to the fact that in IIB every orientifold fixed point contributes four D7--brane giving rise to an $SO(8)$ that is broken by Wilson lines to $SO(4)\times SO(4)\simeq SU(2)^4$. This is consistent with the IIA orientifold that contains eight fixed points, each accompanied by two D6--branes. The symmetry group generated by eight stacks of D6 is therefore $SO(4)^8\simeq SU(2)^{16}$.
Apart from these ``flavor branes'' there are also the usual $N$ D5 (in IIB) or D6 (in IIA) that carry an $SU(N)$ gauge theory on their worldvolume. Both IIA and IIB setups give rise to an $SU(N)$ gauge theory with flavor in the fundamental representation of $SU(2)^{16}$. Possibly more interesting global symmetry groups can be generated by considering a different distribution of the F--theory 7--branes. In \cite{keshavmukhi} it was shown that one can even construct the exceptional gauge groups $E_6$, $E_7$ and $E_8$, which are of particular interest for Grand Unified Theories (GUT's).

One of the remarkable results from \cite{vafa}  was to show that the flux generated superpotential does indeed agree (in lowest order) with the Veneziano--Yankielowicz superpotential for Super--Yang--Mills (SYM) theory. This superpotential has actually obtained corrections from field theory \cite{schwetz, ckl} as well as from string theory \cite{dv} considerations. One question we would like to address is whether generalized topological models (taking the non--K\"ahler structure of the target manifold into account) might be better suited to reproduce these corrections.

Furthermore, we would like to address the additional global symmetry. The field theory analog to Veneziano--Yankielowicz for an $SU(N)$ theory with matter is given by the Affleck--Dine--Seiberg superpotential \cite{affleck}, see also \cite{peskin} for a review. It would be interesting to see if we could reproduce this superpotential or if we would find an extension to it when including the flux due to D7--branes. We would need the precise supergravity solution to see which fluxes are actually turned on. In our setup, the charge of the D7--branes is immediately cancelled by the orientifold planes. We would have to move the orientifold planes away from the flavor branes to observe their effect. This would lead to non--perturbative corrections.

Flux induced superpotentials have been studied in \cite{vafatayl, gukov, fluxpot}. In IIA, the part generated by RR flux would be
\begin{equation}
  W_{IIA}^{RR}\,=\,\int(F_6+J\wedge F_4+J\wedge J\wedge F_2+F_0\,J\wedge J\wedge J)\,.
\end{equation}
Allowing for NS flux leads in general to a complexification of the fluxes \cite{vafa}, so that one finds in addition \cite{jan, gtone}
\begin{equation}
  W_{IIA}^{NS}\,=\,\int(J+iB_{NS})\wedge(F_4+ie^{-\phi}\,d\Omega)\,,
\end{equation}
where $\phi$ is the dilaton as usual.
Note the explicit dependence on $d\Omega$. In \cite{jan} only the real part of $\Omega$ was non--closed, but since we find manifolds with both real and imaginary part non--closed, it was conjectured in \cite{gtone} that the full $d\Omega^{(2,2)}$ will contribute to the superpotential in this way.

Similar remarks hold true for IIB. For a K\"ahler manifold one only expects a superpotential of the form \cite{vafatayl}
\begin{equation}
  W_{IIB} \,=\, \Omega\wedge(F_3+\lambda\,H_{NS})\,,\qquad\qquad\mbox{with}\qquad
    \lambda \,=\, \chi+ie^{-\phi}\,,
\end{equation}
since $dJ=0$ and $d\Omega=0$ on a K\"ahler manifold. For a non--K\"ahler manifold we would expect the same complexification of fluxes such that $H_{NS}\to H_{NS}+idJ$. Obtaining a global supergravity solution in IIB would allow us to calculate those superpotentials in IIB and the IIA mirror accordingly.

The interpretation of the gauge theory in heterotic theory is less clear than in type II. Here, we cannot use the D--brane construction anymore. One would rather expect a gauge theory with gauge group contained in SO(32), but it remains to be seen if this is consistent with U--duality and the gauge theory in IIB. The geometric transition we proposed for type I and heterotic is simply a connection between two metrics as this point and we do not have any proof of a deeper relation on the topological or gauge theory level yet.

We can nevertheless speculate about the superpotential. As pointed out in \cite{realm}, the type I superpotential should be of the form \cite{dasbecksq, carcudallu}
\begin{equation}
  W_{I}\,=\,\int(F_3+i\,dJ)\wedge\Omega + W_{CS}^A + W_{CS}^\omega
\end{equation}
where $W_{CS}$ indicates a Chern--Simons superpotential for the SO(32) gauge field $A$ or spin connection $\omega$, i.e. there is a gauge and a gravitational part to the Chern--Simons potential, which is of the general form
\begin{equation}
  W_{CS}^A \,=\, \int {\rm Tr}\left(A\wedge dA+\frac{2}{3}\,A\wedge A\wedge 
    A\right) \wedge \Omega
\end{equation}
and similarly for $W_{CS}^\omega$. 

The heterotic superpotential should take the form \cite{dasbecksq, carcudallu}
\begin{equation}
  W_{het} \,=\, \int (H+i\,dJ)\wedge\Omega
\end{equation}
with torsion 3--form $H$ that satisfies $dH\,=\,$Tr$\,R\wedge R-\frac{1}{30}\,$Tr$\,F\wedge F$, where $F=dA$ is the gauge fieldstrength and $R$ is the curvature two--form..

\subsubsection*{Further Open Problems}

In order to use geometric transition models for realistic theories one would want to find compact internal manifolds that allow for such transitions. As the analysis in this work was based on local metrics, it should also be applicable to compact manifolds that admit a metric of the type we discussed, e.g. \eqref{startmetric}, in a local patch. In other words, we need compact manifolds that admit locally conical singularities which are resolved or deformed, but they do not have to resemble conifolds globally. 

On such compact spaces one would have to provide a mechanism to cancel the charges of the D5 or D6--branes. As already mentioned in the introduction, this can be done by introducing \=D--branes or orientifold planes \cite{blumenhagen}--\cite{denef}, which would have the further advantage to allow for supersymmetry breaking.

We would also like to give an independent argument for the absence of D--branes after the transition (the orientifold planes and their accompanying flavor branes should be undisturbed by the transition). As observed in \cite{normform}, D6--branes lifted to a $G_2$ holonomy manifold (more precisely a Taub--NUT space) give rise to a normalizable harmonic (1,1) form. If the branes are absent, one would expect this form to become non--normalizable or cease to exist.

In conclusion, we constructed a variety of new (non--compact) string theory backgrounds on non--K\"ahler manifolds in IIA, type I and heterotic SO(32). These backgrounds come in pairs and we argued them to be related by a geometric transition, meaning that one of them contains branes, the other one only flux. A rigorous proof of this claim and the implications for weak--strong coupling dualities in the underlying field theories are left for future work.

\section*{Acknowledgments}

First of all I would like to thank both my advisors, Melanie Becker and Jan Louis for making this joint supervision possible. I am furthermore most grateful towards Keshav Dasgupta who suggested this project to me and was always willing to answer my countless questions.
I would also like to thank all my other collaborators on the various projects that entered into this work as there were Stephon Alexander, Katrin Becker, Sheldon Katz and Radu Tatar.

I am deeply grateful to James Sylvester Gates, Jr., for his support especially during the final year of my thesis. 
Many thanks are also due to my good friends Ken Griggs, Tim Burt and Thom Whittemore. I would also like to thank Dragos Constantin, Axel Krause, William Linch, Willie Merrell and Ram Sriharsha for interesting discussions and for welcoming me at the University of Maryland.
A special source of pride and joy have always been ``my girls'' with whom I had the pleasure to work during the last two years --- Meem Mahmud and Julia Young. 

Although short, my time in Hamburg was nevertheless one of the best I ever had and I would like to thank Iman Benmachiche, Stefan Berge, David Cerdeno, Thomas Grimm, Olaf Hohm, Hans Jockers, Susan Lau, Claudia Lehmann, Fantina Madricardo, Paolo Merlatti, Vanessa Mexner, Andrei Micu, Mathias de Riese, Sakura Sch\"afer--Nameki, Silvia Vaula and Martin Weidner for that.
I would also like to express my gratitude towards my family, especially towards my husband Bastian Angermann for his continuous encouragement, patience and love.

For the time I spent at the University of Maryland I would like to acknowledge support from the German Academic Exchange Service (DAAD), the National Science Foundation, and the University of Maryland Department of Physics and School of Engineering. I would also like to thank the Graduiertenkolleg Hamburg, which is supported by the German Science Foundation (DFG) and the RTN Program MRTN--CT--2004--503369.

\def\theequation{\Alph{section}.\arabic{equation}}
\begin{appendix}

\setcounter{equation}{0}

\section{Geometry and Topology of Conifolds}\label{coni}

The first extensive discussion of the topology and geometry of the Calabi--Yaus known as singular, resolved and deformed conifold was presented in \cite{candelas}. The (singular) conifold is a complex 3--manifold that can be embedded in four dimensional complex space as
\begin{equation}\label{singular}
  \sum_{i=1}^4 (z_i)^2\,=\,0\,,\qquad\qquad z_i\in\mathbb{C}^4\,.
\end{equation}
This space has a conical singularity at $(z_1,z_2,z_3,z_4)=0$ and can be describes as a cone over $S^2\times S^3$ which is also known as $T^{1,1}$.
The singularity can be smoothed in two different ways: via deformation or via resolution, which leads to the other two conifold geometries mentioned above. Asymptotically, they also look like cones over $S^2\times S^3$, but close to the tip of the conifold (the singular point) they are topologically different. It was shown that all three manifolds admit a K\"ahler structure and have vanishing first Chern class, therefore we can pass continously from one geometry to the other.
 
This transition is pictured as starting with a finite size $S^3$, shrinking it to zero size and blowing up an $S^2$ at the singularity, giving rise to deformed, singular and resolved conifold, respectively. 
To see that these spaces are topologically distinct, it suffices to see that their Euler characters differ, since $\chi(S^3)=0$, $\chi($point$)=1$ and $\chi(S^2)=2$.

Consider the singular conifold defined in \eqref{singular}. The base of the cone is a manifold $\mathcal{N}$ given by the intersection of the space of solutions to \eqref{singular} with a sphere of radius $r$ in $\mathbb{C}^4=\mathbb{R}^8$
\begin{equation}\label{base}
  \sum_{i=1}^4 |z_i|^2\,=\,r^2\,.
\end{equation}
Separating $z_i$ into real and imaginary part $z_i=x_i+iy_i$ this equation together with \eqref{singular} becomes
\begin{equation}
  x^i x_i\,=\,\frac{r^2}{2}\,,\quad y^i y_i\,=\,\frac{r^2}{2}\,,\quad x^i y_i
    \,=\,0\,. 
\end{equation}
The first equation defines a three sphere $S^3$ with radius $r/\sqrt{2}$. The others define an $S^2$ fiber over $S^3$. Since all such bundles over $S^3$ are trivial, one finds that $\mathcal{N}$ has topology $S^2\times S^3$. Therefore, the conifold is topologically a cone over $S^2\times S^3$, i.e. the metric of the 3--manifolds reads
\begin{equation}\label{conemetric}
  ds^2 \,=\, dr^2 + r^2 ds_{T^{1,1}}^2\,.
\end{equation}
It was shown in \cite{candelas} that the Ricci flat K\"ahler metric is obtained with
\begin{eqnarray}\label{singbase}
   ds_{T^{1,1}}^2 &=& \frac{1}{9}\,\big(d\psi+\cos\theta_1\,d\phi_1 
    +\cos\theta_2\,d\phi_2\big)^2 + \frac{1}{6}\,\sum_{i=1}^2\big(d\theta_i^2
    +\sin^2\theta_i\,d\phi_i^2\big)\,.
\end{eqnarray}
This can be noted by exploiting the symmetry of $T^{1,1}$ which is given as a coset space $SU(2)\times SU(2)/U(1)$. To see this, define
\begin{equation}
  W\,=\,\frac{1}{\sqrt{2}}\,z^i\sigma_i\,=\,\frac{1}{\sqrt{2}}\, 
    \begin{pmatrix}  z^3+i z^4 & z^1-i z^2\\ z^1+i z^2 & -z^3+i z^4
    \end{pmatrix}
\end{equation}
where $\sigma_i$ are the three Pauli matrices plus the identity. With this definition, equations \eqref{singular} and \eqref{base} become
\begin{eqnarray}
  det\,W &=& -\frac{1}{2}\,\sum_{i=1}^4 (z^i)^2\,=\,0 \\
  Tr\,W^\dagger W &=& \sum_{i=1}^4 |z^i|^2\,=\,r^2\,.
\end{eqnarray}
Defining a matrix $Z=W/r$ these equations read:
det$\,Z=0$ and Tr$Z^\dagger Z=1$.
If $Z_0$ is a particular solution of these equations, say
\begin{equation}
  Z_0\,=\,\begin{pmatrix} 0 & 1 \\ 0 & 0 \end{pmatrix} \,=\, \frac{1}{2}\,
    (\sigma_1+i \sigma_2)
\end{equation}
then it is straightforward to show that a general solution can be written as
$  Z\,=\,L Z_0 R^\dagger$
where $L$ and $R$ belong to SU(2)
\begin{equation}
  L\,=\,\begin{pmatrix} a & -\bar{b}\\ b& \bar{a}\end{pmatrix} \qquad\mbox{and}\qquad
  R\,=\,\begin{pmatrix} k & -\bar{l}\\ l& \bar{k}\end{pmatrix} 
\end{equation}
with $|a|^2+|b|^2=1$ and $|k|^2+|l|^2=1$. Choosing the following parameterization of $L$ and $R$
\begin{eqnarray}\label{eulerangles}\nonumber
  a &=& \cos\frac{\theta_1}{2}\,e^{\frac{i}{2}(\psi_1+\phi_1)}\,,\qquad
    k\,=\,\cos\frac{\theta_2}{2}\,e^{\frac{i}{2}(\psi_2+\phi_2)}\\
  b &=& \sin\frac{\theta_1}{2}\,e^{\frac{i}{2}(\psi_1-\phi_1)}\,,\qquad
    l\,=\,\sin\frac{\theta_2}{2}\,e^{\frac{i}{2}(\psi_2-\phi_2)}
\end{eqnarray}
where $\psi_i,\phi_i,\theta_i$ are the Euler angles of each SU(2) gives rise to coordinates usually used in the literature to discuss conifolds \cite{ pandozayas, papatseyt, ks, mn}.

It is clear that $SU(2)\times SU(2)$ acts transitively on the base $\mathcal{N}$. Certain matrices ($L,R$) leave $Z_0$ fixed. It is easy to check that these are of the form $(L,R)=(\Theta,\Theta^\dagger)$ with
\begin{equation}
  \Theta \,=\, \begin{pmatrix} e^{i\theta} & 0\\0 & e^{-i\theta} \end{pmatrix}\,.
\end{equation}
Therefore, $\mathcal{N}$ can be thought of the set of matrices $(L,R)\sim(L\Theta, R\Theta^\dagger)$, which shows that
\begin{equation}
  \mathcal{N}\,=\,\frac{SU(2)\times SU(2)}{U(1)}\,=\,\frac{S^3\times S^3}{U(1)}
\end{equation}
where the U(1) is generated by $\Theta$. This means that $\mathcal{N}=T^{1,1}$.

For this manifold to be compatible with a K\"ahler structure one requires
\begin{equation}
  g_{\mu\bar{\nu}}\,=\,\partial_\mu \partial_{\bar{\nu}}\mathcal{F}
\end{equation}
where $\mathcal{F}$ is the K\"ahler potential. A K\"ahler potential that is invariant under $SU(2)\times SU(2)$ can be a function of $r^2$ only. In terms of $W$
\begin{equation}\label{kaehlermetric}
  ds^2\,=\,\mathcal{F}'\,Tr(dW^\dagger dW)+\mathcal{F}''\,|Tr\,W^\dagger dW|^2\,.
\end{equation}
The Ricci tensor for a K\"ahler manifold is $R_{\mu\bar{\nu}} = \partial_\mu \partial_{\bar{\nu}}\,\ln\,\sqrt{g}$ with $\sqrt{g}=$det$g_{\mu\bar{\nu}}$. Define a function
\begin{equation}\label{defgamma}
  \gamma(r)\,=\,r^2\,\mathcal{F}'\,,
\end{equation}
then requiring Ricci flatness leads to
\begin{equation}
  \gamma(r)\,=\,r^{4/3}\,.
\end{equation}
After a rescaling $r\to\tilde{r}=\sqrt{3/2}\,r^{2/3}$ one recovers indeed the metric \eqref{conemetric}.

The small resolution is obtained by blowing up an $S^2$ at the tip of the cone. To see this, note first that we could also define the singular conifold after a coordinate redefinition as
\begin{equation}
  wz-uv\,=\,0\,,
\end{equation}
which is equal to the statement that there are non--trivial solutions to
\begin{equation}
  \begin{pmatrix} w & u \\ v & z\end{pmatrix}
  \begin{pmatrix} \xi_1 \\ \xi_2 \end{pmatrix}
  \,=\, 0\,.
\end{equation}
At $(u,v,w,z)=0$ this is solved by any pair $(\xi_1,\xi_2)$, but note that there is an overall scaling freedom $(\xi_1,\xi_2)\sim (\lambda\xi_1,\lambda\xi_2)$, so $(\xi_1,\xi_2)$ actually describe a $\mathbb{CP}^1$ at the tip of the cone. Therefore, the resolved conifold is depicted as $\mathcal{O}(-1)\oplus\mathcal{O}(-1)\to \mathbb{CP}^1$.
We will work in a patch where $\xi_1/\xi_2$ is a good inhomogeneous coordinate on $\mathbb{CP}^1$. Hence
\begin{equation}
  W\,=\,\begin{pmatrix} -u\lambda & u \\ -z\lambda & z \end{pmatrix}\,.
\end{equation}
Equation \eqref{base} becomes
\begin{equation}
  r^2\,=\,Tr\,W^\dagger W \,=\,\sigma\Lambda
\end{equation}
with
$\sigma = |u|^2+|z|^2$ and $\Lambda = 1+|\lambda|^2$.

The K\"ahler potential $\mathcal{K}$ in this case is not simply a function of $r^2$ only, but
\begin{equation}
  \mathcal{K}\,=\,\widetilde{\mathcal{F}}+4 a^2\ln\Lambda
\end{equation}
with $\widetilde{\mathcal{F}}$ being a function of $r^2$ and $a$ is a constant, the resolution parameter. This gives the metric on the resolved conifold
\begin{equation}
  ds^2\,=\,\widetilde{\mathcal{F}}'\,Tr(dW^\dagger dW)+\widetilde{\mathcal{F}}''\,
    |Tr\,W^\dagger dW|^2+4a^2\,\frac{|d\lambda|^2}{\Lambda^2}\,.
\end{equation}
This reduces to the singular conifold metric when $a\to 0$. In terms of the Euler angles \eqref{eulerangles} with $\psi=\psi_1+\psi_2$, this metric was derived in \cite{pandozayas} to be
\begin{eqnarray}\label{resmetricapp}\nonumber
  ds^2 & = & \widetilde{\gamma}'\,dr^2 + \frac{\widetilde{\gamma}'}{4}\,r^2\big(d\psi+\cos\theta_1\,d\phi_1
    +\cos\theta_2\,d\phi_2\big)^2 \\ 
  & &+ \frac{\widetilde{\gamma}}{4}\,\big(d\theta_1^2+\sin^2\theta_1\,d\phi_1^2\big) + 
    \frac{\widetilde{\gamma}+4a^2}{4}\,\big(d\theta_2^2+\sin^2\theta_2\,d\phi_2^2\big)\,,
\end{eqnarray}
with $\widetilde{\gamma}=\widetilde{\gamma}(r)$ going to zero like $r^2$, $a$ is called resolution parameter because it determines the size of the blown up $S^2$ at $r=0$ and $\widetilde{\gamma}'=\partial\widetilde{\gamma}/\partial r^2$. Again, $\widetilde{\gamma}$ is defined as in \eqref{defgamma}. Ricci flatness requires 
\begin{equation}
  \widetilde{\gamma}'\widetilde{\gamma}(\widetilde{\gamma}+4a^2)\,=\,2r^2/3\,,
\end{equation}
which can be solved for $\widetilde{\gamma}(r)$.
It is interesting that there is another metric on the resolved conifold which is related to this one by a flop, basically the exchange of the two $S^2$. 

Let us now turn to the discussion of the deformed conifold. It is obtained by changing \eqref{singular} in a very simple way
\begin{equation}
  \sum_{i=1}^4 (z_i)^2\,=\,\mu^2\,.
\end{equation}
In terms of the matrix $W$ this means
$  det\,W\,=\,-\mu^2/2$
and as before we define a radial coordinate via the relation $r^2=Tr\,(W^\dagger W)$. Splitting the $z_i$ into real and imaginary part we obtain
\begin{equation}
  r^2\,=\,x_i x^i+y_i y^i\,,\qquad \mu^2\,=\,x_i x^i-y_i y^i\,,
\end{equation}
which implies that $r$ ranges from $\mu$ to $\infty$. But it is also apparent that the (deformed) conifold is nothing but the cotangent bundle over a three--sphere, $T^*S^3$.
A particular solution is found to be
\begin{equation}
  W_\mu \,=\, \begin{pmatrix} \frac{\mu}{\sqrt{2}} & \sqrt{r^2-\mu^2}\\
             0 & -\frac{\mu}{\sqrt{2}}\end{pmatrix}
\end{equation}
and the general solution is obtained by setting
$  W\,=\,L\,W_\mu\,R^\dagger$.
For $r\ne\mu$ the stability group is again U(1). So for each $r\ne\mu$ the surfaces $r=$constant are again $S^2\times S^3$. Note however, that for $r=\mu$ the matrix $W_\mu$ is proportional to $\sigma_3$ and is invariant under an entire SU(2). Thus, the ``origin'' of coordinates $r=\mu$ is in fact an $SU(2)=S^3$.

Again, we define a K\"ahler potential $\hat{\mathcal{F}}$ and $\hat{\gamma}=r^2
\hat{\mathcal{F}}$, then the metric is again given by \eqref{kaehlermetric} and the condition for Ricci flat becomes \cite{candelas}
\begin{equation}
  r^2(r^4-\mu^4)(\hat{\gamma}^3)'+3\mu^4\hat{\gamma}^3\,=\,2 r^8\,.
\end{equation}
This can be integrated and one finds that
for $r\to\infty$ the function $\hat{\gamma}$ approaches $r^{4/3}$, which agrees with the singular conifold solution.
In terms of Euler angles the metric is explicitely given as \cite{papatseyt, minasian}
\begin{eqnarray}\label{defmetricapp}\nonumber
  ds^2_{\rm def} &=& \left[\left(r^2 \hat{\gamma}'-\hat{\gamma}\right)
  \,\left(1-\frac{\mu^4}{r^4}\right)+\hat{\gamma}\right]\,\left(\frac{dr^2}{r^2(1-\mu^4/r^4)}
    + \frac{1}{4}\,(d\psi+\cos\theta_1\,d\phi_1+\cos\theta_2\,d\phi_2)^2\right)\\
  &+& \frac{\hat{\gamma}}{4}\,\left[(\sin\theta_1^2\,d\phi_1^2+d\theta_1^2)
    +(\sin\theta_2^2\,d\phi_2^2+d\theta_2^2)\right]\\ \nonumber
  &+& \frac{\hat{\gamma}\mu^2}{2 r^2}\,\left[\cos\psi(d\theta_1
    d\theta_2-\sin\theta_1\sin\theta_2 d\phi_1
    d\phi_2)+\sin\psi(\sin\theta_1 d\phi_1 d\theta_2+\sin\theta_2 d\phi_2
    d\theta_1)\right]\,.
\end{eqnarray}

\section{T--Duality and Buscher Rules}\label{tduality}
\setcounter{equation}{0}

We follow here the conventions from \cite{cvj}, but there are a variety of review articles discussing this topic, e.g. \cite{dabholkar, mtheory, alvarez, skenderis}. 
T--duality can already be observed on the bosonic level. Consider the bosonic part of the ten--dimensional (closed) superstring that is described by $X^\mu(\sigma, \tau)=X_L^\mu(\sigma+\tau)+X_R^\mu(\sigma-\tau)$ with left movers $X_L$ and right movers $X_R$. Compactification on a circle of radius R in the 9th direction, 
\begin{equation}
  X^9 \sim X^9+2\pi R\,.
\end{equation}
leads to a quantization of the center--of--mass momentum, 
as usual in Kaluza--Klein reduction, 
\begin{equation}
  p=\frac{n}{R}\,,\quad n\in\mathbb{Z}\,.
\end{equation}
But additionally, strings can also wind around the compact direction $w$ times, so
\begin{equation}
  X^9(\sigma+2\pi) = X^9(\sigma)+2\pi Rw\,,\quad w\in \mathbb{Z}\,.
\end{equation}
This implies for the momentum, which can also be split into left and right movers $p^9=p^9_L+p^9_R$:
\begin{equation}
  p^9_L=\frac{n}{R}+\frac{wR}{\alpha'}\,,\qquad 
    p^9_R=\frac{n}{R}-\frac{wR}{\alpha'}\,.
\end{equation}
Note, that the mass spectrum 
\begin{equation}
  m^2=\frac{n^2}{R^2}+\frac{w^2 R^2}{\alpha'^2}+\frac{2}{\alpha'}\,(N_L+N_R-2)\,,
\end{equation}
with $N_{R,L}$ being the left and right mover oscillator numbers, is completely symmetric under
\begin{equation}
  R\to R'=\frac{\alpha'}{R} \quad\mbox{and}\quad n\leftrightarrow w\,,
\end{equation}
i.e. the theory compactified on a circle with radius $R$ gives the same spectrum (and Hamiltonian) as another theory compactified on a circle with radius $\alpha'/R$ if one exchanges winding modes $w$ and momentum modes $n$. But this exchange also has the effect 
\begin{equation}\label{tdualp}
  p_L^9 \to p_L^9\,, \qquad p_R^9 \to -p_R^9\,,
\end{equation}
or on the level of $X$: $X^9_L\to X^9_L$ and $X^9_R\to -X^9_R$. The CFT is completely unchanged by choosing $X=X_L-X_R$ instead of $X=X_L+X_R$ apart from the sign change in \eqref{tdualp}. So, T--duality can be viewed as a one--sided parity transformation taking $X^9_R\to -X^9_R$, but leaving the left--movers invariant.
By spacetime supersymmetry, T-duality must also act like a one--sided parity on spacetime fermions, e.g. in IIB with two spinors $S^\alpha$ in Green--Schwarz formalism
\begin{equation}\label{tdualfermion}
  S^\alpha_L \to  S^\alpha_L \quad\mbox{and}\quad S^\alpha_R \to \Gamma^9\Gamma  S^\alpha_R\,. 
\end{equation}
$\Gamma^i$ are the Dirac matrices in d=10 and $\Gamma=\Gamma^1...\Gamma^9$, see \cite{gsw} for a representation of the Clifford algebra in d=10. $\Gamma$ acts as helicity operator and just gives $\pm1$ acting on Ramond states. $\Gamma^9\Gamma$ represents the action of the parity transformation because it anticommutes with $\Gamma^9$ but commutes with all other $\Gamma^i$ ($i\ne9$). This operator changes the chirality of right--moving fermions, so starting with type IIB and two spinors ${\bf 8_s}$ we end up with a theory with one spinor ${\bf 8_s}$ and one conjugate spinor ${\bf 8_c}$ --- type IIA! This duality is not only a symmetry for the free string, but also of the interacting theory. See e.g. \cite{pol}.

T--duality also has to act non--trivially on D--branes if it should consistently exchange IIB with IIA. In fact, since it is a one--sided parity transformation, it exchanges Dirichlet with Neumann boundary conditions, as is obvious in a non--standard formulation of these boundary conditions:
\begin{equation}
  \begin{cases}
    \mbox{Neumann} & \partial_+X=\partial_-X\\
    \mbox{Dirichlet}\quad & \partial_+X=-\partial_-X\,.\\
  \end{cases}
\end{equation}
This means, T--duality along a longitudinal direction of a Dp--brane turns it into a D(p-1) brane; T--duality along a direction transverse to the brane turns it into a D(p+1)--brane.

Let us now consider a theory that includes background fields. Using the sigma model action
\begin{eqnarray}
  S &=& \frac{1}{4\pi\alpha'}\,\int d^2\sigma\sqrt{g}\,\left[\left(g^{ab}G_{\mu\nu}(X)
    +i\epsilon^{ab}B_{\mu\nu}(X)\right)\,\partial_a X^\mu\partial_b X^\nu +  
    \alpha'\phi R\right]
\end{eqnarray}
with background dilaton $\phi$, worldsheet metric $g^{ab}$, worldsheet curvature $R$ and usual graviton and $B_{NS}$ field, one can show that this is invariant under the following T--duality rules \cite{buscher} (the tilde--fields are after T--duality along $y$)
\begin{eqnarray}\label{buscher}\nonumber
  \widetilde{G}_{yy} &=& G_{yy}^{-1}\,,\qquad\qquad\qquad\qquad\qquad\quad
    \widetilde{G}_{\mu y}\,=\, \frac{B_{\mu y}}{G_{yy}}\\
  \widetilde{G}_{\mu\nu} &=& G_{\mu\nu}-\frac{G_{\mu y}G_{\nu y}-B_{\mu y}
    B_{\nu y}}{G_{yy}}\\ \nonumber
  \widetilde{B}_{\mu\nu} &=& B_{\mu\nu}-\frac{B_{\mu y}G_{\nu y}-G_{\mu y}
    B_{\nu y}}{G_{yy}}\,,\qquad \widetilde {B}_{\mu y}\,=\,\frac{G_{\mu y}}{G_{yy}}\\ 
    \nonumber
  e^{2\widetilde{\phi}} &=& G_{yy}^{-1}\,e^{2\phi}\,.
\end{eqnarray}
One can of course perform multiple T--dualities along a number of circles, that form a torus $T^d$ ($d$ being the number of T--duality directions). 

Also taking the RR p--forms $C_p$ into consideration gives the duality transformation \cite{ortin}
\begin{eqnarray}\label{rrtdual}
  \widetilde{C}_{\mu\ldots\nu\alpha y}^{(n)} &=& C_{\mu\ldots\nu\alpha}^{(n-1)}
    -(n-1) G_{yy}^{-1}\,C_{[\mu\ldots\nu\vert y}^{(n-1)} G_{\vert\alpha]y}\\ \nonumber
  \widetilde{C}_{\mu\ldots\nu\alpha\beta}^{(n)} &=& C_{\mu\ldots\nu\alpha\beta y}
    ^{(n+1)}+n\,C_{[\mu\ldots\nu\alpha}^{(n-1)} B_{\beta]y} + n(n-1)\,
    G_{yy}^{-1}\,C_{[\mu\ldots\nu\vert y}^{(n-1)} B_{\vert\alpha\vert y} 
    G_{\vert\beta]y}\,.
\end{eqnarray}
Note that $\alpha,\beta,\mu,\nu\ne y$ and $[\ldots]$ indicates antisymmetrization, where the index $\vert y\vert$ is excluded. 
\pagebreak

\section{Nonlinear Sigma Models and Topological Strings}
\setcounter{equation}{0}

After an introduction to (2,2) supersymmetric sigma models, we will discuss topological sigma models and string theory for
closed three--form $H$ and K\"ahler target and use it to review the Gopakumar--Vafa conjecture in section \ref{conjecture}. In the last section we turn to more general topological models, in particular we are interested in sigma models on generalized K\"ahler targets that also allow for non--vanishing NS flux $H$. An introduction to generalized complex geometry is provided in section \ref{gcg}.

\subsection{Nonlinear Sigma Models}\label{sigma}

String theory is intrinsically linked to sigma models. We can view string theory as the description of a 2--dimensional worldsheet $\Sigma$ propagating through a 10--dimensional target space $M$. The sigma model that describes this theory deals with maps $\phi: \Sigma\to M$. These maps can be promoted to chiral superfields $\Phi$ that have $\phi$ in their lowest component and obey the 2d sigma model action
\begin{equation}\label{sigmaaction}
  S\,=\,-\frac{1}{4}\,\int d\tau\,d\sigma\,d^2\theta\,\left(g_{ij}(\Phi)+
    B_{ij}(\Phi)\right)\, D^\alpha\Phi^i\,D_\alpha\Phi^j\,,
\end{equation}
with indices $i,j=1...d$ parameterizing the target space and
\begin{equation}\label{defD}
  D_\alpha \,=\, \frac{\partial}{\partial\overline{\theta}^\alpha}
    +i\rho^\mu\theta_\alpha\partial_\mu\,,
\end{equation}
where $\rho^\mu$ is a 2d $\gamma$--matrix and $\theta_\alpha$ a two--component Grassmann valued spinor.
Chiral superfields are defined by $\overline{D}^\alpha \Phi^i=0$. See \cite{wess} for an introduction into superspace. Written in terms of superfields, this action has explicit $\mathcal{N}=1$ supersymmetry generated by
\begin{equation}\label{defQ}
  Q_\alpha \,=\, \frac{\partial}{\partial\overline{\theta}^\alpha}
    -i\rho^\mu\theta_\alpha\partial_\mu\,.
\end{equation}
In the case $H=dB=0$ it  has further non--manifest supersymmetry if and only if the target space is K\"ahler \cite{zumino}. The extra supersymmetry transformation is
\begin{equation}\label{susy}
  \delta_\eta\,\Phi^i\,=\,\left(\eta^\alpha D_\alpha\,\Phi^j\right)J^i_j
\end{equation}
where $J^i_j$ is the complex structure of the manifold. If the manifold admits a hyper--K\"ahler structure (i.e. admits three independent K\"ahler structures), then the action \eqref{sigmaaction} admits more than one extra supersymmetry, for a usual K\"ahler structure we find $\mathcal{N}=2$.

Considering a sigma model that does not only contain chiral but also twisted chiral superfields, one can find additional supersymmetry even if the target is {\sl not} K\"ahler. This was proposed in \cite{gates} more than twenty years ago, but only recently it was realized that the constraints one obtains for the target manifolds in this model define a (twisted) generalized K\"ahler structure \cite{gualtieri}, see section \ref{gcg}. In section \ref{gensigma} we will explain how generalized complex geometries can be used to describe generalized topological sigma models.

Let us briefly review the construction from \cite{gates}. The usual non--linear sigma model with chiral superfields \eqref{sigmaaction} is changed by including twisted chiral superfields that satisfy
\begin{eqnarray}\nonumber
  D_+\chi\,=\,\frac{1}{2}\,(1+\gamma_5)\,D\chi\,=\,0 \,,\qquad\qquad
  \overline{D}_-\chi\,=\,\frac{1}{2}\,(1-\gamma_5)\,\overline{D}\chi\,=\,0 
\end{eqnarray}
and similar for their complex conjugates. In d=2 $\gamma_5$ is simply given by the Pauli matrix $\sigma^3$. In theories with only twisted chiral multiplets, $D$ and $\overline{D}$ cannot be distinguished and $\chi$ is equivalent to an ordinary chiral superfield $\Phi$. However, in models with both chiral and twisted chiral superfields, $\chi$ and $\Phi$ are distinct.

The action suggested in \cite{gates} in terms of $\mathcal{N}=1$ real scalar superfields ${\bf \Phi}^i$ $(i=1\ldots d)$ that generalizes \eqref{sigmaaction} is
\begin{equation}
  S\,=\,-\frac{1}{4}\,\int d^2x\,d^2\theta\,\left[g_{ij}({\bf \Phi})\, (D^\alpha{\bf \Phi}^i) 
    \,(D_\alpha{\bf \Phi}^j) +B_{ij}({\bf \Phi})\,(D^\alpha{\bf \Phi}^i) \,(\gamma_5D)_\alpha{\bf \Phi}^j
    \right]\,,
\end{equation}
where ${\bf \Phi}^i$ includes chiral multiplets $\Phi$ as well as twisted chiral multiplets $\chi$. Since it is written in terms of $\mathcal{N}=1$ superfields, this action is manifest supersymmetric, but we are looking for additional supersymmetry. In \cite{gates} the extra supersymmetry transformations are given as
\begin{equation}\label{gensusy}
  \delta_\eta\,{\bf \Phi} \,=\,-i(\eta_+D_-{\bf \Phi}^j)\,(J_+)_j^i+i(\eta_-D_+{\bf \Phi}^j)\,
    (J_-)_j^i
\end{equation}
with two different complex structures $J_+$ and $J_-$. This generalizes \eqref{susy}. The two different complex structures result from the different action of the supersymmetry transformations on chiral and twisted chiral multiplets, respectively. For $\mathcal{N}=2$ chiral multiplets one simply finds
\begin{equation}\label{chiralsusy}
  \delta \Phi\,=\,i\left[\eta^\alpha\hat{Q}_\alpha,\,\Phi\right]\,=\,i\eta^\alpha 
    \hat{D}_\alpha\,\Phi\,,\qquad \delta \overline{\Phi}\,=\,i\left[\eta^\alpha
    \hat{Q}_\alpha,\,\overline{\Phi}\right]\,=\,-i\eta^\alpha \hat{D}_\alpha\,
    \overline{\Phi}\,,
\end{equation}
where the Majorana $\mathcal{N}=1$ spinor derivative $\hat{D}$ and the generator of the non--manifest supersymmetry $\hat{Q}$ are defined in terms of the ordinary $D$ and $Q$ from \eqref{defD} and \eqref{defQ} as
\begin{equation}
  \hat{D}\,=\,\sqrt{\frac{1}{2}}\,\big(D+\overline{D}\big)\,,\qquad 
    \hat{Q}\,=\,\sqrt{\frac{1}{2}}\,\big(D-\overline{D}\big)\,.
\end{equation}
Note that $\{\hat{D},\,\hat{Q}\}=0$. Comparing \eqref{chiralsusy} to \eqref{susy} one finds that the complex structure in this case is given as
\begin{equation}
  \hat{J}^j_i\,=\,\begin{pmatrix} i & 0\\0 &-i\end{pmatrix}\,.
\end{equation}
For $\mathcal{N}=2$ twisted chiral multiplets the supersymmetry transformations read
\begin{equation}\label{twistsusy}
  \delta\chi\,=\,-i(\eta\gamma_5)^\alpha \hat{D}_\alpha\chi\,, \qquad
    \delta\overline{\chi}\,=\,i(\eta\gamma_5)^\alpha \hat{D}_\alpha\overline{\chi}\,.
\end{equation}
In a theory with only twisted chiral multiplets one could absorb $\gamma_5$ into $\eta$, but with also ordinary chiral multiplets present, $\eta$ is already fixed. This transformation is not of the form \eqref{susy}, but one can define a generalized matrix valued complex structure\footnote{This is not to be confused with a generalized complex structure in the spirit of Hitchin and Gualtieri \cite{hitch, gualtieri}.}
\begin{equation}
  \widetilde{J}^j_i\,=\,\begin{pmatrix} -i\gamma_5 & 0\\ 0 & i\gamma_5     
                        \end{pmatrix}
\end{equation}
and write analog to \eqref{susy}
\begin{equation}
  \delta \begin{pmatrix} \chi \\ \overline{\chi} \end{pmatrix}
  \,=\, \big(\eta\widetilde{J}\big)^\alpha\,\hat{D}_\alpha\,
    \begin{pmatrix} \chi \\ \overline{\chi} \end{pmatrix}\,.
\end{equation}
The case $\mathcal{N}=4$ works similar, but is not of interest to us here. 

The two complex structures in \eqref{gensusy} are then found to be
\begin{equation}
  \left(J_\pm\right)^i_j\,=\,\widetilde{J}^i_j\,\pm\,\hat{J}^i_j\,.
\end{equation}
It was shown in \cite{gates} that these two complex structures are indeed integrable. Requiring the commutator of two such generalized supersymmetry transformations \eqref{gensusy} to close (and imposing the equations of motion) leads to
\begin{equation}
  D_+ D_-{\bf\Phi}^i+\Gamma_{+\,jk}^{\;\;i}\,(D_+{\bf\Phi}^j)\,(D_-{\bf\Phi}^k)\,=\,0
\end{equation}
with the affine connections defined in terms of ordinary Christoffel symbols $\Gamma_{jk}^{\;i}$ as
\begin{equation}
  \Gamma_{\pm\,jk}^{\;\;i}\,=\,\Gamma_{jk}^{\;i}\mp H_{jk}^{\;i}\,.
\end{equation}
$H$ is precisely the fieldstrength $H=dB$ and enters as torsion into this relation. This implies that the target manifold is no longer K\"ahler. Note that in above considerations only one torsional connection $\Gamma_{+\,jk}^{\;\;i}$ entered, but not the other one. Furthermore, invariance of the action requires that $g_{ij}$ is Hermitian w.r.t. to both complex structures and the metric is covariantly constant w.r.t. to both torsional connections. 

In summary, it was found that $\mathcal{N}=2$ supersymmetry can be preserved if the manifold possesses a bi--Hermitian structure $(g,B,J_+,J_-)$ (with $H=dB\ne 0$). If one wanted to realize $\mathcal{N}=4$ supersymmetry, one would find two sets of quaternionic structures \cite{gates}.

\subsection{Topological Sigma Models and String Theory}\label{topol}

This section serves the purpose to define topological string amplitudes and explain the difference between open and closed topological string theories. We need this background material to explain the Gopakumar--Vafa conjecture in the next section. In accordance with their observations, we restrict ourselves to the case $H=0$ here. Generalized topological sigma models will be discussed in section \ref{gensigma}.
We will closely follow the review \cite{neitzke}, see e.g. \cite{wittentopol, topol, bcov, marino} for details. 

Topological string theory integrates not only over all maps $\phi$ but also over all metrics on $\Sigma$, this is often called a sigma model coupled to two--dimensional gravity. Classically, the sigma model action depends only on the conformal class of the metric, so the integral over metrics reduces to an integral over conformal (or complex) structures on $\Sigma$. 


The sigma model with K\"ahler target discussed above can be made topological by a procedure called ``twisting'' \cite{wittentopol}, which basically shifts the spin of all operators by 1/2 their R--charge. 
%
There are two conserved supercurrents for the two worldsheet supersymmetries that are nilpotent
\begin{equation}
  (G^\pm)^2\,=\,0\,,
\end{equation}
so one might be tempted to use these as BRST operators and build cohomologies. But they have spin 3/2. The twist shifts their spin by half their R--charge to obtain spin 1 operators 
\begin{equation}
  S_{new}\,=\,S_{old}+\frac{1}{2}\,q
\end{equation}
where $q$ is the U(1) R--charge of the operator in question
Classically, the theory has a vector $U(1)_V$ symmetry and an axial $U(1)_A$ symmetry. Twisting by $U(1)_V$ gives the so--called A--model, twisting by $U(1)_A$ the B--model.  The $U(1)_A$ might suffer from an anomaly unless $c_1(M)=0$, which leads to the requirement that the target must be a Calabi--Yau manifold for the B--model.
One could now define $Q=G^+$ or $Q=G^-$ and use this nilpotent operator as a BRST operator, i.e. restrict one's attention to observables which are annihilated by $Q$.

Before doing so let us note a special feature of $\mathcal{N}=(2,2)$ supersymmetry. Since left and right movers basically decouple, we can split any of the operators $G^\pm$ into 2 commuting copies, one for left and one for right movers. In terms of complex coordinates let us denote the left movers as holomorphic $G^\pm$ and the right movers as antiholomorphic $\overline{G^\pm}$. This makes the (2,2) supersymmetry more apparent. Now twisting can be defined for left and right movers independently and we obtain in principle four models, depending on which we choose as BRST operators:
\begin{eqnarray}\nonumber
  \mbox{A model}:\quad & &(G^+,\,\overline{G}^+)\,,\qquad\qquad\qquad
    \mbox{B model}:\qquad (G^+,\,\overline{G}^-)\\
  \mbox{\={A} model}:\quad & & (G^-,\,\overline{G}^-) \,,\qquad\qquad\qquad
  \mbox{\={B} model}:\qquad  (G^-,\,\overline{G}^+)\,.
\end{eqnarray}
Of these four models, only two are actually independent, since the correlators for A (B) and for \={A} (\={B}) are related by complex conjugation. So we will ignore \={A} and \={B} in the following.

Starting with this setup, one can now discuss observables in topological theories. It turns out, that $Q+\overline{Q}$ in the A--model reduces to the differential operator $d=\partial+\overline{\partial}$ on $M$, i.e. the states of the theory lie in the deRham cohomology. A ``physical state'' constraint requires states to be in $H^{(1,1)}(M)$ only, which corresponds to deformations of the K\"ahler structure on $M$. One can also show that correlators are independent of the complex structure modulus of M, since the corresponding operators are $Q$--exact (they decouple from the computation of string amplitudes). 

In the B--model the relevant cohomology is that of $\overline{\partial}$ with values in $\Lambda^*(TM)$, i.e. the observables are (0,1) forms with values in the tangent bundle $TM$. These correspond to complex structure deformations. One can also show that in this case correlation functions are independent of K\"ahler moduli. So each of the two topological models depends only on half the moduli,
\begin{eqnarray}\nonumber
  \mbox{A model on}\, M: & & \mbox{depends on K\"ahler moduli of}\,M\\ \nonumber
  \mbox{B model on}\, M: & & \mbox{depends on complex structure moduli of}\,M\,.
\end{eqnarray}
In this sense both models describe topological theories, because they only depend on the topology of the target, not its metric.
It can also be shown that the relevant path integral $\int e^{-S}$ simplifies tremendously compared to ordinary field theories. It localizes on $Q$--invariant configurations. These are simply constant maps $\phi:\Sigma\to M$ with $d\phi=0$ for the B--model and holomorphic maps $\overline{\partial}\phi=0$ for the A--model.
In this sense the B--model is simpler than the A--model, because the string worldsheet ``reduces to a point'' on $M$, its correlation functions are those of a field theory on $M$. They compute quantities determined by the periods of the holomorphic 3--form $\Omega^{(3,0)}$, which are sensitive to complex structure deformations. 

The holomorphic maps in the A--model are called ``worldsheet instantons''. Each worldsheet instanton is weighted by
\begin{equation*}
  \exp\left(\int_C (J+i B)\right)
\end{equation*}
where $t=J+i B\;\in H^2(M,\mathbb{C})$ is the complexified K\"ahler parameter and $C$ is the image of the string worldsheet in $M$. Summing over all instantons makes this theory more complicated than the B--model, but only in the sense that it is not local on $M$ and does not straightforwardly reduce to a field theory on $M$. In summery, the A--model moduli are complexified volumes of 2--cycles, while
the B--model moduli are the periods of $\Omega$.

Let us now talk about the relation of these topologically twisted sigma models to string theory. As mentioned before, string theory sums not only over all possible maps $\phi:\,\Sigma\to M$, as discused in the sigma models above, but also over all possible metrics on $\Sigma$. The latter actually reduces to a sum over the moduli space of genus $g$ Riemann surfaces.
The topological string free energy is then defined as a sum over all genera
\begin{equation}
  \mathcal{F}\,=\,\sum_{g=0}^\infty \lambda_s^{2-2g}\,F_g
\end{equation}
with the string coupling $\lambda_s$ and $F_g$ being the amplitude for a fixed genus $g$. The string partition function is given by $\mathcal{Z}=\exp \mathcal{F}$.

The interesting quantities for the topological string theory are therefore the genus $g$ partition functions.
Already at genus zero one finds a lot of interesting information about $M$. In the A--model the genus zero free energy turns out to be
\begin{equation}
  F_0\,=\,\int_M J\wedge J\wedge J + \;\mbox{instanton corrections}\,.
\end{equation}
The first term corresponds to the classical contribution of the worldsheet theory, it gives the leading order contribution in which the string worldsheet just reduces to a point. We have explicitely assumed $M$ to be a complex 3--manifold with the real part of the K\"ahler parameter being $J$. The instanton term contains a sum over all homology classes $H_2(M,\mathbb{Z})$ of the image of the worldsheet, each weighed by the complexified area, and a sum over ``multi--wrappings'' in which the map $\Sigma\to M$ is not one--to--one.

To define the genus zero free energy in the B--model requires a little more effort. We already noted that the relevant moduli are periods of $\Omega\in H^3(M,\mathbb{C})$. This cohomology can be decomposed as
\begin{equation}
  H^3\,=\,H^{3,0}\oplus H^{2,1} \oplus H^{1,2} \oplus H^{0,3}\,.
\end{equation}
For a Calabi--Yau threefold the Hodge numbers are given by $h^{3,0}=h^{0,3}=1$, because there is one unique holomorphic 3--form, and $h^{2,1}=h^{1,2}$, recall the Hodge diamond \eqref{hodgediamond}. Therefore, $H^3(M,\mathbb{C})$ has real dimension $2\,h^{1,2}+2$. It is customary to choose a symplectic basis of 3--cycles $A^i$ and $B_j$ with intersection numbers
\begin{equation}
  A^i \cap A^j=0\,,\quad B_i\cap B_j=0\,,\quad A^i\cap B_j=\delta^i_j\,,\qquad \mbox{with}\,i,j=1,...,h^{1,2}+1\,.
\end{equation}
One can then define homogeneous coordinates on the moduli space of complex structure deformations by
\begin{equation}
  X^i\,:=\,\int_{A^i}\,\Omega\,.
\end{equation}
This gives $h^{1,2}+1$ complex coordinates, although the moduli space only has dimension $h^{1,2}$. This overcounting is due to the fact that $\Omega$ is only unique up to overall rescaling, so the same is true for the coordinates defined this way. Therefore they carry the name ``homogeneous coordinates''.
There are also $h^{1,2}+1$ periods over B--cycles
\begin{equation}
  \hat{F}_i\,:=\,\int_{B^i}\,\Omega\,.
\end{equation}
Due to the relation between A and B cycles, there must be a relation between the periods. In other words, we can express $\hat{F}_i$ as a function of $X^j$.
\begin{equation}
  \hat{F}_i\,=\,\hat{F}_i(X^j)\,.
\end{equation}
One can prove that these satisfy an integrability condition
\begin{equation}
  \frac{\partial}{\partial X^i}\,\hat{F}_j\,=\,\frac{\partial}{\partial X^j}\,\hat{F}_i
\end{equation}
which allows us to define a new function $F$ via
\begin{equation}
  \hat{F}_i\,=\,\frac{\partial}{\partial X^i}\,F
\end{equation}
which is actually nothing but the genus zero free energy of the B-model\footnote{Strictly speaking, $F$ is not a function but rather a section of the line bundle over the moduli space. It depends on the choice of scaling of $\Omega$. Under $\Omega\to \zeta \Omega$ F scales as $F\to \zeta^2 F$, it is homogeneous of degree 2 in the homogeneous coordinates of the moduli space.}. It is given by the simple formula
\begin{equation}
  F\,=\,\frac{1}{2}\,X_iF^i\,.
\end{equation}

In general, the integral over all worldsheets is too hard to carry out explicitely. There are nevertheless some tools that enable one to calculate topological string amplitudes. For example, mirror symmetry between A and B model can be used to compute amplitudes in the model of choice (usually the B--model since it does not obtain instanton corrections) and then extrapolating the result to the mirror theory. We will be more interested in a duality between open and closed strings, which enables one in principle to calculate the free energy at all genera for a particular class of non--compact geometries --- e.g. conifolds. To describe an open topological strings we need to explain what we mean by topological branes that appear as boundaries of $\Sigma$. 

A D--brane corresponds to a boundary condition for $\Sigma$ that is BRST--invariant. In the A--model this implies that the boundary should be mapped to a Lagrangian submanifold\footnote{This means $L$ has half the dimension of $M$ and the K\"ahler form 
restricted to $L$ vanishes.} $L$ of $M$. 
If we allow open strings to end on $L$, we say that the D--branes are wrapped on $L$. Having a stack of N D--branes on $L$ corresponds to including a weighting factor N for each boundary.

We have already discussed how D--branes carry gauge theories in physical strings (we will use ``physical'' for the target space perspective to distinguish it from toplogical strings). The same is true for topological branes. In the A--model it turns out that one can actually compute the exact string field theory, which is again a topological theory: U(N) Chern--Simons theory \cite{witten}. Its action in terms of the U(N) gauge connection $A$ is given by
\begin{equation}
  S\,=\,\int_L \,Tr\left(A\wedge dA+\frac{2}{3}\,A\wedge A\wedge A\right)\,.
\end{equation}
This action might still obtain instanton corrections, but Witten showed that in the special case where $L=S^3$ there are none. This is fascinating, because the $S^3$ in the deformed conifold (which is also $T^*S^3$) is such a Langrangian submanifold.

In physical superstrings, D--branes are sources for RR fluxes. So under what quantity are topological branes charged? The only fluxes available are the K\"ahler 2--form $J$ and the holomorphic 3--form $\Omega$. Wrapping a topological brane on a Lagrangian subspace $L$ of $M$ (in the A model) creates a flux through a 2--cycle $C$ which  ``links'' L. This link means that $C=\partial S$ for some 3--cycle S that intersects L once, so $C$ is homologically trivial in $M$, although it becomes nontrivial if considered as a cycle in $M\setminus L$. This implies that $\int_C J=0$ since $J$ is closed and $C$ trivial.

Wrapping N branes on L has the effect of creating a K\"ahler flux through $C$
\begin{equation}
  \int_C J\,=\,Ng_s\,,
\end{equation}
because the branes act as a $\delta$--function source for the two--form, i.e. $J$ is not closed anymore on L, but $dJ=Ng_s\,\delta(L)$. Similarly, a B--model brane on a holomorphic 2--cycle $Y$ induces a flux of $\Omega$ through the 3--cycle linking $Y$. In principle we could also wrap branes on 0, 4 or 6--cycles in the B--model, but there is no field candidate those branes could be charged under. This suggests a privileged role for 2--cycles.


The A--model branes wrap Lagrangian 3--cycles (whose volume is naturally measured by $\Omega$--the fundamental object of the B--model) whereas B--model branes wrap holomorphic cycles (whose volume is measured by the A model modulus $J$). The full meaning of this interesting connection is still not understood.

\subsection{The Gopakumar--Vafa Conjecture}\label{conjecture}

After all these preliminaries we are now ready to explain the geometric tansition on conifolds. This is a duality between open and closed topological strings (it has been shown that they compute the same string partition function) which has profound physical consequences. The dual gauge theory from the open string sector is $\mathcal{N}=1$ SYM in d=4. The IR dynamics of this gauge theory can be obtained either from the open or from the closed string sector, but the UV dynamics can only be described by open strings, as we cannot generate gauge theories with closed strings. In this sense, both string theory backgrounds are dual, they compute the same superpotential because they have the same topological string partition function. The key to this duality in the gauge theory is to identify parameters from the open string theory with parameters from the closed string theory. In the IR this will be the gluino condensate which is identified either with the K\"ahler or complex structure modulus of a closed or open string theory. 

The geometric transition in question \cite{gopakumar} considers the A model on the deformed conifold $T^*S^3$. As noted by \cite{witten}, the exact partition function of this theory is simply given by U(N) Chern Simons theory. The {\it closed} A--model on this geometry is trivial, because it has no K\"ahler moduli. But the $T^*S^3$ contains a Lagrangian 3--cycle $L=S^3$ on which we can wrap branes in the open A model. This creates a flux $Ng_s$ of $J$ through the 2--cycle C which links L, in this case $C=S^2$. So it is natural to conjecture that this background is dual 
to a background with only flux through this 2--cycle. The resolved conifold is the logical candidate for this dual background as it looks asymptoticall like the deformed conifold, but has a finite $S^2$ at the tip of the cone. This led Gopakumar and Vafa to the following \vspace{5mm}
{\parindent=0.0mm

{\sl Conjecture: The open A model on the deformed conifold $T^*S^3$ with N branes wrapping the $S^3$ is dual to the closed A model on the resolved conifold $\mathcal{O}(-1)\oplus\mathcal{O}(-1)\to \mathbb{CP}^1$ and the size of the $\mathbb{CP}^1$ is determined by $t=Ng_s$.}} \vspace{5mm}

There are no branes anymore in the dual geometry, there is simply no 3--cycle on which they could wrap.
The passage from one geometry to the other is called ``geometric transition'' or ``conifold transition'' in this case. Since the partition function for the open theory is known for all genera (from Chern--Simons theory), this can be used to postulate the full closed string partition function on the resolved conifold, which does have a K\"ahler parameter $J$ and would in general be hard to compute to all orders.

The agreement of the partition function on both sides has been shown in \cite{gopakumar} for arbitrary 't Hooft coupling $\lambda=N g_s$ and to all orders in $1/N$. 
In this sense, this duality is an example of a large N duality which has been suggested by 't Hooft: for large N holes in the Riemann surface of Feynman diagrams are ``filled in'' or ``condensed'', where one takes $N\to\infty$ with $g_s=\,$fixed. 
The authors of \cite{gopakumar} matched the free energy $F_g$ at every genus $g$ via the identification of the 't Hooft coupling
\begin{equation}\label{identifythooft}
  i\lambda\,=\,Ng_s\quad\mbox{(open)} \qquad\longleftrightarrow\qquad
    i\lambda\,=\,t \quad\mbox{(closed)}\,,
\end{equation}
where $t$ is the complexified K\"ahler parameter of the $S^2$ in the resolved conifold and the identification of the 't Hooft coupling for open strings is dictated by the Chern--Simons theory on $S^3$.  

Beyond that, it was also shown that the coupling to gravity (to the metric)\footnote{It might seem contradictory that there can be a coupling to the metric when we are speaking about topological models. The classical Chern--Simons action is indeed independent of the background, but at the quantum level such a coupling can arise. In the closed side there are possible IR divergences, anomalies for non--compact manifolds that depend on the boundary metric of these manifolds.} and Wilson loops take the same form for the open and closed theory.  The two topological string theories described here correspond to the different limits $\lambda\to 0$ and $\lambda\to\infty$, but they are conjectured to describe the same string theory (with the same small $g_s$) only on different geometries. 

\subsubsection*{Embedding in Superstrings and Superpotential}

This scenario has an embedding in ``physical'' type IIA string theory. 
Starting with N D6 branes on the $S^3$ of the deformed conifold we find a dual background with flux through the $S^2$ of the resolved conifold. The Calabi--Yau breaks 3/4 of the supersymmetry (which leaves 8 supercharges), therefore the theory on the worldvolume of the branes has $\mathcal{N}=1$ (the branes break another half of the supersymmetry). There is a U(N) gauge theory on the branes (in the low energy limit of the string theory the U(1) factor decouples and we have effectively SU(N)). As described in the last section, these wrapped branes create flux and therefore a superpotential. This superpotential is computed from topological strings, but we need a gauge theory parameter in which it is expressed. The relevant superfield for $\mathcal{N}=1$ SU(N) is $S$, the chiral superfield with gaugino bilinear in its bottom component. We want to express the free energy $F_g$ in terms of $S$. Since there will be contributions from worldsheet with boundaries, we can arrange this into a sum over holes $h$
\begin{equation}
  F_g(S)\,=\,\sum_{h=0}^\infty F_{g,h}\,S^h\,.
\end{equation}
It turns out that the genus zero term computes the pure gauge theory, i.e. pure SYM, higher genera are related to gravitational corrections.

As discussed above, the open topological string theory is given by Chern Simons on $T^*S^3$, which has no K\"ahler modulus. The superpotential created by the open topological amplitude of genus zero is given by \cite{vafa}
\begin{equation}\label{wopen}
  \lambda_s W^{\rm open}\,=\,\int d^2\theta\,\frac{\partial F_0^\text{open}(S)}{\partial S} 
    +\alpha \,S+\beta
\end{equation}
with $\alpha,\beta=\,$constant, $\alpha\,S$ being the explicit annulus contribution ($h=2$).

Although the topological model is not sensitive to any flux through a 4-- or 6-- cycle, in the superstring theory the corresponding RR forms $F_4$ and $F_6$ can be turned on. In the closed string side this corresponds to a superpotential
\begin{equation}
  \lambda_s W^{\rm closed}\,=\, \int F_2\wedge k\wedge k +i\int F_4\wedge k +\int F_6.
\end{equation}
The topological string amplitude is not modified by these fluxes \cite{vafa}. The genus zero topological string amplitude $F_0$ determines the size of the 4-- and 6--cycle to be $\frac{\partial F_0}{\partial t}$ and $2F_0-t\,\frac{\partial F_0}{\partial t}$, respectively, where $t$ is the usual complexified K\"ahler parameter (of the resolved conifold). If we have $N, L, P$ units of 2--, 4-- and 6--form flux, respectively, the superpotential yields after integration
\begin{equation}
  \lambda_s W^{\rm closed}\,=\,N\,\frac{\partial F_0}{\partial t}+itL+P\,.
\end{equation}
Note that requiring $W=0$ and $\partial_t W=0$ fixes $P$ and $L$ in terms of $N$ and $t$. $N$ is of course fixed by the number of branes in the open string theory.

This looks very similar to the superpotential for the open theory \eqref{wopen}. We have already discussed that the topological string amplitudes agree
\begin{equation}
  F^\text{open} \,=\,F^\text{closed}
\end{equation}
if one identifies the relevant parameters as in \eqref{identifythooft}. In this case we have to identify $S$ with $t$ and $\alpha, \beta$ with the flux quantum numbers $iL, P$. It is clear from the gauge theory side that $\alpha$ (or $L$) is related to a shift in the bare coupling of the gauge theory. In particular, to agree with the bare coupling to all orders we require $iL=V/\lambda_s$, where $V$ is the volume of the $S^3$ that the branes are wrapped on. This gives an interesting relation between the size $V$ of the blown--up $S^3$ (open) and the size $t$ of the blown--up $S^2$ (closed):
\begin{equation}
  \left(e^t-1\right)^N\,=\,\mbox{const}\cdot e^{-V/\lambda_s}\,.
\end{equation}
This indicates that for small $N$ ($Ng_s/V\ll 1$) the D--brane wrapped on $S^3$ description is good (since $t\to 0$), whereas for large $N$ ($Ng_s\gg 1$) the blown--up $S^2$ description is good (since $V\to-\infty$ does not make sense). It should be clear from our discussion that after the $S^3$ has shrunk to zero size there cannot be any D6--branes in the background, but RR fluxes are turned on.

To summarize the superstring picture of the conifold transition: In type IIA we start with N D6--branes on the $S^3$ of the deformed geometry and find as its dual N units if 2--form flux through the $S^2$ of the resolved conifold. In the mirror type IIB, N D5 branes wrapping the $S^2$ of the resolved conifold are dual (in the large N limit) to a background without D--branes but 3--form flux turned on. The geometry after transition is given by the deformed conifold with blown up $S^3$. In both cases we have to identify the complex structure modulus of the deformed conifold with the K\"ahler modulus of the resolved conifold or, roughly speaking, the size of the $S^3$ with the size of the $S^2$.


Let us finish this section with the explicit derivation of the Veneziano--Yankielowicz superpotential in type IIA \cite{vafa}. To lowest order the type IIA superpotential is given by
\begin{equation}\label{superA}
  W(S)\,=\,\int d^2\theta \left(\frac{1}{\lambda_s}\,\tau S+iN^2 \alpha e^{-\tau/N}
    \right)
\end{equation}
where $\tau$ is the chiral superfield with $iC+V/\lambda_s$ in its bottom component, $C$ being the background value of the 3--form gauge field in IIA and $V$ the size of the $S^3$ as before. $C$ takes the role of the $\theta$ angle and the gauge coupling is promoted to a superfield. The second term in \eqref{superA} stems from instanton corrections. It might seem surprising that $V$ enters into this superpotential (although the A model is independent of the complex structure modulus), the reason lies within quantum corrections. But the linear term in S is the only coupling that $V$ has to this theory.

We now use that $\tau$ is actually a dynamical superfield and can be integrated out from its equation of motion. Requiring $\partial_\tau W=0$ gives
\begin{equation}
  S\,=\,i\lambda_s N\alpha e^{-\tau/N}\,.
\end{equation}
Solving this for $\tau$ and plugging the result back into \eqref{superA} gives an effective superpotential for $S$
\begin{equation}
  W_\text{eff}\,=\,-\frac{N}{\lambda_s}\,\left[S\,\log \left( 
    \frac{S}{iN\alpha\lambda_s}\right)-S\right]\,.
\end{equation}
We do indeed recover the Veneziano--Yankielowicz superpotential \cite{vy}. The scale of the gauge theory $\Lambda$ can be identified with $(N\alpha\lambda_s)^{1/3}$. The vacuum of the theory exhibits all the known phenomena of gaugino condensation, chiral symmetry breaking and domain walls. This is a remarkable result and the first example where string theory produces the correct superpotential of a gauge theory.

\subsection{Generalized Complex Geometry}\label{gcg}

Generalized complex geometries (GCG) are a notion introduced rather recently by Hitchin and Gualtieri \cite{hitch, gualtieri} and are of great interest to physicists, because they provide a framework to intertwine the concepts of metric and B--field, something that seems to occur naturally in string theory, as this article should have demonstrated quite explicitely. From the mathematics point if view they were introduced to provide a framework that interpolates between complex and symplectic structures. The basic idea is to view these objects not as linear operations on the tangent bundle of a manifold, but on the direct sum $T\oplus T^*$ of tangent and cotangent bundle. Since the smooth sections of $T\oplus T^*$ have a natural bracket operation, called the 
Courant bracket, there are canonical integrability conditions for these two structures. Therefore, Hitchin defined a generalized complex structure as an almost complex structure $\mathcal{J}$ on $T\oplus T^*$ whose +i--eigenbundle is Courant involutive \cite{hitch}. Schematically, the generalization from complex to generalized complex structures works as follows
\vspace{0.5cm}

\begin{tabular}{p{6cm}p{1cm}p{6.5cm}}
  tangent bundle on manifold $\mathcal M$: $T$ & $\longrightarrow$ & direct sum of 
    tangent and cotangent bundle $T\oplus T^*$\\[1ex]
  almost complex structure & $\longrightarrow$ & generalized almost complex 
    struture\\
  $J$: $T\to T$ with $J^2=-1$ & & $\mathcal{J}$: $T\oplus T^*\,\to\,T\oplus 
    T^*$ with $\mathcal{J}^2=-1$\\[1ex]
  integrability condition w.r.t. Lie & $\longrightarrow$ & 
    integrability condition w.r.t. Courant\\
  bracket & & bracket. 
\end{tabular}
\vspace{0.5cm}

Let $M $ be a real $n$--dimensional manifold. Then $T\oplus T^*$ is $2n$--dimensional and its elements are of the form $X+\xi$ with $X\in T$ a vector field and $\xi\in T^*$ a one--form. There is a natural non--degenerate inner product of signature $(n,n)$ defined by
\begin{equation}
  \langle X+\xi,\,Y+\eta\rangle\,=\,\frac{1}{2}\,\big(\xi(X)+\eta(Y)\big)\,,\qquad
    X,Y\in T\,,\;\xi,\eta\in T^*\,.
\end{equation}
The space $T\oplus T^*$ has a number of symmetries, for example the inner product is invariant under $SO(n,n)$, the one which is of most interest to physicists is the B--field transform.

Let $B:T\to T^*$ with $B^*=-B$, we can therefore view $B$ as a two--form in $\bigwedge^2 T^*$ via $B(X)=\iota_X\,B$, with the interior product $\iota_X: \bigwedge^r(M)\to \bigwedge^{r-1}(M)$ defined as 
\begin{equation}
  \iota_X\,\omega(X_1,\ldots,X_{r-1})\,=\,\omega(X,X_1,\ldots,X_{r-1})
\end{equation}
with vector fields $X_i$. This means, for example, 
\begin{equation}
  \mbox{if}\;X_i=\frac{\partial}{\partial x^i}\qquad \iota_{X_i}\,dx^j\wedge dx^k \,=\,\delta^j_i\,dx^k-\delta^k_i\,dx^j.
\end{equation}
The B--field transform is then defined under the natural splitting $T\oplus T^*$ as a $2n\times 2n$ matrix 
\begin{equation}
  \exp(B)\,=\,\begin{pmatrix} 1 & 0 \\ B & 1 \end{pmatrix}
\end{equation}
and acts as $X+\xi \,\to\, X+\xi+\iota_X B$. This means, it acts as a projection onto $T$ and by a ``shearing'' transformation on $T^*$. 
Consider for example $X+\xi=\partial/\partial x+ dx$ and $B=b\,dx\wedge dy$. The B--field acts on an argument like this as
\begin{equation}
  X+\xi \,\to\, X+\xi+b\,dy\,.
\end{equation}
This is very reminiscent of the action of ``T--duality with B--field along $x$'' we encountered numerous times
throughout this presentation. Our hope is therefore that the non--K\"ahler manifolds we constructed in type IIA (and also heterotic and type I) have a natural interpretations in terms of generalized complex structures. See also \cite{kapustin} for similar interpretations.

After this motivation, let us define the basic quantities needed for generalized complex geometry. We will not be very thorough and not aim for completeness, see \cite{gualtieri} for a complete introduction to the subject. The above mentioned Courant bracket is defined as
\begin{equation}
  \left[X+\xi,\,Y+\eta\right]\,=\,[X,Y]+\mathcal{L}_X\,\eta-\mathcal{L}_Y\,\xi 
    -\frac{1}{2}\,d(\iota_X\,\eta-\iota_Y\,\xi)
\end{equation}
with $X,Y\in T$ and $\xi,\eta \in T^*$. This is a skew--symmetric object but does not satisfy a Jacobi identity.
$\mathcal{L}$ indicates the usual Lie--derivative
\begin{equation}
  \mathcal{L}_X\,=\,d\,\iota_X+\iota_X\,d\,.
\end{equation}
Note that the Courant bracket reduces to the ordinary Lie bracket on vector fields.
The Courant bracket, like the inner product, is not only invariant under diffeomorphisms, but also under the B--field transform. The map $e^B$ is an automorphism of the Courant bracket if and only if B is closed, i.e. $dB=0$.

In physics, we are not only interested in closed B--fields (vanishing field strength). One can, however, defined ``twisted'' quantities that differ from the usual ones by terms involving $H=db$ (with two--form $b$), so the formalism of generalized complex geometries is still applicable. The twisted Courant bracket, for example, is defined in terms of the usual Courant bracket $[\,,\,]$ as
\begin{equation}\label{introH}
  \left[X+\xi,\,Y+\eta\right]_H\,=\,\left[X+\xi,\,Y+\eta\right]+\iota_Y\iota_X\,H
\end{equation}
where $H$ is a real\footnote{$H$ does not necessarily have to be real, but since the Courant bracket is a real quantity, it makes sense to restrict $H$ to be real here.} closed three--form. One can also define a twisted exterior derivative $d_H$ that acts on any form $\eta\in\bigwedge^\bullet T^*$ as
\begin{equation}
  d_H\,\eta \,=\, d\eta+H\wedge\eta\,.
\end{equation}

A Lie algebroid is a vector bundle $L$ on a smooth manifold $M $ equipped with a Lie bracket $[\,,\,]$ on $C^\infty(L)$ and a smooth bundle map $a:L\to T$, called the ``anchor''. The anchor must induce a Lie algebra homomorphism $a:C^\infty(L)\to C^\infty(T)$, i.e.
\begin{equation}\label{liealgebroid}
  a\left([X,Y]\right)\,=\,\left[a(X),a(Y)\right]\qquad \forall X,Y\in C^\infty(L)\,,
\end{equation}
and satisfy a Leibniz rule \cite{mackenzie}. It is useful to think of a Lie algebroid as a generalization of the (complexified) tangent bundle, sine if we take $L=T$ (the tangent bundle) and $a=id$, the bracket reduces to the ordinary commutator of vector fields and both conditions are obviously satisfied. 

%

A complex structure on $M$ is an endomorphism $J:T\to T$ satisfying $J^2=-1$. A symplectic structure on $M$ is a non--degenerate skew form $\omega\in\bigwedge^2T^*$. One may view $\omega$ as a map $T\to T^*$ via the interior product
\begin{equation}
  \omega:\,v\,\to\,\iota_v\,\omega\,,\qquad v\in T\,.
\end{equation}
This implies that a symplectic structure on $T$ can be defined as an isomorphism $\omega:\,T\to T^*$ satisfying $\omega^*=-\omega$, where $\omega^*:\,(T^*)^*=T\,\to \,T^*$. 

A generalized complex structure on $T$ is an endomorphism $\mathcal{J}$ of the direct sum $T\oplus T^*$ which satisfies
\begin{itemize}
  \item $\mathcal{J}$ is complex, i.e. $\mathcal{J}^2=-1$
  \item $\mathcal{J}$ is symplectic, i.e. $\mathcal{J}^*=-\mathcal{J}$\,.
\end{itemize}
The usual complex and symplectic structure are embedded in the notion of generalized complex structure in the following way:
If $J$ is a complex structure on $M$, then the $2n\times 2n$ matrix (written w.r.t. the direct sum $T\oplus T^*$)
\begin{equation}\label{complex}
  \mathcal{J}_J\,=\,\begin{pmatrix} -J & 0 \\ 0 & J^* \end{pmatrix}
\end{equation}
is a generalized complex structure on $T$.
Similarly, if $\omega$ is a symplectic structure on $M$, then
\begin{equation}\label{symplec}
  \mathcal{J}_\omega\,=\,\begin{pmatrix} 0 & -\omega^{-1} \\ \omega & 0 \end{pmatrix}
\end{equation}
is also a generalized complex structure. We therefore observe, that diagonal and anti--diagonal generalized complex structures correspond to complex and symplectic structures, respectively. The interesting aspect of GCG is that it interpolates between the two.

Specifying $\mathcal{J}$ is equivalent to specifying a maximal isotropic subspace of $(T\oplus T^*)\otimes \mathbb{C}$ of real index 0. A subspace $L\subset (T\oplus T^*)\otimes\mathbb{C}$ is isotropic when $\langle X,Y\rangle=0$ for all $X,Y\in L$, it is maximal when its dimension is maximal, i.e. $n$ in our case. Its real index $r$ is given by the complex dimension of $L\cap\overline{L}$. 
Every maximal isotropic in $T\oplus T^*$ corresponds to a pure spinor line bundle. A spinor $\varphi$ is called pure when its null space $L_\varphi=\{v\in T\oplus T^8\,:\,v\cdot\varphi=0\}$ is maximal isotropic. The pure spinor line bundle is generated by 
\begin{equation}\label{purespinor}
  \varphi_L\,=\,\exp(B+i\omega)\,\Omega
\end{equation}
where $B$ and $\omega$ are real two--forms and $\Omega=\theta_1\wedge\ldots\wedge \theta_k$ for some linearly independent one--forms $\{\theta_i\}$. The integer $k$ is called the ``type'' of the maximal isotropic. 
The maximal isotropic is of real index zero if and only if
\begin{equation}
  \omega^{n-k}\wedge\Omega\wedge \overline{\Omega}\,\ne\,0\,.
\end{equation}

The type of a maximal isotropic is the codimension $k$ of its projection onto $T$.
Then any generalized complex structure of type $k=0$ is a B--field transform of a symplectic structure $\mathcal{J}_\omega$ as in \eqref{symplec}, which determines a maximal isotropic $L=\{X-i\omega(X):\;X\in T\otimes\mathbb{C}\}$ and a pure spinor line generated by $\varphi_L=\exp(i\omega)$. The B--field transform gives rise to a generalized complex structure
\begin{equation}
  \mathcal{J}_{k=0}\,=\,e^{-B}\mathcal{J}_\omega e^{B} \,=\,
    \begin{pmatrix} -\omega^{-1}B & -\omega^{-1} \\ \omega+B\omega^{-1}B & 
      B\omega^{-1}
    \end{pmatrix}
\end{equation}
with maximal isotropic $\tilde{L}=e^{-B}L=\{X-(B+i\omega)(X):\;X\in T\otimes \mathbb{C}\}$ and pure spinor line $\varphi_{e^{-B}L}=\exp(B+i\omega)$.

The extremal type $k=n$ is related to complex structures. Note that $\mathcal{J}_J$ as defined in \eqref{complex} determines a maximal isotropic $L=T_{0,1}\oplus T^*_{1,0}$ (where $T_{1,0}=\overline{T_{0,1}}$ is the +i--eigenspace of $J$) and a spinor line generated by $\varphi_L\,=\,\Omega^{n,0}$ (where $\Omega^{n,0}$ is any generator of holomorphic $n$--forms on the $n$--dimensional space $(T,J)$). Then any generalized complex structure of type $k=n$ is the B--field transform of a complex structure, i.e.
\begin{equation}\label{btransformj}
  \mathcal{J}_{k=n}\,=\,e^{-B}\mathcal{J}_J e^{B} \,=\,
    \begin{pmatrix} -J & 0 \\ BJ+J^*B & J^* 
    \end{pmatrix}
\end{equation}
with maximal isotropic $\tilde{L}=e^{-B}L=\{X+\xi-\iota_X B:\;X+\xi\in T_{0,1}\oplus T_{1,0}^*\}$ and pure spinor $\varphi_{e^{-B}L}=\exp(B)\,\Omega^{n,0}$. Note that in this case only the $(0,2)$ component of the real two--form $B$ has any effect. B--field transforms of complex structures are always block--lower--diagonal, an observation used in \cite{kapustin}. 

Let us also note the following integrability considion for generalized (almost) complex structures: A generalized complex structure of type $k=n$ is integrable if and only if the complex structure $J$ is integrable, a generalized complex structure of type $k=0$ is integrable if and only if $d(B+i\omega)=0$.

%

A generalized almost complex structure is said to be a {\sl twisted} generalized complex structure when its +i--eigenbundle is involutive w.r.t. the $H$--twisted Courant bracket, $H$ being the closed real three--form introduced around equation \eqref{introH}. Given any integrable $H$--twisted generalized complex structure $\mathcal{J}$, its conjugate $e^B\mathcal{J}e^{-B}$ is integrable w.r.t. the $H+dB$--twisted Courant bracket, for any smooth two--form $B$. This means that the space of twisted generalized complex structures depends only on the cohomology class $[H]\in H^3(M,\mathbb{R})$.

We close this section we a few remarks about generalized K\"ahler and generalized Calabi--Yau manifolds, as they are of particular importance in string theory.

Since the bundle $T\oplus T^*$ has a natural inner product $\langle\,,\,\rangle$, it has structure group $O(2n,2n)$. The specification of a positive definite metric $G$ ($G^2=1$) that is compatible with this inner product is equivalent to the reduction of the structure group to $O(2n)\times O(2n)$. If the manifold allows for a generalized complex structure $\mathcal{J}$ this means a reduction of the structure group $U(n,n)\subset O(2n,2n)$ to $U(n)\times U(n)$ if the metric $G$ commutes with $\mathcal{J}$. Note that since $G^2=1$ and $\mathcal{J}G=G\mathcal{J}$, the map $G\mathcal{J}$ squares to $-1$, i.e. it defines another generalized complex structure.

Gualtieri \cite{gualtieri} is therefore led to the following definition of a {\sl generalized K\"ahler} manifold: A generalized K\"ahler structure is a pair $(\mathcal{J}_1,\mathcal{J}_2)$ of commuting generalized complex structures such that $G=-\mathcal{J}_1\mathcal{J}_2$ is a positive definite matric on $T\oplus T^*$. In accordance with the observations above this means that the existence of a generalized K\"ahler structure is equivalent to a reduction of the structure group to $U(n)\times U(n)$. This has been exploited to extend string theory on manifolds of $SU(3)$ structure (recall section \ref{torsion}) to manifolds with $SU(3)\times SU(3)$ structure \cite{jangrana}.

A special case of generalized K\"ahler structures is of course the usual K\"ahler structure, since a K\"ahler manifold has both a complex and a symplectic structure. So, we can define two generalized complex structures \eqref{complex} and \eqref{symplec}, which obviously commute, and 
\begin{equation}
  G\,=\,-\mathcal{J}_J\mathcal{J}_\omega\,=\,\begin{pmatrix} 0& g^{-1}\\ g & 0
                                             \end{pmatrix}
\end{equation}
where $g$ is the Riemannian metric on the K\"ahler manifold. 

Any B--field transform of a generalized K\"ahler structure $(\mathcal{J}_1^B,\,\mathcal{J}_2^B)\,=\,(B\mathcal{J}_1B^{-1},\,B\mathcal{J}_2B^{-1})$ is also generalized K\"ahler. In the case with $\mathcal{J}_1=\mathcal{J}_J$ and $\mathcal{J}_2=\mathcal{J}_\omega$, the metric after B--field transform becomes
\begin{equation}
  G^B\,=\,\begin{pmatrix} -g^{-1}B & g^{-1} \\ g-Bg^{-1}B & Bg^{-1}
          \end{pmatrix}\,,
\end{equation}
showing that the generalized K\"ahler metric need not be diagonal. 

According to \cite{gualtieri}, any generalized K\"ahler metric is uniquely specified by the existence of a Riemannian metric $g$ and a two--form $b$, where $b$ does not have to be closed. Therefore, a generalized K\"ahler metric is not simply a B--field transform of a Riemannian metric (for a B--field transform we require $B$ to be closed). The torsion of the generalized K\"ahler structure is given by $h=db$.

Moreover, and this has been exploited in the formulation of generalized topological sigma models (see section \ref{gensigma}), any generalized K\"ahler structure is determined by a bi--Hermitian structure on the manifold $M$. A $U(n)\times U(n)$ structure is equivalent to the specification of $(g,b,J_+,J_-)$ with Riemannian metric $g$, two--form $b$ and two Hermitian 
complex structures $J_\pm$. One could call this a ``bi--Hermitian structure with b--field''. Let $\omega_\pm$ be the two--forms associated to the complex structures $J_\pm$, i.e.
\begin{equation}
  \omega_\pm\,=\,g J_\pm\,.
\end{equation}
Then the maps $b\pm g$ determine the metric G via
\begin{equation}
  G\,=\,\begin{pmatrix} -g^{-1}b & g^{-1} \\ g-bg^{-1}B & bg^{-1}
          \end{pmatrix}
  \,=\,\begin{pmatrix} 1 & 0 \\ b & 1 \end{pmatrix}\,
       \begin{pmatrix} 0 & g^{-1} \\ g & 0 \end{pmatrix}\,
       \begin{pmatrix} 1 & 0 \\ -b & 1 \end{pmatrix}\,.
\end{equation}
The generalized complex structures $\mathcal{J}_{1,2}$ are found to be
\begin{equation}\label{genkaehl}
  \mathcal{J}_{1,2}\,=\,\frac{1}{2}\,\begin{pmatrix} 1 & 0 \\ b & 1 \end{pmatrix}\,
       \begin{pmatrix} J_+\pm J_- & -(\omega_+^{-1}\mp\omega_-^{-1})\\ 
            \omega_+\mp\omega_- & -(J_+^*\pm J_-^*) \end{pmatrix}\,
       \begin{pmatrix} 1 & 0 \\ -b & 1 \end{pmatrix}\,.
\end{equation}
Note that the degenerate case where $J_+=\pm J_-$ corresponds to $(\mathcal{J}_1,\mathcal{J}_2)$ being the B--field transform of a genuine K\"ahler structure.

For a generalized K\"ahler structure the torsion $h$ is of type (2,1)+(1,2) w.r.t. both complex structures $J_\pm$. Also note the following important corollary: For a generalized K\"ahler structure with data $(g,b,J_\pm)$ the following are equivalent
\begin{itemize}
  \item $h=db=0$
  \item $(J_+,g)$ is K\"ahler
  \item $(J_-,g)$ is K\"ahler.
\end{itemize}
In other words, in the absence of torsion the bi--Hermitian structure reduces to a bi--K\"ahler structure. This does not necessarily imply $J_+=J_-$, though.

For a (integrable) generalized K\"ahler structure it is also true that the complex structures $J_\pm$ are covariantly constant w.r.t. to two different torsional connections
\begin{equation}
  \nabla^\pm\,=\,\nabla\pm\frac{1}{2}\,g^{-1}h\,.
\end{equation}
These are metric connections with torsion $T_0(\nabla^\pm)=\pm g^{-1}h$ where $h=db$. This is precisely the geometry first observed in physics \cite{gates}, where it occurred as the target space geometry of $\mathcal{N}=(2,2)$ supersymmetric sigma models with torsion, as we reviewed in section \ref{sigma}.

As before, we can extend the notion of a generalized K\"ahler structure to a twisted generalized K\"ahler structure by requiring integrability of $\mathcal{J}_1$ and $\mathcal{J}_2$ w.r.t. the $H$--twisted Courant bracket. Then $J_\pm$ are still integrable complex structures, but the torsion in the torsional connection is now given by $db+H$, which is still of type (1,2)+(2,1).

In \cite{hitch} also the concept of generalized Calabi--Yau manifolds was introduced. A generalized Calabi--Yau is defined as a generalized K\"ahler structure $(\mathcal{J}_1,\,\mathcal{J}_2)$ where each generalized complex structure has holomorphically trivial canonical bundle.
Analogous, we can define a twisted generalized Calabi--Yau.

\subsection{Generalized Topological Sigma Models}\label{gensigma} 

In section \ref{sigma} we reviewed that $\mathcal{N}=2$ supersymmetry can be preserved if the target manifold of the nonlinear sigma model possesses a bi--Hermitian structure $(g,B,J_+,J_-)$ (with $H=dB\ne 0$). This is exactly the data required for a generalized K\"ahler structure to exist, see section \ref{gcg} and we can find it applying the definitions \eqref{genkaehl}. 
This realization has been used to go beyond the work of \cite{gates} and define sigma models in terms of generalized complex structures only \cite{lindstrom}. Very recently, \cite{lindstrom} has also solved the problem of finding an off--shell formulation for this theory. A formulation for $(2,0)$ models also exists \cite{minalind}, but in the following we will focus on the $(2,2)$ sigma model relevant for type II string theory.

Let us also note that \cite{gates} already realized that when both complex structures commute the manifold reduces locally to prodcut of two K\"ahler structures. The case where $H=0$ has also been analyzed in the GCG framework by \cite{zucchini}. Note that the sigma model of \cite{gates} does not simply reduce to a K\"ahler target, but actually to a bi--K\"ahler target in that case.

The important question is now: Can we use these concepts for string theory? In particular, can we construct a topological string theory based on these generalized sigma models? 
To answer this question 
we focus on the work of Kapustin and Li \cite{kapuli}, which exploits the relation of a bi--Hermitian structure to a generalized K\"ahler structure. They actually define generalized A and B models, i.e. topological models, and give an interpretation of the relevant observables and anomaly cancellation conditions in terms of GCG.

By analogy, one would expect generalized A and B models to depend on only ``half'' the geometric data (recall that the ordinary A model only depends on the K\"ahler structure, the ordinary B model only depends on the complex structure of the target $M$). It turns out that (at least on the classical level) the generalized A and B model depend on only one of the two generalized complex structures that describe the  generalized K\"ahler target.

As already mentioned in section \ref{topol} the topological twist can be performed w.r.t. two different U(1) symmetries: there is an axial $U(1)_A$ and a vector $U(1)_V$ R--symmetry. On a K\"ahler target, both these symmetries exist classically, but $U(1)_A$ has an anomaly unless the manifold has vanishing first Chern class, i.e. is a Calabi--Yau. The A model is defined by twisting the $U(1)_V$, the B model by twisting the $U(1)_A$.

In principle, the same construction can be applied for the ``generalized'' sigma model. 
Recall that the sigma model constructed in \cite{gates} requires a target with a bi--Hermitian structure and non--closed three form $H$.
These two complex structures $J_+$ and $J_-$ induce two different decompositions of the (complexified) tangent bundle
\begin{equation}
  T\otimes\mathbb{C}\;\simeq\; T^{1,0}_+\oplus T^{0,1}_+\;\simeq\; T^{1,0}_-\oplus
    T^{0,1}_-\,
\end{equation}
Under these decompositions the fermionic fields $\psi_\pm$ can be split accordingly into holomorphic and anti--holomorphic parts (recall that the indices $\pm$ indicate left and right movers, so the two complex structures $J_\pm$ are actually interpreted as independent complex structures for left and right movers):
\begin{eqnarray}\label{spinors}\nonumber
  \psi_+& =& \frac{1}{2}\,\big(1-i\,J_+)\,\psi_+ + \frac{1}{2}\,\big(
    1+i\,J_+)\,\psi_+\\
  \psi_-& =& \frac{1}{2}\,\big(1-i\,J_-)\,\psi_- + \frac{1}{2}\,\big(
    1+i\,J_-)\,\psi_-\,.
\end{eqnarray}
The bosons are not charged under U(1) R--symmetry, and classically there are two inequivalent ways to assign U(1) R--charge to the fermions:
\begin{eqnarray}
  U(1)_V\,;: & & q_V\,\left[\frac{1}{2}\,\big(1-i\,J_+)\,\psi_+\right] 
    \,=\, -1\,,\qquad q_V\,\left[\frac{1}{2}\,\big(1-i\,J_-)\,\psi_-\right] 
    \,=\, -1\\
   U(1)_A\,;: & & q_A\,\left[\frac{1}{2}\,\big(1-i\,J_+)\,\psi_+\right] 
    \,=\, -1\,,\qquad q_A\,\left[\frac{1}{2}\,\big(1-i\,J_-)\,\psi_-\right] 
    \,=\, +1\,.
\end{eqnarray}
Then, as described in section \ref{topol}, the inequivalent twist is done by shifting the fermionic spin either by $q_V/2$ or $q_A/2$ and the corresponding models were baptized ``generalized A and B models'' in  \cite{kapuli}. Note that flipping the sign of only $J_-$ exchanges generalized A nd B model.

As for ordinary topological sigma models, one has to ensure anomaly cancellation for the twist to be well-defined. The anomalies in this case can be computed using the Atiyah--Singer index theorem and it was found
\begin{eqnarray}\label{anomaly}\nonumber
  U(1)_V\;:\quad & & c_1\left(T^{1,0}_-\right)-c_1\left(T^{1,0}_+\right)\,=\,0\\
  U(1)_A\;:\quad & & c_1\left(T^{1,0}_-\right)+c_1\left(T^{1,0}_+\right)\,=\,0
\end{eqnarray}
where $c_1$ is the first Chern class of the corresponding tangent bundle. These conditions can be interpreted in terms of GCG. Recall that we had defined a generalized K\"ahler manifold to possess two generalized K\"ahler structures that commute. They define a positive definite metric via $G=-\mathcal{J}_1\,\mathcal{J}_2$ on $T\oplus T^*$.
The generalized complex structure $\mathcal{J}_1$ defined in \eqref{genkaehl} induces two complex structures on $T$. These are precisely the complex structures $J_\pm$ of the bi--Hermitian geometry.

Let $E_1$ and $E_2$ denote the $+i$--eigenbundles of $\mathcal{J}_1$ and $\mathcal{J}_2$, respectively, and let $C_\pm$ indicate the $\pm 1$--eigenbundle of the metric $G$ on $T\oplus T^*$. Since $\mathcal{J}_1$ and $\mathcal{J}_2$ commute (this was one of the conditions for generalized K\"ahler structures), one can decompose $E_1=E_1^+\oplus E_1^-$ and $E_2=E_2^+\oplus E_2^-$, where the superscript $\pm$ labels the eigenvalues $\pm i$ of the {\sl other} generalized complex structure, i.e. $E_1^+$ has eigenvalue $+i$ under $\mathcal{J}_2$. It then follows that
\begin{equation}
  C_\pm\otimes \mathbb{C}\,=\,E_1^\pm \oplus (E_1^\pm)^* \,=\,E_2^\pm \oplus 
    (E_2^\pm)^*\,.
\end{equation}
This can be used to rewrite the anomaly cancellation condition \eqref{anomaly} as
\begin{eqnarray}\nonumber
  U(1)_V\;:\quad & & c_1(E_2)\,=\,0\\
  U(1)_A\;:\quad & & c_1(E_1)\,=\,0\,.
\end{eqnarray}
This might seem to indicate that the U(1) R--symmetry is anomaly free if $\mathcal{J}_2$ or $\mathcal{J}_1$ define a generalized Calabi--Yau structure. This is not quite true, since the generalized Calabi--Yau condition as defined in \cite{hitch} is stronger. But it reverse statement is true: if $\mathcal{J}_1$ is a generalized Calabi--Yau structure, then $c_1(E_1)=0$. Nevertheless, it was shown in \cite{kapuli} that the twisting makes only sense (physically) if the stronger Hitchin--Gualtieri condition is fulfilled, because it also ensures the absence of BRST anomalies. This is similar to a statement in ordinary complex geometry: a manifold $M$ with $c_1(M)=0$ is only a Calabi--Yau if it has nowhere vanishing holomorphic sections of the canonical line bundle (this might be violated if the canonical bundle is not trivial, e.g. if $M$ is not simply connected).

We can therefore state that a generalized A model males sense if $\mathcal{J}_2$ defines a generalized Calabi--Yau, whereas the generalized B model can be defined if $\mathcal{J}_1$ defines a generalized Calabi--Yau. The assignment of $\mathcal{J}_{1,2}$ is not arbitrary, but given in \eqref{genkaehl}. Switching $J_-\to -J_-$ exchanges $\mathcal{J}_1$ and $\mathcal{J}_2$ and therefore also exchanges generalized A and B model. This might be interpreted as a hint towards (generalized) mirror symmetry.

The next immediate question would be: What are the relevant BRST operators and observables? Let us follow the discussion of the generalized B--model in \cite{kapuli}, since the generalized A--model is obtained by flipping the sign of $J_-$. Let $Q_\pm$ denote the usual supersymmetry generators in light cone coordinates and $\widetilde{Q}$ the generator of the extra, non--manifest supersymmetry. One can then define two operators
\begin{equation}\label{genQs}
  Q_L\,=\,\frac{i}{2}\,\left(Q_++i \widetilde{Q}_+\right)\,,\qquad
    Q_R\,=\,\frac{i}{2}\,\left(Q_-+i \widetilde{Q}_-\right)
\end{equation}
that are nilpotent and anticommute. The twist shifts the spin of these operators, so that we can define a good BRST operator for the generalized B--model as
\begin{equation}
  Q_{BRST}\,=\,Q_L+Q_R\,.
\end{equation}

It was than further shown in \cite{kapuli} that the BRST complex (for the B--model) coincides with the cohomology of the Lie algebroid $L$ (see definition around equation \eqref{liealgebroid}) associated to $E_1$, the $+i$--eigenbundle of $\mathcal{J}_1$.
The natural choice for the complex Lie algebroid in this case is to take $L$ to be the $-i$--eigenbundle of $\mathcal{J}_1$. The bracket on $L$ is then induced by the Courant bracket on $T\oplus T^*$ and the anchor is the projection $a:\,L\to T\otimes\mathbb{C}$. The associated complex controls the deformations of the twisted generalized complex structure $\mathcal{J}_1$ (with $H=db$ fixed).

One can furthermore consider the cohomology of states. In the usual sigma model with K\"ahler target, there is a well--known state--operator isomorphism identifying both cohomologies. In general, the cohomology of operators is given by the chiral ring whereas the cohomology of states is interpreted as the zero--energy states in the RR sector. The isomorphism between the two spaces is given by the spectral flow \cite{lerche}.

Even if the U(1) R--symmetry is anomalous in the usual sigma model and the twisting does not make sense, one can nevertheless define the chiral ring and the space of RR ground states. They are now (in general) non--isomorphic. E.g. for the K\"ahler target with $H=0$ the chiral ring is given by $H^\bullet(\bigwedge^\bullet T(X))$, whereas the RR ground states are given by the deRham cohomology $H^\bullet(\Omega^\bullet(X))$. Only for a Calabi--Yau target $X$ both spaces are isomorphic. For the generalized B--model the chiral ring is given by Lie algebroid cohomology associated with $\mathcal{J}_1$ \cite{kapuli} (this holds true even is the twist cannot be defined). 

To discuss the cohomology of states, one defines an operator $Q=Q_++iQ_-$ which turns out to be\footnote{up to a numerical factor of $-\sqrt{2}i$} a twisted deRham operator
\begin{equation}
  d_H\,=\,d-H\wedge\,.
\end{equation}
Note that this reduces to the ordinary deRham operator when $H=0$. Therefore, the supersymmetric ground states for the generalized B--model are given by the $d_H$--cohomology. The BRST operator $Q_{BRST}$ is related to $Q$ via
\begin{equation}
  Q_{BRST} \,=\, \frac{1}{2}\,\big(Q+[J_R,Q]\big)
\end{equation}
where $J_R$ is the Noether current associated to the $U(1)_R$--symmetry we twisted with \cite{kapuli}. This actually implies
\begin{equation}
  Q_{BRST} \,=\, \overline{\partial}_H
\end{equation}
with the twisted generalized Dolbeault--operator as defined by Gualtieri \cite{gualtieri}. It then follows that the BRST--cohomology of operators is isomorphic to the Lie algebroid cohomology of $L$ (the $-i$--eigenbundle of $\mathcal{J}_1$). On the quantum level this isomorphism may be changed due to worldsheet instantons.

Above consideration were basically a generalization of the closed topological string theory presented in section \ref{topol}. For open strings (worldsheet with boundaries), we would also need to introduce the concept of generalized topological A and B branes, see \cite{kapustin, zabzine, kapubrane}. 

Thus, we have learned that (with some effort) one can define a topological string theory for generalized K\"ahler targets, if the manifold is generalized Calabi--Yau w.r.t. to one of the generalized complex structures. Then the topological observables depend only on the cohomology of this generalized complex structure, but not on both.

\end{appendix}




\begin{thebibliography}{150}
\addcontentsline{toc}{section}{References}
  \bibitem{gtone}
    M.~Becker, K.~Dasgupta, A.~Knauf and R.~Tatar,
    ``Geometric transitions, flops and non-Kaehler manifolds. I,''  
    Nucl.\ Phys.\ B {\bf 702} (2004) 207 [arXiv:hep-th/0403288].
  \bibitem{realm}
    S.~Alexander, K.~Becker, M.~Becker, K.~Dasgupta, A.~Knauf and R.~Tatar,
    ``In the realm of the geometric transitions,'' 
    Nucl.\ Phys.\ B {\bf 704} (2005) 231 [arXiv:hep-th/0408192].
  \bibitem{gttwo}  M.~Becker, K.~Dasgupta, S.~Katz, A.~Knauf and R.~Tatar,
    ``Geometric transitions, flops and non-Kaehler manifolds. II,''
    Nucl.\ Phys.\ B {\bf 738} (2006) 124 [arXiv:hep-th/0511099].
  \bibitem{ks} I.~R.~Klebanov and M.~J.~Strassler,
    ``Supergravity and a confining gauge theory: Duality cascades and
    $\chi_{SB}$-resolution of naked singularities,''
    JHEP {\bf 0008} (2000) 052 [arXiv:hep-th/0007191].
  \bibitem{vafa} C.~Vafa,
    ``Superstrings and topological strings at large N,''
    J.\ Math.\ Phys.\  {\bf 42}, 2798 (2001) [arXiv:hep-th/0008142].
  \bibitem{gsw} M.~B.~Green, J.~H.~Schwarz, E.~Witten, {\it Superstring Theory} 
    (2 volumes), Cambridge University Press (1987).
  \bibitem{pol} J.~Polchinski, {\it String Theory} (2 volumes), Cambridge University 
    Press (1998).
  \bibitem{kaku} M.~Kaku, {\it Introduction to Superstrings}, Springer Verlag (1988).
  \bibitem{kirit} E.~Kiritsis, ``Introduction to superstring theory,'' Leuven 
    University Press (1998) [arXiv:hep-th/9709062].
  \bibitem{polwit} E.~Witten, ``String theory dynamics in various dimensions,''
    Nucl.\ Phys.\ B {\bf 443} (1995) 85 [arXiv:hep-th/9503124]; 
  J.~Polchinski and E.~Witten, ``Evidence for Heterotic - Type I String Duality,''
    Nucl.\ Phys.\ B {\bf 460} (1996) 525 [arXiv:hep-th/9510169].
  \bibitem{dabhol} A.~Dabholkar, ``Ten-dimensional heterotic string as a soliton,'' 
    Phys.\ Lett.\ B {\bf 357} (1995) 307 [arXiv:hep-th/9506160].
  \bibitem{branes}  J.~Polchinski, ``Dirichlet-Branes and Ramond-Ramond Charges,''
    Phys.\ Rev.\ Lett.\  {\bf 75} (1995) 4724 [arXiv:hep-th/9510017].
  \bibitem{cosmo}  S.~Kachru, R.~Kallosh, A.~Linde and S.~P.~Trivedi,
    ``De Sitter vacua in string theory,'' Phys.\ Rev.\ D {\bf 68}, 046005 (2003)
    [arXiv:hep-th/0301240]; 
  S.~Kachru, R.~Kallosh, A.~Linde, J.~Maldacena, L.~McAllister and S.~P.~Trivedi,
    ``Towards inflation in string theory,'' JCAP {\bf 0310}, 013 (2003)
    [arXiv:hep-th/0308055]; 
  A.~Linde, ``Prospects of inflation,''
    Phys.\ Scripta {\bf T117}, 40 (2005) [arXiv:hep-th/0402051]; 
  V.~Balasubramanian, ``Accelerating universes and string theory,'' 
    Class.\ Quant.\ Grav.\  {\bf 21}, S1337 (2004) [arXiv:hep-th/0404075];  
  C.~P.~Burgess, ``Inflatable string theory?,'' Pramana {\bf 63}, 1269 (2004)
    [arXiv:hep-th/0408037].
  \bibitem{dabholkar} A.~Dabholkar,
    ``Lectures on orientifolds and duality,'' arXiv:hep-th/9804208.
  \bibitem{blumenhagen} R.~Blumenhagen, L.~Gorlich and B.~Kors,
    ``Supersymmetric 4D orientifolds of type IIA with D6-branes at angles,''
    JHEP {\bf 0001}, 040 (2000) [arXiv:hep-th/9912204]; 
  R.~Blumenhagen, L.~Gorlich and T.~Ott,
    ``Supersymmetric intersecting branes on the type IIA T**6/Z(4) orientifold,''
    JHEP {\bf 0301}, 021 (2003) [arXiv:hep-th/0211059].
  \bibitem{kachruori} S.~Kachru, M.~B.~Schulz and S.~Trivedi,
    ``Moduli stabilization from fluxes in a simple IIB orientifold,''
    JHEP {\bf 0310} (2003) 007 [arXiv:hep-th/0201028];
  \bibitem{orienticomp} B.~Acharya, M.~Aganagic, K.~Hori and C.~Vafa,
    ``Orientifolds, mirror symmetry and superpotentials,'' arXiv:hep-th/0202208;
    
  I.~Brunner and K.~Hori, ``Orientifolds and mirror symmetry,''
    JHEP {\bf 0411}, 005 (2004) [arXiv:hep-th/0303135]; 
  I.~Brunner, K.~Hori, K.~Hosomichi and J.~Walcher,
    ``Orientifolds of Gepner models,'' arXiv:hep-th/0401137.
  \bibitem{hans} M.~Grana, T.~W.~Grimm, H.~Jockers and J.~Louis,
    ``Soft supersymmetry breaking in Calabi-Yau orientifolds with D-branes  and 
    fluxes,'' Nucl.\ Phys.\ B {\bf 690}, 21 (2004) [arXiv:hep-th/0312232]; 
  T.~W.~Grimm and J.~Louis, ``The effective action of N = 1 Calabi-Yau orientifolds,''
    Nucl.\ Phys.\ B {\bf 699}, 387 (2004) [arXiv:hep-th/0403067]; 
    ``The effective action of type IIA Calabi-Yau  orientifolds,'' 
    Nucl.\ Phys.\ B {\bf 718}, 153 (2005) [arXiv:hep-th/0412277]; 
  H.~Jockers and J.~Louis,
    ``The effective action of D7-branes in N = 1 Calabi-Yau orientifolds,''
    Nucl.\ Phys.\ B {\bf 705} (2005) 167 [arXiv:hep-th/0409098]; 
  H.~Jockers, ``The effective action of D-branes in Calabi-Yau orientifold
    compactifications,'' Fortsch.\ Phys.\  {\bf 53} (2005) 1087 
    [arXiv:hep-th/0507042]; 
  T.~W.~Grimm, ``The effective action of type II Calabi-Yau orientifolds,''
    Fortsch.\ Phys.\  {\bf 53}, 1179 (2005) [arXiv:hep-th/0507153].
  \bibitem{vaula} R.~D'Auria, S.~Ferrara, F.~Gargiulo, M.~Trigiante and S.~Vaula,
    ``N = 4 supergravity Lagrangian for type IIB on T**6/Z(2) in presence of
    fluxes and D3-branes,'' JHEP {\bf 0306}, 045 (2003) [arXiv:hep-th/0303049];
  C.~Angelantonj, S.~Ferrara and M.~Trigiante,
    ``New D = 4 gauged supergravities from \\ N = 4 orientifolds with fluxes,''
    JHEP {\bf 0310}, 015 (2003) [arXiv:hep-th/0306185].
  \bibitem{kors} M.~Berg, M.~Haack and B.~Kors,
    ``An orientifold with fluxes and branes via T-duality,''
    Nucl.\ Phys.\ B {\bf 669}, 3 (2003) [arXiv:hep-th/0305183].
  \bibitem{susi} D.~Lust, S.~Reffert and S.~Stieberger,
    ``Flux-induced soft supersymmetry breaking in chiral type IIB orientifolds
    with D3/D7-branes,'' Nucl.\ Phys.\ B {\bf 706}, 3 (2005) [arXiv:hep-th/0406092];
    
  D.~Lust, S.~Reffert, W.~Schulgin and S.~Stieberger,
    ``Moduli stabilization in type IIB orientifolds. I: Orbifold limits,''
    arXiv:hep-th/0506090.
  \bibitem{denef} F.~Denef, M.~R.~Douglas, B.~Florea, A.~Grassi and S.~Kachru,
    ``Fixing all moduli in a simple F-theory compactification,'' arXiv:hep-th/0503124.
  \bibitem{braneworlds} L.~E.~Ibanez, F.~Marchesano and R.~Rabadan,
    ``Getting just the standard model at intersecting branes,''
    JHEP {\bf 0111} (2001) 002 [arXiv:hep-th/0105155].
  \bibitem{intersecsm}  R.~Blumenhagen, B.~Kors, D.~Lust and T.~Ott,
    ``The standard model from stable intersecting brane world orbifolds,''
    Nucl.\ Phys.\ B {\bf 616} (2001) 3 [arXiv:hep-th/0107138].
  \bibitem{koko} C.~Kokorelis,``Exact standard model structures from intersecting 
    D5-branes,'' Nucl.\ Phys.\ B {\bf 677} (2004) 115 [arXiv:hep-th/0207234].
  \bibitem{gabi} G.~Honecker, ``Chiral supersymmetric models on an orientifold 
    of Z(4) x Z(2) with intersecting D6-branes,'' Nucl.\ Phys.\ B {\bf 666} (2003) 175
    [arXiv:hep-th/0303015].
  \bibitem{lustig} D.~Lust, ``Intersecting brane worlds: A path to the standard 
    model?,'' Class.\ Quant.\ Grav.\  {\bf 21} (2004) S1399 [arXiv:hep-th/0401156].
  \bibitem{patisala} M.~Cvetic, T.~Li and T.~Liu,
    ``Supersymmetric Pati-Salam models from intersecting D6-branes: A road to
    the standard model,'' Nucl.\ Phys.\ B {\bf 698} (2004) 163 [arXiv:hep-th/0403061].
  \bibitem{nakahara} Mikio Nakahara, ``Geometry, Topology and Physics,'' Graduate 
    Student Series in Physics, Adam Hilger, Bristol, England (1990).
  \bibitem{candel} P.~Candelas, ``Lectures on Complex Manifolds,'' Proceedings of the 
    Trieste Spring School 1987; 
  T.~Eguchi, P.~B.~Gilkey and A.~J.~Hanson,
    ``Gravitation, Gauge Theories And Differential Geometry,'' 
    Phys.\ Rept.\  {\bf 66} (1980) 213.
  \bibitem{compact} M.~Bodner, A.~C.~Cadavid and S.~Ferrara,
    ``(2,2) Vacuum Configurations For Type IIA Superstrings: N=2 Supergravity
    Lagrangians And Algebraic Geometry,'' Class.\ Quant.\ Grav.\  {\bf 8} (1991) 789.
  \bibitem{mirror}  S.~Ferrara and S.~Sabharwal, ``Quaternionic Manifolds For Type II 
    Superstring Vacua Of Calabi-Yau Spaces,''
    Nucl.\ Phys.\ B {\bf 332} (1990) 317;
  M.~Bodner and A.~C.~Cadavid, ``Dimensional Reduction Of Type IIB 
    Supergravity And Exceptional Quaternionic Manifolds,'' 
    Class.\ Quant.\ Grav.\  {\bf 7} (1990) 829;
  R.~B\"ohm, H.~G\"unther, C.~Herrmann and J.~Louis, ``Compactification of type IIB 
    string theory on Calabi-Yau threefolds,'' Nucl.\ Phys.\ B {\bf 569}, 229 (2000)
    [arXiv:hep-th/9908007].
  \bibitem{strominger} A.~Strominger, ``Superstrings With Torsion,''
    Nucl.\ Phys.\ B {\bf 274} (1986) 253.
  \bibitem{dewit} B.~de Wit, D.~J.~Smit and N.~D.~Hari Dass,
    ``Residual Supersymmetry Of Compactified D = 10 Supergravity,''
    Nucl.\ Phys.\ B {\bf 283} (1987) 165.
  \bibitem{hull} C.~M.~Hull, 
    ``Superstring Compactifications With Torsion And Space-Time Supersymmetry,''
    Print-86-0251 (CAMBRIDGE).
  \bibitem{granaflux} M.~Grana,
    ``Flux compactifications in string theory: A comprehensive review,''
    Phys.\ Rept.\  {\bf 423} (2006) 91 [arXiv:hep-th/0509003].
  \bibitem{typeiiflux} J.~Polchinski and A.~Strominger, 
    ``New Vacua for Type II String Theory,'' Phys.\ Lett.\ B {\bf 388} (1996) 736
    [arXiv:hep-th/9510227]; 
  J.~Michelson, ``Compactifications of type IIB strings to four dimensions with  
    non-trivial classical potential,'' Nucl.\ Phys.\ B {\bf 495}, 127 (1997)
    [arXiv:hep-th/9610151]; 
  G.~Curio, A.~Klemm, D.~Lust and S.~Theisen, ``On the vacuum structure of type II 
    string compactifications on  Calabi-Yau spaces with H-fluxes,'' 
    Nucl.\ Phys.\ B {\bf 609}, 3 (2001) [arXiv:hep-th/0012213].
  \bibitem{vafatayl} T.~R.~Taylor and C.~Vafa, 
    ``RR flux on Calabi-Yau and partial supersymmetry breaking,'' 
    Phys.\ Lett.\ B {\bf 474}, 130 (2000) [arXiv:hep-th/9912152].
  \bibitem{fluxstable}  
  J.~F.~G.~Cascales and A.~M.~Uranga, ``Chiral 4d string vacua with D-branes and 
    moduli stabilization,'' arXiv:hep-th/0311250; 
  J.~F.~G.~Cascales, M.~P.~Garcia del Moral, F.~Quevedo and A.~M.~Uranga,
    ``Realistic D-brane models on warped throats: Fluxes, hierarchies and  moduli
    stabilization,'' JHEP {\bf 0402}, 031 (2004) [arXiv:hep-th/0312051].
  \bibitem{becksis} K.~Becker and M.~Becker, 
    ``Supersymmetry breaking, M-theory and fluxes,''
    JHEP {\bf 0107} (2001) 038 [arXiv:hep-th/0107044].
  \bibitem{blumentayl} R.~Blumenhagen, D.~Lust and T.~R.~Taylor, 
    ``Moduli stabilization in chiral type IIB orientifold models with fluxes,'' 
    Nucl.\ Phys.\ B {\bf 663}, 319 (2003) [arXiv:hep-th/0303016]. 
  \bibitem{kachruflux} S.~Kachru and A.~K.~Kashani-Poor,
    ``Moduli potentials in type IIA compactifications with RR and NS flux,''
    JHEP {\bf 0503}, 066 (2005) [arXiv:hep-th/0411279].
  \bibitem{giddings} S.~B.~Giddings, S.~Kachru and J.~Polchinski,
    ``Hierarchies from fluxes in string compactifications,''
    Phys.\ Rev.\ D {\bf 66} (2002) 106006 [arXiv:hep-th/0105097].
  \bibitem{landscape} L.~Susskind, ``The anthropic landscape of string theory,''
     arXiv:hep-th/0302219; 
  M.~R.~Douglas, ``The statistics of string / M theory vacua,''
    JHEP {\bf 0305} (2003) 046 [arXiv:hep-th/0303194]; S.~Ashok and M.~R.~Douglas,
    ``Counting flux vacua,'' JHEP {\bf 0401}, 060 (2004) 
    [arXiv:hep-th/0307049];
  T.~Banks, M.~Dine and E.~Gorbatov, 
    ``Is there a string theory landscape?,'' JHEP {\bf 0408} (2004) 058
    [arXiv:hep-th/0309170].
  \bibitem{flori} F.~Gmeiner, R.~Blumenhagen, G.~Honecker, D.~Lust and T.~Weigand,
    ``One in a billion: MSSM-like D-brane statistics,'' JHEP {\bf 0601}, 004 (2006)
    [arXiv:hep-th/0510170].
  \bibitem{mtheory} J.~H.~Schwarz, ``Lectures on superstring and M theory dualities,''
    Nucl.\ Phys.\ Proc.\ Suppl.\  {\bf 55B}, 1 (1997) [arXiv:hep-th/9607201];
  J.~P.~Gauntlett, ``M-theory: Strings, duality and branes,''
    Contemp.\ Phys.\  {\bf 39} (1998) 317;
  A.~Sen, ``An introduction to non-perturbative string theory,'' 
    arXiv:hep-th/9802051.
  \bibitem{obers} N.~A.~Obers and B.~Pioline, ``U-duality and M-theory,''
    Phys.\ Rept.\  {\bf 318}, 113 (1999) [arXiv:hep-th/9809039].
  \bibitem{alvarez} E.~Alvarez, L.~Alvarez-Gaume and Y.~Lozano,
    ``An introduction to T duality in string theory,''
    Nucl.\ Phys.\ Proc.\ Suppl.\  {\bf 41}, 1 (1995) [arXiv:hep-th/9410237].
  \bibitem{dixon} J. Dixon and D. Gepner, unpublished.
  \bibitem{lerche} W.~Lerche, C.~Vafa and N.~P.~Warner, 
    ``Chiral Rings In N=2 Superconformal Theories,'' 
    Nucl.\ Phys.\ B {\bf 324} (1989) 427.
  \bibitem{aspinwall} P.~S.~Aspinwall, C.~A.~Lutken and G.~G.~Ross,
    ``Construction And Couplings Of Mirror Manifolds,''
    Phys.\ Lett.\ B {\bf 241} (1990) 373.
  \bibitem{plesser} B.~R.~Greene and M.~R.~Plesser,
    ``Duality In Calabi-Yau Moduli Space,''
    Nucl.\ Phys.\ B {\bf 338} (1990) 15.
  \bibitem{candelpark} P.~Candelas, X.~C.~De La Ossa, P.~S.~Green and L.~Parkes,
    ``A Pair Of Calabi-Yau Manifolds As An Exactly Soluble Superconformal Theory,''
    Nucl.\ Phys.\ B {\bf 359} (1991) 21.
  \bibitem{syz} A.~Strominger, S.~T.~Yau and E.~Zaslow,
    ``Mirror symmetry is T-duality,'' Nucl.\ Phys.\ B {\bf 479}, 243 (1996)
    [arXiv:hep-th/9606040].
  \bibitem{vafaf} C.~Vafa, ``Evidence for F-Theory,''
    Nucl.\ Phys.\ B {\bf 469} (1996) 403 [arXiv:hep-th/9602022].
  \bibitem{sen} A.~Sen, ``F-theory and Orientifolds,''
    Nucl.\ Phys.\ B {\bf 475} (1996) 562 [arXiv:hep-th/9605150]; 
    ``Orientifold limit of F-theory vacua,'' Phys.\ Rev.\ D {\bf 55} (1997) 7345
    [arXiv:hep-th/9702165]; 
  \bibitem{candelas} P.~Candelas and X.~C.~de la Ossa,
    ``Comments On Conifolds,'' Nucl.\ Phys.\ B {\bf 342} (1990) 246.
  \bibitem{klebanovwitten} I.~R.~Klebanov and E.~Witten,
    ``Superconformal field theory on threebranes at a Calabi-Yau singularity,''
    Nucl.\ Phys.\ B {\bf 536} (1998) 199 [arXiv:hep-th/9807080].
  \bibitem{klebanekrasov} I.~R.~Klebanov and N.~A.~Nekrasov,
    ``Gravity duals of fractional branes and logarithmic RG flow,''
    Nucl.\ Phys.\ B {\bf 574} (2000) 263 [arXiv:hep-th/9911096].
  \bibitem{klebanovtseyt} I.~R.~Klebanov and A.~A.~Tseytlin,
    ``Gravity duals of supersymmetric SU(N) x SU(N+M) gauge theories,''
    Nucl.\ Phys.\ B {\bf 578} (2000) 123 [arXiv:hep-th/0002159].
  \bibitem{vy} G.~Veneziano and S.~Yankielowicz,
    ``An Effective Lagrangian For The Pure N=1 Supersymmetric Yang-Mills Theory,''
    Phys.\ Lett.\ B {\bf 113} (1982) 231.
  \bibitem{schwetz} G.~R.~Farrar, G.~Gabadadze and M.~Schwetz,
    ``On the effective action of N = 1 supersymmetric Yang-Mills theory,''
    Phys.\ Rev.\ D {\bf 58} (1998) 015009 [arXiv:hep-th/9711166].
  \bibitem{ckl} D.~G.~Cerdeno, A.~Knauf and J.~Louis,
    ``A note on effective N = 1 super Yang-Mills theories versus lattice
    results,'' Eur.\ Phys.\ J.\ C {\bf 31} (2003) 415
    [arXiv:hep-th/0307198].
  \bibitem{dv} R.~Dijkgraaf and C.~Vafa,
    ``Matrix models, topological strings, and supersymmetric gauge theories,''
    Nucl.\ Phys.\ B {\bf 644} (2002) 3 [arXiv:hep-th/0206255].
  \bibitem{gopakumar}  R.~Gopakumar and C.~Vafa,
    ``On the gauge theory/geometry correspondence,''
    Adv.\ Theor.\ Math.\ Phys.\  {\bf 3} (1999) 1415 [arXiv:hep-th/9811131].
  \bibitem{hori} K.~Hori, A.~Iqbal and C.~Vafa, ``D-branes and mirror symmetry,'' 
    arXiv:hep-th/0005247.
  \bibitem{aganagic} M.~Aganagic, A.~Karch, D.~Lust and A.~Miemiec,
    ``Mirror symmetries for brane configurations and branes at singularities,''
    Nucl.\ Phys.\ B {\bf 569} (2000) 277 [arXiv:hep-th/9903093].
  \bibitem{flop} M.~Atiyah, J.~M.~Maldacena and C.~Vafa,
    ``An M-theory flop as a large N duality,''
    J.\ Math.\ Phys.\  {\bf 42} (2001) 3209 [arXiv:hep-th/0011256].
  \bibitem{branecascade} K.~Dasgupta, K.~Oh and R.~Tatar,
    ``Geometric transition, large N dualities and MQCD dynamics,''
    Nucl.\ Phys.\ B {\bf 610}, 331 (2001) [arXiv:hep-th/0105066]; 
    ``Open/closed string dualities and Seiberg duality 
    from geometric transitions in M-theory,'' JHEP {\bf 0208}, 026 (2002)
    [arXiv:hep-th/0106040].
  \bibitem{oh} K.~Dasgupta, K.~h.~Oh, J.~Park and R.~Tatar,
    ``Geometric transition versus cascading solution,''
    JHEP {\bf 0201} (2002) 031 [arXiv:hep-th/0110050];.
  \bibitem{kachru} S.~Kachru, M.~B.~Schulz, P.~K.~Tripathy and S.~P.~Trivedi,
    ``New supersymmetric string compactifications,''
    JHEP {\bf 0303} (2003) 061 [arXiv:hep-th/0211182].
  \bibitem{jan} S.~Gurrieri, J.~Louis, A.~Micu and D.~Waldram,
    ``Mirror symmetry in generalized Calabi-Yau compactifications,''
    Nucl.\ Phys.\ B {\bf 654}, 61 (2003)
    [arXiv:hep-th/0211102].
  \bibitem{mn} J.~M.~Maldacena and C.~Nunez,
    ``Towards the large N limit of pure N = 1 super Yang Mills,''
    Phys.\ Rev.\ Lett.\  {\bf 86} (2001) 588
    [arXiv:hep-th/0008001].
  \bibitem{micu} J.~Louis and A.~Micu,
    ``Type II theories compactified on Calabi-Yau threefolds in the presence  of
    background fluxes,'' Nucl.\ Phys.\ B {\bf 635} (2002) 395
    [arXiv:hep-th/0202168].
  \bibitem{greg} R.~Gregory, J.~A.~Harvey and G.~W.~Moore,
     ``Unwinding strings and T-duality of Kaluza-Klein and H-monopoles,''
     Adv.\ Theor.\ Math.\ Phys.\  {\bf 1} (1997) 283 [arXiv:hep-th/9708086].
  \bibitem{pandozayas} L.~A.~Pando Zayas and A.~A.~Tseytlin,
    ``3-branes on resolved conifold,'' JHEP {\bf 0011} (2000) 028
    [arXiv:hep-th/0010088].
  \bibitem{papatseyt} G.~Papadopoulos and A.~A.~Tseytlin,
    ``Complex geometry of conifolds and 5-brane wrapped on 2-sphere,''
    Class.\ Quant.\ Grav.\  {\bf 18} (2001) 1333 [arXiv:hep-th/0012034].
  \bibitem{minasian} R.~Minasian and D.~Tsimpis,
    ``On the geometry of non-trivially embedded branes,''
    Nucl.\ Phys.\ B {\bf 572} (2000) 499 [arXiv:hep-th/9911042].
  \bibitem{gtthree} K.~Dasgupta, M.~Grisaru, R.~Gwyn, S.~Katz, A.~Knauf and R.~Tatar,
  ``Gauge-Gravity Dualities, Dipoles and New Non-Kahler Manifolds,''
  arXiv:hep-th/0605201.
  \bibitem{dasmukhi} K.~Dasgupta and S.~Mukhi,
    ``Brane constructions, conifolds and M-theory,''
    Nucl.\ Phys.\ B {\bf 551} (1999) 204 [arXiv:hep-th/9811139]; 
    ``Brane constructions, fractional branes and anti-de Sitter domain walls,''
    JHEP {\bf 9907} (1999) 008 [arXiv:hep-th/9904131].
  \bibitem{uranga} A.~M.~Uranga,
    ``Brane configurations for branes at conifolds,''
    JHEP {\bf 9901} (1999) 022 [arXiv:hep-th/9811004].
  \bibitem{ohta} K.~Ohta and T.~Yokono,
    ``Deformation of conifold and intersecting branes,''
    JHEP {\bf 0002} (2000) 023 [arXiv:hep-th/9912266].
  \bibitem{cvetic} M.~Cvetic, G.~W.~Gibbons, H.~Lu and C.~N.~Pope,
    ``Ricci-flat metrics, harmonic forms and brane resolutions,''
    Commun.\ Math.\ Phys.\  {\bf 232} (2003) 457
    [arXiv:hep-th/0012011].
  \bibitem{keha} A.~Kehagias, 
    ``New type IIB vacua and their F-theory interpretation,''
    Phys.\ Lett.\ B {\bf 435} (1998) 337 [arXiv:hep-th/9805131].
  \bibitem{wecht}  A.~Flournoy, B.~Wecht and B.~Williams,
    ``Constructing nongeometric vacua in string theory,''
    Nucl.\ Phys.\ B {\bf 706}, 127 (2005) [arXiv:hep-th/0404217];
  J.~Shelton, W.~Taylor and B.~Wecht,
    ``Nongeometric flux compactifications,''
    JHEP {\bf 0510}, 085 (2005) [arXiv:hep-th/0508133].
  \bibitem{simeon} S.~Hellerman, J.~McGreevy and B.~Williams,
    ``Geometric constructions of nongeometric string theories,''
    JHEP {\bf 0401}, 024 (2004) [arXiv:hep-th/0208174].
  \bibitem{nongeomet} A.~Dabholkar and C.~Hull,
    ``Generalised T-duality and non-geometric backgrounds,'' arXiv:hep-th/0512005.
  \bibitem{hitch} N.~Hitchin, ``Generalized Calabi-Yau manifolds,''
    Quart.\ J.\ Math.\ Oxford Ser.\  {\bf 54} (2003) 281 [arXiv:math.dg/0209099].
  \bibitem{gualtieri} M.~Gualtieri, ``Generalized Complex Geometry,'' Oxford 
    University DPhil thesis [arXiv:math.DG/0401221].
  \bibitem{kapuli} A.~Kapustin and Y.~Li,
    ``Topological sigma-models with H-flux and twisted generalized complex
    manifolds,'' arXiv:hep-th/0407249.
  \bibitem{kapustin} A.~Kapustin,
    ``Topological strings on noncommutative manifolds,''
    Int.\ J.\ Geom.\ Meth.\ Mod.\ Phys.\  {\bf 1} (2004) 49 [arXiv:hep-th/0310057].
  \bibitem{minas}  M.~Grana, R.~Minasian, M.~Petrini and A.~Tomasiello,
    ``Supersymmetric backgrounds from generalized Calabi-Yau manifolds,''
    JHEP {\bf 0408} (2004) 046 [arXiv:hep-th/0406137]; 
    ``Type II strings and generalized Calabi-Yau manifolds,''
    Comptes Rendus Physique {\bf 5} (2004) 979 [arXiv:hep-th/0409176]; 
    ``Generalized structures of N = 1 vacua,'' JHEP {\bf 0511} (2005) 020
    [arXiv:hep-th/0505212].
  \bibitem{minasfidanza} P.~Grange and R.~Minasian,
    ``Modified pure spinors and mirror symmetry,'' 
    Nucl.\ Phys.\ B {\bf 732} (2006) 366 [arXiv:hep-th/0412086];
  S.~Fidanza, R.~Minasian and A.~Tomasiello,
    ``Mirror symmetric SU(3)-structure manifolds with NS fluxes,''
    Commun.\ Math.\ Phys.\  {\bf 254} (2005) 401 [arXiv:hep-th/0311122].
  \bibitem{jangrana} M.~Grana, J.~Louis and D.~Waldram,
    ``Hitchin functionals in N = 2 supergravity,'' arXiv:hep-th/0505264; 
  I.~Benmachiche and T.~W.~Grimm, ``Generalized N=1 Orientifold Compactifications and 
    the Hitchin functionals,'' arXiv:hep-th/0602241.
  \bibitem{lindstrom} U.~Lindstrom, M.~Rocek, R.~von Unge and M.~Zabzine,
    ``Generalized Kaehler geometry and manifest N = (2,2) supersymmetric
    nonlinear sigma-models,''
    JHEP {\bf 0507} (2005) 067 [arXiv:hep-th/0411186];
    ``Generalized Kaehler manifolds and off-shell supersymmetry,'' 
    arXiv:hep-th/0512164;
  A.~Bredthauer, U.~Lindstrom and J.~Persson,
    ``First-order supersymmetric sigma models and target space geometry,''
    arXiv:hep-th/0508228.
  \bibitem{minalind} U.~Lindstrom, R.~Minasian, A.~Tomasiello and M.~Zabzine,
    ``Generalized complex manifolds and supersymmetry,''
    Commun.\ Math.\ Phys.\  {\bf 257} (2005) 235 [arXiv:hep-th/0405085].
  \bibitem{zucchini} R.~Zucchini,
    ``A topological sigma model of biKaehler geometry,'' arXiv:hep-th/0511144.
  \bibitem{kachrutomas} W.~y.~Chuang, S.~Kachru and A.~Tomasiello,
    ``Complex / symplectic mirrors,'' arXiv:hep-th/0510042.
  \bibitem{marchesano} F.~Marchesano, ``D6-branes and torsion,''
    arXiv:hep-th/0603210.
  \bibitem{salamon} S. Chiossi, S. Salamon,
    ``The intrinsic torsion of $SU(3)$ and $G_2$ structures,''
    Proc. conf. Differential Geometry Valencia 2001 [math.DG/0202282].
  \bibitem{beckbeck} 
    K.~Becker and M.~Becker, ``M-Theory on Eight-Manifolds,''
    Nucl.\ Phys.\ B {\bf 477} (1996) 155 [arXiv:hep-th/9605053].
  \bibitem{keshavmukhi} K.~Dasgupta and S.~Mukhi,
    ``F-theory at constant coupling,'' Phys.\ Lett.\ B {\bf 385} (1996) 125
    [arXiv:hep-th/9606044].
  \bibitem{sav} K.~Dasgupta, G.~Rajesh and S.~Sethi,
    ``M theory, orientifolds and G-flux,''
    JHEP {\bf 9908}, 023 (1999), hep-th/9908088.
  \bibitem{schubert} S. Katz and S. A. Stromme, ``Schubert: a maple package for 
    intersection theory'', http://www.mi.uib.no/schubert/
  \bibitem{ouyang} P.~Ouyang,
    ``Holomorphic D7-branes and flavored N = 1 gauge theories,''
    Nucl.\ Phys.\ B {\bf 699} (2004) 207 [arXiv:hep-th/0311084].
 \bibitem{cvetlupope} M.~Cvetic, G.~W.~Gibbons, H.~Lu and C.~N.~Pope,
    ``Cohomogeneity one manifolds of Spin(7) and G(2) holonomy,''
    Phys.\ Rev.\ D {\bf 65} (2002) 106004 [arXiv:hep-th/0108245]; 
    ``M-theory conifolds,'' Phys.\ Rev.\ Lett.\  {\bf 88} (2002) 121602
    [arXiv:hep-th/0112098]; ``A G(2) unification of the deformed and resolved 
    conifolds,'' Phys.\ Lett.\ B {\bf 534} (2002) 172 [arXiv:hep-th/0112138].
  \bibitem{brand} A.~Brandhuber, J.~Gomis, S.~S.~Gubser and S.~Gukov,
    ``Gauge theory at large N and new G(2) holonomy metrics,''
    Nucl.\ Phys.\ B {\bf 611} (2001) 179 [arXiv:hep-th/0106034].
  \bibitem{grana} A.~Butti, M.~Grana, R.~Minasian, M.~Petrini and A.~Zaffaroni,
    ``The baryonic branch of Klebanov-Strassler solution: A supersymmetric family
    of SU(3) structure backgrounds,''
    JHEP {\bf 0503} (2005) 069, hep-th/0412187.
  \bibitem{minakas} P. Kaste, Ruben Minasian, M. Petrini, A. Tomasiello
    ``Nontrivial RR two-form field strength and SU(3)-structure''
    Fortsch.Phys. {\bf 51} 764 (2003), hep-th/0412187.
  \bibitem{gauntlett} J.~P.~Gauntlett, D.~Martelli and D.~Waldram,
    ``Superstrings with intrinsic torsion,''
    Phys.\ Rev.\ D {\bf 69}, 086002 (2004) [arXiv:hep-th/0302158];
    \{the former\} and S.~Pakis ``G-structures and wrapped NS5-branes,'' 
    Commun.\ Math.\ Phys.\  {\bf 247}, 421 (2004) [arXiv:hep-th/0205050].
  \bibitem{lust} G.~L.~Cardoso, G.~Curio, G.~Dall'Agata, D.~Lust, P.~Manousselis and 
    G.~Zoupanos, ``Non-Kaehler string backgrounds and their five torsion classes,''
    Nucl.\ Phys.\ B {\bf 652} (2003) 5 [arXiv:hep-th/0211118].
  \bibitem{frey} A.~R.~Frey, ``Notes on SU(3) structures in type IIB supergravity,''
    JHEP {\bf 0406}, 027 (2004) [arXiv:hep-th/0404107].
  \bibitem{dallagata} G.~Dall'Agata,
    ``On supersymmetric solutions of type IIB supergravity with general fluxes,''
    Nucl.\ Phys.\ B {\bf 695}, 243 (2004) [arXiv:hep-th/0403220].
  \bibitem{behrndt} K.~Behrndt, M.~Cvetic and P.~Gao,
    ``General type IIB fluxes with SU(3) structures,''
    Nucl.\ Phys.\ B {\bf 721}, 287 (2005) [arXiv:hep-th/0502154].
  \bibitem{gstructure} K.~Yano, ``Differential Geometry on Complex and Almost Complex 
    Spaces,'' Macmillan, New York, 1965; 
  M.~Falcitelli, A.~Farinola and S.~Salamon, 
    ``Almost--Hermitian Geometry'', Diff. Geo {\bf 4} (1994) 259;
  D.~Joyce, ``Compact Manifolds with Special Holonomy'', Oxford University Press, 
    Oxford, 2000;
  T.~Friedrich and S.~Ivanov, ``Parallel spinors and connections with skew-symmetric 
    torsion in string theory,'' arXiv:math.dg/0102142;
  S.~Salamon, ``Alomost Parallel Structures,'' 
    Contemp. Math. {\bf 288} (2001), 162-181 [math.DG/0107146].
  \bibitem{sala} S. Salamon, {\it Riemannian Geometry and Holonomy Groups}, 
    Pitman Research Notes in Mathematics 201, Longman, Harlow (1989). 
  \bibitem{hitchin} N. ~Hitchin, ``Stable forms and special metrics'',
    Contemp. Math., {\bf 288}, Amer. Math. Soc. (2000) [math.DG/0102128].
  \bibitem{misra} A.~Misra,
    ``Flow equations for uplifting half-flat to Spin(7) manifolds,''
    J.\ Math.\ Phys.\  {\bf 47}, 033504 (2006) [arXiv:hep-th/0507147];
  A.~Franzen, P.~Kaura, A.~Misra and R.~Ray, ``Uplifting the Iwasawa,''
    Fortsch.\ Phys.\  {\bf 54}, 207 (2006) [arXiv:hep-th/0506224].
  \bibitem{daskat} K.~Becker and K.~Dasgupta, ``Heterotic strings with torsion,''
    JHEP {\bf 0211} (2002) 006 [arXiv:hep-th/0209077].
  \bibitem{dasgreenbeck} K.~Becker, M.~Becker, K.~Dasgupta and P.~S.~Green,
    ``Compactifications of heterotic theory on non-Kaehler complex manifolds. I,''
    JHEP {\bf 0304} (2003) 007 [arXiv:hep-th/0301161]; \{the former\} and E.~Sharpe,
   ``Compactifications of heterotic strings on non-Kaehler complex  manifolds. II,''
    Nucl.\ Phys.\ B {\bf 678} (2004) 19 [arXiv:hep-th/0310058].
  \bibitem{andrei} S.~Gurrieri, A.~Lukas and A.~Micu,
    ``Heterotic on half-flat,'' Phys.\ Rev.\ D {\bf 70} (2004) 126009
    [arXiv:hep-th/0408121];
  A.~Micu, ``Heterotic compactifications and nearly-Kaehler manifolds,''
    Phys.\ Rev.\ D {\bf 70}, 126002 (2004) [arXiv:hep-th/0409008].
  \bibitem{hullwit} C.~M.~Hull and E.~Witten,
    ``Supersymmetric Sigma Models And The Heterotic String,''
    Phys.\ Lett.\ B {\bf 160}, 398 (1985); ``Compactifications Of The Heterotic 
    Superstring,'' Phys.\ Lett.\ B {\bf 178}, 357 (1986).
  \bibitem{dasbecksq} K.~Becker, M.~Becker, K.~Dasgupta and S.~Prokushkin,
    ``Properties of heterotic vacua from superpotentials,''
    Nucl.\ Phys.\ B {\bf 666} (2003) 144 [arXiv:hep-th/0304001].
  \bibitem{carcudallu} G.~L.~Cardoso, G.~Curio, G.~Dall'Agata and D.~Lust,
    ``BPS action and superpotential for heterotic string compactifications  with
    fluxes,'' JHEP {\bf 0310} (2003) 004 [arXiv:hep-th/0306088]; ``Heterotic string 
    theory on non-Kaehler manifolds with H-flux and  gaugino condensate,''
    Fortsch.\ Phys.\  {\bf 52} (2004) 483 [arXiv:hep-th/0310021].
  \bibitem{ntwofivebrane} J.~P.~Gauntlett, N.~Kim, D.~Martelli and D.~Waldram,
      ``Wrapped fivebranes and ${\cal N} = 2$ super Yang-Mills theory,''
      Phys.\ Rev.\ D {\bf 64}, 106008 (2001), hep-th/0106117;
    F.~Bigazzi, A.~L.~Cotrone and A.~Zaffaroni,
      ``${\cal N} = 2$ gauge theories from wrapped five-branes,''
      Phys.\ Lett.\ B {\bf 519}, 269 (2001), hep-th/0106160;
    P.~Di Vecchia, A.~Lerda and P.~Merlatti,
      ``${\cal N} = 1$ and ${\cal N} = 2$ super Yang-Mills theories from wrapped 
      branes,''Nucl.\ Phys.\ B {\bf 646}, 43 (2002), hep-th/0205204.
  \bibitem{gates} S.~J.~Gates, C.~M.~Hull and M.~Rocek,
    ``Twisted Multiplets And New Supersymmetric Nonlinear Sigma Models,''
    Nucl.\ Phys.\ B {\bf 248} (1984) 157.
  \bibitem{wittentop} E.~Witten, ``Mirror manifolds and topological field theory,''
    in {\it Essays on Mirror Manifolds}, ed. S.~T.~Yau (International Press, 1992),
    arXiv:hep-th/9112056.
  \bibitem{malikov}F. ~Malikov, V. ~Schechtman, A. ~Vaintrob,
    ``Chiral de Rham complex'', math.AG/9803041; 
  F. ~ Malikov, V. ~Schechtman,
    ``Chiral de Rham complex. II,'' math.AG/9901065; 
    ``Chiral Poincar\'e duality,'' math.AG/9905008; 
  V.~ Gorbounov, F. ~Malikov, V. ~Schechtman,
    ``Gerbes of chiral differential operators,'' math.AG/9906117.
  \bibitem{wittwo} E.~Witten,
    ``Two-dimensional models with (0,2) supersymmetry: Perturbative aspects,''
    hep-th/0504078.
  \bibitem{katz} S.~Katz and E.~Sharpe, 
    ``Notes on certain (0,2) correlation functions,''
    Commun.\ Math.\ Phys.\  {\bf 262}, 611 (2006) [arXiv:hep-th/0406226]; 
   E.~Sharpe, ``Notes on correlation functions in (0,2) theories,''
    arXiv:hep-th/0502064.
  \bibitem{kapu} A.~Kapustin,
    ``Chiral de Rham complex and the half-twisted sigma-model,'' arXiv:hep-th/0504074.
  \bibitem{affleck} I.~Affleck, M.~Dine and N.~Seiberg,
    ``Dynamical Supersymmetry Breaking In Supersymmetric QCD,'' 
    Nucl.\ Phys.\ B {\bf 241} (1984) 493.
  \bibitem{peskin} M.~E.~Peskin,
    ``Duality in supersymmetric Yang-Mills theory,'' arXiv:hep-th/9702094.
  \bibitem{gukov} S.~Gukov, C.~Vafa and E.~Witten, 
    ``CFT's from Calabi-Yau four-folds,'' Nucl.\ Phys.\ B {\bf 584}, 69 (2000)
    [Erratum-ibid.\ B {\bf 608}, 477 (2001)] [arXiv:hep-th/9906070]; 
  S.~Gukov, ``Solitons, superpotentials and calibrations,'' 
    Nucl.\ Phys.\ B {\bf 574}, 169 (2000) [arXiv:hep-th/9911011].
  \bibitem{fluxpot} B.~S.~Acharya and B.~J.~Spence,
    ``Flux, supersymmetry and M theory on 7-manifolds,'' arXiv:hep-th/0007213; 
  C.~Beasley and E.~Witten, ``A note on fluxes and superpotentials in M-theory 
    compactifications on manifolds of G(2) holonomy,''
    JHEP {\bf 0207}, 046 (2002) [arXiv:hep-th/0203061]; 
  K.~Behrndt and C.~Jeschek,
    ``Superpotentials from flux compactifications of M-theory,''
    Class.\ Quant.\ Grav.\  {\bf 21}, S1533 (2004) [arXiv:hep-th/0401019].
  \bibitem{normform} Y.~Imamura, ``Born-Infeld action and Chern-Simons term from 
    Kaluza-Klein monopole in M-theory,'' Phys.\ Lett.\ B {\bf 414}, 242 (1997)
    [arXiv:hep-th/9706144]; 
  A.~Sen, ``Dynamics of multiple Kaluza-Klein monopoles in M and string theory,''
    Adv.\ Theor.\ Math.\ Phys.\  {\bf 1}, 115 (1998) [arXiv:hep-th/9707042]; 
    ``A note on enhanced gauge symmetries in M- and string theory,''
    JHEP {\bf 9709}, 001 (1997) [arXiv:hep-th/9707123].
  \bibitem{cvj} C.~V.~Johnson, ``D-brane primer,'' arXiv:hep-th/0007170.
  \bibitem{skenderis} K.~Skenderis, 
    ``Black holes and branes in string theory,''
    Lect.\ Notes Phys.\  {\bf 541}, 325 (2000) [arXiv:hep-th/9901050].
  \bibitem{buscher} T.~H.~Buscher,
    ``A Symmetry Of The String Background Field Equations,''
    Phys.\ Lett.\ B {\bf 194} (1987) 59; 
    ``Path Integral Derivation Of Quantum Duality In Nonlinear Sigma Models,''
    Phys.\ Lett.\ B {\bf 201} (1988) 466.
  \bibitem{ortin} E.~Bergshoeff, C.~M.~Hull and T.~Ortin,
    ``Duality in the type II superstring effective action,''
    Nucl.\ Phys.\ B {\bf 451}, 547 (1995) [arXiv:hep-th/9504081]; 
  P.~Meessen and T.~Ortin,
    ``An Sl(2,Z) multiplet of nine-dimensional type II supergravity theories,''
    Nucl.\ Phys.\ B {\bf 541} (1999) 195 [arXiv:hep-th/9806120].
  \bibitem{wess} J.~Wess and J.~Bagger, {\it Supersymmetry and Supergravity}, 
    Princeton University Press (1983).
  \bibitem{zumino} B.~Zumino,
    ``Supersymmetry And Kahler Manifolds,'' Phys.\ Lett.\ B {\bf 87} (1979) 203.
  \bibitem{neitzke} A.~Neitzke and C.~Vafa,
    ``Topological strings and their physical applications,''
    arXiv:hep-th/0410178.
  \bibitem{wittentopol} E.~Witten,
    ``Topological Sigma Models,'' Commun.\ Math.\ Phys.\  {\bf 118} (1988) 411.
  \bibitem{topol}  E.~Witten, ``Topological Quantum Field Theory,''
    Commun.\ Math.\ Phys.\  {\bf 117}, 353 (1988); 
  T.~Eguchi and S.~K.~Yang, ``N=2 Superconformal Models As Topological Field 
    Theories,'' Mod.\ Phys.\ Lett.\ A {\bf 5} (1990) 1693.
  \bibitem{bcov} M.~Bershadsky, S.~Cecotti, H.~Ooguri and C.~Vafa, ``Kodaira-Spencer 
    theory of gravity and exact results for quantum string amplitudes,''
    Commun.\ Math.\ Phys.\  {\bf 165}, 311 (1994) [arXiv:hep-th/9309140].
  \bibitem{marino}  M.~Marino, ``Chern-Simons theory and topological strings,''
    Rev.\ Mod.\ Phys.\  {\bf 77} (2005) 675 [arXiv:hep-th/0406005].
  \bibitem{witten} E.~Witten, ``Chern-Simons gauge theory as a string theory,''
    Prog.\ Math.\  {\bf 133} (1995) 637 [arXiv:hep-th/9207094]; 
    ``Noncommutative Geometry And String Field Theory,'' 
    Nucl.\ Phys.\ B {\bf 268} (1986) 253.
  \bibitem{mackenzie} M.~Mackenzie, ``Lie groupoids and Lie algebroids in 
    differential geometry'', volume 124 of London Mathematical Society Lecture Note 
    Series, Cambridge University Press, Cambridge (1987).
  \bibitem{zabzine} M.~Zabzine,
    ``Geometry of D-branes for general N = (2,2) sigma models,''
    Lett.\ Math.\ Phys.\  {\bf 70} (2004) 211 [arXiv:hep-th/0405240].
  \bibitem{kapubrane} A.~Kapustin and Y.~Li,
    ``Open string BRST cohomology for generalized complex branes,''
    arXiv:hep-th/0501071; 
  A.~Kapustin, ``A-branes and noncommutative geometry,''
    arXiv:hep-th/0502212.
\end{thebibliography}
\end{document}